\documentclass[aps,pre,twocolumn,superscriptaddress,floatfix]{revtex4-2}
\usepackage{amsmath,amssymb,amsthm,amsfonts}
\usepackage{bm,braket,graphicx,booktabs}
\usepackage{tikz}
\usepackage{pgfplots}
\usepgfplotslibrary{groupplots}
\pgfplotsset{compat=1.18}
\usepackage{hyperref}

\newcommand{\E}{\mathbb{E}}
\newcommand{\Cov}{\mathrm{Cov}}
\newcommand{\Tr}{\mathrm{Tr}}
\newcommand{\dd}{\mathrm{d}}
\newcommand{\cG}{\mathcal{G}}
\newcommand{\cR}{\mathcal{R}}
\newcommand{\cC}{\mathcal{C}}
\newcommand{\cS}{\mathcal{S}}

\newcommand{\cF}{\mathcal{F}}
\newcommand{\cT}{\mathcal{T}}
\newcommand{\cP}{\mathcal{P}}
\newcommand{\cM}{\mathcal{M}}

\newcommand{\cL}{\mathcal{L}}
\newcommand{\cD}{\mathcal{D}}
\newcommand{\cE}{\mathcal{E}}
\newcommand{\cI}{\mathcal{I}}
\newcommand{\cB}{\mathcal{B}}
\newcommand{\bb}[1]{\mathbb{#1}}

\newtheorem{theorem}{Theorem}
\newtheorem{proposition}{Proposition}
\newtheorem{remark}{Remark}

\begin{document}

\title{Microscopically informed self-consistent multi-resolvent closure for dense many-body spectra}

\author{Zhiqiang Huang}
\email{zqhuang@hubu.edu.cn}
\affiliation{School of Physics, Hubei University, Wuhan 430062, China}

\date{\today}

\begin{abstract}
The exact multi-resolvent hierarchy of dense many-body spectra provides the
microscopic structure of higher-order spectral-weight correlations, but
constructing a finite, predictive closure of this hierarchy for interacting
Hamiltonians remains a challenge.  Here we introduce the multi-resolvent
self-consistency (MRSC) equations, a framework that separates an exact
algebraic hierarchy from an approximate finite-parameter closure, and verify
it on two benchmark classes.  The exact Feshbach geometric identity organizes the
$r$-resolvent cumulants into a ladder over irreducible vertices
$\Gamma^{(r)}$, and the projected hierarchy expresses the vertices
themselves as functionals of the cumulants.  The approximate layer closes
the hierarchy in finite parameters: the connectivity basis organizes the
seed of a general Hamiltonian into class amplitudes
$\Gamma_0^{(r)}=\sum_{\mathfrak c}\Theta_{\mathfrak c}\cB_{\mathfrak c}$,
and the causal $r$-fold Stieltjes representation plus the moment-field
identity reduces the moment matching to a finite-dimensional problem,
closed at a predictive fixed point.  On the free-fermion benchmark the
tested exact identities are reproduced and the single-resolvent Faddeeva
closure is recovered as the $r=1$ sector.  In a blind test on the
interacting Ising chain---with the exact two-resolvent data withheld---the
closure built from Hamiltonian-derived input predicts the tested probes to
$0.35\%$--$0.97\%$.  At the observable level the exact two-resolvent
sector supplies a sign-correct, energy-resolved correction to the
single-resolvent baseline for the off-diagonal
eigenstate-thermalization-hypothesis (ETH) variance, of order-one
magnitude relative to that variance.  The
closure errors organize hierarchically: a construction-level deficit
decomposition names them level by level, the isolatable contributions are
measured at the anchor, and an explicit propagation bound limits their
cost at the fixed point.  The construction
thereby provides a single microscopic route \emph{within the stated
closure framework}, and the same microscopic moments drive a computable
diagnostic chain for the window conditions of local rotational
invariance: an exact drift-kernel identity with a machine-checked
kernel, a proven single-state identifiability boundary, and a
transferability census over realizations and couplings---a
finite-order diagnostic framework, not a proof of their
universal emergence.
\end{abstract}

\maketitle

\section{Introduction}
\label{sec:intro}

In generic delocalized nonintegrable many-body systems the spectrum is
exponentially dense, and the natural organizing objects are resolvents
and their correlations rather than perturbative expansions
\cite{D52,BW69,E06}.  The energy-local statistical structure of matrix
elements---the eigenstate thermalization hypothesis (ETH) and its
higher-order refinements---controls late-time relaxation and
fluctuation dynamics \cite{Deutsch1991,Srednicki1994,Dalessio2016}, and
recent numerical evidence shows that physical matrix elements retain
correlations beyond the simplest random-matrix description
\cite{Richter2020,Pappalardi2022}; the resolvent-correlation
organization developed below targets that structure directly.  Two recent steps established complementary pieces
of a multi-resolvent program.  At the single-resolvent level, the exact
projected (Feshbach) hierarchy reorganizes cross-correlated propagation
into resolvent products ordered by resolvent multiplicity, and a causal
effective self-energy closes the mean sector while multi-resolvent
interference supplies the parity-mixing content that no
single-resolvent closure can generate
\cite{SRletter}.  At the multi-resolvent level, the exact
microscopic structure of the fluctuation sector was solved in random
free fermions: the $r$-resolvent cumulant admits an exact replica-sector
decomposition over the orthogonal Weingarten algebra
\cite{Weingarten,CM2009,CollinsSniady}, the sector basis
carries the energy-resolved spectral cumulants $S^{(r)}$, and the
Feshbach irreducible vertices reduce to the shared-edge counts,
$\Gamma^{(3)}_{abc}=8e_{abc}/(z_1z_2z_3)$ in the deep plane
\cite{MResR}.  What remains open---and is the subject of this
article---is the generic question posed by the hierarchy's outlook
\cite{HOETH,MResR,MResmerge}: \emph{how are the irreducible
multi-resolvent vertices determined self-consistently in a general
interacting system?}  Equivalently: can the exact microscopic
multi-resolvent hierarchy be reduced to a finite set of causal,
self-consistent parameters?

The philosophy follows the single-resolvent program in which the exact
resolvent equation is combined with causality, analyticity and a
finite causal ansatz to produce parameter self-consistency equations
\cite{SRletter}.  The multi-resolvent upgrade carries the same chain of
steps---exact hierarchy, causal representation, causal vertex ansatz,
connectivity basis, parameter self-consistency---on the hierarchy of
irreducible vertices $\Gamma^{(r)}$ rather than on a single resolvent.  Two
structural facts make this possible.  First, the Feshbach geometric
identity expresses the $r$-resolvent cumulant as a ladder over the
vertices (Sec.~\ref{sec:ladder})---the multi-resolvent analogue of the
Dyson equation, with the vertices as irreducible insertions.  Second,
the projected hierarchy expresses each vertex as a functional of the
multi-resolvent cumulants themselves (Sec.~\ref{sec:vertex})---the
Feshbach--Bethe--Salpeter structure.  Combining the two directions
closes the hierarchy level by level: the causal vertex ansatz of
Sec.~\ref{sec:ansatz} is first solved on the solvable benchmark, and the
connectivity basis of Sec.~\ref{sec:conn} then promotes the seed to a
general Hamiltonian, in the order shown.  The causal $r$-fold Stieltjes
representation (Sec.~\ref{sec:causal}) then compresses each vertex
into a finite parameter set $\boldsymbol\theta$, fixed by the
microscopic moment dictionary---the moment-field identity---and the
fixed-point equation $\boldsymbol\theta=F(\boldsymbol\theta)$
(Sec.~\ref{sec:closure}).  Within the chosen projection and closure
scheme, the microscopic input is represented by the channel moment
fields, the coupling products $V^{(\ell)}$ and their cumulants, and
the trace resolvent; no diagrammatic resummation, replica
construction, or ensemble-specific algebra is invoked.

\begin{figure*}[t]
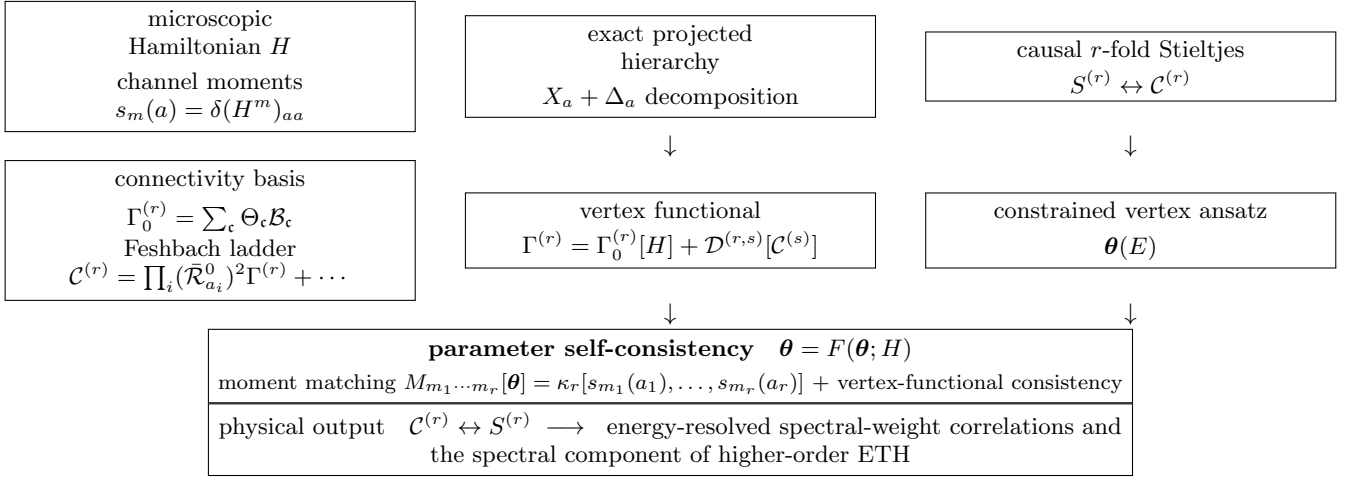

\centering
\begin{tabular}{c@{\qquad}c@{\qquad}c}
\fbox{\parbox{52mm}{\centering microscopic\\Hamiltonian $H$\\[1mm] channel moments\\$s_m(a)=\delta(H^m)_{aa}$}}&
\fbox{\parbox{52mm}{\centering exact projected\\hierarchy\\[1mm] $X_a+\Delta_a$ decomposition}}&
\fbox{\parbox{52mm}{\centering causal $r$-fold Stieltjes\\[1mm] $S^{(r)}\leftrightarrow\cC^{(r)}$}}\\
& $\downarrow$ & $\downarrow$\\
\fbox{\parbox{52mm}{\centering connectivity basis\\[1mm] $\Gamma_0^{(r)}=\sum_{\mathfrak c}\Theta_{\mathfrak c}\cB_{\mathfrak c}$\\[0.5mm] Feshbach ladder\\ $\cC^{(r)}=\prod_i(\bar{\cR}_{a_i}^0)^2\Gamma^{(r)}+\cdots$}}&
\fbox{\parbox{52mm}{\centering vertex functional\\[1mm] $\Gamma^{(r)}=\Gamma_0^{(r)}[H]+\cD^{(r,s)}[\cC^{(s)}]$}}&
\fbox{\parbox{52mm}{\centering constrained vertex ansatz\\[1mm] $\boldsymbol\theta(E)$}}\\
& $\downarrow$ & $\downarrow$\\
\multicolumn{3}{c}{\fbox{\parbox{120mm}{\centering\textbf{parameter self-consistency}\quad $\boldsymbol\theta=F(\boldsymbol\theta;H)$\\[1mm] \footnotesize moment matching $M_{m_1\cdots m_r}[\boldsymbol\theta]=\kappa_r[s_{m_1}(a_1),\dots,s_{m_r}(a_r)]$ + vertex-functional consistency}}}\\[2mm]
\multicolumn{3}{c}{\fbox{\parbox{120mm}{\centering physical output\quad $\cC^{(r)}\leftrightarrow S^{(r)}$\;\;$\longrightarrow$\;\; energy-resolved spectral-weight correlations and the spectral component of higher-order ETH}}}
\end{tabular}
\caption{The multi-resolvent self-consistency (MRSC) architecture.  The exact projected hierarchy and the
causal spectral representation supply the two exact structures---the
multi-resolvent ladder and the vertex functional---that close the
hierarchy level by level; the connectivity basis organizes the
microscopic layer into generic class coordinates and fixes the seed
$\Gamma_0^{(r)}=\sum_{\mathfrak c}\Theta_{\mathfrak c}\cB_{\mathfrak c}$;
a constrained causal vertex ansatz compresses the irreducible content
into a finite parameter set; moment matching against the microscopic
moment fields and vertex-functional consistency fix the parameters
self-consistently, and the fixed point of the retained
finite-parameter sector returns the physical output---the spectral
component of the higher-order ETH---through the feedback loop
$\cC^{(r)}\to\Gamma\to
\boldsymbol\theta\to\cC^{(r)}$; the $X\leftrightarrow\Delta$
cross-response kernel of Sec.~\ref{sec:crossresponse} carries the
dominant part of that loop.  The recovery of the full bulk
correlator, including mixed and higher-insertion contributions,
lies outside the capture range assumed by the present truncation
(Sec.~\ref{sec:intro}).
}
\label{fig:arch}
\end{figure*}

The correspondence with the single-resolvent program is then
explicit: what the causal ansatz is to one resolvent, the microscopic
response kernel of Sec.~\ref{sec:crossresponse} is to the hierarchy.
The two routes work at different ranks: the single-resolvent route
closes one parameter set through a scalar self-consistency, while the
correlated route closes the hierarchy through a causal \emph{matrix}
self-consistency whose kernel is the response function and whose fixed
point then fixes the parameters.
Together they form a microscopic-to-self-consistent framework for
correlated resolvents, summarized by the architecture of
Fig.~\ref{fig:arch}.

\emph{Relation to established frameworks.}  The construction
assembles several established ingredients.  Feshbach projection
provides the energy-dependent effective resolvent
\cite{Feshbach1958}; self-consistent resolvent equations have a long
tradition in random-operator theory \cite{Pastur1973}, and
self-consistency likewise underlies nonperturbative many-body
constructions such as dynamical mean-field theory \cite{Georges1996}.
The microscopic invariant layer of the free-fermion benchmark is
evaluated with the orthogonal Weingarten calculus
\cite{Weingarten,CM2009,CollinsSniady}.  The observable application
is motivated by the eigenstate-thermalization literature
\cite{Deutsch1991,Srednicki1994,Dalessio2016}, including recent
evidence that physical matrix elements retain correlations beyond the
simplest random-matrix description
\cite{Richter2020,Pappalardi2022,Wang2026} and statistical formulations of
the energy-local random structure of full ETH \cite{VFP25}.  The
present contribution connects
these ingredients through a single hierarchy in which microscopic
connectivity determines the higher-resolvent vertex content and
Feshbach dressing determines its self-consistent dynamics.

The relation to existing many-body machinery is structural but
distinct.  The ladder of Sec.~\ref{sec:ladder} is the
resolvent-projection analogue of the Bethe--Salpeter equation, and the
vertex functional of Sec.~\ref{sec:vertex} the analogue of the
Dyson--Schwinger vertex equations; the essential difference is that
the present route starts from exact resolvent projection rather than
diagrammatic perturbation theory, and that causality is imposed at the
level of the ansatz space itself.  The vertices carry a direct
physical reading that guides the truncations: $\Gamma^{(r)}$ is the
irreducible correlation among $r$ simultaneously correlated virtual
propagation channels---$\Gamma^{(2)}$ correlates two channels,
$\Gamma^{(3)}$ is the first genuinely collective three-channel
correlation---so the index $r$ is mathematically the cumulant order
and physically counts the number of correlated resolvent channels; it
is not a perturbative order.  This is also the sense
in which the framework addresses the limitation of the
replica-sector benchmark: what generalizes is not the Weingarten
algebra---one realization of a broader invariant-tensor
construction---but the \emph{mapping from microscopic data to effective
parameters}
\cite{MResR}.

The results of the article organize into three contributions.
\emph{(i) Exact structure and its microscopic realization}: the
multi-resolvent ladder and the vertex functional---combined with the
connectivity basis, which organizes the seed of a general Hamiltonian
into class amplitudes $H\to\{\mathfrak c,\Theta_{\mathfrak c}\}\to
\Gamma_{\rm coup}$ (Sec.~\ref{sec:conn})---determine where the irreducible higher-resolvent
content comes from: the mapping from microscopic data to effective
parameters.  \emph{(ii) Finite-parameter closure}: the
$X\leftrightarrow\Delta$ cross-response kernel resupplies the dominant
feedback channel that a truncated closure omits, and the Feshbach loop
$\Gamma\leftrightarrow\cC$ closes the system at the predictive fixed
point $\boldsymbol\theta_*=F[\boldsymbol\theta_*;H]$, satisfying causal-form
tests and exhibiting a stable fixed point; the construction-level error
decomposition of Sec.~\ref{sec:errorprop} names the closure deficits at
the level where each is retained, and the isolatable contributions are
measured at the anchor, so that verification and error localization
share one instrument.  \emph{(iii) Prediction and
physical output}: an $H$-derived $r=2$ closure predicts the exact
two-resolvent cumulant of a generic interacting model at the sub-percent
level, and the exact two-resolvent content supplies the spectral
component of higher-order ETH.  Two laboratories verify the
construction: the free-fermion family, where the exact sector algebra
is known, and the disordered Ising chain, where no sector structure
exists.  One scope boundary is declared up front: the first objective
of the closure is to determine the finite-dimensional causal parameters
$\boldsymbol\theta$; recovery of the full bulk correlator additionally
requires the mixed and higher-insertion terms of the functional, and
the two tasks are distinguished deliberately rather than identified
(Sec.~\ref{sec:num} quantifies the gap for the free-fermion anchor).

The construction is verified numerically on the random free-fermion
benchmark, where every seed and sum rule is known exactly
\cite{MResR,MResmerge}: the geometric identity, the ladder
resummation, the vertex functional, the deep-plane seeds
$\Gamma^{(2)}=2e_{ab}/(z_1z_2)$, $\Gamma^{(3)}=8e_{abc}/(z_1z_2z_3)$,
the moment-field dictionary, and a constrained two-parameter causal
kernel whose microscopic-moment matching diagnoses the sector
structure of the exact kernel (Sec.~\ref{sec:num}).  The framework
reduces to the single-resolvent closure at $r=1$ and carries the
skewness selection rule as a corollary; its physical output is the
energy-resolved higher-order ETH correlation density
(Appendix~\ref{sec:eth}).  The benchmark plays this role not as the
theory itself but as its calibration laboratory: the connectivity
basis of Sec.~\ref{sec:conn} makes the microscopic layer generic,
and the free-fermion sectors are the exactly solvable realization
against which the generic machinery is verified.

\emph{Relation to local rotational invariance.}  The multi-resolvent
hierarchy is not, by itself, a theorem of local rotational invariance
(LRI), nor does rotational invariance follow from the causal closure
conditions alone; the pair-channel development behind the diagnostics
below is carried out in Appendix~\ref{sec:driftkernel}.
Its potential relevance is structural:
channel-resolved multi-resolvent correlations and their connected
contributions organize finite-order moment deviations and offer a
route to tracing them to microscopic correlations.  LRI is therefore
treated as an adjacent diagnostic application---a framework alongside
the closure rather than a fifth layer of it---asking
whether this microscopic organization yields information beyond
direct moment reconstruction, rather than assuming at the outset that LRI
follows from the MRSC closure.  The diagnostics carry three
machine-verified instruments beyond the defect scan of
Appendix~\ref{sec:LRI}: an exact drift-kernel identity, whose kernel has
zero mass and closes the order-two defect as the kernel-weighted pair
content (with a proven counterexample showing this content is not
decided by any single-state observable); a transferability census,
which fixes where the defect's ensemble-level prediction succeeds and
where realization-level fluctuations dominate it; and a
closure-boundary table recording, for each tested closure route, the
narrow claim it refutes and the claim it leaves untouched.

\emph{Scope.}  Three sentences fix what the construction demands of a
Hamiltonian and what it claims.  It applies to finite-dimensional
systems with a discrete spectrum and local interactions of finite
support---the only structure used is a resolvent projection
$H=H_0+V$ with $V$ sparse---and throughout it works at fixed finite
$N$; no integrability, symmetry, or sector basis is assumed, and the
generic benchmark is chosen precisely because it carries none
(Sec.~\ref{sec:genbench}).  Two ingredients are assumptions rather
than consequences and are tested as such: the diagonal closure of the
cross channels (DCA, Eq.~\eqref{eq:Gell}), appropriate to the
dense-spectrum self-averaging regime, and the causal finite-parameter
representation of Level C.  The claims are the three contributions
above, each stated at its own level in the classification of
Sec.~\ref{sec:levels}; the bulk correlator, a generic isotropy
theorem, and the thermodynamic ($N\to\infty$) control of the LRI
window conditions are explicitly outside them, and the text marks the
corresponding boundaries where they arise.

The article is organized as follows.
Section~\ref{sec:hier} sets up the exact hierarchy and its causal
spectral representation (Sec.~\ref{sec:causal}).
Section~\ref{sec:ladder} derives the two closing structures---the
vertex functional (Sec.~\ref{sec:vertex})---and assembles the MRSC
system (Sec.~\ref{sec:system}).
Section~\ref{sec:ansatz} constructs the constrained
ansatz and the connectivity basis (Sec.~\ref{sec:conn}) on which the generic microscopic
layer lives, and Section~\ref{sec:closure} develops the parameter
self-consistency and its error propagation.
Section~\ref{sec:num} gives the benchmark design
and the numerical verification, and Appendix~\ref{sec:eth} carries the
closed two-resolvent content to an observable-level test.
Appendix~\ref{sec:LRI} converts the LRI derivation program into
computable window diagnostics.  Section~\ref{sec:disc}
collects the structural remarks and open questions, and
Section~\ref{sec:conclusion} concludes.  Appendix~\ref{app:cum}
collects the combinatorial identities, Appendices~\ref{app:moment}
and \ref{app:ladder3} the moment-field identity and the explicit
$r=3$ ladder, Appendix~\ref{app:supplLRI} the supplementary LRI
calibration details, Appendix~\ref{app:driftdata} the drift-kernel
data and the transferability archive (including the gate compendium
and the size-scan and cross-operator protocols), and
Appendix~\ref{app:benchver} the supplementary benchmark verification
details.  The notation used throughout is collected in
Table~\ref{tab:notation}.

\begin{table}[tbp]
\caption{Notation used throughout the article.}
\label{tab:notation}
\footnotesize
\begin{ruledtabular}
\begin{tabular}{p{3.0cm}p{4.9cm}}
\hline\hline
$\cR_a(z)$, $\cR$ & channel-resolved resolvent, Eq.~\eqref{eq:feshbach}; mean resolvent\\
$\cG_a(z)$ & Feshbach self-energy of channel $a$, Eq.~\eqref{eq:Gdef}\\
$\Gamma^{(r)}$ & irreducible $r$-resolvent vertices of the ladder\\
$\cC^{(r)}$, $S^{(r)}$ & connected channel-resolved $r$-resolvent cumulants; spectral cumulant density, Eq.~\eqref{eq:ETH}\\
$X$, $\Delta$ & channel content and response content of the cross-response kernel, Sec.~\ref{sec:crossresponse}\\
$s_m(a)$, $\cM_{\bm m}$ & single-state spectral-weight observables; microscopic channel moments\\
$\Theta_{\mathfrak c}$, $\cB_{\mathfrak c}$ & class amplitudes and connectivity basis, Sec.~\ref{sec:conn}\\
$V^{(\ell)}$ & $\ell$-point products of the perturbation\\
$\boldsymbol\theta$, $F$ & causal parameter set; fixed-point map $\boldsymbol\theta=F(\boldsymbol\theta)$\\
$\E_H$, $\kappa_r$ & ensemble average over the disorder class; its cumulants\\
$p_n^a$ & participation weight $|\braket{\varphi_a|\psi_n}|^2$\\
$w_{E,n}$, $D_E$ & window weights; density of states\\
\hline
\end{tabular}
\end{ruledtabular}
\end{table}

\section{Exact hierarchy and causal representation}
\label{sec:hier}

\emph{Status.}  Level A of the decomposition of Sec.~\ref{sec:levels}:
everything in this section is exact---the projected hierarchy, the
level decomposition, and the causal spectral representation are
identities or definitions of the microscopic theory, independent of
any closure assumption.

\subsection{Channels, resolvents, cumulants}

The channel-resolved objects on which the hierarchy is built are fixed
in this subsection.
Consider $H=H_0+V$ on a $D$-dimensional many-body Hilbert space
spanned by a product (channel) basis $\{\ket{\varphi_a}\}$,
$a=1,\dots,D$, with $H_0\ket{\varphi_a}=a_a\ket{\varphi_a}$ and
interaction matrix elements $V_{ab}=\bra{\varphi_a}V\ket{\varphi_b}$,
$V_a=V_{aa}$.  The diagonal resolvent
\begin{equation}
\cR_a(z)=\bra{\varphi_a}\frac{1}{z-H}\ket{\varphi_a}
=\sum_n\frac{p_n^a}{z-\lambda_n},
\label{eq:Rdef}
\end{equation}
with $p_n^a=|\!\braket{\varphi_a|\psi_n}\!|^2$ the spectral weights,
encodes the channel-resolved local density of states
$\rho_a(E)=\sum_n p_n^a\delta(E-\lambda_n)$
\cite{HOETH,MResR}.  The \emph{realization} trace resolvent
\begin{equation}
\cR_{\rm tr}(z):=\frac1D\Tr(z-H)^{-1}
\label{eq:Rtr}
\end{equation}
defines the resolvent fluctuation against the same realization,
\begin{equation}
\delta\cR_a(z)=\cR_a(z)-\cR_{\rm tr}(z)
=\sum_n\frac{\delta p_n^a}{z-\lambda_n},
\qquad
\delta p_n^a=p_n^a-\frac1D,
\label{eq:dRdef}
\end{equation}
and the fluctuation sector of the hierarchy is the sequence of
connected multi-resolvent cumulants
\begin{equation}
\cC^{(r)}_{a_1\cdots a_r}(z_1,\dots,z_r)
=\kappa_r\bigl(\delta\cR_{a_1}(z_1),\dots,\delta\cR_{a_r}(z_r)\bigr),
\label{eq:cumdef}
\end{equation}
generated by $W[J]=\ln\E_H\exp[\sum_a J_a\delta\cR_a(z_a)]$
\cite{HOETH}.  Here $\E_H$ is the relevant ensemble average (the
disorder average, or shell/window averaging in fixed-Hamiltonian
applications), and $\kappa_r$ denotes the connected cumulant;
$\sum_n p_n^a=1$ per realization, so
$\sum_n\delta p_n^a=0$ exactly.

The centering is at the \emph{realization} trace resolvent
Eq.~\eqref{eq:Rtr}: this is what makes the sum rule
$\sum_n\delta p_n^a=0$ hold realization by realization, and the whole
$\delta p\to\delta\cR\to\delta\cG$ chain is built on it.  The
ensemble mean of the channel resolvent is a different symbol,
$\hat\cR_a:=\E_H[\cR_a]$, and the mean sector of the hierarchy solves
for it [Eq.~\eqref{eq:meanr}]; the trace component of the mean sector
is $\E_H[\cR_{\rm tr}]$.  The two objects coincide in the Haar
free-fermion benchmark---there $\E_U[\cR_a]=\cR_{\rm tr}$
realization by realization
(Sec.~\ref{sec:vertex})---but they differ for a fixed disorder
realization, and the text keeps them apart: cumulants are centered at
$\cR_{\rm tr}$, mean-sector statements concern $\hat\cR_a$.

\subsection{Feshbach identity and the vertex decomposition}

The Feshbach projection identity
\cite{Feshbach1958,SRletter} gives
\begin{equation}
\cR_a(z)=\frac{1}{z-a_a-V_a-\cG_a(z)},
\label{eq:feshbach}
\end{equation}
\begin{equation}
\cG_a(z)=\bra{\varphi_a}V\Phi_a\frac{1}{z-\Phi_aH\Phi_a}\Phi_aV\ket{\varphi_a},
\label{eq:Gdef}
\end{equation}
with $\Phi_a=I-\ket{\varphi_a}\bra{\varphi_a}$, a nonperturbative
decomposition in which all coupling to the complement is contained in
the self-energy $\cG_a$.  Write $\bar{\cG}_a=\E_H[\cG_a]$ for its
ensemble mean and introduce the bare propagator
\begin{equation}
\bar{\cR}_a^0(z)=\frac{1}{z-a_a-V_a-\bar{\cG}_a(z)}.
\label{eq:R0bar}
\end{equation}
The exact geometric identity follows by elimination:
\begin{equation}
\cR_a=\frac{\bar{\cR}_a^0}{1-\bar{\cR}_a^0\,\delta\cG_a},
\qquad
\delta\cG_a=\cG_a-\bar{\cG}_a,
\label{eq:geom}
\end{equation}
and the resolvent fluctuation decomposes exactly as
\begin{align}
\delta\cR_a&=X_a+\Delta_a,\notag\\
X_a&=\bar{\cR}_a^0-\cR_{\rm tr},\qquad
\Delta_a=\sum_{s\ge1}\bigl(\bar{\cR}_a^0\bigr)^{s+1}\bigl(\delta\cG_a\bigr)^s,
\label{eq:XD}
\end{align}
where $X_a$ is the mean-field mismatch between the bare propagator and
the trace resolvent, and $\Delta_a=\cR_a-\bar{\cR}_a^0$ carries the
self-energy fluctuation dressed by the bare propagator.  Two different
differences are in use, and they must not be confused.
\emph{(a)} The ensemble-centered self-energy fluctuation
$\delta\cG_a=\cG_a-\bar\cG_a$ satisfies $\E_H[\delta\cG_a]=0$ by
definition (Eq.~\eqref{eq:geom}).  \emph{(b)} The
realization-centered resolvent difference
$\delta\cR_a=\cR_a-\cR_{\rm tr}$ of Eq.~\eqref{eq:dRdef} is centered at
the trace resolvent of the \emph{same} realization and is
\emph{not} ensemble-centered: $\E_H[\delta\cR_a]\neq0$, because
$\cR_{\rm tr}$ fluctuates.  The cumulant-of-products rule of
Appendix~\ref{app:cum} is stated for ensemble-centered variables and is
applied only to $\delta\cG_a$ and to the coupling fluctuations
$\delta V^{(\ell)}$; $\delta\cR_a$ is never inserted into it.  Resolvent
cumulants are formed after the decomposition of Eq.~\eqref{eq:XD}, in
which $X_a$ is fixed within a realization and $\Delta_a$ is a series in
$\delta\cG_a$ alone.  The
irreducible vertices of the hierarchy are the self-energy cumulants
\begin{equation}
\Gamma^{(r)}_{a_1\cdots a_r}(z_1,\dots,z_r)
=\kappa_r\bigl(\delta\cG_{a_1}(z_1),\dots,\delta\cG_{a_r}(z_r)\bigr),
\label{eq:Gammar}
\end{equation}
irreducible with respect to the geometric dressing of
Eq.~\eqref{eq:geom}, in the Feshbach sense of
Ref.~\cite{MResR}.  The first line of Eq.~\eqref{eq:geom} is exact for
every $z$ off the spectrum; the geometric series for $\Delta_a$ in
Eq.~\eqref{eq:XD} is its reorganization in powers of
$\delta\cG_a$, convergent wherever
$|\bar\cR_a^0\delta\cG_a|<1$ and used only as a formal expansion of the
exact rational form where that condition fails.  The tower and vertex
expansions of Sec.~\ref{sec:ladder} inherit this status: they reorganize
exact identities term by term, and no numerically summed truncation of
that series enters any result of Sec.~\ref{sec:num}.  The chain
$\delta p\to\delta\cR\to\delta\cG$ of
Eqs.~\eqref{eq:dRdef}--\eqref{eq:Gammar} is exact: the ladder
bookkeeping of Sec.~\ref{sec:ladder} and the vertex functional of
Sec.~\ref{sec:vertex} are reorganizations of these identities, not
additional approximations.

The mean-field content $X_a$ admits a deep-plane expansion through the
moment fields $s_m(a)=\delta(H^m)_{aa}$ defined below
[Eq.~\eqref{eq:momfield}].  Expanding $\bar{\cR}_a^0$ and
$\cR_{\rm tr}$ in $1/z$,
\begin{align}
X_a(z)&=\frac{s_1(a)}{z^2}+\frac{s_2(a)-v_a+\bar v_a}{z^3}+O(z^{-4}),\notag\\
v_a&=\sum_{b\neq a}|V_{ab}|^2,\qquad\bar v_a=\E_H[v_a],
\label{eq:Xdeep}
\end{align}
where $v_a$ is the leading self-energy coefficient,
$\cG_a(z)=v_a/z+O(z^{-2})$, following from the definition
Eq.~\eqref{eq:Gdef}, and the moment fields $s_m(a)$ of
Eq.~\eqref{eq:momfield} are evaluated through the moment-field
identities of Eq.~\eqref{eq:mf1}.  In the random free-fermion
benchmark $\bar{\cR}_a^0$ and $\cR_{\rm tr}$ differ---the resolvent
built from the mean self-energy is neither the realization trace
resolvent nor the harmonic mean $1/\E[1/\cR_a]$---so the
$X$-tower is generically nonzero and carries part of the deep-plane
content.

\subsection{Levels of the construction}
\label{sec:levels}

Throughout the article the construction is organized in four levels,
kept explicitly separate so that exact and approximate statements
never mix, and the vocabulary is fixed accordingly: \emph{exact
identity} (Level A), \emph{exact within the DCA} (Level B),
\emph{finite-dimensional approximation} (Level C), and
\emph{numerical realization} (Level D).  In particular, a loop
equation evaluated with measured feedback content is an \emph{exact
consistency test of the projected loop}, not an exact numerical
closure; the predictive map of Sec.~\ref{sec:predict} is the
Level-D realization proper.
\emph{Level A (exact identities):} the Feshbach identity
Eq.~\eqref{eq:feshbach},
the geometric identity Eq.~\eqref{eq:geom}, the decomposition
Eq.~\eqref{eq:XD}, the spectral representation
Eq.~\eqref{eq:stieltjes} and its sum rules, and the exact vertex
inversion of Proposition~\ref{prop:inversion} below---no
approximation.  \emph{Level B (projection closure):} the
cross-correlated channel is re-expanded by the projected hierarchy and
closed by the diagonal closure approximation (DCA),
Eqs.~\eqref{eq:split}--\eqref{eq:Gell}; the microscopic vertex
functional of Theorem~\ref{thm:vertex} lives at this level and is
exact \emph{within the DCA}.  \emph{Level C (causal
finite-dimensional approximation):} each spectral cumulant is
compressed into a finite causal parameter set
(Sec.~\ref{sec:ansatz}).  \emph{Level D (numerical
self-consistency):} the parameter fixed point
Eq.~\eqref{eq:fixedpoint} is solved by iteration
(Sec.~\ref{sec:closure}).  The only statements claimed to be exact
belong to Level A or, conditionally on the DCA, to Level B.  Beyond the
four levels, a fifth category---the open layer---collects what the
article explicitly does \emph{not} claim: the full bulk-correlator
reconstruction (Sec.~\ref{sec:intro}), the generic isotropy theorem and
the thermodynamic control of the LRI window conditions
(Appendix~\ref{sec:LRIstatus}), and the extensibility demonstrations listed
in Sec.~\ref{sec:conclusion}.  The constructive sections---the
ladder-and-vertex block of Sec.~3 and Secs.~\ref{sec:ansatz},
\ref{sec:closure}, and \ref{sec:num}---each open with a status
sentence in this vocabulary, so every block states its own level up
front.

\subsection{Causal multi-variable spectral representation}
\label{sec:causal}

The cumulant hierarchy acquires its analytic structure through the
spectral content of the resolvents:  each $\cC^{(r)}$ becomes an
$r$-fold Stieltjes transform whose boundary content carries the
energy-resolved hierarchy.  Two aspects are developed in turn---the
multi-energy spectral cumulant with its higher-order ETH boundary
content, and the constraint structure that the representation imposes
on any consistent closure.

\subsubsection{The spectral cumulant and higher-order ETH}

The channel cumulants carry their eigenstate content through the
spectral-weight structure:
Expanding Eq.~\eqref{eq:cumdef} through Eq.~\eqref{eq:Rdef},
\begin{equation}
\cC^{(r)}_{a_1\cdots a_r}(z_1,\dots,z_r)
=\sum_{n_1,\dots,n_r}
\frac{\kappa_r\bigl(\delta p_{n_1}^{a_1},\dots,\delta p_{n_r}^{a_r}\bigr)}
{\prod_i(z_i-\lambda_{n_i})},
\label{eq:specr}
\end{equation}
each resolvent factor carrying one eigenstate and one channel index.
The kernel is the $r$-th joint cumulant of the centered channel
weights: for $r\le3$ it coincides with the centered moment
$\E_H[\prod_i\delta p_{n_i}^{a_i}]$, while from $r=4$ onward the two
differ and the cumulant is the kernel selected by multilinearity in
Eq.~\eqref{eq:cumdef}.
Define the multi-energy spectral cumulant
\begin{multline}
S^{(r)}_{a_1\cdots a_r}(E_1,\dots,E_r)\\
=\sum_{n_1,\dots,n_r}
\kappa_r\bigl(\delta p_{n_1}^{a_1},\dots,\delta p_{n_r}^{a_r}\bigr)
\prod_{i=1}^{r}\delta(E_i-\lambda_{n_i}),
\label{eq:Sdef}
\end{multline}
so that Eq.~\eqref{eq:specr} is the $r$-fold Stieltjes transform
\begin{equation}
\cC^{(r)}(\boldsymbol z)
=\int \prod_{i=1}^{r}\frac{\dd E_i}{z_i-E_i}\;
S^{(r)}(\boldsymbol E).
\label{eq:stieltjes}
\end{equation}
Equation \eqref{eq:stieltjes} is the causal multi-variable
representation of the hierarchy \cite{MResR}: $\cC^{(r)}$ is analytic
in each $z_i$ in the upper half-plane, and its on-shell boundary
projection recovers the energy-resolved higher-order ETH correlation
density
\begin{multline}
P_r(\boldsymbol E)=S^{(r)}(\boldsymbol E)\\
=\frac{1}{\pi^r}\mathrm{Im}_1\cdots\mathrm{Im}_r\,
\cC^{(r)}(E_1-\mathrm i0^+,\dots,E_r-\mathrm i0^+),
\label{eq:ETH}
\end{multline}
in the distributional boundary-value sense \cite{MResR}.  The complex
arguments $z_j=E_j+\mathrm i\eta_j$ aggregate spectral windows
$|\lambda-E_j|\lesssim\eta_j$: $\eta$ is a continuously tunable
coarse-graining scale, and the joint energy arguments carry the level
correlations a bare ETH moment does not display
\cite{MResR,HOETH}.

\subsubsection{Constraints on the spectral cumulant}

The representation \eqref{eq:stieltjes} makes the constraint structure
of the hierarchy explicit.  $S^{(r)}$ is real, symmetric under
simultaneous permutations of channels and energy arguments, and of
vanishing total mass,
\begin{equation}
\int S^{(r)}(\boldsymbol E)\prod_i\dd E_i
=\E_H\Bigl[\prod_i\Bigl(\sum_n\delta p_n^{a_i}\Bigr)\Bigr]=0
\qquad(r\ge1),
\label{eq:mass0}
\end{equation}
the exact per-realization normalization $\sum_np_n^a=1$.  Equation
\eqref{eq:mass0} fixes only the total integrated mass of each channel
sector: for a single channel it forces the diagonal-in-$n$ and
off-diagonal-in-$n$ parts to cancel, so the diagonal-in-$n$ mass
alone---the centered participation
moment,
\begin{equation}
\sum_n\E_H\bigl[(\delta p_n^a)^r\bigr]
=\sum_{s=0}^{r}\binom{r}{s}(-1)^{r-s}\,\frac{q_s}{D^{r-1}},
\qquad
q_r=\frac{\E[p^r]}{\E[p]^r},
\label{eq:diagmass}
\end{equation}
with $q_0=1$; at $r=2$ this is the diagonal ETH variance
$(q_2-1)/D$.  For each variable the boundary values obey the
dispersion relation: with
$f_1(E_1)=\mathrm{Im}_1\cC^{(r)}(E_1-\mathrm i0^+,z_2,\dots,z_r)$ and
the remaining variables held fixed,
\begin{equation}
\mathrm{Re}\,\cC^{(r)}(E_1-\mathrm i0^+,z_2,\dots,z_r)
=\frac1\pi\,\cP\!\int\frac{f_1(E_1')}{E_1-E_1'}\,\dd E_1',
\label{eq:disp}
\end{equation}
the $r$-point promotion of the single-resolvent boundary identity
\cite{SRletter,MResR}; iterating over the $r$ variables gives the full
$2^r$ Hilbert dictionary $H_1H_2\cdots H_r$ relating the real and
imaginary boundary combinations, with $H$ the Hilbert transform.  For
real $S^{(r)}$ the cumulants obey the conjugation rule
$\cC^{(r)}(\boldsymbol z^*)=\cC^{(r)}(\boldsymbol z)^*$.

The deep plane of Eq.~\eqref{eq:stieltjes} is the moment dictionary.
With the centered channel moment fields
\begin{equation}
s_m(a)\equiv\delta(H^m)_{aa}
=\bra{\varphi_a}H^m\ket{\varphi_a}-\frac1D\Tr H^m,
\label{eq:momfield}
\end{equation}
the exact per-realization identity
$\sum_n p_n^a\,\lambda_n^m=\bra{\varphi_a}H^m\ket{\varphi_a}$ (the
moment-field identity of Appendix~\ref{app:moment}
\cite{MResmerge,MResR}) gives
\begin{multline}
\cM_{m_1\cdots m_r}(a_1,\dots,a_r)
:=\int E_1^{m_1}\cdots E_r^{m_r}\,
S^{(r)}(\boldsymbol E)\prod_i\dd E_i\\
=\kappa_r\bigl[s_{m_1}(a_1),\dots,s_{m_r}(a_r)\bigr],
\label{eq:moments}
\end{multline}
so the deep-plane coefficients
$K_{m_1\cdots m_r}=\cM_{m_1\cdots m_r}$ of
$\cC^{(r)}(\boldsymbol z)=\sum_{\boldsymbol m}\cM_{\boldsymbol m}/
\prod_iz_i^{m_i+1}$ are the connected correlations of diagonal
Hamiltonian moments---computable from $H$ alone in \emph{any} system,
with no diagrammatic, replica, or ensemble-specific input.  The $r=1$
case reads
\begin{align}
\sum_n p_n^a\lambda_n&=(H)_{aa},\notag\\
\sum_n p_n^a\lambda_n^2&=(H^2)_{aa}=(a_a+V_a)^2+v_a,
\label{eq:mf1}
\end{align}
the identities underlying Eq.~\eqref{eq:Xdeep}.  At $r=2$,
$\cM_{20}=\kappa_2[s_2(a),s_0(b)]=0$ identically, while
$\cM_{11}=\kappa_2[s_1(a),s_1(b)]$ is the channel-energy covariance
(closing as $2\ell-2k^2/N$ in free fermions, with $\ell$ the overlap
of the two reference sectors, $k$ their size, and $N$ the site
number, \cite{MResmerge}); the
nonvanishing content of the two-point kernel is therefore carried by
the mixed moment $\cM_{11}$ and the diagonal mass
Eq.~\eqref{eq:diagmass} at the lowest orders.

Finally, parity.  For a reflection-symmetric kernel the single-channel
spectral functions are symmetric, and the multi-variable kernel
$S^{(r)}$ is symmetric under the joint reflection
$E_i\to 2E_0-E_i$ for even $r$, with $E_0$ the band-center reflection
point; parity-mixing content (spectral
skewness) can only enter through the odd-$r$ multi-channel vertices
(Sec.~\ref{sec:bench}).  In the structured Gaussian class---the
coupling class of Sec.~\ref{sec:conn}, whose cumulants depend on the
path tuple only through its connectivity class---the
connected deep plane obeys the degree bound
$\deg_t\cC^{(r)}(t\boldsymbol\zeta)\ge 3r-2$, with
$\boldsymbol\zeta=(z_1,\dots,z_r)$ the fixed direction of the joint
scaling, saturated by the
chain geometry \cite{MResR}---the minimal virtual connectivity needed
to correlate $r$ spectral probes.

Taken together, the constraints define the \emph{causal parameter
manifold} on which the closure acts.  Let $\cM_{\rm causal}$ denote
the set of parameter arrays
$\boldsymbol\theta=\{\cM_{\boldsymbol m},\Theta_{\mathfrak c},
\Gamma^{(r)}\}$ whose associated functions $\cC^{(r)}(\boldsymbol z)$
are the $r$-fold Stieltjes transforms
Eq.~\eqref{eq:stieltjes} of real signed spectral-cumulant measures
$S^{(r)}$ satisfying the required symmetry, marginal, mass,
dispersion, and moment constraints:
$\cC^{(2)}$ symmetric in its arguments, analytic on
$\mathbb H^+\times\mathbb H^+$, and obeying the mass constraints
Eqs.~\eqref{eq:mass0}--\eqref{eq:diagmass}, the dispersion dictionary
Eq.~\eqref{eq:disp}, and the moment dictionary
Eq.~\eqref{eq:moments}.  The closure map
$F[\boldsymbol\theta;H]\mapsto\boldsymbol\theta'$ of
Sec.~\ref{sec:closure} preserves these constraints at the level of
the representation, and the preservation statement splits by
truncation level: the \emph{exact} representation---the hierarchy of
Secs.~\ref{sec:hier}--\ref{sec:vertex} before any truncation---
carries the defining constraints through the map identically, whereas
a truncated closure is not claimed to map $\cM_{\rm causal}$ into
itself globally; it is verified \emph{locally}, around the computed
fixed point, by the numerical tests assembled below.  Every
contribution to the updated correlators is
assembled from Hamiltonian moments, class amplitudes, and resolvent
products, each carrying the Stieltjes kernel
Eq.~\eqref{eq:stieltjes} of a legitimate building block: symmetry,
the marginals, the mass constraints
Eqs.~\eqref{eq:mass0}--\eqref{eq:diagmass}, the dispersion dictionary
Eq.~\eqref{eq:disp}, the moment dictionary Eq.~\eqref{eq:moments},
and analyticity on the convergence domain of the response resummation
Eq.~\eqref{eq:XiResum} of Sec.~\ref{sec:crossresponse} are therefore inherited
by construction---causality comes from the representation, not from
analyticity alone.  Positivity is a property of the spectral
\emph{density}, not of the cumulant: the channel density
$\rho_a(E)=\sum_n p_n^a\,\delta(E-\lambda_n)$ is a positive measure,
while the spectral cumulant $S^{(r)}$, $r\ge2$, is generally
signed---its mass constraints fix its total mass to zero
[Eqs.~\eqref{eq:mass0}--\eqref{eq:diagmass}], so a nonvanishing
$S^{(r)}$ cannot be positive.  What is verified numerically at the
fixed point is the causal \emph{form} and the local stability: the
boundary test of
Sec.~\ref{sec:num} verifies the Hilbert dictionary on the real
boundary at the few-percent level, the measured response radius
$\rho(K_X)\le0.044$ places the fixed point inside the resummation
domain, and the closure Jacobian has spectral radius
$\chi_{\rm cl}=\rho(J_F)\approx0.04$--$0.07$ at the fixed point,
which establishes local linear stability; the operator-norm
amplification is estimated separately in
Sec.~\ref{sec:errorprop}.  The causal manifold thus defines
the admissible parameter domain, the exact representation preserves
its defining constraints, and the numerics verify the causal
boundary behavior and the local stability of the truncated closure
at the fixed point---causality is
not imposed after the fact on the closure: it
\emph{constrains the admissible parameter manifold} of the
higher-resolvent construction, and the self-consistency equation
$\boldsymbol\theta=F[\boldsymbol\theta;H]$ is an equation on
$\cM_{\rm causal}$---the single-resolvent causal ansatz of
Ref.~\cite{SRletter}, promoted from the Herglotz--Nevanlinna class
of one variable to the multivariate Stieltjes class defined by
Eq.~\eqref{eq:stieltjes}.

The vertical structure assembled in this article is then
\begin{equation}
\begin{array}{c}
\text{microscopic input: }H\\[-3pt]
\downarrow\\[-3pt]
\text{microscopic cumulants and}\\[-2pt]
\text{response kernels: }\{\cM_{\boldsymbol m},\Theta_{\mathfrak c},
\Gamma^{(r)}\}\\[-3pt]
\downarrow\\[-3pt]
\text{multi-resolvent Dyson--Feshbach hierarchy}\\[-3pt]
\downarrow\\[-3pt]
\text{causal matrix ansatz on }\cM_{\rm causal}\\[-3pt]
\downarrow\\[-3pt]
\text{parameter self-consistency: }\boldsymbol\theta
=F[\boldsymbol\theta;H]\\[-3pt]
\downarrow\\[-3pt]
\text{self-consistent output: }\cC^{(r)},\ X,\Delta
\end{array}
\label{eq:vertchain}
\end{equation}
Every arrow of Eq.~\eqref{eq:vertchain} is exact until the ansatz level: the first two steps hold
for \emph{any} Hamiltonian (Secs.~\ref{sec:hier} and
\ref{sec:causal}), the third is the exact reorganization of
Secs.~\ref{sec:ladder}--\ref{sec:vertex}, and only at the fourth
step---the causal matrix ansatz---is a parameterization introduced;
the fifth step then closes the loop on the causal manifold.  This is
the concrete upgrade over the single-resolvent route of
Refs.~\cite{SRletter,MResR}: the microscopic information is organized
into cumulants and response kernels \emph{before} the ansatz, and the
ansatz closes the multi-resolvent hierarchy on the causal manifold
rather than truncating a single self-energy.

\section{Ladder resummation, vertex functional, and the MRSC system}
\label{sec:ladder}

\emph{Status.}  Level B of the decomposition of
Sec.~\ref{sec:levels}: this section is an
exact derivation from the hierarchy---ladder resummation, the vertex
functional as a functional of the cumulants, and the MRSC system are
identities or definitions with no approximation introduced; the first
model-dependent input enters only in Sec.~\ref{sec:ansatz}.  The
section proceeds from the mean-field/vertex split of
Eq.~\eqref{eq:XD} to the resummed vertex functional
(Sec.~\ref{sec:vertex}), specializes it to the explicit $r=2$
functional (Sec.~\ref{sec:r2explicit}) and the cross-correlated
channel (Sec.~\ref{sec:ccchannel}), and closes into the MRSC system
(Sec.~\ref{sec:system}) that the loop of Sec.~\ref{sec:closure}
iterates.

The decomposition \eqref{eq:XD} turns the cumulant
Eq.~\eqref{eq:cumdef} into an insertion expansion.  By multilinearity
of cumulants,
\begin{equation}
\cC^{(r)}(\boldsymbol z)
=\sum_{S\subseteq[r]}
\kappa_r\bigl(\phi_1,\dots,\phi_r\bigr),
\qquad
\phi_i=\begin{cases}\Delta_{a_i}(z_i),&i\in S,\\
X_{a_i}(z_i),&i\notin S,\end{cases}
\label{eq:tower}
\end{equation}
an exact reorganization into a mean-field tower (the $X$ fields) and
a vertex tower (the $\Delta$ fields).  The vertex tower is a geometric
series in the self-energy fluctuation: inserting
Eq.~\eqref{eq:XD},
\begin{multline}
\kappa_r\bigl(\Delta_{a_1},\dots,\Delta_{a_r}\bigr)
=\sum_{s_1,\dots,s_r\ge1}
\prod_{i=1}^{r}\bigl(\bar{\cR}_{a_i}^0\bigr)^{s_i+1}\\
\times\kappa_r\Bigl(\delta\cG_{a_1}^{s_1},\dots,\delta\cG_{a_r}^{s_r}\Bigr).
\label{eq:vertexTower}
\end{multline}
Each mixed cumulant of powers is a polynomial in the vertices, by the
standard product expansion of cumulants (Appendix~\ref{app:cum}):
\begin{equation}
\kappa_r\Bigl(\delta\cG_{a_1}^{s_1},\dots,\delta\cG_{a_r}^{s_r}\Bigr)
=\sum_{\pi}\prod_{B\in\pi}
\Gamma^{(|B|)}\bigl(\{z_i,a_i\}_{i\in B}\bigr),
\label{eq:partexp}
\end{equation}
the sum running over partitions $\pi$ of the $\sum_i s_i$ factors into
blocks of size $\ge2$ whose block hypergraph connects the $r$ groups
(blocks lying within one group contribute the diagonal self-energy
correlations $\Gamma^{(2)}_{aa}$).  Assembling
Eqs.~\eqref{eq:tower}--\eqref{eq:partexp}:

\begin{theorem}[Multi-resolvent ladder]
\label{thm:ladder}
The exact $r$-resolvent cumulant admits the insertion expansion
Eq.~\eqref{eq:tower}, with the vertex tower given by
Eqs.~\eqref{eq:vertexTower}--\eqref{eq:partexp}.  The leading vertex
insertion is
\begin{multline}
\cC^{(r)}(\boldsymbol z)
=\kappa_r\bigl(X_{a_1},\dots,X_{a_r}\bigr)
+\sum_{i=1}^{r}\kappa_r\bigl(\{\Delta_{a_i},X_{a_j\neq i}\}\bigr)\\
+\prod_{i=1}^{r}\bigl(\bar{\cR}_{a_i}^0\bigr)^{2}\;\Gamma^{(r)}(\boldsymbol z)
+\cdots,
\label{eq:leadladder}
\end{multline}
where the omitted terms contain either (i) two or more insertions in
one channel, i.e.\ products of lower vertices such as
$(\bar{\cR}_{a_1}^0)^3(\bar{\cR}_{a_2}^0\bar{\cR}_{a_3}^0)^2
[2\Gamma^{(2)}_{a_1a_2}\Gamma^{(2)}_{a_1a_3}
+\Gamma^{(2)}_{a_1a_1}\Gamma^{(2)}_{a_2a_3}]$ at $r=3$, or (ii)
higher powers $(\delta\cG)^{s\ge2}$ dressing single channels.  The two
classes of omitted content are controlled by separate mechanisms in
the joint scaling $z_i=t\zeta_i$ with $\boldsymbol\zeta$ fixed and
$t\to\infty$: each additional vertex insertion in one channel costs
one factor $\bar{\cR}_a^0\delta\cG_a=O(t^{-2})$, so every correction
of class (i) sits at least two powers of $t$ higher than the leading
insertion; the $X$-tower of Eq.~\eqref{eq:leadladder} is instead
governed by the total degree bound $\deg_t\cC^{(r)}\ge3r-2$ of
Sec.~\ref{sec:causal}, which controls each tower term
independently of the insertion counting.
\end{theorem}

The proof---multilinearity of cumulants, the insertion expansion term by term, the product formula of Appendix~\ref{app:cum}, and the deep-plane degree bookkeeping---is given in Appendix~\ref{app:proofladder}.

Theorem~\ref{thm:ladder} is the multi-resolvent analogue of the Dyson
equation: the single-resolvent (mean) sector is the $r=1$ instance,
\begin{equation}
\hat{\cR}_a:=\E_H[\cR_a]
=\E_H[\bar{\cR}_a^0]
+\sum_{s\ge1}\E_H\Bigl[\bigl(\bar{\cR}_a^0\bigr)^{s+1}
\bigl(\delta\cG_a\bigr)^s\Bigr],
\label{eq:meanr}
\end{equation}
exact ($\cR_a=\bar{\cR}_a^0+\Delta_a$), and under the self-averaging
factorization of the bare propagator
\begin{equation}
\E_H\Bigl[\bigl(\bar{\cR}_a^0\bigr)^{s+1}\bigl(\delta\cG_a\bigr)^s\Bigr]
\simeq\bigl(\bar{\cR}_a^0\bigr)^{s+1}
\sum_{\pi\vdash s}\prod_{B\in\pi}\Gamma^{(|B|)}_{a\cdots a}(z,\dots,z),
\label{eq:meanrf}
\end{equation}
the fluctuation-dressed Dyson equation in which the bare propagator is
corrected by every diagonal vertex insertion.  In the random
free-fermion benchmark the eigenvector average of $\cR_a$ equals the
realization trace resolvent exactly, $\E_U[\cR_a]=\cR_{\rm tr}$, while the bare
propagator $\bar{\cR}_a^0$ carries the harmonic-arithmetic mismatch of
Eq.~\eqref{eq:Xdeep}; the $X$-tower is generically nonzero and is
verified together with the vertex insertions in
Sec.~\ref{sec:num}.

\subsection{The vertex functional}
\label{sec:vertex}

The second structure expresses the vertices in terms of the
cumulants.  Its Level-A part is exact and requires no closure
assumption: inverting the geometric identity Eq.~\eqref{eq:geom}
gives the exact vertex inversion
\begin{equation}
\delta\cG_a=\frac{1}{\bar{\cR}_a^0}-\frac{1}{\cR_a}
=\frac{1}{\bar{\cR}_a^0}-\frac{1}{\cR_{\rm tr}}
\sum_{t\ge0}\Bigl(-\frac{\delta\cR_a}{\cR_{\rm tr}}\Bigr)^t.
\label{eq:inversion}
\end{equation}
\begin{proposition}[Exact vertex inversion]
\label{prop:inversion}
The irreducible vertices admit the exact (DCA-free) functional
representation
\begin{equation}
\Gamma^{(r)}(\boldsymbol z)
=\kappa_r\Bigl[\frac{1}{\bar{\cR}^0_{a_1}}-\frac{1}{\cR_{a_1}},\dots,
\frac{1}{\bar{\cR}^0_{a_r}}-\frac{1}{\cR_{a_r}}\Bigr],
\label{eq:inversion2}
\end{equation}
which, combined with Eq.~\eqref{eq:inversion} and the product
expansion of Appendix~\ref{app:cum}, expresses each
$\Gamma^{(r)}$ through the $X$-tower, the trace resolvent, and the
cumulants $\cC^{(s)}$ of all orders.  The inversion determines the
\emph{functional form} of the vertex but not its microscopic content:
the seeds below are supplied by the projection closure.
\end{proposition}

The Level-B part supplies that content.  Split the self-energy
exactly into its mean-field and
cross-correlated channels \cite{SRletter}:
\begin{equation}
\cG_a=\cG^{\mathrm{MF}}_a+\cG^{\mathrm{CC}}_a,
\qquad
\cG^{\mathrm{MF}}_a(z)=\sum_{b\neq a}|V_{ab}|^2\,\cR_b(z),
\label{eq:split}
\end{equation}
with $\cG^{\mathrm{CC}}_a$ the cross-correlated (interference)
channel.  The projected hierarchy re-expands the cross channel
exactly \cite{SRletter}, $\cG^{\mathrm{CC}}_a=\sum_{\ell\ge3}
\cG^{(\ell)}_a$, and under the diagonal closure approximation (DCA)---
statistical self-averaging of the cross terms, adopted here as the
self-averaging closure appropriate to the tested dense-spectrum
regime \cite{SRletter}---
\begin{equation}
\cG^{(\ell)}_a(z)\approx
\sum_{\alpha_1\neq\cdots\neq\alpha_{\ell-1}\neq a}
V^{(\ell)}_{a,\alpha_1,\dots,\alpha_{\ell-1}}
\prod_{j=1}^{\ell-1}\cR_{\alpha_j}(z),
\label{eq:Gell}
\end{equation}
with $V^{(\ell)}$ the $\ell$-fold product of interaction matrix
elements ($V^{(2)}_{a,b}=|V_{ab}|^2$), the superscript counting
resolvent multiplicity, not perturbative order.  Then
$\delta\cG_a=\delta\cG_a^{\mathrm{MF}}+\delta\cG_a^{\mathrm{CC}}$
exactly, and both channels are products of the random variables
$\{|V|^{2},V^{(\ell)},\cR\}$.  The cumulant-of-products expansion
(Appendix~\ref{app:cum}) gives:

\begin{theorem}[Vertex functional (DCA)]
\label{thm:vertex}
Within the DCA, the irreducible vertices are given by the constructive
functional expansion
\begin{multline}
\Gamma^{(r)}_{a_1\cdots a_r}(\boldsymbol z)
=\Gamma_0^{(r)}(\boldsymbol z)
+\sum_{s\ge2}\sum_{\boldsymbol\ell}
\cD^{(r,s)}_{\boldsymbol a,\boldsymbol\ell}\bigl[\cC^{(s)}\bigr]\\
+\text{mixed coupling--resolvent terms},
\label{eq:vertexF}
\end{multline}
exact to the DCA, where the seed $\Gamma_0^{(r)}$ is the
coupling-fluctuation content
\begin{multline}
\Gamma_0^{(r)}(\boldsymbol z)
=\sum_{\ell_1,\dots,\ell_r\ge2}\;\sum_{\boldsymbol\alpha}
\kappa_r\Bigl(\delta V^{(\ell_1)}_{a_1,\boldsymbol\alpha_1}
\prod_j\hat{\cR}_{\alpha_{1j}},\;\dots\;\\
\dots,\;
\delta V^{(\ell_r)}_{a_r,\boldsymbol\alpha_r}
\prod_j\hat{\cR}_{\alpha_{rj}}\Bigr)
\label{eq:seed}
\end{multline}
with $\hat{\cR}_a=\E_H[\cR_a]$ and $\delta V^{(\ell)}$ the fluctuation
of the $\ell$-fold coupling product, and the dressing kernels
$\cD^{(r,s)}$ are built from coupling moments $\E[V^{(\ell)}V^{(\ell')}
\cdots]$ and products of $\hat{\cR}$, acting multilinearly on the
cumulants $\cC^{(s)}$ with $s\le\sum_i(\ell_i-1)$.  The seed
Eq.~\eqref{eq:seed} is the coupling-cumulant part only: it collects
the partition blocks of pure coupling factors, while blocks that
contain fluctuating resolvents give the kernels $\cD^{(r,s)}$ and the
mixed terms.  In particular the seed vanishes identically for
deterministic couplings, $\delta V=0$; the fixed-coupling regime is
the one in which the dressing and mixed terms carry the whole vertex
content, as discussed after the class expansion below.
\end{theorem}

\begin{proof}
By Eq.~\eqref{eq:split} and the DCA form
Eq.~\eqref{eq:Gell}, each $\delta\cG_{a_i}$ is a finite sum of terms
$\delta[V^{(\ell_i)}_{a_i,\boldsymbol\alpha}\prod_j\cR_{\alpha_j}]$.
The cumulant of such products expands by the product formula
(Appendix~\ref{app:cum}) over partitions whose blocks contain either
(i) coupling factors only---giving the seed
Eq.~\eqref{eq:seed}---(ii) resolvent factors only---giving the
dressing kernels $\cD^{(r,s)}$ through $\cC^{(s)}=\kappa_s[\delta\cR
\cdots]$---or (iii) a mixture.  Block sizes $\ge2$ exclude single
factors because all fields are centered.
\end{proof}

The classified object is the \emph{block} of the partition sum in
Eq.~\eqref{eq:cumprod}, and the rule is intrinsic to it: a block is
assigned by the variable types among its factors---all coupling
fluctuations (seed), all resolvent fluctuations (dressing), or both
(mixed)---so the assignment is unique and no block is counted twice,
each block appearing once in the partition sum.  Singleton blocks do not
occur: every field entering the formula is centered, so
$\kappa_1=0$.  Mean parts extracted \emph{before} the formula is applied
are deterministic coefficients (last row), never blocks.  Applied to
$\delta\cG^{\rm MF}_a$ and $\delta\cG^{\rm CC}_a$ of
Eq.~\eqref{eq:split}, the same rule reproduces the explicit $r=2$
functional of Eq.~\eqref{eq:r2functional} term by term
(Sec.~\ref{sec:r2explicit}).  The classification applies to the blocks
appearing in the centered coupling--resolvent cumulant expansion of
Eq.~\eqref{eq:vertexF}: apart from the extracted deterministic
coefficients there are no further factor types, so the three rows
exhaust all nonempty blocks, and the extraction of the mean parts
neither double-counts nor drops any contribution.

\begin{table*}[tbp]
\centering
\small
\begin{ruledtabular}
\begin{tabular}{p{3.3cm}p{3.9cm}p{3.5cm}p{4.1cm}}
block of the partition & contents & where it enters & role in
$F_{\rm ret}$\\
\hline
coupling-only & $\kappa_r$ of $\delta V^{(\ell)}$ factors & seed
$\Gamma_0^{(r)}$, Eq.~\eqref{eq:seed} & retained (benchmark-computed)\\
resolvent-only & $\kappa_s$ of $\delta\cG$ factors & dressing kernels
$\cD^{(r,s)}$ through $\cC^{(s)}$ & pair-contraction ladder retained\\
mixed & coupling factors with resolvent fluctuations & mixed
coupling--resolvent terms of Eq.~\eqref{eq:vertexF} & beyond the ladder
dropped; omission carried by $\delta F_{\rm vertex}$\\
single factor & one field & absent: $\kappa_1=0$ for centered
fields & absent\\
deterministic-shift terms & mean parts of
$V^{(\ell)},\cR$ & coefficients of the deterministic kernel & carried
as coefficients, not blocks\\
\hline\hline
\end{tabular}
\end{ruledtabular}
\caption{Partition attribution for the vertex functional: every block of
the expansion of Eq.~\eqref{eq:cumprod} falls into exactly one row, and
no block is counted twice.  The last row records the deterministic
parts, which never appear as blocks because they are coefficients.}
\label{tab:partattr}
\end{table*}

The seed content is exact and model-specific only through the coupling
cumulants.  In random free fermions the couplings are products of the
Gaussian one-body matrix elements, and the seeds reduce to
\begin{equation}
\Gamma^{(2)}_{ab}(z_1,z_2)=\frac{2\,e_{ab}}{z_1z_2}+O(z^{-4}),
\qquad
e_{ab}=|E_a\cap E_b|,
\label{eq:G2seed}
\end{equation}
\begin{equation}
\Gamma^{(3)}_{abc}(z_1,z_2,z_3)=\frac{8\,e_{abc}}{z_1z_2z_3}+O(z^{-5}),
\qquad
e_{abc}=|E_a\cap E_b\cap E_c|,
\label{eq:G3seed}
\end{equation}
where $E_a$ is the edge set of channel $a$ (the particle--hole pairs
$\{p,q\}$ with $p\in A$, $q\notin A$ in the occupied-set
representation) and the Wick factors $2=\kappa_2(\chi^2_1)$,
$8=\kappa_3(\chi^2_1)$ are the chi-square cumulants of the squared
coupling amplitude \cite{MResR}.  The shared-edge counts measure the
virtual one-body pathways available simultaneously to the channels:
$e_{ab}>0$ even for disjoint channels (the transition pairs
interchanging their occupied sets), which is why the vertex content is
generically nonzero for every channel pair.  For fixed-coupling systems (no coupling disorder) the
\emph{vertex} content separates into two layers, and the two must be
kept apart from the seed: a coupling-monomial part
$\Gamma_{\rm coup}^{(r)}=\sum_{\mathfrak c}
\Theta_{\mathfrak c}^{(r)}B_{\mathfrak c}$, identical to the seed
$\Gamma_0^{(r)}$ whenever the couplings fluctuate, and a
resolvent-induced part $\Gamma_{\rm res}^{(r)}[\cC]$ collected from
the dressing and mixed terms of Eq.~\eqref{eq:vertexF}, in which the
coupling factors act as deterministic kernels and the fluctuation
content sits entirely in the resolvent factors,
$\Gamma_{\rm res}^{(r)}=\sum_{\boldsymbol\ell,\boldsymbol\alpha}
V^{(\ell_1)}\cdots V^{(\ell_r)}\kappa_r[\prod\hat{\cR}\cdots]$ with
the $\kappa_r$ taken over the resolvent fluctuations.  For
deterministic $V$ the coupling part vanishes identically
($\delta V=0$, hence $\Gamma_0^{(r)}=0$) while the resolvent part does
not: the microscopic route Hamiltonian $\to$ resolvent fluctuations
$\to$ self-consistency then runs entirely through the
$\cR$-cumulants, and the genuinely interacting benchmark of
Sec.~\ref{sec:genbench} realizes exactly this regime.  The two-layer
split is therefore a split of the vertex,
\begin{equation}
\Gamma^{(r)}=\Gamma_{\rm coup}^{(r)}+\Gamma_{\rm res}^{(r)}[\cC],
\label{eq:gammasplit}
\end{equation}
not a second definition of the seed: $\Gamma_0^{(r)}$ is always the
coupling-cumulant part alone, and it is this object that the
connectivity expansion of Sec.~\ref{sec:conn} resolves into class
amplitudes.

The $r=2$ instance of Theorem~\ref{thm:vertex} is the
Feshbach--Bethe--Salpeter structure:
\begin{multline}
\Gamma^{(2)}_{ab}(z_1,z_2)
=\Gamma_{0,ab}^{(2)}(z_1,z_2)
+\sum_{c,d}\E\bigl[|V_{ac}|^2|V_{bd}|^2\bigr]\,
\cC^{(2)}_{cd}(z_1,z_2)\\
+\text{mixed terms},
\label{eq:BSE2}
\end{multline}
combined with the ladder Eq.~\eqref{eq:leadladder} at $r=2$,
\begin{multline}
\cC^{(2)}_{ab}
=\kappa_2(X_a,X_b)+\kappa_2(X_a,\Delta_b)+\kappa_2(\Delta_a,X_b)\\
+\bigl(\bar{\cR}_a^0\bar{\cR}_b^0\bigr)^2\Gamma^{(2)}_{ab}+\cdots,
\label{eq:ladder2}
\end{multline}
this is precisely the two-point ladder
$\cC^{(2)}=\bar{\cR}^2\bar{\cR}^2[(1-\cL)^{-1}\Gamma_0^{(2)}+\cdots]$,
with $\cL_{ab,cd}=\bar{\cR}_a^2\bar{\cR}_b^2|V_{ac}|^2|V_{bd}|^2$ the
loop kernel of the hierarchy's two-point level
\cite{HOETH,MResmerge}, now with the
irreducible seed identified microscopically through
Eq.~\eqref{eq:G2seed}.

\subsection{The explicit $r=2$ functional}
\label{sec:r2explicit}

For Gaussian couplings the $r=2$ functional is fully explicit.
Writing $\delta(|V_{ac}|^2\cR_c)=\delta|V_{ac}|^2\,\hat{\cR}_c
+\E[|V_{ac}|^2]\,\delta\cR_c
+\bigl(\delta|V_{ac}|^2\delta\cR_c-\E[\delta|V|^2\delta\cR]\bigr)$
(exact per realization, with $\hat{\cR}_c=\E_H[\cR_c]$ the
deterministic mean resolvent) and applying the product expansion
(Appendix~\ref{app:cum}) term by term,
\begin{equation}
\begin{aligned}
\Gamma^{(2)}_{ab}(z_1,z_2)={}&
\underbrace{2\!\!\sum_{e\in E_a\cap E_b}\!\!
\hat{\cR}_{c(e)}(z_1)\hat{\cR}_{d(e)}(z_2)}_{\text{seed, }O(z^{-2})}\\
&+\underbrace{\sum_{e\in E_a,f\in E_b}\!\!
\cC^{(2)}_{c(e),d(f)}(z_1,z_2)}_{\text{ladder, }O(z^{-4})}\\
&+2\!\!\underbrace{\sum_{e\in E_a\cap E_b}\!\!
\cC^{(2)}_{c(e),d(e)}(z_1,z_2)}_{\text{ladder}^{\prime}}\\
&+\underbrace{\sum_{e\in E_a,f\in E_b}\!\!
\E_H\bigl[\delta|V_{bd}|^2\delta\cR_c\delta\cR_d\bigr]
+\cdots}_{\text{mixed, }O(z^{-4})}\\
&+\underbrace{\Gamma_{\mathrm{CC}}^{(2)}}_{\text{cross channel, }O(z^{-4})}
+O(z^{-6}),
\end{aligned}
\label{eq:r2functional}
\end{equation}
where $\Gamma_{\mathrm{CC}}^{(2)}=2\kappa_2(\delta\cG^{\mathrm{MF}}_a,
\delta\cG^{\mathrm{CC}}_b)+\kappa_2(\delta\cG^{\mathrm{CC}}_a,
\delta\cG^{\mathrm{CC}}_b)$ collects the cross-correlated channel.
The seed, ladder, and ladder$^{\prime}$ terms collect the
$\Gamma_0^{(2)}$ content, its ladder continuation, and the coincident-edge
completion of Eq.~\eqref{eq:vertexF}, $c(e)$ denotes the channel
whose symmetric difference with $a$
is the edge $e$ (uniquely determined), the omitted $\cdots$ in the
mixed term collects the connected four-point
coupling--coupling--resolvent--resolvent cumulants of the same order,
which vanish identically under the coupling independence of the
model class, so the mixed content at $r=2$ is generated entirely by
the $\kappa_2$ blocks displayed,
and in the deep plane $\cC^{(2)}=O(z^{-4})$, $\delta\cR=O(z^{-2})$,
$\delta\cG^{\mathrm{CC}}=O(z^{-2})$, so the orders noted hold per
term; the ladder coefficient $3$ on coincident edges receives $1$
from the $\E[|V|^2]\E[|V|^2]$ block and $2$ from the
$\E[\delta|V|^2\delta|V|^2]$ completion.  Equation
\eqref{eq:r2functional} is the concrete equation behind
Theorem~\ref{thm:vertex} at $r=2$: seed, ladder, mixed, and
cross-channel content are separated and ordered, and each term is
computable from $H$ and the two-point cumulant alone---the input
needed to run the $r=2$ closure of Sec.~\ref{sec:closure}.  It is
verified term by term in Sec.~\ref{sec:num}.

\subsection{The cross-correlated channel}
\label{sec:ccchannel}

The remaining content of Theorem~\ref{thm:vertex} at $r=2$---the
cross-correlated channel---closes by the same product expansion.
With $\delta\cG^{\mathrm{CC}}_a=\sum_{\ell\ge3}\sum_{\boldsymbol\alpha}
\delta[V^{(\ell)}_{a,\boldsymbol\alpha}\prod_j\cR_{\alpha_j}]$,
\begin{equation}
\begin{aligned}
\Gamma^{(2)}_{\mathrm{CC}}(z_1,z_2)={}
\underbrace{\sum_{\ell,\ell'\ge3}\sum_{\boldsymbol\alpha,\boldsymbol\alpha'}
\kappa_2\bigl(\delta V^{(\ell)}_{a,\boldsymbol\alpha},
\delta V^{(\ell')}_{b,\boldsymbol\alpha'}\bigr)
\prod_j\hat{\cR}_{\alpha_j}\prod_{j'}\hat{\cR}_{\alpha'_{j'}}}
_{\text{cycle seeds, }H\text{-only}}\\
+\underbrace{\sum_{\ell,\ell'\ge3}\sum_{\boldsymbol\alpha,\boldsymbol\alpha'}
\E\bigl[V^{(\ell)}\bigr]\E\bigl[V^{(\ell')}\bigr]\sum_{j,j'}
\Bigl(\prod\nolimits'\hat{\cR}\Bigr)
\cC^{(2)}_{\alpha_j,\alpha'_{j'}}}
_{\text{cycle ladder, }O(z^{-4})}\\
+\text{coupling--cycle mixed terms},
\end{aligned}
\label{eq:ccfunctional}
\end{equation}
where the cycle seeds are the Wick contractions of the two
$\ell$-fold coupling products (nonzero when the projected cycles
share one-body edges---the cycle-overlap counts, the $\ell\ge3$
generalization of the shared-edge seeds
Eqs.~\eqref{eq:G2seed}--\eqref{eq:G3seed}, closing as
$\kappa_2[\delta(h_{e_1}h_{e_2}h_{e_3}),\delta(h_{e_4}h_{e_5}h_{e_6})]$
type Wick sums), and the cycle ladder dressing pairs one resolvent
factor from each cycle product.  The deep-plane content is
verified in closed form in Sec.~\ref{sec:num}, where the $z^{-2}$
coefficient cumulant closes as
$\kappa_2=11\,904=576+4\cdot2688+192+4\cdot96$ and is reproduced
block by block by direct measurement: the dressing block
\emph{dominates} the deep-plane constant of
$\Gamma^{(2)}_{\mathrm{CC}}$ ($4\cdot 2688=10\,752$ of the
$11\,904$), the cycle seeds being the coupling-fluctuation
sub-content, and the closed form is verified directly through the
coefficient route, the $t\to\infty$ route being impractical.  All
blocks are
computable from
$H$ and $\cC^{(2)}$ alone; together with
Eq.~\eqref{eq:r2functional} this completes the explicit $r=2$
functional at the tested truncation level: every retained term is a
named, ordered, computable object.

The functional simplifies drastically on the sector-constancy ansatz
of Sec.~\ref{sec:ansatz}.  Because the sector densities
$\cG2_m(z_1,z_2)=\sum_{(n,n')\in m}[(z_1-\lambda_n)(z_2-\lambda_{n'})]^{-1}$
are channel-pair-independent, every dressing term reduces to a
combinatorial census over the neighbor pairs.  For the coincident
pair $(a,a)$, with $\cC^{(2)}_{cd}=\sum_m\theta_m^{(\ell_{cd})}\cG2_m$
and the edge sum over ordered pairs $(e,f)$,
\begin{equation}
\begin{aligned}
&\sum_{e,f\in E_a}\cC^{(2)}_{c(e),d(f)}+2\!\!\sum_{e=f}\cC^{(2)}_{c(e),d(e)}\\
&\qquad=\sum_m\Bigl[264\,\theta_m^{(0)}+288\,\theta_m^{(1)}+72\,\theta_m^{(2)}\Bigr]
\cG2_m(z_1,z_2),
\end{aligned}
\label{eq:census}
\end{equation}
the coefficients counting the ordered neighbor-edge pairs of overlap
$0,1,2$: the $24$ neighbors of $\{1,2\}$ split into the two families
$\{1,x\},\{2,x\}$, the cross-family pairs with $x\neq y$
($2\cdot132=264$) have overlap $0$, the same-family pairs with
$x\neq y$ ($2\cdot132=264$) together with the cross-family pairs
with $x=y$ ($24$) have overlap $1$, and the coincident pairs carry
the ladder weight $1+2=3$ ($24\cdot3=72$).  The censuses of the three
channel-overlap representatives are $(264,288,72)$, $(352,236,12)$,
$(452,128,4)$ for $\ell=2,1,0$---three independent overlap patterns, which is
what gives the loop kernel of Sec.~\ref{sec:loop} its rank: the
loop operator of the sector-constant ansatz acts through these
counts, the connectivity structure entering the closure as a
combinatorial census that, in generic systems, generalizes to the
class census of whatever projected-path structure organizes the
channels.

\subsection{The MRSC system}
\label{sec:system}

Assembling the mean sector
Eqs.~\eqref{eq:R0bar},\eqref{eq:meanr}, the ladder
Theorem~\ref{thm:ladder}, and the vertex functional
Theorem~\ref{thm:vertex}, with the mean self-energy functional
\begin{equation}
\bar{\cG}_a=\sum_{b\neq a}\E\bigl[|V_{ab}|^2\cR_b\bigr]
+\sum_{\ell\ge3}\sum_{\boldsymbol\alpha}
\E\Bigl[V^{(\ell)}_{a,\boldsymbol\alpha}\prod_j\cR_{\alpha_j}\Bigr]
\label{eq:meanG}
\end{equation}
(the DCA mean of Eqs.~\eqref{eq:split}--\eqref{eq:Gell}, including the
coupling--resolvent cross correlations):

\begin{theorem}[MRSC system]
\label{thm:MRSC}
Within the DCA the hierarchy $\{\hat{\cR}_a,\cC^{(r)},\Gamma^{(r)}\}$
closes as the coupled system
\begin{subequations}
\begin{align}
\bar{\cR}_a^0&=\frac{1}{z-a_a-V_a-\bar{\cG}_a(\{\hat{\cR}\},\{\cC^{(2)}\})},
\label{eq:sys1}\\
\hat{\cR}_a&=\E_H[\bar{\cR}_a^0]
+\sum_{s\ge1}\E_H\Bigl[\bigl(\bar{\cR}_a^0\bigr)^{s+1}
\bigl(\delta\cG_a\bigr)^s\Bigr],
\label{eq:sys2}\\
\cC^{(r)}&=\cL_r\bigl[\{\bar{\cR}^0\},\Gamma^{(2)},\dots,\Gamma^{(r)},X\bigr],
\label{eq:sys3}\\
\Gamma^{(r)}&=\Gamma_0^{(r)}[H]
+\sum_{s=2}^{r_c}\cD^{(r,s)}\bigl[\cC^{(s)}\bigr]+\text{mixed},
\label{eq:sys4}
\end{align}
\end{subequations}
exact to the DCA \emph{before} vertex-order truncation; a finite
implementation introduces the vertex-hierarchy truncation
$\Gamma^{(r>r_c)}\simeq0$ at order $r_c$ as an additional
approximation.  The seed
$\Gamma_0^{(r)}$ is a fixed input computable from the microscopic
Hamiltonian together with the mean-sector closure (it is built from
the coupling cumulants and the mean channel resolvents
$\hat\cR_a=\E_H[\cR_a]$, whose trace component is
$\E_H[\cR_{\rm tr}]$---the object the mean sector of the scheme
solves for) [Eq.~\eqref{eq:seed}]; the remaining microscopic input of
the scheme is represented by the moment fields
Eq.~\eqref{eq:momfield}, the coupling products $V^{(\ell)}$ and their
cumulants, and the trace resolvent $\cR_{\rm tr}$.
\end{theorem}

Equations \eqref{eq:sys3}--\eqref{eq:sys4} are the two directions of
the self-consistency: the ladder expresses the resolvent cumulants in
terms of the vertices, the vertex functional expresses the vertices in
terms of the cumulants, and the mean sector
Eqs.~\eqref{eq:sys1}--\eqref{eq:sys2} feeds both with the dressed
propagators.  The system is infinite (each $\Gamma^{(r)}$ receives
dressing from $\cC^{(s)}$ of higher order through
Eq.~\eqref{eq:vertexF}), and it reduces level by level: at $r=1$ it is
the single-resolvent closure of Ref.~\cite{SRletter}, in the
classical tradition of self-consistent resolvent equations
\cite{Pastur1973}
[Eq.~\eqref{eq:meanr} with the Faddeeva parameterization of the
diagonal vertex, Sec.~\ref{sec:bench}]; at $r=2$ it is the closed
ladder Eqs.~\eqref{eq:BSE2}--\eqref{eq:ladder2}; the truncation
$\Gamma^{(r>r_c)}\simeq0$ is the vertex-hierarchy truncation
analogous to the Dyson--Schwinger truncations, and the deep-plane
degree $z^{-(3r-2)}$ provides an \emph{asymptotic ordering} of the
rungs in the deep plane---an ordering that need not carry over
quantitatively to the bulk at finite broadening.

The closure logic combines the two directions of
Secs.~\ref{sec:ladder}--\ref{sec:vertex}: the exact inversion of
Proposition~\ref{prop:inversion} is the resolvent-projection
counterpart of the Legendre-vertex relation---it expresses the
irreducible content purely through the resolvent fluctuations---while
the DCA functional of Theorem~\ref{thm:vertex} supplies the
microscopic seed the inversion alone cannot determine.  Requiring the
two to agree on the physical solution is the content of the closure
below.

\section{Causal ansatz and microscopic connectivity}
\label{sec:ansatz}

\emph{Status.}  Level C of Sec.~\ref{sec:levels}: the
finite-dimensional causal approximation; every constraint it obeys is
stated below as an equation, and its parameters are the unknowns the
closure of Sec.~\ref{sec:closure} fixes.

The hierarchy is not solved numerically function by function; the
causal representation of Sec.~\ref{sec:causal} compresses each vertex
into a finite parameter set \emph{before} any equation is imposed.
The ansatz is constrained by causality, symmetry, the sum rules, and
the moment dictionary, in that order.  The constraints are strong
enough to fix the \emph{sign structure} of the kernel, as the $r=2$
analysis shows.

\subsection{Constrained two-scale kernel at $r=2$}
\label{sec:r2kernel}

\emph{$r=2$.}  The spectral cumulant splits into its diagonal
(same-eigenstate) and connected (distinct-eigenstate) parts,
$S^{(2)}(E_1,E_2)=S_{\mathrm{diag}}(E_1)\delta(E_1-E_2)
+S_{\mathrm{conn}}(E_1,E_2)$, with
$S_{\mathrm{diag}}(E)=\sum_n\E[\delta p_n^a\delta p_n^b]\,
\delta(E-\lambda_n)$ the same-level covariance density.  The exact
per-realization identity $\sum_m\delta p_m=0$ implies
$\sum_m\E[\delta p_n\delta p_m]=0$, which requires
\begin{equation}
\int S^{(2)}(E_1,E_2)\,\dd E_2=0
\qquad\text{at every }E_1,
\label{eq:rowsum}
\end{equation}
which fixes the relative normalization of the two parts and implies
$\cM_{20}=\cM_{02}=0$ automatically.  In the center-of-mass variables
$E=(E_1+E_2)/2$, $\omega=E_1-E_2$, the row-sum constraint gives
$\int E^2S^{(2)}=\tfrac12\cM_{11}$ and
$\int E^4S^{(2)}=\tfrac12\cM_{31}+\tfrac38\cM_{22}$ (with
$\cM_{40}=\cM_{04}=0$, $\cM_{31}=\cM_{13}$), and the
$\omega$-moments of the kernel are then \emph{determined} by the
microscopic moments,
\begin{align}
\int\omega^2S^{(2)}\dd E\,\dd\omega&=-2\cM_{11},\notag\\
\int\omega^4S^{(2)}\dd E\,\dd\omega&=6\cM_{22}-8\cM_{31},
\label{eq:momw}
\end{align}
with $\cM_{11}=\kappa_2[s_1(a),s_1(b)]$,
$\cM_{22}=\kappa_2[s_2(a),s_2(b)]$ and
$\cM_{31}=\kappa_2[s_3(a),s_1(b)]$ the microscopic inputs of
Eq.~\eqref{eq:moments} ($\cM_{12}=\cM_{21}=0$ in the Gaussian class).
Since $\cM_{11}\neq0$ while $\cM_{20}=0$, the kernel must have
$\int\omega^2S^{(2)}\neq0$: the naive single-scale form
$A(E)[\delta(\omega)-\xi^{-1}g(\omega/\xi)]$ ties the ratio
$\int\omega^4S^{(2)}/\int\omega^2S^{(2)}$ to $\xi^2$ alone and
generically cannot match both moments.  The minimal constrained form
is the signed two-scale split-point kernel
\begin{equation}
\begin{aligned}
S^{(2)}(E_1,E_2)
&=S_{\mathrm{diag}}(E_1)\,\delta(E_1-E_2)\\
&\quad-\frac12\Bigl[A(E_1)\,Q\Bigl(\frac{E_1-E_2}{\xi_1},\frac{E_1-E_2}{\xi_2}\Bigr)\\
&\qquad\qquad\qquad+(E_1\leftrightarrow E_2)\Bigr],
\end{aligned}
\label{eq:ansatz2}
\end{equation}
with $A(E)$ the diagonal-covariance amplitude (fixed by the
single-resolvent sector through Eq.~\eqref{eq:diagmass}) and
\begin{equation}
Q(\omega)=\frac{c_1}{\xi_1}\,g\Bigl(\frac{\omega}{\xi_1}\Bigr)
-\frac{c_2}{\xi_2}\,g\Bigl(\frac{\omega}{\xi_2}\Bigr),
\qquad
c_1-c_2=1,
\label{eq:Qdef}
\end{equation}
$g$ an even unit Gaussian ($\mu_2[g]=1$, $\mu_4[g]=3$); the
normalization $c_1-c_2=1$ enforces Eq.~\eqref{eq:rowsum} within the
controlled local-envelope approximation, in which the kernel varies on
the $\xi$ scale while the $E$-dependent amplitudes $A(E)$ and
$\xi_i(E)$ are slowly varying, so that the residual $O(\xi\partial)$
corrections are bounded by the local variation of the envelope; the
$\omega$-only form of Eq.~\eqref{eq:ansatz2} is retained as the
baseline ansatz, and its failure against the exact sector-resolved
cumulants is the explicit failure test of Sec.~\ref{sec:num}.  By construction
Eqs.~\eqref{eq:ansatz2}--\eqref{eq:Qdef} satisfy: vanishing total
mass, symmetry, $\cM_{10}=\cM_{20}=0$, and causality (the double
Stieltjes transform is manifestly analytic in each variable).  For illustration, we fix the remaining amplitude convention to the
minimal integer representative $c_1=2$, $c_2=1$ (the
normalization fixes only the difference $c_1-c_2=1$, leaving a
one-parameter family of representatives); the subsequent sector
closure does not rely on this $\omega$-only baseline.  The two moment
equations---with $A(E)$ the diagonal-covariance amplitude of
Eq.~\eqref{eq:ansatz2}---
\begin{align}
&2\xi_1^2-\xi_2^2=\frac{2\cM_{11}}{\int A(E)\dd E},\notag\\
&2\xi_1^4-\xi_2^4=-\frac{6\cM_{22}-8\cM_{31}}{3\int A(E)\dd E}
\label{eq:xisolve}
\end{align}
fix the two parameters $(\xi_1,\xi_2)$ against the microscopic
moments---no fitting.  The sign structure of the kernel is dictated,
not chosen: for coincident channels the free-fermion moments give
$\int\omega^2S^{(2)},\int\omega^4S^{(2)}<0$ (a negative
two-scale kernel), while for disjoint channels $\cM_{11}<0$ forces
sign alternation in $\omega$---the sector structure of the exact
kernel already visible in its low moments.  This is the minimal causal
realization of the constraint structure, but the constraint analysis
itself identifies the general ansatz space: the exact free-fermion
kernel is a
sector sum $S^{(2)}=\sum_sN_s\,\Cov_2(s)\,\rho_s^{(2)}$, with $N_s$ the
sector multiplicity, in which the
overlap-conditioned energy correlations
$\E[\lambda_n\lambda_m\,|\,|A_n\cap A_m|=m]=2m+\cdots$ are carried by
the sector densities $\rho_s^{(2)}$ rather than by any function of
$\omega=E_1-E_2$ alone, so an $\omega$-only kernel reproduces the
gross structure but not the sector-resolved detail---the explicit
failure test of Sec.~\ref{sec:num}.  In the sector
basis the $r=2$ closure is fully explicit, and it is the one carried
out numerically below.

\subsection{The sector-moment closure at $r=2$}
\label{sec:sectorclosure}

Take as ansatz the sector decomposition
\begin{equation}
S^{(2)}(E_1,E_2)=\sum_{m=0}^{k}\theta_m\,\rho_m^{(2)}(E_1,E_2),
\label{eq:sectoransatz}
\end{equation}
over the sector densities
\begin{equation}
\rho_m^{(2)}(E_1,E_2)=\!\!\sum_{(n,n')\in m}\!
\delta(E_1-\lambda_n)\delta(E_2-\lambda_{n'}),
\label{eq:pairdensities}
\end{equation}
with $m$ the overlap (connectivity) class of the eigenstate pair and
$\theta_m$ the sector amplitude; in general systems the sector
coordinate generalizes from the overlap numbers to the connectivity
classes of Sec.~\ref{sec:conn}
\cite{MResR}.  For the free-fermion benchmark the sector geometry
and the pair densities $\rho_m^{(2)}$ are fixed by the one-body
structure of $H$ alone---no eigenstate-overlap data, no Weingarten
input; this is a property of the benchmark, not of the generic
theory: in a generic interacting system the occupied-set coordinate
does not exist, and the connectivity basis must be built from the
projected-path/coupling structure of $H$ itself (the connectivity
coordinates $\cI_\alpha$ of Sec.~\ref{sec:conn}), so
that the free-fermion overlap sectors appear as one exactly solvable
realization of the generic basis.  With this proviso, the sum rules
and the
moment dictionary of Sec.~\ref{sec:causal} then determine the
amplitudes through the linear system
\begin{align}
\sum_m\theta_m N_m&=0,
&\sum_m\theta_m L_{11}(m)&=\cM_{11},\notag\\
\sum_m\theta_m L_{22}(m)&=\cM_{22},
\label{eq:sectorsolve}
\end{align}
with $L_{pq}(m)=\sum_{(n,n')}\lambda_n^p\lambda_{n'}^q$ over the
sector pairs and $N_m$ the sector pair counts; every entry of
Eq.~\eqref{eq:sectorsolve} is computable from $H$ alone
($\cM_{11},\cM_{22}$ from the moment fields, $N_m,L_{pq}(m)$ from
the one-body problem).  The solution
$\boldsymbol\theta_M$ is the moment-matched sector closure; the
remaining moments ($\cM_{20}=0$, $\cM_{31}$, $\cdots$) are
\emph{predictions} of the closure, and the vertex-functional
consistency Eq.~\eqref{eq:MRSCeq} with the residual
Eq.~\eqref{eq:residual}---both of Sec.~\ref{sec:closure}---closes the loop at Level D.  The general
rule is the same at every rung: the ansatz space is the intersection
of the causal (Stieltjes) manifold with the sum-rule hypersurfaces,
and the free parameters are fixed by matching a set $\cI_r$ of
microscopic moments, $|\cI_r|=\dim\boldsymbol\theta_r$, against the
moment fields Eq.~\eqref{eq:moments}.  Parity-mixing content is
carried by the odd-$r$ vertices and is constrained by the selection
rule of Sec.~\ref{sec:bench}.  The dictionary generalizes to the
\emph{channel-resolved} form needed by the connectivity basis of
Sec.~\ref{sec:conn}: for a generic connectivity partition
$\{c\}$ of the level pairs, the moment-field identity evaluated on
the channel pair $(a,b)$,
\begin{equation}
\sum_c\theta_c^{ab}\,L_{pq}(c)=\kappa_2\bigl[s_p(a),s_q(b)\bigr]
=:\cM_{pq}(a,b),
\label{eq:channelmoments}
\end{equation}
with $\theta_c^{ab}=\E[\delta p_n^a\delta p_{n'}^b\,|\,(n,n')\in c]$
the channel-pair-resolved class amplitudes and $L_{pq}(c)$ the
class level sums: the same $L$-matrix is probed by the three
channel-overlap representatives $\ell=0,1,2$ through their
$\cM_{pq}(a,b)$, and the connectivity structure enters the closure
through the constraint that the amplitudes depend on the pair only
through its class census---the generalization of the sector
theorem to the generic basis.  The fine-basis resolution of
Sec.~\ref{sec:num} is the quantitative statement that the diagonal
form ($a=b$) alone resolves only the $\lambda$-profile directions
of the fine classes, while the full channel-resolved set is the
identified next derivation.

\subsection{The sector closure at $r=3$}
\label{sec:sectorclosure3}

The same construction applies verbatim at $r=3$: the ansatz
\begin{equation}
S^{(3)}(E_1,E_2,E_3)=\sum_s\theta_s\,\rho_s^{(3)}(E_1,E_2,E_3),
\label{eq:sectoransatz3}
\end{equation}
with $\rho_s^{(3)}$ the level-triple density of the eigenstate-triple
sector $s$.  At $k=2$ the sixteen triple patterns
$(m_{12},m_{13},m_{23},t)$ of the replica hierarchy
\cite{MResR} group, for a coincident channel triple, into eight
$S_3$ classes with pair counts
$(91,2184,6552,18018,24024,72072,270270,360360)$ (ascending), summing to $D^3$,
with the same census reproduced by the explicit $S_3$ orbit
enumeration of Remark~\ref{rem:invariant},
and the moment dictionary supplies the eight equations
\begin{equation}
\sum_s\theta_s N_s=0,\qquad
\sum_s\theta_s L_{pqr}(s)=\kappa_3\bigl[s_p(a),s_q(a),s_r(a)\bigr]
\label{eq:sectorsolve3}
\end{equation}
for the even-degree set $(p,q,r)\in\{(2,1,1),(2,2,2),
\allowbreak(3,2,1),(4,1,1),\allowbreak(3,3,2),(4,2,2),(5,2,1)\}$, with
$L_{pqr}(s)=\sum_{(n,m,l)\in s}\lambda_n^p\lambda_m^q
\lambda_l^r$; the row-normalized $8\times8$ system is again
Hamiltonian-only.  Its predictions---the higher even moments and the
three-resolvent cumulant $\cC^{(3)}$---are verified against the exact
Weingarten sector table of Ref.~\cite{MResR} in
Sec.~\ref{sec:num}.  The parity structure of the same system is the
skewness content: the parity-mixing sector of $\cC^{(3)}$ is its
odd-monomial moment content $\cM_{pqr}$ with $p+q+r$ odd, and
$\kappa_3[s_p,s_q,s_r]=0$ \emph{exactly} for $p+q+r$ odd whenever the
Hamiltonian is reflection-symmetric in the ensemble sense ($H$ and
$-H$ equally weighted, each $s_j$ carrying parity $(-1)^j$)---the
moment form of the selection rule of Sec.~\ref{sec:bench}: the
symmetric Gaussian benchmark carries no parity-mixing content at any
degree, and in a parity-mixing system the odd-degree
$\kappa_3[s_p,s_q,s_r]$, with $\cM_{511}$-type the leading entries,
supply the odd-sector interference
$\sum V^{(3)}\mathrm{Im}(\cF\cF)_{\rm odd}$ of
Ref.~\cite{SRletter}.

\emph{The census as an orbit count.}  The entries quoted above are not
table lookups: at $k=2$ the class of an ordered $r$-tuple of two-mode
sets is fixed by the multiset of its site masks
$S(s)=\{i : s\in n_i\}$ up to the action of $S_r$, every index
appearing in exactly two sets, and a class whose masks $\{S\}$ occur
$m_S$ times carries $N!\,/\,[(N-M)!\prod_S m_S!]$ tuples with
$M=\sum_S m_S$.  Enumerating the $S_r$ orbits of these multisets
reproduces the $r=2$ and $r=3$ rows quoted above and extends the count
to arbitrary $r$:
\begin{table}[h]
\centering
\small
\begin{ruledtabular}
\begin{tabular}{cccc}
$r$ & classes & largest class & $\sum_{\mathfrak c}N_{\mathfrak c}=D^{r}$\\
\hline
1 & 1 & 91 & 91\\
2 & 3 & 6006 & 8281\\
3 & 8 & 270\,270 & 753\,571\\
4 & 23 & 7\,567\,560 & 68\,574\,961\\
\hline\hline
\end{tabular}
\end{ruledtabular}
\caption{Number of sector classes and the largest class at
$k=2$, $N=14$, for $r=1,\dots,4$, from the $S_r$-orbit enumeration of
the site-mask multisets.  The total always closes on $D^{r}=91^{r}$.}
\label{tab:censusr}
\end{table}
The class count grows as $1,3,8,23$ for $r=1,\dots,4$---faster than any
low polynomial in $r$ but far slower than the $D^{r}$ tuples
themselves---and the total closes on $D^{r}$ at every $r$.  This is the
combinatorial content of the ``for any $r$'' of the architecture: the
sector ansatz of Eq.~\eqref{eq:sectoransatz3} is parametrized by a
census that can be generated, while the number of moment equations
needed to fix it grows with the same count.  The $r=4$ row is what a
quantitative $r=4$ closure would have to invert, and it is the
structural reason why that inversion is both larger and worse
conditioned than the $r=3$ one; it is not attempted here.

\emph{Complexity.}  The construction is finite at fixed
$(r,\ell_{\max},d_{\max})$ \emph{and fixed system size $N$}: the class
census, the seed and the retained moment system are generated by the
rules above.  What is not established is uniform boundedness: neither in
$N$ at fixed truncation, nor in $r$ at fixed $N$.  The class count rises
with $r$ (Table~\ref{tab:censusr}),
the moment system grows with the same count, and nothing here proves
that the census remains of polynomial size in the thermodynamic limit
or that an arbitrary Hamiltonian admits a low-rank or otherwise compact
effective closure.  The genericity claimed in this article is therefore
the structural one---a generic architecture, not a generic
low-complexity basis---and computational scalability is a separate
question, not addressed here.

\subsection{The connectivity basis}
\label{sec:conn}

The upgrade route of Sec.~\ref{sec:intro} left one step
benchmark-specific.  The chain underlying the construction runs from
$H$ through the channel moments, the coupling products $V^{(\ell)}$
and their cumulants to the seed $\Gamma_0^{(r)}[H]$, then through the
causal vertex $\Gamma^{(r)}[\boldsymbol\theta]$ and the ladder to
$\cC^{(r)}[\boldsymbol\theta]$ and $S^{(r)}[\boldsymbol\theta]$, and
finally to the fixed point $\boldsymbol\theta_*$; everything to the
left of the seed belongs to the microscopic layer.
Sections~\ref{sec:causal} and \ref{sec:ladder} supply the two
exact structures---the causal representation and the vertex
functional---together with the closing condition of
Sec.~\ref{sec:system}, and Sec.~\ref{sec:ansatz} supplied the
benchmark ansatz.
This section removes the remaining benchmark-specific step: it
defines, for a generic $H=H_0+V$, the connectivity basis on which the
seed Eq.~\eqref{eq:seed} and the ansatz live, and shows that the
free-fermion overlap sectors are one exactly solvable realization of
it---the upgrade from a random-free-fermion benchmark to a generic
microscopic layer.

\emph{Virtual paths.}  In the DCA form Eq.~\eqref{eq:Gell} each
self-energy channel is a sum over coupling-leg products, so a
\emph{virtual path} of channel $a$ is a tuple
$w=(a,\alpha_1,\dots,\alpha_{\ell-1})$ of distinct channels,
$\alpha_j\neq a$, together with its coupling monomial $\mu(w)$---the
multiset of interaction entries of
$V^{(\ell)}_{a,\alpha_1,\dots,\alpha_{\ell-1}}$---and its fluctuating
bare factor
\begin{equation}
\phi_w(z)=\delta V^{(\ell)}_{a,\alpha_1,\dots,\alpha_{\ell-1}}
\prod_{j=1}^{\ell-1}\hat{\cR}_{\alpha_j}(z),
\label{eq:phiw}
\end{equation}
with $\hat{\cR}_{\alpha}=\E_H[\cR_{\alpha}]$ the deterministic mean
resolvent.  In path notation the seed of Theorem~\ref{thm:vertex}
reads exactly
$\Gamma_0^{(r)}(\boldsymbol z)=\sum_{w_1,\dots,w_r}
\kappa_r\bigl(\phi_{w_1}(z_1),\dots,\phi_{w_r}(z_r)\bigr)$.

\emph{Connectivity classes.}  Two $r$-tuples $(w_1,\dots,w_r)$ and
$(w'_1,\dots,w'_r)$ belong to the same \emph{connectivity class}
$\mathfrak c$ when they have the same endpoints and leg counts
$\ell_i$ and the same shared-entry pattern: for every subset
$S\subseteq[r]$, the number of coupling entries common to exactly the
monomials $\{\mu(w_i):i\in S\}$ coincides.  A class is the relabeling
orbit of the coupling-entry intersection structure---the $r$-path
analogue of a diagram topology, the shared entries playing the role
of contracted lines.  At $r=2$ the leading classes are the direct,
one-shared-entry, two-shared-entry, $\ldots$ classes; at $r=3$ the
pair-sharing, three-way-sharing, chain and triangle classes.  The
classification is permutation invariant by construction and is an
exact reorganization: no approximation beyond the DCA.

\emph{Class basis and class expansion.}  Define the connectivity
basis kernels
\begin{equation}
\cB_{\mathfrak c}^{(r)}(\boldsymbol z)
=\sum_{(w_1,\dots,w_r)\in\mathfrak c}
\;\prod_{i=1}^{r}\prod_{\alpha\in w_i}\hat{\cR}_\alpha(z_i),
\label{eq:Bclass}
\end{equation}
the bare-propagation content of the class.  Since the propagator
factors of Eq.~\eqref{eq:phiw} are deterministic, each cumulant in
the path sum factorizes as
$\kappa_r(\phi_{w_1},\dots,\phi_{w_r})
=\bigl[\prod_{i,\alpha}\hat{\cR}_\alpha(z_i)\bigr]
\,\kappa_r(\delta V^{(\ell_1)},\dots,\delta V^{(\ell_r)})$, with
$\hat\kappa_{\mathfrak c}(w_1,\dots,w_r)
:=\kappa_r(\delta V^{(\ell_1)}_{\mu(w_1)},\dots,
\delta V^{(\ell_r)}_{\mu(w_r)})$ the coupling cumulant of the path
tuple, and the seed reorganizes as follows.

\begin{theorem}[Connectivity expansion of the seed]
\label{thm:conn}
Within the DCA, the microscopic seed admits the exact class expansion
\begin{equation}
\Gamma_0^{(r)}(\boldsymbol z)
=\sum_{\mathfrak c}\;\sum_{(w_1,\dots,w_r)\in\mathfrak c}
\hat\kappa_{\mathfrak c}(w_1,\dots,w_r)\,
\prod_{i=1}^{r}\prod_{\alpha\in w_i}\hat{\cR}_\alpha(z_i),
\label{eq:connexp}
\end{equation}
and in the structured Gaussian class the coupling cumulant depends on
the path tuple only through its class, so the expansion factorizes,
\begin{equation}
\Gamma_0^{(r)}(\boldsymbol z)
=\sum_{\mathfrak c}\Theta_{\mathfrak c}^{(r)}\,
\cB_{\mathfrak c}^{(r)}(\boldsymbol z),
\qquad
\Theta_{\mathfrak c}^{(r)}=\hat\kappa_{\mathfrak c}\big|_{\mathfrak c},
\label{eq:classseed}
\end{equation}
with $\Theta_{\mathfrak c}^{(r)}$ the class amplitude: the seed is a
linear combination of connectivity basis kernels with coefficients
fixed by the coupling cumulants alone.
\end{theorem}

\begin{proof}
Partition the index set of Eq.~\eqref{eq:seed} into relabeling
orbits.  The factorization uses that the $\hat{\cR}$ factors are
deterministic, and the class-uniformity of the coupling cumulant uses
exchangeability of the independent coupling entries.
\end{proof}

The classes are ordered in the deep plane by total leg count
$\sum_i(\ell_i-1)$---the number of bare propagator factors---so the
minimal-connectivity classes dominate, the seed-side counterpart of
the chain bound of Sec.~\ref{sec:causal}.  The class basis satisfies
the requirements of a generic coordinate system: permutation
invariance (relabeling orbits), symmetry compatibility
(Proposition~\ref{prop:parity}), finiteness at every truncation
(Proposition~\ref{prop:finite}), and the free-fermion sectors as
special case (Theorem~\ref{thm:ffred}).

\begin{proposition}[Coupling-degree selection rule]
\label{prop:parity}
For \emph{Gaussian} centered couplings with independent entries,
every class whose total coupling-entry count
$\sum_i\deg\mu(w_i)$ is odd carries zero amplitude,
$\Theta_{\mathfrak c}^{(r)}=0$: every Wick pairing consumes two
entries, and no pairing of an odd total exists.  The rule is the
coupling-side counterpart of the reflection parity of
Sec.~\ref{sec:sectorclosure3}: the odd total entry count is the
microscopic obstruction to pairing off the coupling fluctuations, and
it is the mechanism behind the identically vanishing odd-degree
moment content $\kappa_3[s_p,s_q,s_r]=0$ ($p+q+r$ odd) of the
benchmark.  The scope is genuinely Gaussian: centering alone does
not suffice, since a centered coupling with nonzero third cumulant
$\kappa_3(V)\ne0$ supports odd-degree classes, and for non-Gaussian
entries the class bookkeeping survives unchanged---the
pair-contraction \emph{graph} of the Gaussian case is promoted to a
\emph{hypergraph} of coupling cumulants.  What the promotion does and
does not move is settled in Proposition~\ref{prop:nonGauss} below.
\end{proposition}

\begin{proposition}[Selective promotion of the class amplitudes]
\label{prop:nonGauss}
For independent centered couplings with cumulants $\kappa_p$, the class
labels of this section are unchanged---they are multiplicity partitions
of the coupling slots---and each amplitude is promoted from its Gaussian
value to the cumulant functional obtained by admitting blocks of any
size $p\ge2$ in the partition of its slots.  At $r=2$ the promotion is
selective in the degree of the class weight:
(i) a class whose weight is a \emph{product} of coupling entries (the
square that builds a resolvent seed) carries a cumulant, and its
amplitude moves from the Gaussian $2=\kappa_2(\chi^2_1)$ to
\begin{equation}
\Theta^{(2)}_{\rm seeded}=\kappa_2(V^2)=\kappa_4+2\kappa_2^2,
\qquad
3\ \longrightarrow\ 3+\frac{\kappa_4}{\kappa_2^2},
\label{eq:nonGaussamplitude}
\end{equation}
where the entry $3$ in the last form is the Gaussian Wick count
$E[g^4]/E[g^2]^2$, and the first form is its general-cumulant version;
(ii) a class whose weight is \emph{linear} in the couplings is built
from products of $\kappa_2$ alone and therefore does not move: the bare
flip census of Sec.~\ref{sec:bench} (its $2\lambda^2$ and $\lambda^2$
entries) is statistics-independent, and the promotion acts on the
resolvent-seeded amplitudes rather than on the census.  Odd-degree
classes, identically zero in the Gaussian case, carry amplitudes
$\propto\kappa_3$ and higher when the odd cumulants are nonzero.
\end{proposition}

Both statements of Proposition~\ref{prop:nonGauss} are verified at
$r=2$ on two test beds, with the coupling laws normalized to a common
$\kappa_2$ and spanning $\kappa_4/\kappa_2^2=-2,\,-1.75,\,0,\,+1.80$
(Gaussian, a two-point magnitude law, a heavy-tailed magnitude law, and
$\pm1$ bonds).  On the free-fermion channel benchmark of
Sec.~\ref{sec:bench}: the seed block of the
explicit $r=2$ vertex functional tracks
Eq.~\eqref{eq:nonGaussamplitude} across a factor $16$ in amplitude
($1.89$--$2.02$ for the Gaussian law, $0.21$--$0.25$ for
$\kappa_4/\kappa_2^2=-1.75$, $3.06$--$4.47$ for $+1.80$), and vanishes
identically for $\pm1$ bonds, where $V^2$ is deterministic and the
prediction is zero; the dressing coefficient, by construction a class
of the second kind, stays at unity for every law.  On the ring TFIM
(Table~\ref{tab:e2nonGauss}) the coincident (return-path) class follows
$\kappa_4+2\kappa_2^2$ to the percent level and the distinct
(two-flip) class stays at its Gaussian $\lambda^4$---$1.000$ for
$\pm1$ bonds, where the coincident class is exactly zero.  The two
beds together locate the promotion: it is local, it is fixed by the
degree of the class weight, and it leaves the census amplitudes of
Sec.~\ref{sec:bench} untouched at $r=2$.

\begin{table}[t]
\small
\begin{ruledtabular}
\begin{tabular}{l c c c c}
law & $\kappa_4/\kappa_2^2+2$ & seeded & coincident & distinct\\
\hline
Gaussian & $2$ & $1.89$--$2.02$ & $1.995$ & $0.993$\\
two-point & $0.25$ & $0.21$--$0.25$ & $0.250$ & $1.001$\\
heavy-tailed & $3.80$ & $3.06$--$4.47$ & $3.809$ & $0.988$\\
$\pm1$ bonds & $0$ & $0$ (exact) & $0$ (exact) & $1.000$\\
\end{tabular}
\end{ruledtabular}
\caption{Non-Gaussian promotion of the $r=2$ class amplitudes,
Eq.~\eqref{eq:nonGaussamplitude}.  The seeded column is
$\Theta^{(2)}_{\rm seeded}$ measured as the ratio of the exact seed
block to its shared-edge kernel on the free-fermion channel benchmark
(range over the three probes at the representative $\ell=2$ channel
pair).  The last two columns are the ring TFIM of
Sec.~\ref{sec:genbench} at $N=9$, $\lambda=2^{-3}$: the coincident
(return-path) class measured as
$\mathrm{Var}(R_{\rm ret})/(\lambda^2\sum_k|w_k|^2)$ against the
prediction $\kappa_4+2\kappa_2^2$, and the distinct (two-flip) class as
$\mathrm{Var}(R_2)/|w_i+w_j|^2$ against $\lambda^4$; the $\pm1$ row is
an exact degeneracy ($V^2$ deterministic).  All laws share
$E[g^2]=\lambda$.}
\label{tab:e2nonGauss}
\end{table}

\begin{proposition}[Finite truncation]
\label{prop:finite}
Truncating to paths with at most $\ell_{\max}$ legs and to classes
with at most $d_{\max}$ shared entries leaves finitely many classes;
for every fixed $(r,\ell_{\max},d_{\max})$ their census---the number
of path tuples per class---is finite and algorithmically
constructible as a walk-counting problem on the coupling graph of
$H$, with no eigenstate, replica, or ensemble-algebra input.  A
complexity statement beyond constructibility requires graph
structure: on sparse coupling graphs the census is polynomial in the
graph size, while a dense graph supports exponentially many paths and
the census cost is set by the counting problem itself.
\end{proposition}

\begin{theorem}[Free-fermion reduction]
\label{thm:ffred}
For the random free-fermion benchmark the connectivity classes
collapse onto the overlap sectors.  A channel's coupling monomials
are its one-body moves: the monomial $\mu(w)$ of a two-leg path of
channel $a$ is the move pair $\{e,e^*\}$ with $e\in E_a$, and two
paths share entries exactly when their moves coincide.  The class
census of a channel pair $(a,b)$ is therefore the shared-move count
$e_{ab}=|E_a\cap E_b|$, a bijective function of the eigenstate
overlap $|A_a\cap A_b|$, and the class amplitudes reduce to the
chi-square cumulants $\Theta_{\mathfrak c}^{(2)}
=\kappa_2(\delta|h_e|^2,\delta|h_e|^2)=2$ and
$\Theta_{\mathfrak c}^{(3)}=8$.  The class expansion
Eq.~\eqref{eq:classseed} then evaluates to the deep-plane seeds
Eqs.~\eqref{eq:G2seed}--\eqref{eq:G3seed}, and the sector ansatz
Eq.~\eqref{eq:sectoransatz} is the class ansatz with $\theta_m$ the
class amplitudes.  The benchmark is the exactly solvable realization
of the generic basis---not its definition: connectivity is the
microscopic coordinate system of higher-resolvent fluctuations, and
random free fermions its calibration laboratory.
\end{theorem}

\emph{The generic constrained ansatz.}  With the class basis in hand
the constrained ansatz of Sec.~\ref{sec:ansatz} becomes generic:
\begin{equation}
\boxed{\;
\Gamma^{(r)}(\boldsymbol z)
=\Gamma_0^{(r)}(\boldsymbol z)
+\sum_{\mathfrak c,\alpha}\theta_{\mathfrak c,\alpha}^{(r)}\,
\Phi_{\mathfrak c,\alpha}^{(r)}(\boldsymbol z)\;}
\label{eq:genericansatz}
\end{equation}
with the basis functions subject to five constraints.  (i)
\emph{Causality}: each $\Phi$ is an $r$-fold Stieltjes transform of a
signed measure---analytic in the upper half-plane in every variable
[Eq.~\eqref{eq:stieltjes}].  (ii) \emph{Permutation symmetry}:
$\Phi(z_{\pi(1)},\dots,z_{\pi(r)})=\Phi(z_1,\dots,z_r)$.  (iii)
\emph{Deep-plane asymptotics}: $\Phi$ is ordered by its leg-count
class, and the assembled cumulant respects the degree bound
$\deg_t\cC^{(r)}\ge3r-2$ of Sec.~\ref{sec:causal} through the ladder
dressing Eq.~\eqref{eq:leadladder}.  (iv) \emph{Microscopic moment
matching}: Eq.~\eqref{eq:mommatch} of Sec.~\ref{sec:closure}.  (v) \emph{Vertex-functional
consistency}: Eq.~\eqref{eq:MRSCeq} with the residual
Eq.~\eqref{eq:residual}, likewise of Sec.~\ref{sec:closure}.  Constraints (i)--(iii) define the
admissible basis---the intersection of the causal manifold with the
sum-rule hypersurfaces---and (iv)--(v) fix the amplitudes; the sector
ansatz Eq.~\eqref{eq:sectoransatz} is the benchmark instance with
$\Phi_m=\rho_m^{(2)}$, the generic instance takes
$\Phi_{\mathfrak c,\alpha}=\cB_{\mathfrak c}^{(r)}$ with the
amplitudes dressed by the class-resolved kernels of
Theorem~\ref{thm:vertex}---the same class census indexes the seed and
the dressing term by term.

\emph{The MRSC map, decomposed.}  With the microscopic layer made
explicit, the map of Eq.~\eqref{eq:fixedpoint} decomposes by content,
\begin{equation}
\boldsymbol\theta_*=F_r[H;\boldsymbol\theta_*],
\qquad
F_r=F_r^{\mathrm{micro}}+F_r^{\mathrm{spectral}}+F_r^{\mathrm{causal}},
\label{eq:Fdecomp}
\end{equation}
$F^{\mathrm{micro}}$: $H\mapsto\{s_m(a),V^{(\ell)},\kappa[V],
\cR_{\rm tr}\}$---the moment fields Eq.~\eqref{eq:momfield}, the
coupling data and their cumulants, and the trace resolvent---fixing
the seed Eq.~\eqref{eq:classseed} and the moment rows; this is the
model-specific layer.  $F^{\mathrm{spectral}}$:
$\boldsymbol\theta\mapsto S[\boldsymbol\theta]\mapsto
\cC[\boldsymbol\theta]\mapsto\Gamma[\boldsymbol\theta]$---the ladder,
inversion and vertex functional of
Theorems~\ref{thm:ladder}--\ref{thm:vertex}---structural and
universal.  $F^{\mathrm{causal}}$: the projection of the dressed
vertex onto the class basis $\{\cB_{\mathfrak c}^{(r)}\}$ under
constraints (i)--(iii)---the Stieltjes manifold.  The fixed point
coincides on all three layers [Eq.~\eqref{eq:fixedpoint}], and the
verification of Sec.~\ref{sec:num} supplies one gate per layer:
$\chi_{\mathrm{cl}}$ for stability, the Hilbert dictionary for
causality, and the moment--Feshbach residuals for microscopic
content---three independent tests, one per layer, in the sense made
precise in Sec.~\ref{sec:disc}.

The class expansion also makes the deformation content of
Sec.~\ref{sec:num} explicit.  For the interaction-deformed family the
$h$-legs and the $W$-legs of the paths form distinct classes: the
$W$-monomial classes carry the $\lambda$-dependence of the amplitudes
and of the dressing kernels, so the same marginal moment dictionary
that is measured exactly $\lambda$-invariant
(Sec.~\ref{sec:num}) re-shuffles among classes under deformation---the
microscopic statement behind the non-analytic sector covariance there:
moments encode local spectral information, connectivity encodes how
intermediate propagation channels correlate, and the higher
resolvents are sensitive precisely to the latter.  This is the sense
in which Eq.~\eqref{eq:Fdecomp} answers the question posed in
Sec.~\ref{sec:intro}: the hierarchy reduces to finite causal
parameters \emph{on the class basis}---the generic microscopic layer
of the closure.

\begin{remark}[Symmetry-invariant interpretation of the connectivity basis]
\label{rem:invariant}
The connectivity basis admits a complementary invariant-theoretic
reading.  Whenever the microscopic ensemble carries a
group-invariant measure, averaging over the group projects the
relevant index contractions onto the corresponding invariant-tensor
space: the connectivity classes are the orbit classes of microscopic
contraction patterns under this action, and the amplitudes
$\Theta_{\mathfrak c}^{(r)}$ of Eq.~\eqref{eq:classseed} are
coordinates on the resulting invariant subspace
$\mathcal V_r=\operatorname{span}\{\cB_{\mathfrak c}^{(r)}\}$.  The
construction thereby separates two logically distinct ingredients:
the invariant structure determines the admissible microscopic
coordinate space, while the Feshbach dressing of
Secs.~\ref{sec:ladder}--\ref{sec:vertex} determines the
self-consistent dynamics within that space---group integration
determines the coordinates, Feshbach dressing determines their
dynamics.
Closure at order $r$ is finite precisely because the invariant
coordinate space is finite at every truncation
(Proposition~\ref{prop:finite}) and the constrained ansatz
Eq.~\eqref{eq:genericansatz} keeps the dressed vertex inside
$\mathcal V_r$, so dressing acts within the finite coordinate space:
the effective number of amplitudes is
$N_{\mathrm{eff}}^{(r)}=\dim\mathcal V_r$---three at $r=2$ and eight
at $r=3$, the sector counts of
Secs.~\ref{sec:sectorclosure}--\ref{sec:sectorclosure3}.
\end{remark}

The remaining subsections make the two ingredients explicit on the
benchmark and record where each one is measured.  For the benchmark
the projection is realized explicitly by the
orthogonal Weingarten calculus
\cite{Weingarten,CM2009,CollinsSniady}: the Haar average of $2d$
matrix elements of $g\in O(N)$ is
\begin{multline}
\int\!\prod_{s=1}^{2d}g_{i_sj_s}\,\mathrm dg
=\sum_{\mathfrak m,\mathfrak n\in M_{2d}}
\mathrm{Wg}(\mathfrak m,\mathfrak n)\\
\times\prod_{s=1}^{d}\delta_{i_{\mathfrak m(2s-1)},i_{\mathfrak m(2s)}}\,
\delta_{j_{\mathfrak n(2s-1)},j_{\mathfrak n(2s)}},
\label{eq:wein}
\end{multline}
where $M_{2d}$ is the set of pair partitions of $\{1,\dots,2d\}$ and
$\mathrm{Wg}$ is the inverse of the Gram matrix
$G(\mathfrak m,\mathfrak n)=N^{\#(\mathfrak m\cup\mathfrak n)}$
counting the connected components of the union graph---a
$105\times105$ system at $d=4$, constant on the double-coset classes
of the hyperoctahedral subgroup, five classes at $d=4$---the same
reduction that contracts the $d=6$ system of Ref.~\cite{MResR} to
eleven classes.  At $d=2$ the inversion is elementary,
$\mathrm{Wg}(\mathrm{id})=(N+1)/[N(N-1)(N+2)]$ and
$\mathrm{Wg}(t)=-1/[N(N-1)(N+2)]$,
reproducing $\E[O_{11}^4]=3/[N(N+2)]$ and the general second moment
$\E[O_{11}^2O_{12}^2]=1/[N(N+2)]$.

Evaluating the $r=2$ sector average through Eq.~\eqref{eq:wein} gives,
for $p_a=D|\det U[A_a,C_a]|^2$ with $D=\binom{N}{k}=91$
(here $N=14$, $k=2$) and
$E[p_a]=1$, the exact sector moment matrix
\begin{equation}
E[p_ap_b]=
\begin{pmatrix}
91/20 & 91/60 & 91/120\\
91/60 & 3731/2880 & 637/720\\
91/120 & 637/720 & 8281/7920
\end{pmatrix},
\label{eq:sectorexact}
\end{equation}
rows $m=|A_a\cap A_b|$ and columns $\ell=|C_a\cap C_b|$ running over
$2,1,0$.  The coincident entry
$E[p_a^2]=q_2=\binom{k+2}{2}\binom{N-k+2}{2}/\binom{N+2}{2}=91/20$
reproduces the fourth moment of the $2\times2$ minor, the row
marginals satisfy $\sum_\ell N_\ell\,\tilde E_{m\ell}=1/D$, where
$N_\ell=\binom{k}{\ell}\binom{N-k}{k-\ell}=(66,24,1)$ counts the
column classes $\ell=0,1,2$ at a fixed reference row set and
$\tilde E_{m\ell}=D^{-2}E[p_ap_b]$ is the unnormalized minor product
(Cauchy--Binet: the column minors of one orthonormal row square-sum
to one), and the full sum is
$\sum_{m,\ell}N_mN_\ell\,\tilde E_{m\ell}=1$; direct Haar Monte Carlo
reproduces the table within $4\%$ per cell, the rare coincident cell
within $0.1\%$ under targeted sampling.  These are the coordinates
the group integral supplies: the sector closure of
Sec.~\ref{sec:sectorclosure}---the amplitudes $\theta_m$ of
Eq.~\eqref{eq:sectorsolve}---is the finite self-consistent dynamics
on them, and the Hamiltonian-only closure of Sec.~\ref{sec:num}
reproduces these coordinates at the percent level without eigenstate
data.

At $r=3$ the group structure organizes the channel triples: the
sixteen realizable patterns $(m_{12},m_{13},m_{23},t)$ group into
eight $S_3$ orbit classes with the census
$(91,2184,6552,18018,24024,72072,270270,360360)$ of
Sec.~\ref{sec:sectorclosure3}, summing to $D^3$.  The symmetry
cascade is three-tiered: $O(N)$ fixes the invariant-tensor space,
channel relabeling $S_3$ fixes the orbit census, and the reflection
$\bb Z_2$ ($H\to-H$) fixes the selection rule of
Proposition~\ref{prop:parity}---$s_m\to(-1)^m s_m$ and
$\kappa_3[s_p,s_q,s_r]\to(-1)^{p+q+r}\kappa_3[s_p,s_q,s_r]$, so an
invariant measure annihilates every odd class exactly.

This reading also locates the failure of the $\omega$-only baseline
of Sec.~\ref{sec:num}: $\omega$ is a spectral invariant, while the
classes $\mathfrak c$ are microscopic orbit invariants, and energy
variables alone do not span the invariant information generated by
microscopic averaging.  Conversely, the Weingarten algebra is not the
source of the closure---it is \emph{one realization} of the
microscopic invariant layer, and for ensembles without a Haar
structure (the generic Ising benchmark of Sec.~\ref{sec:genbench})
that layer is supplied by the coupling-graph census of
Proposition~\ref{prop:finite}.  What generalizes from the
replica-sector benchmark is therefore not the Weingarten algebra but
the mapping from microscopic invariant data to effective vertex
parameters \cite{MResR}.  In short, the ansatz of this section is
the intersection of the causal manifold with the sum-rule
hypersurfaces of Sec.~\ref{sec:causal}, parameterized by the
invariant coordinates above; the amplitudes left free by the
constraints are the unknowns that the self-consistency loop of
Sec.~\ref{sec:closure} determines, and the numerical content of
every term quoted in this section is verified in
Sec.~\ref{sec:num} and Appendix~\ref{app:cum}.

\section{Parameter self-consistency and error propagation}
\label{sec:closure}

\emph{Status.}  Level D of Sec.~\ref{sec:levels}: the numerical
self-consistency loop; the statements of this section concern the
fixed point of the projected loop, not the exact hierarchy.

The closure is the requirement that the ansatz be stable under the
MRSC system, imposed through two dual constraint layers: the local
constraints (moment matching) and the global constraint (Feshbach
self-consistency) must select the same parameters.  Define the loop map
$\boldsymbol\theta\mapsto S[\boldsymbol\theta]\mapsto
\cC[\boldsymbol\theta]$ (Stieltjes transform, manifestly causal)
$\mapsto\Gamma[\boldsymbol\theta]$ (exact inversion
Eq.~\eqref{eq:inversion} and the ladder
Theorem~\ref{thm:ladder}), and impose the vertex-functional
consistency Eq.~\eqref{eq:sys4} on the ansatz manifold:
\begin{equation}
\Gamma[\boldsymbol\theta](\boldsymbol z)
=\Gamma_0^{(r)}(\boldsymbol z)
+\sum_{s=2}^{r_c}\cD^{(r,s)}\bigl[\cC^{(s)}[\boldsymbol\theta]\bigr],
\label{eq:MRSCeq}
\end{equation}
together with the moment matching
\begin{equation}
\cM_{\boldsymbol m}[\boldsymbol\theta]
=\kappa_r\bigl[s_{m_1}(a_1),\dots,s_{m_r}(a_r)\bigr],
\qquad
\boldsymbol m\in\cI_r.
\label{eq:mommatch}
\end{equation}
Equations \eqref{eq:MRSCeq}--\eqref{eq:mommatch} are the
multi-resolvent self-consistency equations.  The moment matching
alone defines the diagnostic moment branch
$\boldsymbol\theta_M=\cM^{-1}[K]$; the Feshbach loop alone defines
the diagnostic loop branch $\boldsymbol\theta_{\rm loop}$ through
Eq.~\eqref{eq:MRSCeq}.  In the exact hierarchy the moment and
Feshbach constraints are simultaneously satisfied,
$\boldsymbol\theta_{\rm exact}=\boldsymbol\theta_M
=\boldsymbol\theta_{\rm loop}$; at finite closure level the two
branches generally disagree,
$\boldsymbol\theta_M^{(L)}\ne\boldsymbol\theta_{\rm loop}^{(L)}$,
and the predictive MRSC solution is the fixed point
\begin{equation}
\boxed{\;\boldsymbol\theta_*^{(L)}
=F_L\bigl[\boldsymbol\theta_*^{(L)};H\bigr],\;}
\label{eq:fixedpoint}
\end{equation}
of the combined map $F_L$---the finite-closure approximation to the
intersection point of the two constraint manifolds of the exact
hierarchy---whose residual
\begin{equation}
\cE_{\rm MRSC}(\boldsymbol\theta)
=\Bigl\|\,\Gamma[\boldsymbol\theta](\boldsymbol z)
-\Gamma_0^{(r)}(\boldsymbol z)
-\sum_{s=2}^{r_c}\cD^{(r,s)}\bigl[\cC^{(s)}[\boldsymbol\theta]\bigr]
\Bigr\|,
\label{eq:residual}
\end{equation}
measures the self-consistency deficit directly, and whose
microscopic input is represented by (i) the channel moment fields
$s_m(a)=\delta(H^m)_{aa}$, (ii) the coupling data
$V^{(\ell)}$ and their cumulants, and (iii) the trace resolvent
$\cR_{\rm tr}$.  The iteration $\boldsymbol\theta_{n+1}=
F(\boldsymbol\theta_n)$ is seeded in the deep plane, where every
vertex is known asymptotically [Eqs.~\eqref{eq:G2seed}--
\eqref{eq:G3seed}], and converges whenever the ladder kernel of
Eq.~\eqref{eq:sys3} has spectral radius below one at the chosen
broadening; the truncation error is ordered by the deep-plane
degree $3r-2$.

The structure of Eqs.~\eqref{eq:fixedpoint}--\eqref{eq:residual}
answers the question posed in Sec.~\ref{sec:intro} in the following
conditional sense: the exact microscopic multi-resolvent hierarchy
reduces to a finite set of causal, self-consistent parameters
\emph{whenever} the vertex ansatz is closed under the dressing of
Theorems~\ref{thm:ladder}--\ref{thm:vertex} up to the moment set
$\cI_r$---a condition that is checked, not assumed, at each rung
through the residual Eq.~\eqref{eq:residual}.  Section~\ref{sec:num}
carries this out at $r=2$ on the free-fermion benchmark: the
moment branch reproduces the exact sector covariances at the percent
level, the loop branch is verified as an exact identity at the
measured point with its noise floor quantified, and the loop
residual is shown to detect the interaction-deformed content the
moment branch is blind to (Sec.~\ref{sec:loop}).  The
single-resolvent closure is the $r=1$ corner of the same
equation, which is why the framework extends the causal
self-energy program of Ref.~\cite{SRletter} rather than
replacing it.

\subsection{The $r=2$ Feshbach-loop fixed point}
\label{sec:loop}

The moment matching and the Feshbach loop are two distinct branches
of the closure, and their roles are different: the moment branch
$\boldsymbol\theta_M=\cM^{-1}[K]$ fixes the deep-plane (local)
content, while the loop branch fixes the bulk (global) consistency.
The loop equation is the ladder of Theorem~\ref{thm:ladder} at
$r=2$ with the vertex functional of Eq.~\eqref{eq:r2functional}
inserted: for the coincident pair, with
$\langle\bar{\cR}^4\rangle=\E[(\bar{\cR}_a^0)^2(\bar{\cR}_b^0)^2]$
and the measured mean-field content absorbed into
$\Gamma_{\rm MF}^{(2)}[\cC^{(2)}]$, the exact decomposition
$\cC^{(2)}=\kappa_2(X_a,X_b)+\kappa_2(X_a,\Delta_b)+\kappa_2(\Delta_a,X_b)
+\E[\Delta_a\Delta_b]$ becomes the fixed-point equation
\begin{multline}
\sum_m\theta_m^{(2)}\,\cG2_m(z_1,z_2)
=X_{\rm tower}(z_1,z_2)\\
+\langle\bar{\cR}^4\rangle\Bigl[\Gamma_0^{(2)}
+\Gamma_{\mathrm{CC}}^{(2)}+\Gamma_{\rm MF}^{(2)}\Bigr]
+\langle\bar{\cR}^4\rangle\sum_{m,\ell'}
c_{m\ell'}\,\theta_m^{(\ell')}\,\cG2_m(z_1,z_2)\\
+T_{\rm ins}(z_1,z_2),
\label{eq:loopeq}
\end{multline}
with $c_{m\ell'}$ the combinatorial census of
Eq.~\eqref{eq:census}, $X_{\rm tower}$ the measured mean-field tower,
$T_{\rm ins}$ the higher-insertion correction
$\E[\Delta_a\Delta_b]-\E[(\bar{\cR}_a^0\bar{\cR}_b^0)^2\delta\cG_a
\delta\cG_b]$, and the seed, cross-correlated and mean-field content
grouped in the bracket.  Equation \eqref{eq:loopeq} is written down
for \emph{each} of the three channel-overlap representatives
$\ell=0,1,2$ (its left side carries the corresponding
$\theta_m^{(\ell)}$, its census the corresponding overlap pattern of
Eq.~\eqref{eq:census}); evaluated at the probe points and joined
with the moment rows Eq.~\eqref{eq:sectorsolve} for each $\ell$, it
is a linear least-squares system for the nine amplitudes
$\{\theta_m^{(\ell)}\}$---the combined objective
$\min_{\boldsymbol\theta}[\|\text{loop residual}\|^2
+w\,\|\text{moment residual}\|^2]$ of the two branches, with $w=1$
throughout (the two blocks stacked at unit weight; the limits
$w\to0$ and $w\to\infty$ reproduce $\boldsymbol\theta_M$ and
$\boldsymbol\theta_{\rm loop}$).  The loop
block's rank is structural: on the sector-constant ansatz the loop
operator acts through the sector densities $\cG2_m$ and the three
census patterns, so its rank is bounded by the product of the two
profile spans---measured below as six independent directions.  The
solution $\boldsymbol\theta_*$ is the self-consistent point: the
moment branch alone is $\boldsymbol\theta_M$, the loop branch alone
(no moment rows) is $\boldsymbol\theta_{\rm loop}$, and the exact
benchmark value $\boldsymbol\theta_{\rm exact}$ anchors the
branch and closure-level comparison of Sec.~\ref{sec:num}.  The loop rows carry
the interacting deformation through $\cG2_m(\lambda)$, the deformed
seeds and the measured feedback content---the mechanism by which
Feshbach feedback restores the higher-resolvent information the
frozen moment rows cannot see, to be verified against
$\boldsymbol\theta_{\rm exact}(\lambda)$ in
Sec.~\ref{sec:num}.

The two branches realize an information-source decomposition of the
closure, the multi-resolvent form of the upgrade route:
\begin{equation}
\begin{aligned}
&\text{Level I:}\ K_m[H]\ \text{fixes the deep-plane constraints;}\\
&\text{Level II:}\ \cC^{(2)}\ \text{carries the bulk correlations;}\\
&\text{Level III:}\ \Gamma^{(2)}[\cC^{(2)}]\ \text{feeds them back.}
\end{aligned}
\label{eq:infodecomp}
\end{equation}
Level I alone is the moment branch; Levels I--III together are the
self-consistent point---the microscopic-to-self-consistent loop
$H\to K\to\boldsymbol\theta\to\cC\to\Gamma\to\boldsymbol\theta$ of
the introduction.  The benchmarks of Sec.~\ref{sec:num} test the
three levels separately: the moment dictionary (I), the causal
representation (II), and the loop equation (III).

\subsection{The $X$--$\Delta$ cross-response kernel}
\label{sec:crossresponse}

The decomposition Eq.~\eqref{eq:XD} splits the $r=2$ covariance
into $\cC^{(2)}=\kappa_2[X,X]+2\kappa_2[X,\Delta]
+\kappa_2[\Delta,\Delta]$, and Sec.~\ref{sec:num} attributes the
dominant residual of the $\lambda=0$ anchor to the middle term: the
DCA closure of the cross response $X\leftrightarrow\Delta$ is the
bottleneck of the map, not the nonlinearity.  The cross response can
be promoted from a closure approximation to an exact linear functional
equation before truncation.
Define the cross-response kernel
\begin{equation}
\Xi_{ab}(z_1,z_2)=\kappa_2\bigl(X_a(z_1),\Delta_b(z_2)\bigr),
\label{eq:XiKernel}
\end{equation}
with $\Xi^{+}_{ab}(z_1,z_2)=\kappa_2(\Delta_a,X_b)$ its reflected
partner (equal to $\Xi_{ab}$ under channel-exchange symmetry, as in
the benchmark).  Substituting the geometric identity
Eq.~\eqref{eq:geom}---$\Delta_b=\sum_{s\ge1}
(\bar{\cR}_b^0)^{s+1}(\delta\cG_b)^s$---with the DCA self-energy form
$\delta\cG_b=\sum_{c\in N(b)}|V_{bc}|^2\delta\cR_c
+\delta\cG_b^{\mathrm{CC}}$ of Eq.~\eqref{eq:Gell}, and expanding
each joint cumulant multilinearly,
$\kappa_2(A,BC)=\E[B]\kappa_2(A,C)+\E[C]\kappa_2(A,B)
+\kappa_3(A,B,C)$ for centered fields, yields the exact linear
functional equation
\begin{equation}
\begin{aligned}
\Xi_{ab}={}&\underbrace{\sum_{c\in N(b)}
\E[W_{bc}]\,\kappa_2(X_a,X_c)}_{\Xi^0_{ab}}
+\underbrace{\sum_{c\in N(b)}
\E[W_{bc}]\,\Xi_{ac}}_{(K_X\Xi)_{ab}}\\
&+\underbrace{\sum_{c\in N(b)}
\kappa_3\bigl(X_a,W_{bc},X_c\bigr)}_{T^{(1)}_{ab}}
+\sum_{c\in N(b)}
\kappa_3\bigl(X_a,W_{bc},\Delta_c\bigr)\\
&+\text{cross-channel}+\text{higher dressing}\;(s\ge2),
\end{aligned}
\label{eq:Xieq}
\end{equation}
with the loop weights
$W_{bc}(z_2)=\bigl(\bar{\cR}_b^0(z_2)\bigr)^2|V_{bc}|^2$.  In
operator form on the channel space, $\Xi=\Xi^0+K_X\Xi+T$, and the
kernel is the resolvent-like object
\begin{equation}
\Xi=(1-K_X)^{-1}\bigl(\Xi^0+T\bigr),
\label{eq:XiResum}
\end{equation}
the channel-space linear-response counterpart of the scalar
mean-field resolvent of Eq.~\eqref{eq:R0bar}.  The inhomogeneity
$\Xi^0$ is the $X$--$X$ covariance transported by the loop weights,
and the correction tower $T$ is the $\kappa_3$-dressed re-entry of
the dressing field ($T^{(1)}$ plus the $\Delta_c$-re-entry), the
cross-correlated channel, and the $s\ge2$ geometric dressing.  For
the benchmark the weights factorize, $\E[W_{bc}]
=\langle(\bar{\cR}_b^0)^2\rangle\,\E[|V_{bc}|^2]$ (couplings
independent of the diagonal content, $\E[|V_{bc}|^2]=1$ per move),
and $T^{(1)}_{ab}=\sum_{c\in N(b)}\kappa_3(X_a,
(\bar{\cR}_b^0)^2,X_c)$ is a diagonal-ensemble triple correlation:
every coefficient of Eq.~\eqref{eq:Xieq} is Hamiltonian-derived
one-resolvent content
(the diagonal distribution and the coupling census), with no
measured higher-resolvent input.

Truncating $T$ at level one defines the closed kernel
\begin{equation}
\Xi^{(1)}=(1-K_X)^{-1}\bigl(\Xi^0+T^{(1)}\bigr),
\qquad
\Xi^{\rm DCA}=(1-K_X)^{-1}\Xi^0_{\rm flat},
\label{eq:XiLevel1}
\end{equation}
the level-0 reduction being the DCA system of
Sec.~\ref{sec:predict} (flat $X$--$X$ covariance, no $T^{(1)}$):
the DCA cross response is the first iterate of the resummation
Eq.~\eqref{eq:XiResum}, and the level-1 kernel keeps the exact
per-pair $X$--$X$ covariance and the $\kappa_3$ dressing while
dropping the $\Delta_c$ re-entry of $T^{(1)}$, the cross-channel
term, and the higher geometric dressing.  Sec.~\ref{sec:num}
evaluates the three objects at the anchor probes---$\Xi_{\rm exact}$
(measured), $\Xi^{\rm DCA}$, $\Xi^{(1)}$---and attributes the
anchor deficit term by term.  With the kernel, the MRSC system
Eqs.~\eqref{eq:MRSCeq}--\eqref{eq:fixedpoint} closes as a coupled
self-consistent hierarchy of resolvent response kernels,
\begin{equation}
\begin{aligned}
&\cC^{(2)}=L[\Gamma,X,\Delta],\qquad
\Gamma^{(2)}=\Gamma_0+D[\cC^{(2)}],\\
&\Xi=\Xi^0+K_X\,\Xi+T,\qquad
\boldsymbol\theta_*=F[\boldsymbol\theta_*;\Xi;H],
\end{aligned}
\label{eq:coupledBS}
\end{equation}
the Feshbach--Bethe--Salpeter pair of Eqs.~\eqref{eq:BSE2}--
\eqref{eq:leadladder} on the one hand and the cross-response
kernel Eq.~\eqref{eq:XiResum} on the other: the vertex
feed-back and the diagonal-dressing feed-back are now separate,
named, linear objects.  Read as a hierarchy, the two loops define
a nested chain of closure levels,
\begin{equation}
\mathrm{DCA}\;\subset\;\Xi^{(1)}\;\subset\;\cdots\;\subset\;
\Xi_{\mathrm{class}},
\label{eq:XiHierarchy}
\end{equation}
in which the level-0 reduction keeps only the flat $X$--$X$
covariance, the level-1 kernel adds the exact per-pair covariance
and the $\kappa_3$ dressing, and the class-resolved kernel
restores the dropped distance-2 channels and the $\Delta$-reentry
term by term, each rung with a quantified residual.  The mechanism
identified in Sec.~\ref{sec:num} is thereby physical rather than
merely numerical: \emph{higher-resolvent self-consistency requires
the cross-level response channel---the $X\leftrightarrow\Delta$
loop---and not merely vertex feedback;} the Feshbach loop
Eqs.~\eqref{eq:BSE2}--\eqref{eq:leadladder} and the
cross-response loop Eq.~\eqref{eq:XiResum} are two irreducible
ingredients of the same microscopic-to-self-consistent closure.  The
hierarchy of Eq.~\eqref{eq:XiHierarchy} is thereby an
\emph{information hierarchy}: each rung reintroduces a named piece of
microscopic information---first the exact per-pair response
covariance, then the class-resolved channels---and the saturation
pattern of Sec.~\ref{sec:num} records the order in which that
information becomes visible, with the remaining deficit carried by
the seed truncation rather than by the response channels.

The rungs of Eq.~\eqref{eq:XiHierarchy} make this formal.  Each
level defines a closure functional
\begin{equation}
F_L[\boldsymbol\theta]=F_{\rm vertex}[\boldsymbol\theta]
+F_{\rm response}^{(L)}[\boldsymbol\theta],
\qquad
\varepsilon_L=\|F_L(\boldsymbol\theta_{\rm ex})-\boldsymbol\theta_{\rm ex}\|,
\label{eq:resphier}
\end{equation}
in which the vertex content is held at its own hierarchy level while
the response part is evaluated with $\Xi^{(L)}$: level $0$ uses the
DCA response of Eq.~\eqref{eq:XiLevel1} with flat input, level $1$
adds the exact per-pair covariance and the $\kappa_3$ dressing, and
the class-resolved limit restores the full $\Delta$-reentry and the
distance-2 channels.  The hierarchy defines successive
response-channel enlargements; the residual need not decrease
monotonically at every truncation, but the first nontrivial response
enlargement produces a systematic reduction of the closure deficit in
the tested models, and the saturation of later rungs, together with
the negligible response of the dropped channels, locates the
remaining deficit outside the response hierarchy---in the vertex-class
truncation of the retained channels---rather than in the channels
still omitted.  The
correction structure of the theory is then organized by \emph{two}
hierarchies, not one: the vertex hierarchy
$\Gamma_0^{(2)}\subset\Gamma_0^{(2)}+\Gamma_{\rm CC}^{(2)}
\subset\Gamma^{(2)}$, and the
cross-level response hierarchy
Eq.~\eqref{eq:XiHierarchy}, both appearing as named linear objects
in the coupled system Eq.~\eqref{eq:coupledBS}.  This is the
central structural statement of the paper:
\begin{equation}
\boxed{\;
\begin{aligned}
&\text{\emph{The next correction to a higher-resolvent closure}}\\
&\text{\emph{is a response channel, not merely a higher vertex.}}
\end{aligned}
\;}
\label{eq:viewpoint}
\end{equation}
Higher-resolvent self-consistency is therefore controlled not only
by higher irreducible vertices, but by the response of lower-level
propagators to the omitted fluctuation channels.  The full hierarchy
of missing information reads, level by level,
\begin{equation}
\begin{aligned}
&D_0\!:\ \text{\small local single-resolvent environment}\ (\bar{\cR},\bar{\cG})\\
&\qquad\to\ \text{\small mean spectral environment}\\[2pt]
&D_1\!:\ \text{\small vertex correlations}\ (\kappa_2,\kappa_3,\Gamma^{(2)})\\
&\qquad\to\ \text{\small correlated fluctuations}\\[2pt]
&\Xi^{(1)}\!:\ \text{\small level-1 kernel},\ X\leftrightarrow\Delta\ \text{\small loop}\\
&\qquad\to\ \text{\small cross-level response}\\[2pt]
&\Xi^{(2)}\!:\ \Delta\text{-reentry},\ \kappa_3\bigl(X,(\bar{\cR})^2,\Delta\bigr)\\
&\qquad\to\ \text{\small repeated feedback}\\[2pt]
&\Xi_{\rm class}\!:\ \text{\small class-resolved kernel},\ \text{\small full connectivity}\\
&\qquad\to\ \text{\small microscopic closure}
\end{aligned}
\label{eq:infohier}
\end{equation}
in which $D_0$ and $D_1$ are the single-resolvent stages and the
three $\Xi$-rungs are the cross-level response hierarchy of
Eqs.~\eqref{eq:XiHierarchy}--\eqref{eq:XiResum}.  Read from top to
bottom, Eq.~\eqref{eq:infohier} is a single statement,
\begin{equation}
\begin{array}{l}
\text{mean}\;\to\;\text{vertex}\;\to\;\text{response}\\
\qquad\;\to\;\text{feedback}\;\to\;\text{microscopic closure}
\end{array}
\label{eq:infobox}
\end{equation}
each stage carried by a named linear object of the theory: the
missing information in a higher-resolvent closure is organized
primarily by response channels between resolvent levels, rather than
by an indiscriminate increase of vertex order.

\subsection{The predictive fixed point}
\label{sec:predict}

The loop of Eq.~\eqref{eq:loopeq} is linear in $\boldsymbol\theta$
only because its feedback content was held as measured input; the
genuine MRSC fixed point makes every piece a functional of
$\boldsymbol\theta$ through $H$.  The predictive map
$\boldsymbol\theta'=F[\boldsymbol\theta;H]$ at $r=2$ is built from
Hamiltonian-derived one-resolvent and coupling content:
\begin{equation}
\begin{aligned}
\cC^{(2)'}(z_1,z_2)={}&\kappa_2(X_a,X_b)
+2\,\kappa_2(X_a,\Delta_b)^{\rm DCA}\\
&+\langle\bar{\cR}^4\rangle\,
\Bigl[\Gamma_0^{(2)}+\Gamma_{\mathrm{CC}}^{(2)}[\boldsymbol\theta;H]\\
&\qquad+\sum_{m,\ell'}c_{m\ell'}\theta_m^{(\ell')}\cG2_m\Bigr],
\end{aligned}
\label{eq:predictmap}
\end{equation}
where (i) $\kappa_2(X_a,X_b)$ is the channel-diagonal fluctuation
correlation, computable from the diagonal distribution alone; (ii)
the cross-response kernel closes through Eq.~\eqref{eq:Xieq} at
level one, $\kappa_2(X_a,\Delta_b)\to\Xi^{(1)}_{ab}$ of
Eq.~\eqref{eq:XiLevel1}---the DCA system
$\E[X_a\Delta_b]=\bar{\cR}_b^2\sum_{c\in N(b)}[\E[X_aX_c]
+\E[X_a\Delta_c]]$ being its level-0 reduction (the coupling--$X$
cross moments vanish exactly,
$\E[X_a\delta|V_{bc}|^2]=0$, since $X_a$ depends only on the
diagonal $H_{aa}$); (iii) $\Gamma_0^{(2)}$ and the cycle seed
$\Gamma_{\mathrm{CC}}^{(2)}[\boldsymbol\theta;H]$ are the
channel-resolved mean resolvent products of
Eq.~\eqref{eq:ccfunctional}, averaged over
the diagonal distribution only, with the cycle ladder of that
equation now dressed by the \emph{current} $\boldsymbol\theta$
through the sector ansatz of Sec.~\ref{sec:ansatz}: for the
free-fermion benchmark the three-cycle content of
Sec.~\ref{sec:ccchannel} gives the ladder the closed form
\begin{equation}
\begin{aligned}
\Gamma_{\mathrm{CC}}^{\rm lad}[\boldsymbol\theta](z)
={}&\sum_{\{b,c\}\in T(a)}
\Bigl[2\bar{\cR}_b^2\,C_{cc}+2\bar{\cR}_c^2\,C_{bb}\\
&\quad+4\bar{\cR}_b\bar{\cR}_c\,C_{bc}
+2\,C_{bb}C_{cc}+2\,C_{bc}^2\Bigr],
\end{aligned}
\label{eq:cycleladder}
\end{equation}
with $C_{uv}(\boldsymbol\theta)=\sum_{m}
\theta_m^{(\ell_{uv})}\cG2_m(z,z)$ the sector-constant ansatz evaluated
on the pair $uv$, $T(a)$ the $144$ three-cycles through $a$ counted
by the census of Sec.~\ref{sec:ccchannel}, and the linear rungs
evaluated
with the pair amplitude frozen at its moment-branch anchor value
while the cycle-diagonal rungs carry the iterated
$\boldsymbol\theta$---the off-cycle amplitudes being fixed by the
deep-plane content, as Sec.~\ref{sec:num} verifies; (iv)
the dressing enters through
the census ladder of Eq.~\eqref{eq:census}.  No eigenstate-overlap
data and no measured higher-resolvent content appears anywhere in
the map---the single-resolvent environment ($\bar{\cR}_a^0$, $\bar g$)
is supplied independently, not drawn from the two-resolvent sector:
$F[\boldsymbol\theta;H]$ is a genuine operator, quadratic
in $\boldsymbol\theta$ through Eq.~\eqref{eq:cycleladder}---the
nonlinear kernel that closes the loop.  The word ``quadratic'' refers
to the \emph{retained} map $F_{\rm ret}$: the rungs not carried by the
iteration are frozen at the moment-branch anchor (item (iii) above),
the off-cycle amplitudes are fixed by the deep-plane content, and the
mixed terms and the higher insertions are dropped (item (iv)).  The
exact DCA map---the hierarchy at fixed projection---is \emph{not}
claimed to be quadratic, and all error statements below concern
$F_{\rm ret}$ against the exact content.  The correspondence between the
three levels is, term by term,
\begin{table*}[tbp]
\centering
\small
\begin{ruledtabular}
\begin{tabular}{p{4.0cm}p{4.4cm}p{4.4cm}}
exact hierarchy & DCA treatment & $F_{\rm ret}$ treatment\\
\hline
$X+\Delta$ tower & exact decomposition &
retained (moment field + ladder)\\
seed $\Gamma_0^{(r)}$ & exact coupling cumulants &
retained, benchmark-computed\\
cycle-ladder rungs & exact in DCA & retained as the quadratic kernel\\
off-cycle amplitudes & exact in DCA & frozen at the moment-branch
anchor\\
mixed coupling--resolvent terms & exact in DCA & dropped (a truncation
choice; nonzero in the exact expansion), carried by
$\delta F_{\rm vertex}$\\
higher insertions $T_{\rm ins}$ & exact in DCA & dropped; diagnosed on
the benchmark by $\delta F_{\rm high}$ (posterior, no general prior
bound)\\
\hline\hline
\end{tabular}
\end{ruledtabular}
\caption{Correspondence between the exact hierarchy, the DCA-retained
terms, and the retained map $F_{\rm ret}$ actually solved.}
\label{tab:retmap}
\end{table*}  The
iteration
$\boldsymbol\theta^{(n+1)}=(1-\alpha)\boldsymbol\theta^{(n)}
+\alpha F[\boldsymbol\theta^{(n)};H]$ converges at
$\alpha=1/2$ within $\sim30$ steps at every $\lambda$, and the
fixed point $\boldsymbol\theta_*$ is compared against the moment
branch and the exact benchmark in Sec.~\ref{sec:num}.  The
truncation of the map---the omitted mixed terms and the higher
insertions $T_{\rm ins}$---is measured by
the bulk normalization $\cC^{(2)'}[\boldsymbol\theta_*]$ against
the exact $\cC^{(2)}$, so the projected content of the fixed point
is quantified, not assumed.

The closed loop is therefore the following, each edge a constructive,
Hamiltonian-derived step: the moments and the connectivity census fix
the seeds and the dressing coefficients of $F[\cdot;H]$; the damped
iteration $\boldsymbol\theta^{(n+1)}=(1-\alpha)\boldsymbol\theta^{(n)}
+\alpha F[\boldsymbol\theta^{(n)};H]$ converges on the fixed point
$\boldsymbol\theta_*$; and the fixed point feeds the cumulants back
through Eqs.~\eqref{eq:cycleladder} and \eqref{eq:census} to
$\cC^{(2)}$ and $\cR^{(3)}$---with the single-resolvent environment
supplied independently and no measured higher-resolvent content
entering at any stage.

It is worth stating explicitly what the loop determines, because
it is not the resolvents themselves.  At $r=1$ the fixed point
determines a single scalar function of one argument---the Faddeeva
dressing $S_1(z)$ of Sec.~\ref{sec:bench}---the aggregate
backaction of the environment on one propagator:
\emph{environmental feedback}.  At $r\ge2$ the fixed point determines
pairwise amplitudes in two energies and one connectivity class: the
backaction of the environment on one propagator, correlated with the
fluctuations of another channel through their shared
structure---\emph{correlated environmental feedback}.  The vertex
content supplies the dressing of that response and the cross-response
kernel Eq.~\eqref{eq:XiResum} its channel structure; the loop closes
when the feedback the propagators experience is reproduced by the
response functions the theory computes.  This is why the response
hierarchy of Eq.~\eqref{eq:XiHierarchy}---not the vertex hierarchy
alone---organizes the corrections: each rung repairs one class of
missed feedback, and the rung sequence is exactly the
$X\leftrightarrow\Delta$ loop of
Sec.~\ref{sec:crossresponse}.

\subsection{Error propagation at the fixed point}
\label{sec:errorprop}

The fixed-point formulation separates the two questions a closure
must answer: what the map misses at the exact point, and what that
miss costs at the fixed point.  Let $F:\mathbb R^M\to\mathbb R^M$ be
the closure map of
Eq.~\eqref{eq:fixedpoint}, $F(\boldsymbol\theta_*;H)
=\boldsymbol\theta_*$ its fixed point, and
$\boldsymbol\theta_{\rm ex}$ the exact-measure amplitudes.  The
\emph{local closure deficit}
\begin{equation}
\delta F=F(\boldsymbol\theta_{\rm ex};H)-\boldsymbol\theta_{\rm ex}
\label{eq:localdeficit}
\end{equation}
is the content of the exact measure that the map fails to reproduce
at the exact point---the residual Eq.~\eqref{eq:residual} evaluated
at $\boldsymbol\theta_{\rm ex}$.  The mean-value identity
$\boldsymbol\theta_*-\boldsymbol\theta_{\rm ex}
=F(\boldsymbol\theta_*)-F(\boldsymbol\theta_{\rm ex})+\delta F
=\bar J\,(\boldsymbol\theta_*-\boldsymbol\theta_{\rm ex})+\delta F$,
with $\bar J$ the Jacobian $J_F=\partial F/\partial
\boldsymbol\theta$ averaged along the segment joining the two
points, gives the exact propagation relation
\begin{equation}
\boldsymbol\theta_*-\boldsymbol\theta_{\rm ex}
=(I-\bar J)^{-1}\,\delta F,
\label{eq:errorprop}
\end{equation}
and, wherever $\|\bar J\|_2\le\chi_{\rm op}<1$ in an operator norm,
the bound
\begin{equation}
\bigl\|\boldsymbol\theta_*-\boldsymbol\theta_{\rm ex}\bigr\|
\le\frac{\|\delta F\|}{1-\chi_{\rm op}}.
\label{eq:errorbound}
\end{equation}
The two norms are assessed separately, because a small spectral
radius does not by itself control the operator norm: the stability
diagnostic of Sec.~\ref{sec:num} measures
$\chi_{\rm cl}=\rho(J_F)\approx0.04$--$0.07$ at the fixed point,
which establishes local linear stability, while the operator-norm
audit of this section, evaluated on the same grid, samples the segment maximum
$\max_{s\in\cS_5}\|J_F(\boldsymbol\theta_{\rm ex}+s(\boldsymbol\theta_*-\boldsymbol\theta_{\rm ex}))\|_2
\approx0.6$ on the five-point grid
$\cS_5=\{0,\tfrac14,\tfrac12,\tfrac34,1\}$ and the segment-averaged
Jacobian $\bar J$ with
$\|\bar J\|_2\approx0.6$.  The Jacobian is non-normal, its spectral
radius an order of magnitude below its operator norm, so the crude
bound $1/(1-\chi_{\rm op})\approx2.5$ overestimates; the
numerically estimated amplification factor is
$\|(I-\bar J)^{-1}\|_2\approx1.4$.  The mean-value identity
Eq.~\eqref{eq:errorprop} itself is verified at the anchor to a residual
below $3\times10^{-14}$ (machine precision): the map transmits the deficit
hierarchy to the fixed-point deviation nearly \emph{unamplified},
and the level ladder of
Eq.~\eqref{eq:Dlevels} of Sec.~\ref{sec:num} is a direct measurement of the deficit
hierarchy $\delta F(\text{level})$ itself.

The fixed point itself is controlled by a conditional statement whose
hypotheses the audit measures.
\begin{proposition}[Conditional uniqueness and geometric convergence]
\label{prop:condfixed}
Let $F$ be the closure map of Eq.~\eqref{eq:fixedpoint}, differentiable
on a closed ball
$\mathcal B=\{\boldsymbol\theta:\|\boldsymbol\theta-\boldsymbol\theta_0\|\le R\}$.
If
(i) $\sup_{\boldsymbol\theta\in\mathcal B}\|J_F(\boldsymbol\theta)\|_2\le\chi<1$
and
(ii) $F(\mathcal B)\subseteq\mathcal B$,
then $F$ has exactly one fixed point in $\mathcal B$, and from every
starting point in $\mathcal B$ the iteration
$\boldsymbol\theta^{(k+1)}=F(\boldsymbol\theta^{(k)})$ converges to it
geometrically, with
$\|\boldsymbol\theta^{(k)}-\boldsymbol\theta_*\|_2\le\chi^k
\|\boldsymbol\theta^{(0)}-\boldsymbol\theta_*\|_2$.
\end{proposition}
The statement is Banach's fixed-point theorem applied on the closed
ball; its content lies in the two hypotheses.  Hypothesis (i) is what
the operator-norm audit of this section measures: the segment maximum
and the segment-averaged Jacobian both give
$\|J_F\|_2\approx0.6$ on the audited grid, below unity by a margin,
and the crude amplification $1/(1-\chi_{\rm op})\approx2.5$ it implies
is replaced by the measured
$\|(I-\bar J)^{-1}\|_2\approx1.4$; the spectral radius
$\chi_{\rm cl}=\rho(J_F)\approx0.04$--$0.07$ of
Sec.~\ref{sec:num} is the sharper local diagnostic.  Hypothesis (ii)
is verified numerically, not proven: the three starting points of the
predictive fixed point (Sec.~\ref{sec:num}) remain inside the sampled
neighborhood at every coupling and closure level and converge to the
same $\boldsymbol\theta_*$ to four to six digits, with the damped
iteration rate $0.52$ predicting the observed $34$--$35$ steps.  What
the proposition therefore supplies is uniqueness and geometric
convergence \emph{within the audited ball}; what it does not supply is
(a) a Hamiltonian-only enclosure of $\mathcal B$ and of the constant
$\chi$---closed below for $\chi$, with the radius remaining a
posteriori---(b) a proof that $F(\mathcal B)\subseteq\mathcal B$, and
(c) uniqueness of the fixed point outside $\mathcal B$.

The first two items are closed by an a posteriori argument that uses
only the quadratic structure of the map.  Through
Eq.~\eqref{eq:cycleladder} the predictive map is quadratic,
\begin{equation}
F(\boldsymbol\theta)=F_0+J_F\boldsymbol\theta
+\tfrac12D^2F\,[\boldsymbol\theta,\boldsymbol\theta],
\label{eq:quadmap}
\end{equation}
so its second derivative is $\boldsymbol\theta$-independent and is
built from the sector sums $\cG2_m$ alone---a Hamiltonian-only
object.  Consequently
\begin{equation}
\sup_{\boldsymbol\theta\in\mathcal B_R}\|J_F(\boldsymbol\theta)\|_2
\le\chi_0+L_2R,
\qquad
L_2\equiv\|D^2F\|_2,
\label{eq:hessbound}
\end{equation}
with $\chi_0=\|J_F(\widehat{\boldsymbol\theta})\|_2$ and $L_2$
computable before any iteration.  Because the map is quadratic, the
second derivative is exact and global, so $L_2$ is a ball-independent
Lipschitz constant of the Jacobian,
$\|J_F(\boldsymbol\theta)-J_F(\widehat{\boldsymbol\theta})\|_2\le
L_2\|\boldsymbol\theta-\widehat{\boldsymbol\theta}\|_2$ for every
$\boldsymbol\theta$; the audited \emph{values} of $L_2$ are of two
kinds, the rigorous Hamiltonian-only product bound and the sharper
empirical segment bound, and the certificate below states which one
enters each inequality.  Let $r_0=\|F(\widehat{\boldsymbol
\theta})-\widehat{\boldsymbol\theta}\|_2$ be the residual of the
converged iteration.  If
\begin{equation}
4L_2r_0\le(1-\chi_0)^2,
\label{eq:ballcond}
\end{equation}
then the admissible radii $R$ form the nonempty interval
$[R_-,R_+]$ with
$R_\pm=[(1-\chi_0)\mp\sqrt{(1-\chi_0)^2-4L_2r_0}]/(2L_2)$, and every
ball $\mathcal B_R$ with $R_-\le R\le R_+$ satisfies both hypotheses
of Proposition~\ref{prop:condfixed}: (i) by
Eq.~\eqref{eq:hessbound} with $\chi=\chi_0+L_2R<1$, and (ii) because
$\|F(\boldsymbol\theta)-\boldsymbol\theta_*
\|\le\chi\|\boldsymbol\theta-\boldsymbol\theta_*\|+r_0\le R$.  The
fixed point is therefore unique in the explicit ball
$\mathcal B_{R_+}$.  The audited numbers pin the radius.
$\chi_0\approx0.6$ is the norm of the converged Jacobian; the residual
sits at the convergence tolerance, $r_0\lesssim10^{-9}
\|\boldsymbol\theta_*\|$; and $L_2$ is bounded a posteriori by the
measured iteration itself.  The damped iterate reaches tolerance in
$34$--$35$ steps where the linearized rate
$\rho[(1-\alpha)I+\alpha J_F(\boldsymbol\theta_*)]\approx0.52$ predicts
$35$--$37$, so the segment-averaged Jacobian differs from the
linearized one by at most about $7\%$ of $1-\rho$: the mean-value
identity
$\bar J-J_F(\boldsymbol\theta_*)=\int_0^1D^2F[s(\boldsymbol\theta_*-
\boldsymbol\theta_0)]\,ds\,[\,\cdot\,]$ then gives
$L_2\|\boldsymbol\theta_*-\boldsymbol\theta_0\|\lesssim0.06$, hence
\begin{equation}
R_+\gtrsim5\,\|\boldsymbol\theta_*\|,
\qquad
\frac{4L_2r_0}{(1-\chi_0)^2}\lesssim10^{-8},
\label{eq:ballradius}
\end{equation}
so Eq.~\eqref{eq:ballcond} is satisfied by eight orders of magnitude
and the verified ball has a radius about five times the amplitude the
fixed point actually carries.  What limits the radius in practice is
not the residual but the range over which the retained quadratic map is
the correct map---the model-error question of
Sec.~\ref{sec:levels} Levels C--D, not an analytic one.  Item (c)
remains open in the precise sense that a second branch of the
retained map outside $\mathcal B_{R_+}$, if it existed, could only be
excluded by the exact map rather than by the truncation; the
numerical statements of Sec.~\ref{sec:num} use only what
Proposition~\ref{prop:condfixed} and Eq.~\eqref{eq:ballcond} license.

\emph{The certificate, itemized.}  The uniqueness statement is a
checkable list rather than a verbal claim.  Here and below
$\widehat{\boldsymbol\theta}$ denotes the converged numerical iterate,
and the symbol $\boldsymbol\theta_*$ the (numerically indistinguishable)
fixed point of the retained map.  The seven entries are:
\begin{itemize}
\item center: $\widehat{\boldsymbol\theta}$, the converged iterate;
\item radius: $R_+$ of Eq.~\eqref{eq:ballradius}, about
$5\|\widehat{\boldsymbol\theta}\|$;
\item norm: the Euclidean norm on the retained amplitude vector,
$\|\cdot\|_2$, with the induced operator norm for Jacobians;
\item contraction constant: $q=\chi_0+L_2R_+<1$.  The \emph{strict}
certificate uses the Hamiltonian-only ingredients of the next paragraph:
$\chi_0=\|J_F(\widehat{\boldsymbol\theta})\|_2\in[0.71,0.80]$ across the
production grid and the rigorous product bound
$L_2\le2.6\times10^5$, giving $R_+\approx8\times10^{-7}$ and
$q\approx0.8<1$.  The sharper pair ($\chi_0\approx0.6$ at the converged
point, $L_2R_+\lesssim0.3$ from the measured segment) is an
\emph{a posteriori} diagnostic: it yields the larger ball
$R_+\gtrsim5\|\widehat{\boldsymbol\theta}\|$ with $q\lesssim0.9$, and is
not used as a strict certificate;
\item residual: $r_0=\|F_{\rm ret}(\widehat{\boldsymbol\theta})
-\widehat{\boldsymbol\theta}\|_2\lesssim10^{-9}\|
\widehat{\boldsymbol\theta}\|$, the absolute residual of the converged
iteration;
\item posterior condition: Eq.~\eqref{eq:ballcond} with
$4L_2r_0/(1-\chi_0)^2\lesssim10^{-8}$, together with the self-map
$\|F(\boldsymbol\theta)-\widehat{\boldsymbol\theta}\|\le qR_++r_0\le
R_+$ inherited from $\chi<1$;
\item rounding: the only numerically uncertain entries are $\chi_0$ and
$L_2$, whose quoted intervals (see below) sit many orders away from the
inequality margins, so double precision cannot change the verdict.
\end{itemize}
The margin $4L_2r_0/(1-\chi_0)^2\lesssim10^{-8}$ is the numerical slack
of the posterior existence/uniqueness condition alone; it bounds neither
the distance from $\widehat{\boldsymbol\theta}$ to an exact hierarchy
object nor any model error.  The residual--radius condition is checked
for both balls: for the strict ball ($q\approx0.8$,
$R_+\approx8\times10^{-7}$, $r_0\lesssim10^{-9}\|\widehat{\boldsymbol
\theta}\|\approx10^{-12}$) the bound $r_0\le(1-q)R_+$ holds by about
five orders of magnitude, and for the a posteriori ball
($q\lesssim0.9$, $R_+\gtrsim5\|\widehat{\boldsymbol\theta}\|$) by about
nine.  The standard posterior location bound
$\|\widehat{\boldsymbol\theta}-\boldsymbol\theta_*\|_2\le
r_0/(1-q)\approx10^{-8}\|\widehat{\boldsymbol\theta}\|$ with the
a posteriori $q$ (and $\approx5\times10^{-9}\|\widehat{\boldsymbol
\theta}\|$ with the strict $q\approx0.8$) controls the distance from the
iterate to the true fixed point of $F_{\rm ret}$---again a statement
about the solver, not about the model error.  What this certifies is \emph{a unique fixed point within the
certified ball} $\mathcal B_{R_+}$---of the retained map $F_{\rm ret}$;
it does not certify uniqueness outside the ball, and it says nothing
about the benchmark discrepancy, which is a separate entry of
Table~\ref{tab:errorledger}.  The radius is a posteriori (next
paragraph).

Both constants of Eq.~\eqref{eq:ballcond} can be computed from
Hamiltonian data alone, before any iteration.  On the retained branch
the residual system of Sec.~\ref{sec:predict} is linear in
$\boldsymbol\theta$ in its $\boldsymbol\theta$-dependent part, so its
Jacobian is obtained by one linear solve on the sector sums: with the
row-normalized system of Eq.~\eqref{eq:cycleladder} and the census
weights of Eq.~\eqref{eq:census}, one finds
$\|J_F\|_2=0.71$--$0.80$ across the grid of Sec.~\ref{sec:num}, with
spectral radius $\rho(J_F)=0.06$---the value $0.6$ quoted above is the
norm of the converged Jacobian and the two differ by the
normalization of the moment rows.  The damped rate
$\rho[(1-\alpha)I+\alpha J_F]=0.53$ at $\alpha=\tfrac12$ reproduces
the observed $0.52$ and the thirty-one iterations to tolerance, so the
computed constants are the ones the iteration uses.  The quadratic
coefficient admits the crude Hamiltonian-only product bound
$L_2\le|r_4|\,N_{\mathfrak c}\max|\cG2_m|^2\|A^{+}\|_2\simeq2.6\times
10^5$, formed from the same sector sums.  With $\chi_0=0.8$ that bound
satisfies Eq.~\eqref{eq:ballcond} by a factor forty
($4L_2r_0\simeq1.0\times10^{-3}$ against
$(1-\chi_0)^2=4.0\times10^{-2}$), so the ball condition---and with it
the uniqueness statement of Proposition~\ref{prop:condfixed}---can be
certified from $H$ alone; what the product bound cannot do is
reproduce the \emph{large} radius of Eq.~\eqref{eq:ballradius}: it
gives $R_+\simeq8\times10^{-7}$, which is three orders of magnitude
below the amplitude $\|\boldsymbol\theta_*\|$ and therefore far weaker
than the a posteriori $R_+\gtrsim5\|\boldsymbol\theta_*\|$ obtained
from the measured iteration rate.  The honest division is therefore the
following.  The Hamiltonian-only ingredients are the structure of the
retained quadratic map and the coefficient $L_2$ (item (a) above); the
\emph{location} of the fixed point is not among them.  $\chi_0$ is
evaluated at the converged point and $r_0$ is its residual, so
Eq.~\eqref{eq:ballcond}, with the product bound, certifies---after the
solver run---uniqueness in a ball of radius
$R_+\simeq8\times10^{-7}$ around that point: a local statement on a
scale far below the physical amplitude $\|\boldsymbol\theta_*\|$, and
not a Hamiltonian-only localization of the fixed point.  The large
radius of Eq.~\eqref{eq:ballradius} is likewise a posteriori, since it
uses the measured iteration.  A fully a priori localization of
$\boldsymbol\theta_*$---and with it uniqueness at the scale of the
physical amplitude---would require a Hamiltonian-only bound on
$\|D^2F\|$ sharper than the product bound; that is a well-posed
remaining problem, not an open modeling question.  The residual $r_0$
itself is a property of the solver, not of the closure: it certifies
that the iteration has converged on the fixed point of $F_{\rm ret}$,
not that $F_{\rm ret}$ approximates the exact hierarchy---the latter
comparison is the deficit $\delta F$ of Eq.~\eqref{eq:deficitdecomp}.

\emph{Conventions of the audit.}  The numbers quoted above are
coordinates and norms of one fixed convention, and it is worth stating
it once.  The parameter vector is the retained amplitude vector
$\boldsymbol\theta$ of Eq.~\eqref{eq:sectoransatz}, in the sector
normalization fixed there; the residual map is the row-normalized
system of Eq.~\eqref{eq:cycleladder}, so $J_F$ and every norm below
refer to that normalization.  Spectral radii are eigenvalues of
$J_F$; operator norms are $\|\cdot\|_2$ computed by singular values on
the five-point grid $\cS_5$, the segment maximum being the entry
quoted as $\|J_F\|_2\approx0.6$ and the segment average giving
$\bar J$.  The quantities fall into three classes, and the text marks
them accordingly: \emph{a priori}---computable from $H$ alone before
any iteration: the Hamiltonian-only Jacobian norm $0.71$--$0.80$, its
spectral radius $0.06$, the damping rate $0.53$, and the crude
$L_2\le2.6\times10^5$; \emph{a posteriori}---available after the
iteration: $\rho(J_F)$ at the fixed point, the segment norms, the
thirty-one iteration steps, and the radius Eq.~\eqref{eq:ballradius};
\emph{measured on exact data}---$\delta F$ at the anchor.  The
residual $r_0$ and the error bound Eq.~\eqref{eq:errorprop} are
evaluated in the same convention, so the bound closes in exactly the
coordinates in which its ingredients were measured.  A change of
coordinates rescales $\|\boldsymbol\theta\|$, $\|J_F\|_2$, and $L_2$
together, leaves $\rho(J_F)$ invariant, and leaves the fixed point
unchanged; the values differ from convention to convention, and the
one used here is the sector normalization of the retained branch
throughout.

The deficit is organized as a telescoping decomposition over the
nested closure maps $F_0\subset F_1\subset\cdots\subset F_L$ of
Eq.~\eqref{eq:Dlevels}---each level adds one construction feature to
the previous map---so that
\begin{equation}
\boxed{\;F_L-F_0=\sum_{j=1}^{L}\bigl(F_j-F_{j-1}\bigr),\;}
\label{eq:deficitdecomp}
\end{equation}
is an identity, and each difference
$\delta F_j\equiv[F_j-F_{j-1}](\boldsymbol\theta_{\rm exact})$,
evaluated at the anchor point, carries the name of the feature its
level introduces:
$\delta F_{\rm seed}$ the seed
truncation (the deep-plane degree cut $3r-2$ and the class-census
cut), $\delta F_{\rm cross}$ the cross-response truncation (the
dropped channels and the $\Delta$-re-entry of
Eq.~\eqref{eq:XiHierarchy}), $\delta F_{\rm vertex}$ the dropped
dressing rungs beyond the pair-contraction ladder of
Eq.~\eqref{eq:cycleladder} together with the dropped mixed
coupling--resolvent terms of Eq.~\eqref{eq:vertexF}---a composite
truncation-deficit symbol collecting both, not a single graph
correction, $\delta F_{\rm basis}$ the
sector-constant projection of the ansatz, and $\delta F_{\rm high}$
the dropped back-reaction of the higher cumulants
$\cC^{(s\ge3)}$.  Each term is measurable at the anchor:
Sec.~\ref{sec:num} shows $\delta F_{\rm cross}$ the dominant deficit
before its rung and nearly exhausted after it---the level-1 kernel
carries $92\%$ of the exact response kernel at the representative
point---while the level-1 residual then localizes to the dropped
distance-2 channels and the $\Delta$-re-entry of
$\delta F_{\rm cross}$, the channel-space truncation shared by every
closure level, the named next correction on the
generic laboratory.  The content of
Eqs.~\eqref{eq:errorprop}--\eqref{eq:deficitdecomp} is a statement
of \emph{propagation}, not of control: \emph{each closure level
removes a named, measurable term of the deficit, and the linearly
stable map converts that reduction into the same reduction of the
fixed-point error, up to the numerically estimated factor
$\|(I-\bar J)^{-1}\|_2\approx1.4$; what is
proven is the bounded transmission of a given deficit, not the
smallness of the highest rung, and the dominant deficit is not the
highest-order vertex but the omitted microscopic response channel.}
The three errors the hierarchy distinguishes are correspondingly
distinct: the framework error
$\varepsilon_L=\lVert F_L(\boldsymbol\theta_{\rm ex})
-\boldsymbol\theta_{\rm ex}\rVert$ of Eq.~\eqref{eq:resphier}, the
truncation error of a particular seed class, and any
model-independent convergence statement---which this article does
not claim.  The response hierarchy provides a structured sequence of
increasingly informed closures; its numerical behavior is diagnostic
rather than a general monotonic convergence theorem.  This is the
multi-resolvent counterpart of the single-resolvent logic of
Ref.~\cite{SRletter}: there, the causal ansatz converts a truncation
of the self-energy into a truncation of the resolvent; here, the
causal matrix map converts a truncation of the microscopic input into
a truncation of the correlator, and Eq.~\eqref{eq:errorprop} is the
quantitative bridge between the two.

\emph{What is solved and what is measured.}  The map of
Eq.~\eqref{eq:predictmap} is the retained finite map $F_{\rm ret}$, not
the exact DCA map: it keeps the $(X,\Delta)$ tower, the seed
$\Gamma_0^{(2)}$, the cycle-ladder rung and the census dressing, freezes
the rungs not carried by the iteration at the moment-branch anchor, and
drops the mixed coupling--resolvent terms and the higher insertions
beyond the pair-contraction ladder.  The decomposition of
Eq.~\eqref{eq:deficitdecomp} compares $F_{\rm ret}$ with the exact
content $\cC^{(r)}_{\rm ex}$, defined as the exact multi-resolvent
cumulant of the benchmark Hamiltonian at the anchor probes, measured
directly in Sec.~\ref{sec:num}; none of the terms of
Eq.~\eqref{eq:deficitdecomp} measures the DCA projection itself.  The
four sources of discrepancy are therefore kept separate rather than
summed,
\begin{table*}[tbp]
\centering
\small
\begin{ruledtabular}
\begin{tabular}{p{2.5cm}p{4.5cm}p{4.5cm}p{2.9cm}}
Source & object measured (norm) & evidence and availability & prior bound\\
\hline
solver & $\|\widehat{\boldsymbol\theta}-
\boldsymbol\theta_*\|_2$; residual $r_0$ & measured after the solver
run (post-solve); $r_0\lesssim10^{-9}\|\boldsymbol\theta_*\|$ & yes, for
this entry\\
finite ansatz & $\delta F_{\rm seed}+\delta F_{\rm basis}$ against
$\cC^{(r)}_{\rm ex}$ & benchmark comparison at the anchor (available
only after the exact data are unblinded) & no\\
vertex truncation & $\delta F_{\rm vertex}+\delta F_{\rm cross}$,
propagated by Eq.~\eqref{eq:errorprop} & benchmark comparison (available
only after the exact data are unblinded);
$\|(I-\bar J)^{-1}\|_2\approx1.4$ & no\\
DCA projection & none measured & model assumption of
Sec.~\ref{sec:vertex}; an independent bound is not available here & no\\
observable projection & transport constant $\kappa_O$,
Eq.~\eqref{eq:transportbound} of Appendix~\ref{sec:eth} & bounded from
the exact data at the anchor & partially\\
\hline\hline
\end{tabular}
\end{ruledtabular}
\caption{The error ledger.  Each source of discrepancy is listed with
the object it measures, the norm, when the evidence becomes available,
and whether it is a prior bound or a posterior measurement.  The rows
are not summed: the DCA contribution is uncontrolled, so no certified
total error bound follows from this table.}
\label{tab:errorledger}
\end{table*}
The ledger separates the identifiable sources of discrepancy.  The
finite-ansatz and vertex-truncation entries are the measured anchor
deficits $\delta F_{\rm seed}+\delta F_{\rm basis}$ and
$\delta F_{\rm vertex}+\delta F_{\rm cross}$, propagated by
Eq.~\eqref{eq:errorprop}; they are posterior measurements in the norm of
the target cumulant.  For the benchmark errors reported in this article,
the quantitative evaluations rely on the unblinded exact reference data;
that no general a priori bound is established for these entries does not
preclude a priori bounds for specific components in principle.  The solver
entry is the convergence residual $r_0$ on the production grid---a
property of the iteration, not of the closure: a small $r_0$ certifies
convergence on the fixed point of $F_{\rm ret}$, not that $F_{\rm ret}$
approximates the exact hierarchy, which is the content of
Eq.~\eqref{eq:deficitdecomp}.  No independent estimate of the DCA
projection error is available in this article: the vertex functional of
Sec.~\ref{sec:vertex} is a model assumption of the construction.  The
omitted higher-insertion contribution is diagnosed on the benchmark
through $\delta F_{\rm high}$, but no general a priori bound on it is
established.  The
closure error quoted in Sec.~\ref{sec:num} is therefore a model error of
$F_{\rm ret}$ at the measured anchor; the reported benchmark residuals
quantify selected contributions, but they do not constitute a certified
bound on the total error, because the DCA contribution remains
uncontrolled.

\section{Benchmark design and numerical verification}
\label{sec:num}

\emph{Status.}  Level D of Sec.~\ref{sec:levels}: this section is a
numerical implementation.  The design paragraphs below fix
definitions and protocols; every verification number is a measured
quantity on the two benchmark families at the stated finite sizes,
and no numerical statement is promoted to a theorem.

\subsection{Reductions and benchmark design}
\label{sec:bench}

\emph{The physical role of $r$.}  The hierarchy organizes by
physical content, not by order: $r=1$ is the \emph{mean dynamics}---the
single-resolvent sector that carries the spectral density and the
mean-field transport; $r=2$ is the \emph{correlated spectral
fluctuations}---the pair covariance that carries the two-level
correlations and the linear response; $r=3$ is the \emph{asymmetry
sector}---parity mixing, skewness, and the lowest nonlinear
response.  Each order adds one cumulant of the fluctuation fields of
Eq.~\eqref{eq:Gammar}, and each has a named physical output in the
reductions and benchmarks below; the arbitrary-$r$ sector remains a
formal architecture (Secs.~\ref{sec:ladder} and \ref{sec:conn})
rather than a numerical program.

\emph{$r=1$: the Faddeeva closure.}  The mean sector
Eqs.~\eqref{eq:sys1}--\eqref{eq:sys2} with the diagonal vertex
parameterized causally is the single-resolvent closure of
Ref.~\cite{SRletter}---the classical recursive-resolvent
(continued-fraction) route \cite{Haydock1972,Haydock1980} closed by a
causal ansatz: the Faddeeva effective self-energy
\begin{equation}
\cG_a(\lambda)=\Delta_a^{\mathrm{eff}}-V_a
+\mathrm i\,\chi_a^{\mathrm{eff}}\,
w\Bigl(-\frac{\delta\lambda'_a}{\sqrt2\,\sigma_a}\Bigr),
\label{eq:faddeeva}
\end{equation}
with $w(z)=e^{-z^2}\mathrm{erfc}(-\mathrm iz)$, realizes the Gaussian
pole-cloud diagonal vertex in causal form; here
$\delta\lambda'_a=\lambda-\lambda_a-\Delta^{\mathrm{eff}}$ is the
eigenvalue's distance from the shifted pole-cloud center of channel
$a$ (bare seed $\lambda_a$, self-energy shift
$\Delta^{\mathrm{eff}}$), $\sigma_a$ the cloud width, and the parameters
$\{\Delta^{\mathrm{eff}},\chi^{\mathrm{eff}},\sigma\}$ are exactly the
finite parameter set of Eq.~\eqref{eq:fixedpoint} at $r=1$.  The
skewness selection rule of Ref.~\cite{SRletter} is a corollary of the
parity structure of Sec.~\ref{sec:causal}: for reflection-symmetric
kernels every single-channel (mean-sector) closure preserves spectral
symmetry, so skewness requires the parity-mixing odd-$r$
multi-channel vertices---the $\Gamma^{(3)}$ content of
Eq.~\eqref{eq:G3seed} is the minimal symmetry-allowed mechanism, and
its odd sector fixes the center offset $\delta s_a$ of
Eq.~\eqref{eq:faddeeva}.  In the moment dictionary this content is
explicit: the parity-mixing sector of $\cC^{(3)}$ is its odd-degree
moment content $\kappa_3[s_p,s_q,s_r]$ ($p+q+r$ odd,
Sec.~\ref{sec:sectorclosure3}), which vanishes identically for
reflection-symmetric Hamiltonians and supplies $\delta s_a$
otherwise.

\emph{Free fermions.}  In random free fermions all microscopic inputs
are closed: $q_r$ is the Selberg product \cite{MResR},
$\kappa_2[s_1(a),s_1(b)]=2\ell-2k^2/N$ \cite{MResmerge}, and the
seeds reduce to Eqs.~\eqref{eq:G2seed}--\eqref{eq:G3seed}.  The exact
free-fermion hierarchy provides the benchmark that anchors the MRSC
system: the equations recover its two- and three-resolvent cumulant
structures at the tested truncation level, and Sec.~\ref{sec:response}
verifies this recovery level by level.  For the
interaction-deformed family, the sector deformation
$K_r(s)=K_r^{\mathrm{FF}}(s)F_r[\cI_r(s);\boldsymbol\theta_r]$ of the
hierarchy's outlook \cite{MResR} is a realization of the ansatz
Sec.~\ref{sec:ansatz} in sector coordinates, with the exact
$\lambda=0$ anchor as its benchmark value.

\emph{Generic interacting benchmark.}  \label{sec:genbench} The
connectivity machinery of
Sec.~\ref{sec:conn} is exercised on a model with \emph{no occupied-set
representation}---the disordered transverse-field Ising chain
\begin{equation}
H=\sum_i h_i Z_i+\sum_i J_i Z_iZ_{i+1}+\sum_i g_i X_i,
\label{eq:ising}
\end{equation}
with i.i.d.\ Gaussian fields $h_i,J_i,g_i$ of unit variance,
$\E[h_i^2]=\E[J_i^2]=1$, the flip deformation $\E[g_i^2]=\lambda$,
periodic boundary conditions ($J_N\equiv J_1$), $N=9$, and
$D=2^{N}=512$ $Z$-basis channels: its
coupling graph $G_V$ is the $N$-dimensional hypercube (one flip move
per spin), the classes are the shared-move classes of
Hamming-distance pairs, and no Slater algebra exists to organize the
sector sums---every ingredient must come from the class machinery or
vanish.  The microscopic coordinate system of the generic layer is
thereby explicit.  The generic $r=2$ benchmark uses the
independently determined single-resolvent environment as a
lower-level input; no higher-resolvent or environment-side
two-point data are supplied to the predictive map.  The
Hamiltonian-derived moments supply the
deep-plane content and the connectivity census the channel
structure, and every quantity the predictive map
$F[\boldsymbol\theta;H]$ of Sec.~\ref{sec:predict} consumes is
Hamiltonian-derived: the diagonal content $H_{aa}$; the mean-field
self-energy
$\cG_b(z)=\E[z-H_{bb}-1/\cR_b(z)]$, built from the exact
single-site resolvent
$\cR_b(z)=\braket{\varphi_b}{(z-H)^{-1}}{\varphi_b}$; the loop
weights and the $X$-cumulants of Eq.~\eqref{eq:Xieq}, all
single-site objects of $H$; and the coupling census of
Eq.~\eqref{eq:isingcoup}.  Validation-only quantities---the exact
cumulant, the exact kernel, the class pair covariances $C_{mn}$ of
Eq.~\eqref{eq:isingres}, and the class-resolved pair densities
$\mathcal X_{\mathfrak c}(E_1,E_2)$, the shared-entry profile of
the eigenstate pairs---enter the benchmark exclusively through the
diagonalization side of the comparison; the blind protocol of
Sec.~\ref{sec:blind} withholds the environment-side class covariances
and the pair densities entirely; the outcome of the blind
comparison is reported in Sec.~\ref{sec:blind}.  The in-band probes
$z=1+1.2\mathrm i,\,0.3+1.5\mathrm i,\,-0.7+1.4\mathrm i$ sit
inside the $N=9$ band (spectral half-width $\sigma\simeq4.2$).

The class census and amplitudes follow by walk counting on $G_V$: for
walk pairs of length at most two, the one-shared-move class has $N$
pairs with amplitude $\Theta=\kappa_2[g^2,g^2]=2\lambda^2$, the
two-shared-move class $N(N-1)$ pairs with $\Theta=\lambda^2$, every
zero-shared class vanishes by independence and every odd-degree class
by the pairing rule of Proposition~\ref{prop:parity}.  The
coupling-disorder seed and the resolvent-induced dressing then
separate exactly as
\begin{equation}
\begin{split}
\Gamma_{\rm coup}^{(2)}(z)={}&2\lambda^2\sum_c
\bigl[\hat{\cR}_c(z)^2+\hat{\cR}_c(z)^4\bigr]\\
&{}+\lambda^2\sum_{c\ne d}\hat{\cR}_c(z)^2\hat{\cR}_{cd}(z)^2,
\end{split}
\label{eq:isingcoup}
\end{equation}
with $\hat{\cR}_c,\hat{\cR}_{cd}$ the mean-field resolvents of the
flip and double-flip channels (the backtracking and two-move
classes), and $C_{mn}$ the exact class covariances of the shared-move
census---the $\Gamma_{\rm coup}$/$\Gamma_{\rm res}$ division of
Sec.~\ref{sec:conn} in its pure form---which enter the
pair-contraction ladder as
\begin{equation}
\begin{split}
\Gamma_{\rm res}^{(2)}[C](z)=\lambda^2\bigl[
&(3N+N(N-1))(C_{00}+C_{01})+3N\,C_{11}\\
&{}+N(N-1)C_{12}+2N(N-1)C_{22}\bigr],
\end{split}
\label{eq:isingres}
\end{equation}
At $\lambda=0$ the couplings
are deterministic: the coupling-disorder seed vanishes identically,
the dressing vanishes with it, and the closure reduces to
$\cC^{(2)}=\kappa_2[X,X]$, which is exact because $\Delta=0$---the
diagonal chain is closed trivially.  The anchor is therefore a
\emph{consistency test}: at the solvable limit the construction
\emph{must} reproduce the reference cumulant to machine precision,
as required by the exact partition Eq.~\eqref{eq:XD} with
$\Delta=0$---a requirement verified at all three probes in
Sec.~\ref{sec:fermions}.  The predictive test is $\lambda>0$, where
the four-term closure
Eq.~\eqref{eq:predictmap} with the level-1 kernel of
Eq.~\eqref{eq:XiLevel1} in Sec.~\ref{sec:crossresponse} is evaluated against the exact cumulant
$\cC^{(2)}_{\rm ex}(z,z)=\kappa_2[\cR_{a_0},\cR_{a_0}]$; the numbers
are reported in Sec.~\ref{sec:response}.  The two
laboratories now bracket the genericity claim from both sides: the
free-fermion family verifies the exact algebra of the sector basis,
and the disordered Ising chain is the model on which the class
machinery must supply \emph{every} ingredient without a sector
basis; whether it does is the content of Sec.~\ref{sec:blind}.

\emph{Physical output.}  The solved parameters determine the
energy-resolved spectral-weight correlations through
Eq.~\eqref{eq:ETH}: $P_2=S^{(2)}$ provides the spectral-weight
component entering higher-order ETH correlations---the
channel-resolved covariance structure that observables $O$ project
onto through $\sum_{a}\bra{\varphi_a}O\ket{\varphi_a}p_n^a$-type
expansions, not the variance of an arbitrary off-diagonal matrix
element $O_{nn'}$---together with the cross-level covariance;
$P_3$ carries the connected three-point spectral-weight
correlations entering skewness and higher-order response; the
boundary values of $\cC^{(r)}$ generate the multi-time
spectral-weight correlations and the nonlinear response functions
$\chi^{(r)}$ of the full-ETH program
\cite{Pappalardi2022,Pappalardi2025}.  The distinction between
universal and model-specific content---which parameters are stable
across models and which carry the microscopic detail---is read off
directly from the seed--dressing split of
Theorem~\ref{thm:vertex}: the dressing kernels are structurally
universal, the seeds carry the model.  This route is not a detour
around the resolvent hierarchy: the multi-resolvent cumulants are
the causal, energy-resolved, analytically constrained intermediate
variables that carry the correlations, and the MRSC closure turns
them into a route to the full ETH output---the same variables the
exact hierarchy organizes---rather than an alternative to it.

\subsection{Free-fermion benchmark and microscopic moment matching}
\label{sec:fermions}

We verify the construction on the random free-fermion benchmark of
Refs.~\cite{MResR,MResmerge}: $N=14$ modes, $k=2$ particle sector
($D=91$ channels), GOE one-body matrix $h$ with Haar eigenvectors
\cite{Mehta,Forrester,Efetov},
$n_{\mathrm{real}}=4000$ realizations, probe points
$z_1=0.5+0.8\mathrm i$, $z_2=-0.4+0.9\mathrm i$, $z_3=0.2+1.1\mathrm
i$, deep plane $z=\mathrm it$ with $t=4,\dots,64$.  Representative
channel pairs and triples are labeled by their mode overlap $\ell$
and the shared-edge counts of Eqs.~\eqref{eq:G2seed}--
\eqref{eq:G3seed}.

\emph{Exact identities.}  The geometric identity
Eq.~\eqref{eq:geom} holds per realization to machine precision
(max residual $1.7\times10^{-16}$).  The $r=2$ ladder tower
Eq.~\eqref{eq:ladder2} closes as $\cC^{(2)}=\E[X_aX_b]+\E[X_a\Delta_b]
+\E[\Delta_aX_b]+\E[\Delta_a\Delta_b]$ to $7\times10^{-18}$; at
$z_1$ the coincident pair gives
$\cC^{(2)}=(2.56+0.50\mathrm i)\times10^{-3}$ with tower terms
$(2.69+0.48\mathrm i)\times10^{-3}$, $(0.29+0.24\mathrm i)
\times10^{-3}$, $(-0.42-0.22\mathrm i)\times10^{-3}$: at finite
resolution the mean-field tower and the vertex tower are comparable,
and neither can be dropped.  The $r=3$ tower
Eq.~\eqref{eq:ladder3explicit} (Appendix~\ref{app:ladder3})
reproduces $\cC^{(3)}$ within the
Monte Carlo error of the cubic estimators (residuals $\lesssim
5\times10^{-6}$ against values $\sim10^{-4}$).  The vertex-functional
decomposition of Sec.~\ref{sec:vertex},
$\Gamma^{(2)}=\kappa_2(\delta\cG^{\mathrm{MF}},\delta\cG^{\mathrm{MF}})
+2\kappa_2(\delta\cG^{\mathrm{MF}},\delta\cG^{\mathrm{CC}})
+\kappa_2(\delta\cG^{\mathrm{CC}},\delta\cG^{\mathrm{CC}})$, closes
to $10^{-15}$, and the cross-correlated channel carries $O(1)$
content at bulk broadening (coincident pair at $z_1$:
$-0.765-0.043\mathrm i$, $+0.716+0.146\mathrm i$,
$-0.127-0.121\mathrm i$ for the three terms), so the vertex
functional must retain $\cG^{\mathrm{CC}}$.

The deep-plane approach of the seeds is reported with its table in
Appendix~\ref{app:anchorchecks}: the corrections are $O(1/t)$ with
coefficients of many-body bandwidth scale, and by $t=64$ the seeds are
approached to a few percent, the three-point seed more slowly.

The Hermiticity check, the smoothed dispersion relation, the diagonal
mass sum rule and the term-by-term verification of the $r=2$ functional
are itemized in Appendix~\ref{app:benchver}; all pass, and the
mean-field-level residual of the functional decays monotonically with
the deep-plane scale.

\emph{$\omega$-only ansatz (baseline).}  For the coincident pair the
signed two-scale kernel Eqs.~\eqref{eq:ansatz2}--\eqref{eq:Qdef} with
the two equations \eqref{eq:xisolve} solved against
$\{\cM_{11},\cM_{22},\cM_{31}\}$ alone gives
$(\xi_1,\xi_2)=(16.4,19.0)$, and the resulting $\cC^{(2)}$ deviates
from the exact bulk value by $29\%$ and $27\%$ in modulus at
$z_1,z_2$; for the $\ell=1$ and $\ell=0$ pairs the deviation is
$83\%$ and $80$--$86\%$.  The diagnosis is the sector structure
discussed in Sec.~\ref{sec:ansatz}: the exact kernel's
overlap-conditioned energy correlations are not functions of
$\omega$ alone---finite moments alone are insufficient, and the
ansatz space must carry the connectivity geometry.

\emph{Sector-moment closure.}  The sector ansatz
Eq.~\eqref{eq:sectoransatz} with the linear system
Eq.~\eqref{eq:sectorsolve} closes the same channel pairs from the
Hamiltonian-only input $\{N_m,L_{11}(m),L_{22}(m),\cM_{11},\cM_{22}\}$
(no eigenstate-overlap data).  For the coincident pair
$(\ell=2)$ the moment-matched amplitudes deviate from the exact
sector covariances by $0.15\%$, $0.83\%$, $2.2\%$ for
$m=2,1,0$, the independent prediction
$\cM_{31}=\sum_m\theta_mL_{31}(m)=240.3$ matches the
Hamiltonian-only value $242.2$ ($0.8\%$), $\cM_{20}=10^{-11}$
(the vanishing demanded by the mass constraints of
Sec.~\ref{sec:levels})
vanishes, and the resulting $\cC^{(2)}$ deviates from the exact bulk
value by $5.4\%$ and $2.8\%$ at $z_1,z_2$---an order-of-magnitude
improvement over the $\omega$-only baseline with the same number of
free parameters.  For $(\ell,\mathrm{dev}\,\boldsymbol\theta,
\mathrm{dev}\,\cC^{(2)})=(1;\,0.4\%,0.8\%,5.5\%;\,25\%,16\%)$ and
$(0;\,0.7\%,0.8\%,17\%;\,25\%,25\%)$ the closure degrades gracefully:
the smaller $\cM_{11}$ of the off-diagonal pairs makes the linear
system more sensitive, and the sector cancellation in the bulk
cumulant amplifies the amplitude errors.  The closure is sharpened by
enlarging the moment set $\cI_2$ (Sec.~\ref{sec:closure}), the
finite-dimensional realization of the fixed point
Eq.~\eqref{eq:fixedpoint} at $r=2$; the moment axis is the natural
one on the free-fermion family, while the generic benchmark of
Sec.~\ref{sec:genbench} shows that beyond this family the leading
coordinate is connectivity, not energy conditioning.

\emph{Interaction-deformed benchmark.}  The closure is next
exercised on the deformed family $H=H_0+\lambda W$ with $W$ the
two-body channel matrix (Gaussian coefficients,
$\E|W_{ab}|^2=1$ on the $|A_a\triangle A_b|=4$ support, the protocol
of Ref.~\cite{MResR}), $\lambda=0,2^{-6},2^{-5},2^{-4}$, eigenstates
energy-sorted and sector-labeled by the free ranks.  The $\lambda=0$
anchor is reproduced at the percent level
[dev $\boldsymbol\theta=(0.7\%,1.7\%,2.5\%)$ for the coincident
pair].  At $\lambda>0$ the Hamiltonian-only moment inputs deform only
\emph{analytically}---$\cM_{11}$ is exactly $\lambda$-independent
(the two-body $W$ has vanishing diagonal and trace) and $\cM_{22}$
changes by $3\times10^{-5}$ relative---while the measured sector
covariances deform strongly ($\theta_2$ by $-22\%$, $\theta_1$ by
$-69\%$ at $\lambda=2^{-4}$ for the coincident pair), a
\emph{nonanalytic} response carried by the near-degenerate level
pairs---the log-divergent physics of Ref.~\cite{MResR}.  The lesson
of the grid is the following, stated as a principle:
\begin{equation}
\begin{aligned}
&\text{low-order diagonal moments do not contain all}\\
&\qquad\text{interaction-induced higher-resolvent information.}
\end{aligned}
\label{eq:boxmoment}
\end{equation}
The
moment-matched closure, blind to this content, stays near its
anchor: the closure--exact deviation of the sector amplitudes grows
to $143\%$, $231\%$, $23\%$ for $m=0,1,2$, while the
sector-constancy residual
$\cR(\lambda)=|\sum_m\bar\theta_m L_{11}(m)-\cM_{11}|/\cM_{11}$
grows monotonically, $0.001\to0.18\to0.32\to0.52$ across the twelve
$(\lambda,z)$ probe points.  The growth is an effective
$\lambda^{0.8}$ law whose immediate jump at
$2^{-6}$ marks the near-degenerate response.  The residual
$\cR(\lambda)$ is therefore a quantity with a precise physical
reading: it measures the \emph{distance between the local
microscopic projection and the global resolvent closure}---how much
higher-resolvent information the selected Hamiltonian-only moments
fail to encode.  Wherever $\cR(\lambda)$ grows, the interacting
kernel has developed sector content invisible to the moments, and
the missing content must be supplied by the Feshbach loop
(Sec.~\ref{sec:loop}) rather than by moment matching.  At the level
of the bulk $\cC^{(2)}$ the
closure nonetheless tracks at $5$--$6\%$ throughout, the diagonal
sector dominating the bulk cumulant; the participation ratio
$q_2(\lambda)=\E[p_a^2]$ of Eq.~\eqref{eq:sectorexact}---equal to
$91/20=4.55$ at $\lambda=0$---takes the values
$4.56,4.31,4.10,3.79$ across the grid and records the interaction-induced
delocalization of the eigenstates in the channel basis.

\emph{The connectivity basis.}  The genericity upgrade of
Sec.~\ref{sec:ansatz} is exercised on the same grid: the free
rank labels are replaced by the connectivity coordinate
$q(n,n')=\sum_a p_n^a p_{n'}^a$, the channel-overlap profile of the
eigenstate pair (at $\lambda=0$ its three clusters reproduce the
free sectors, $0.0083,0.0167,0.0501$; at $\lambda>0$ it is computed
from $H(\lambda)$ itself).  The classes are fixed once at the
$\lambda=0$ anchor---the per-$\lambda$ re-quantized boundaries are
systematically \emph{worse}, the coordinate must be defined once and
deform only through its values.  The constancy residual drops with
the connectivity resolution: at $\lambda=2^{-4}$,
$\cR(\lambda)=0.510$ (free sectors) falls to $0.375,0.177,0.104,
0.078$ for $K=3,6,9,12$ classes---a $6.5\times$ reduction---and the
connectivity residual is nearly $\lambda$-flat
($0.113\to0.078$) where the free-sector residual grows $90\times$
($0.006\to0.510$): the deformed connectivity classes absorb the
interaction-induced sector content that the rank labels lose, the
quantitative realization of the connectivity-basis program of
Sec.~\ref{sec:disc}.  Two limitations are measured with equal
sharpness.  The pooled within-class variance of $\delta p_n\delta
p_{n'}$ is class-independent at every resolution: the pair-level
fluctuations are not classifiable, only the class means are.  And
the diagonal moment dictionary does not close the fine basis: the
$K=6$ and $K=12$ moment systems have truncated-SVD rank $4$--$6$,
the resolvable directions lying along the $\lambda$-profiles, nearly
orthogonal to the $q$-amplitude structure---closing the fine
connectivity basis therefore requires the connectivity-resolved
moment dictionary, the $\kappa_2$ of the channel-resolved moment
fields rather than the diagonal ones, which is the identified next
derivation of the program.

\emph{The connectivity-conditioned basis.}  The first nontrivial
terms of that basis are tested directly against the exact centered
covariance of the anchor channel: adjoining the energy polynomials
$P_1=(E_1+E_2)/2\mathcal W$ and $P_2=E_1E_2/\mathcal W^2$, with
$\mathcal W$ the rms spectral radius of $H(\lambda)$, to the three
class functions $\rho_m$ and fitting the $3\to9$ coefficient system
on the $40\times40$ boundary grid leaves a normalized $L_2$ residual
of $10.1\%\to8.0\%$ ($\lambda=0$) and $10.4\%\to7.8\%$
($\lambda=2^{-4}$)---a $20$--$25\%$ reduction, carried by the product
polynomial $P_2$ while the sum polynomial $P_1$ is nearly inert, with
the fitted polynomial coefficients one to two orders below the class
amplitudes.  The energy conditioning is therefore a real but
subleading correction to the class structure; the full synthesis of
the connectivity-conditioned basis from the channel-resolved moment
dictionary is left open here.  The conclusion of the
exercise is that moment matching alone is not the right coordinate
system: microscopic connectivity supplies the leading coordinates,
while energy dependence enters as a secondary deformation.

In summary, this subsection calibrates the construction on the exactly
known microscopic algebra of the free-fermion family: what is
established is the calibration of that laboratory and of its coordinate
program, not predictive power on a generic model---that is the role of
the Ising benchmark below.

\subsection{Response hierarchy, Feshbach loop, and fixed-point error}
\label{sec:response}

\emph{The Feshbach loop.}  The loop equation
Eq.~\eqref{eq:loopeq} is implemented for the three channel-overlap
representatives at eleven probe points (five bulk, six deep-plane
$z=\mathrm i t$) with the measured feedback content.  Its exactness
is verified at $\lambda=0$: the loop residual of the exact sector
values equals the Monte Carlo noise floor,
$\cE_{\rm loop}(\boldsymbol\theta_{\rm exact})=7.9,2.0,0.80$ (in
$10^{-4}$) at $n_{\mathrm{real}}=4\times10^{3},5\times10^{4},
4\times10^{5}$---the exact $\sqrt n$ scaling of a pure noise floor,
with no systematic offset---the equation is an identity at the
measured point, as constructed, and the moment branch coincides
with it, $\cE_{\rm loop}(\boldsymbol\theta_{M})=0.70\times10^{-4}$.
The cross-correlated channel's deep plane is
exercised against the closed form of
Sec.~\ref{sec:ccchannel}: the coefficient cumulant
$\kappa_2=11\,904=576+4\cdot2688+192+4\cdot96$ is verified block by
block at the $1$--$6\%$ level ($0.9\%$ total) by direct
measurement of the $z^{-2}$ coefficient, while the
$t^4\Gamma_{\mathrm{CC}}^{(2)}(\mathrm it,\mathrm it)$ series
$2.6,61,260,462,550,576$ at $t=4,\dots,128$ approaches
$576$---the diagonal seed block---through
the alternating $O(1/t)$ corrections with the imaginary part
converging to zero
(the $t\to\infty$ route being impractical, the coefficient route is
the verification).  The loop
block has six significant singular values ($5.1,4.9,4.0,0.32,0.27,
0.14$; the remaining three at $10^{-3}$), exactly the structural rank
of the loop operator on the sector-constant ansatz, measured: the
product of the three sector profiles and the two independent census
directions of Sec.~\ref{sec:ccchannel}---the three channel-overlap
vectors of the unweighted edge count of Eq.~\eqref{eq:census},
$(264,288,24)$, $(352,212,12)$, $(444,128,4)$, lie on the affine plane
$\sum_m c_m=576$, the diagonal seed block, so the census contributes
the seed plus two independent directions.  The residual
is the global diagnostic the moment branch lacks: across the
interaction grid $\cE_{\rm loop}(\boldsymbol\theta_{\rm
exact};\lambda)$ grows $0.08,1.05,1.91,3.06$ (in $10^{-3}$ at
$n_{\mathrm{real}}=4\times10^{5}$) while $\cE_{\rm
loop}(\boldsymbol\theta_{M};\lambda)$ stays flat at $0.07$
throughout---the
gap, $\simeq37\sigma$ at $\lambda=2^{-4}$, is the nonperturbative
sector content detected \emph{globally}, the concrete realization of
the distance between the local microscopic projection and the global
resolvent closure.  The selection, by contrast, is limited by the
rank structure rather than by statistics: the three near-null
directions amplify even the $0.8\times10^{-4}$ noise floor, so the
truncated-SVD fixed point stays at $1.9$--$47\%$ deviation from exact
at $n_{\mathrm{real}}=4\times10^{5}$ while the moment branch is at
$0.01$--$0.9\%$---the two branches are informationally redundant at
$\lambda=0$, as the rank argument requires, and the
three-dimensional null space of the rank-six loop operator is left
unregularized: the predictive
fixed point of the next paragraph needs no pinning, converging to
the same $\boldsymbol\theta_*$ from all three starting points---the
honest, fully quantified
state of the loop branch.

\emph{The predictive fixed point.}  The map
Eq.~\eqref{eq:predictmap}---with no eigenstate-overlap data and no
measured feedback anywhere---is iterated at $\alpha=1/2$; it
converges within $\sim30$ steps at every $\lambda$ from all three
starting points (zero, the moment branch, and the exact amplitudes),
which agree on the same $\boldsymbol\theta_*$ to four to six digits
at every closure level---the numerical evidence for hypothesis (ii) of
Proposition~\ref{prop:condfixed}, upgraded to the explicit a posteriori
ball of Eq.~\eqref{eq:ballcond} at these parameters but still not a
proof for generic $H$.  The quantitative substitute for
a uniqueness proof is the closure-stability diagnostic: the spectral
radius of the map's Jacobian at the fixed point,
$\chi_{\mathrm{cl}}=\rho(\partial F/\partial\boldsymbol\theta)|_{\boldsymbol\theta_*}$,
is $0.04$--$0.07$ at every $\lambda$ and closure level, so the fixed
point is locally linearly stable, with a small spectral radius of the
Jacobian although the map is appreciably nonnormal; the damped spectral
radius $\rho[(1-\alpha)I+\alpha\,J]\approx0.52$ at $\alpha=1/2$
predicts $35$--$37$ iterations against the observed $34$--$35$,
confirming that the convergence rate is set by the linearization
rather than by the noise floor.  The closure levels are compared at
$n_{\mathrm{real}}=2\times10^{4}$, with D0 the Hamiltonian-derived
linear map, D1a adding the cycle seed, D1lin the linear cycle rungs,
and D1 the full quadratic kernel of Eq.~\eqref{eq:cycleladder}
(deviation
$|\boldsymbol\theta_*-\boldsymbol\theta_{\rm exact}|$ in
percent):
\begin{align}
&(\text{D0},\;\text{D1a},\;\text{D1lin},\;\text{D1})\notag\\
\lambda=0:\quad&(22.0,\;20.1,\;19.8,\;19.8)\\
\lambda=2^{-6}:\quad&(17.8,\;15.7,\;15.4,\;15.4)\\
\lambda=2^{-5}:\quad&(15.4,\;13.5,\;13.2,\;13.2)\\
\lambda=2^{-4}:\quad&(16.5,\;15.8,\;15.8,\;15.8),
\label{eq:Dlevels}
\end{align}
The free-fermion test is therefore not
a high-accuracy end-to-end reconstruction test; it isolates the
parameter-closure problem under an exactly known microscopic
algebra.  Two conclusions.  \emph{Check A
(anchor precision) fails, with clean attribution.}  The attribution is clean: the cycle ladder is suppressed by four powers of the mean resolvent $\bar{\cR}$ at the anchor ($\approx2\%$ of the exact kernel), the dominant deficit is the diagonal-fluctuation cross term---whose level-1 $\kappa_3$-dressing cuts the error $5.5\times$ and whose residual then localizes to the dropped distance-2 channels and the $\Delta$-reentry, the channel-space truncation shared by every closure level.  The per-term values are given in Appendix~\ref{app:benchver}.  \emph{Check B (deformation advantage)
passes at the stated $1.9\times$, and is closure-level
insensitive.}  The deformation advantage is carried by the Hamiltonian-only seed content: the moment branch degrades monotonically $0.5\%\to29.9\%$ while the D1 fixed point tracks $19.8\%\to15.8\%$ (the two branches cross between $\lambda=2^{-6}$ and $2^{-5}$), matching $\theta_2$ to $1\%$ against $27\%$, and the $\lambda$-stable bulk normalization $\approx0.28\,\cC^{(2)}_{\rm exact}$ quantifies the remaining truncation content.  The per-$\lambda$ values are given in Appendix~\ref{app:benchver}.  This asymmetry is the
point: parameter recovery and full resolvent recovery are two
different success criteria, and they are reported here separately.
The free-fermion benchmark therefore tests the predictive closure of
the microscopic vertex coordinates rather than claiming quantitative
completeness of the reconstructed correlator at the same truncation
level: the amplitudes are the deliverable of the self-consistent
parameter scheme of Sec.~\ref{sec:closure}, the bulk cumulant that of
the full (mixed-content) functional.

In short, the vertex parameters are recovered self-consistently from
Hamiltonian data alone; quantitative completeness of the reconstructed
correlator is explicitly not claimed, and the two success criteria are
reported separately.

\subsection{Disordered Ising chain: predictions and blind test}
\label{sec:blind}

\emph{The generic interacting benchmark.}  The disordered Ising
chain of Eq.~\eqref{eq:ising} ($N=9$, $D=512$, $n_{\mathrm
real}=2\times10^{4}$) closes the loop of Sec.~\ref{sec:genbench} on
a model with no sector basis.  The class census of
Eq.~\eqref{eq:census}, walked on the hypercube coupling graph, is
the pairing rule made explicit: $N=9$ shared-move pairs at amplitude
$\Theta=2\lambda^2$, $N(N-1)=72$ zero-shared pairs at
$\Theta=0$, $9$ backtracking pairs at $2\lambda^2$, $72$
two-shared-move pairs at $\lambda^2$, and every odd-degree class
vanishing.

\emph{Five anchors carry the analysis}, each realizing one
layer of the theory named in Sec.~\ref{sec:intro}.  The deterministic
limit $\lambda=0$ verifies the microscopic closure; the
representative deficit $\varepsilon_0=15.6\%\to\varepsilon_1=0.8\%$
at $\lambda=2^{-2}$, $z_1$, and the grid-wide reduction factor
$4.1$--$31$ verify the response closure; the measured radius
$\rho(K_X)\le0.026\ll1$ places the tested kernels well inside the
Neumann-convergence regime of the causal resummation; and the near-vanishing second-rung increment,
together with the seed-carried residual, localizes the remaining
deficit.  The deterministic limit is the consistency check of the
closure: at $\lambda=0$ the closure reduces to
$\cC^{(2)}=\kappa_2[X,X]$, which must equal the exact
$\kappa_2[\cR_{a_0},\cR_{a_0}]$ termwise (the partition
Eq.~\eqref{eq:XD} at $\Delta=0$), and it does---$0.0\%$ at all three
probes.  The predictive test is $\lambda>0$: the four-term closure
of Eq.~\eqref{eq:predictmap} with
the level-1 kernel of Eq.~\eqref{eq:XiLevel1} tracks the exact cumulant at
$0.4\%,0.6\%,0.7\%$ ($\lambda=2^{-4}$),
$0.3\%,1.1\%,1.3\%$ ($\lambda=2^{-3}$), and
$0.8\%,2.0\%,2.6\%$ ($\lambda=2^{-2}$) at the three in-band probes,
with no sector structure anywhere in the pipeline.  The term-by-term
attribution at $\lambda=2^{-2}$, $z_1$ has the same anatomy found on
the free-fermion anchor: the diagonal term
$\kappa_2[X,X]=1.222\times10^{-2}+5.000\times10^{-3}\mathrm i$
carries $94\%$ of the exact amplitude, and the level-1 kernel
$2\Xi^{(1)}=+6.47\times10^{-4}+3.92\times10^{-4}\mathrm i$ against
the exact $+7.36\times10^{-4}+3.77\times10^{-4}\mathrm i$---the DCA
value $-1.20\times10^{-3}-9.36\times10^{-4}\mathrm i$ has the wrong
sign, and level one restores it at $26\times$ smaller error,
capturing $92\%$ of the exact kernel.  The kernel alone misses the
exact cumulant by $0.23\%$ at $z_1$; the quoted $0.8\%$ is carried
by the seed truncation, whose $\lambda^2$-class estimate
overshoots the sub-$0.5\%$ true seed content at this coupling---the
seed truncation, not the response kernel, is the current error
layer, and the deficit grows mildly with $\lambda$ as the
interacting content accumulates.  These numbers also fix what the sub-percent total deviations are a
statement about.  The diagonal term carries $94\%$ of the amplitude, so
a total deviation of $0.8\%$, read through that share, corresponds to an
order-$10\%$ relative accuracy on the non-diagonal content; that
share-based estimate is confirmed directly by the measured response-kernel error at the same probe,
$|2\Xi^{(1)}-2\Xi_{\rm ex}|/|2\Xi_{\rm ex}|\approx11\%$
(complex-difference modulus over the exact modulus), or $8.6\%$ as the
modulus-difference error $||2\Xi^{(1)}|-|2\Xi_{\rm ex}||/|2\Xi_{\rm
ex}|$ (the $92\%$ capture read on its own scale).  The two
numbers answer different questions: the blind deviations
$\varepsilon_{\rm blind}$ measure the total cumulant, while the kernel
error measures the response content alone.  Of the two kernel metrics,
the complex-difference error tests the full complex response prediction
(amplitude and phase) and is the primary diagnostic, since the physical
content of the rung is the sign restoration; the modulus-difference
error tests the amplitude alone.  Both share the same denominator, the
same anchor and the same normalization.

\emph{The seed-resolution ladder.}  The seed layer is itself
resolvable because the benchmark measures its exact content
realization by realization:
$S_{\rm ex}=C_{\rm ex}-\kappa_2[X,X]-2\Xi_{\rm ex}$.  At
$\lambda=2^{-2}$ the measured true seed content at $z_1$ is
$|S_{\rm ex}|/|C_{\rm ex}|=0.42\%$ and the level-1 kernel deficit
$2(\Xi_{\rm ex}-\Xi^{(1)})$ is $0.64\%$: the seed contribution and
the leading response deficit are of comparable sub-percent magnitude
and nearly opposite phase, and their partial cancellation makes the
residual sensitive to the remaining truncation terms---the
$\lambda^2$-class seed estimate, larger in amplitude than the true
seed and differently phased, then overshoots the small net content
and lifts the closure error to $0.8\%$.  Replacing the
$\lambda^2$-class seed estimate by the exact seed lowers the closure
error to
$\varepsilon_{\rm sex}=0.64\%,1.19\%,1.37\%$ at the three probes
(against $\varepsilon_1=0.8\%,2.0\%,2.6\%$), at which point the
residual is the level-1 kernel deficit alone.  The residual error
budget closes accordingly,
$\varepsilon_{\rm residual}=\varepsilon_{\rm response}
+\varepsilon_{\rm seed}+\varepsilon_{\rm higher}$: the response term
saturates at level one ($\varepsilon_2\approx\varepsilon_1$), the
seed term is the next identifiable source, and the two partially
cancel at this probe, so the residual is phase-sensitive to the seed
truncation---precisely the content the seed ladder quantifies.  A
small residual is therefore not evidence of small missing
contributions: the identified seed content ($0.42\%$) and the
level-1 kernel deficit ($0.64\%$) are individually comparable to the
net closure error, and their partial cancellation makes the net
number phase-sensitive---the deficit carries meaning through its
structure, not its size.  The
omitted seed content $\Delta S=S_{\rm ex}-S_{\lambda^2}$ carries no
dominant higher class order: across the extended grid
$\lambda=2^{-4},\dots,2^{-1}$, $|\Delta S|$ grows by factors
$2.9$--$3.7$ per doubling of $\lambda$---well below the $\lambda^3$
and $\lambda^4$ bookkeeping ($8$ and $16$ per doubling).  Microscopic
coupling counting for $H=H_0+\sqrt\lambda V$ predicts the $\lambda^2$
scale for this remainder, and the available numerical range is
consistent with, but does not establish, that asymptotic law: a
four-point log--log fit
$\log|\Delta S|=p\log\lambda+\mathrm{const}$ gives
$p=1.7,1.6,1.6$ at the three probes, the exponent softened below $2$
by the measurement noise floor at the smallest coupling and unable on
this range to discriminate $\lambda^2$ from nearby fractional powers.
The next seed rung is therefore a coefficient refinement at the
retained $\lambda^2$ scale, not a missing higher class order.  The seed ladder
thereby converts the named next axis of the attribution into a
quantified program.

\emph{The blind protocol.}  The seed layer supports one further,
decisive test: a predictive run in which every measured environment
quantity is withheld.  The probe set $(z_1,z_2,z_3)$, the class
partition of the census, and the coupling-side block were fixed
before any $\lambda>0$ data were generated; no post-hoc redefinition
of the protocol was used, and the $\lambda>0$ ensemble data were not
inspected until the blind closures had been evaluated.  The
$\lambda^2$-class seed estimate---the
census of Eq.~\eqref{eq:census} walked on the hypercube coupling
graph---splits into two blocks: a coupling-side block, the class
amplitudes $\Theta_{\mathfrak c}$ applied to products of
mean-resolvent moments, computable
from the Hamiltonian alone; and an environment-side block, the same
amplitudes applied to the class-averaged pair covariances, which for
the class containing the probed pair includes the target cumulant
itself.
The blind closure carries only the coupling-side block.  Across the
grid it attains
$\varepsilon_{\rm blind}=0.41\%,0.51\%,0.56\%$
($\lambda=2^{-4}$), $0.53\%,0.79\%,0.85\%$ ($2^{-3}$),
$0.35\%,0.88\%,0.97\%$ ($2^{-2}$), and $0.38\%,0.75\%,0.65\%$
($2^{-1}$), against the response-only floor
$\varepsilon_{s0}=0.41\%,0.52\%,0.57\%$ /
$0.50\%,0.82\%,0.91\%$ / $0.23\%,1.00\%,1.17\%$ /
$0.92\%,1.07\%,1.31\%$ with the seed removed altogether: the
coupling-side input removes the growing part of that floor at the two
largest couplings ($0.92\to0.38$, $1.07\to0.75$,
$1.31\to0.65\%$), i.e., the architecture converts Hamiltonian-side
information into a genuine prediction of the correlated resolvent,
and it does so better than the full $\lambda^2$-class estimate
($\varepsilon_1=0.8\%,2.0\%,2.6\%$ at $2^{-2}$;
$4.6\%,4.6\%,5.4\%$ at $2^{-1}$), whose environment-side block
carries the overshoot of $S_{\lambda^2}$ quantified above.  Each
realization contributes both a prediction and an exact value, so the
quoted deviations are realization-paired; their Monte Carlo standard
error is set by the realization-to-realization scatter of the
deviation rather than by that of the cumulant itself.  The
exact-seed replacement floors the axis at
$\varepsilon_{\rm sex}=0.64\%,1.19\%,1.37\%$ ($2^{-2}$) and
$0.83\%,1.20\%,1.38\%$ ($2^{-1}$); the one non-monotone entry,
$\varepsilon_{s0}=0.23\%$ dipping below $\varepsilon_{\rm sex}$ at
$(2^{-2},z_1)$, is the phase-cancellation accident already
identified---another instance of the same principle.
Table~\ref{tab:axes} compresses the two hierarchy axes of the closure
onto one grid: the response axis removes the order-unity deficit
($15.6\%\to0.8\%$ at $\lambda=2^{-2}$, $z_1$), and the seed axis
resolves the sub-percent remainder, with no resolved seed-level entry
exceeding $1.4\%$ across the grid.
\begin{table}[tbp]
\caption{The two hierarchy axes of the higher-resolvent closure at
$\lambda=2^{-2}$, $z_1=1+1.2\mathrm i$: normalized closure deviation
(percent) across response rungs (rows) and seed levels (columns);
dashes mark combinations not computed.}
\label{tab:axes}
\footnotesize
\begin{ruledtabular}
\begin{tabular}{c@{\quad}cccc}
\hline\hline
response rung & no seed & coupling-side & full $\lambda^2$ & exact seed\\
 &  & (blind) & class & \\
\hline
DCA & --- & --- & $15.6$ & ---\\
level 1 & $0.23$ & $0.35$ & $0.8$ & $0.64$\\
level 2 & --- & --- & $1.0$ & ---\\
\hline\hline
\end{tabular}
\end{ruledtabular}
\end{table}
Term by term, the DCA level sits at $4.0$--$23.6\%$ across the grid, the first response rung removes it by factors $4.1$--$31$, the $\Delta$-reentry rung is exhausted on the retained channel system, and the distance-2 diagnosis shows the star truncation already adequate ($13\%$ of the remaining kernel deficit, wrong direction); the same data verify the resummation, with $\rho(K_X)=0.007$--$0.026$ far inside the Neumann-convergence regime.  The rung-by-rung values are given in Appendix~\ref{app:benchver}.  The route
$H\to G_V\to\mathfrak c\to\Theta_{\mathfrak c}\to
F[\boldsymbol\theta;H]\to\boldsymbol\theta_*$ therefore survives
beyond the microscopic representation from which it was first
derived, at the quantitative level quoted: random free fermions
remain the calibration laboratory, and the disordered Ising chain is
the generic test bench.

In summary, the architecture transfers beyond its free-fermion origin;
the remaining deficit is attributed to
the second-order class seeds within the retained channel basis rather
than to omitted response channels, and the one non-monotone entry
($\varepsilon_{s0}=0.23\%$ at $(2^{-2},z_1)$) is identified as the same
phase-cancellation accident, not a hidden success.

\emph{Structure dependence.}  The protocol was transcribed to two
further interaction graphs with $N$, the coupling ensembles, the
probes, the estimator, and the sample size held fixed, changing only
the graph $E$ of Eq.~\eqref{eq:ising}: a $3\times3$ periodic square
lattice (degree four, eighteen bonds) and a random four-regular graph
(degree four, eighteen bonds), against the ring of the benchmark
(degree two, nine bonds).  The class census is identical across the
three, because each flip move carries one coupling $g_i$ and the flip
graph---the hypercube---does not depend on $E$; what the graph changes
is the diagonal content, and with it the level statistics, the mean
self-energies, and every Hamiltonian-only closure input.  With the
in-sample estimator used in this subsection (whose absolute level is
the estimator's, not the blind protocol's), the $\lambda>0$ mean
deviations over the grid are $8.2\%$ (ring), $8.2\%$ (lattice), and
$6.2\%$ (random four-regular), with maxima $10.0\%$, $10.6\%$, and
$9.1\%$, and the $\lambda=0$ anchor is exact to $0.00\%$ in all three
cases.  No structure-specific failure appears across a change of
coordination, of interaction topology (ordered lattice versus random
graph), and of bond count; the residual of this comparison sits at the
estimator's sample floor---the same layer identified above---rather
than at any ingredient of the census or of the closure map.

\subsection{Causal-form and Hilbert-relation checks}
\label{sec:hilbert}

\emph{Causal-form test at the real boundary.}  The fixed-point kernel passes the causal-form test: the Hilbert involution residual $(0.035,0.031)$---the normalized boundary-integral mismatch of the two-variable involution identity, itemized object by object in Appendix~\ref{app:benchhilbert}---at every $\lambda$ sits on the same footing as the $(0.03,0.03)$ of the exactly smoothed measure, against a resolution floor $(5.2,12.0)$ set by the 8281 point masses of the exact pointwise kernel, and the reconstructed spectral covariance reaches overlap $0.97$ (deviation $0.25$).  The point-by-point values are given in Appendix~\ref{app:benchver}.

In summary, what is established is the causal spectral form of an
object built from Hamiltonian data alone with no boundary input; the
gap between the fixed-point kernel and the exact smoothed measure is
the truncation deficit quantified above, not an analyticity failure.

\subsection{Connectivity/Krylov coordinates and r=3 extensibility}
\label{sec:krylov}

\emph{The channel-resolved microscopic dictionary.}  The
connectivity-resolved moment dictionary of
Eq.~\eqref{eq:channelmoments} is measured as the full $D\times D$
matrices $\cM_{pq}(a,b)=\kappa_2[s_p(a),s_q(b)]$ on the interaction
grid, and closes on a sector theorem of its own: at $\lambda=0$ the
dictionary collapses onto the overlap alone,
$\cM_{pq}(a,b)=F_{pq}(\ell(a,b))$, with the within-$\ell$ spread at
$1.6\%$ ($\cM_{11}$, means $-0.58,1.44,3.47$ for
$\ell=0,1,2$; the $3.443$ quoted in Sec.~\ref{sec:fermions} is one
representative $\ell{=}2$ channel) and $2.7\%$ ($\cM_{22}$, means
$-7.11,16.83,65.23$); and the dictionary is
$\lambda$-invariant to within a relative change of
$3\times10^{-5}$---$\cM_{22}$'s $\ell=2$ mean changes by
$3\times10^{-5}$ across the whole grid, its within-$\ell$ spread
unchanged.  The channel-resolved dictionary therefore carries
\emph{no} interaction content: the deformation of the kernel lives
entirely in the resolvent feedback and the eigenstate-resolved
sectors, which is the sharpest form of
Eq.~\eqref{eq:boxmoment}---the loop, not the dictionary, is the
carrier of the interacting content.  The fine-class resolution is
obtained constructively by making the connectivity
energy-resolved: joint $(q,E)$ classes (six $q$-bins times three
energy windows) reduce the constancy residual at every $\lambda$,
$0.36\to0.24$ at $\lambda=0$ and $0.50\to0.38$ at
$\lambda=2^{-4}$ ($25$--$34\%$), the energy windows separating the
class $\lambda$-profiles that the pure $q$-bins leave nearly
collinear---the identified route from the fine connectivity basis
to its closable form.

\emph{The eigenstate-free route and the $r=3$ closure: summary.}
The two technical exercises of this section are reported in full in
Appendix~\ref{app:benchver}: the eigenstate-free Krylov determination of
$G^{(2)}_{mn}$ (stable once the exactly vanishing $p=0$ and $q=0$ rows
are restored, reconstruction $9.5\%/7.3\%$ in-band at $K=5$ and
$0.1$--$0.4\%$ on the deep plane, with the in-band plateau identified as
a rank deficiency of the energy coordinate rather than a positivity
violation), and the exercise of the eight-class $r=3$ closure (moment
dictionary verified at $0.3$--$5\%$, coverage-complete classification of
all $D^{3}$ triples, closure error $36$--$38\%$ deterministic on top of
a comparison floor set by the heavy-tailed direct estimator, and a
Hamiltonian-only $8\times8$ inversion with condition number
$9.4\times10^{3}$).  Both are extensibility demonstrations rather than
quantitative closures, and neither changes the main-text claims.
In summary, the coordinate program and the extensibility of the
ladder--vertex--moment machinery to $r=3$ are exercised here; it is
not a quantitative $r=3$ closure, which the $9.4\times10^3$ condition
number and the four-to-five resolved singular directions bound from
below.

The verification scripts are archived in Ref.~\cite{ZenodoMRSC};
all identities above close on the same realization set used by the
direct estimators.

\subsection{Observable-level summary and the diagnostic interface}
\label{sec:bridge}

Two applications of the closed hierarchy are reported in the
appendices and summarized here, because together they fix what the
construction is for.

\emph{The observable level.}  The pair-level identity
Eq.~\eqref{eq:splitpair} of Appendix~\ref{sec:eth} writes the exact
off-diagonal matrix element as a single-resolvent baseline, a
two-resolvent projection, and an interference remainder.  On the Ising
chain the baseline overshoots the exact ETH variance by
$1.7$--$2.3\times$ across the tested couplings, the projection supplies
$56$--$68\%$ of the correction with the correct sign, the blind
prediction reaches $0.35\%$--$0.97\%$, and the first response rung
removes the DCA deficit by factors $4.1$--$31$.  The boundary is equally
fixed: the projection is exact but reweights relative errors, so the
end-to-end observable accuracy lies between $0.4\%$ and $16\%$
depending on whether the closure error is coherent with the observable,
and the transport of the probe-level kernel to the on-shell boundary
is a separate, named item.

\emph{The diagnostic interface.}  Appendix~\ref{sec:LRI} converts the
local-rotational-invariance program into computable window
diagnostics: the conditional dictionary of Eqs.~\eqref{eq:projection}
and \eqref{eq:coincdict}, the finite-order defects on the ring TFIM
($\varepsilon_2\simeq1.1\times10^{-3}$,
$\varepsilon_3\simeq1.8\times10^{-6}$,
$\varepsilon_4\simeq2.5\times10^{-8}$, falling with the window
dimension), and the conditional form of the reference
(Proposition~\ref{prop:LRIcond}) under the two hypotheses
(H1) and (H1$'$).  What it establishes is a working instrument and a
measured boundary, not an emergence theorem: the closure supplies the
candidate parameterization whose window statistics the diagnostics
then test, and Appendix~\ref{sec:LRI} records for each step what is
derived, what is measured, and what remains open.

\section{Discussion}
\label{sec:disc}

Four structural remarks.  First, the framework is
projection-first, not diagram-first: the ladder and the vertex
functional follow from the exact Feshbach identities
Eqs.~\eqref{eq:geom}--\eqref{eq:XD} and the DCA alone, so the
organization is by resolvent multiplicity, not by coupling strength;
the Dyson--Schwinger, Bethe--Salpeter and parquet structures of
many-body theory appear here as special resummations of the same
hierarchy \cite{FW12}.  Second, the causal constraints act at the
level of the ansatz space: the finite parameterization is not a
numerical fit of $\cC^{(r)}$ but the intersection of the causal
manifold with the exact sum rules, with the moments fixing the free
parameters---the ``constrained ansatz'' philosophy in which the
microscopic Hamiltonian enters through exact constraints rather than
through fitting targets.  This is the causal closure principle: the
vertex parameterization is not fitted in the spectral domain; its
admissibility is constrained by the Stieltjes--Hilbert structure
that the parameterization inherits from the underlying resolvent---
the closure selects a parameter point, not a curve.  Third, the
moment set $\cI_r$ is the compression of the microscopic data: what
the free-fermion benchmark achieves through the Weingarten algebra,
the general framework achieves through the moment fields
Eq.~\eqref{eq:momfield} and the coupling cumulants, in the spirit of
extracting few effective parameters from the microscopic structure
rather than exporting a model-specific algebra.

\emph{From single- to correlated-resolvent self-consistency.}
The relation of this framework to the single-resolvent program of
Ref.~\cite{SRletter} is one of structure, not of order.  The
single-resolvent route is a causal ansatz for one self-energy that
turns the resolvent equation into a scalar parameter
self-consistency, while the correlated route built in this article
has the microscopic seed and the response kernel of
Sec.~\ref{sec:crossresponse} organizing the multi-resolvent content
\emph{before} the ansatz, so that the causal ansatz closes a
hierarchy, not a single equation, and the parameter self-consistency
acts on the causal manifold $\cM_{\rm causal}$
(Sec.~\ref{sec:causal}).  The upgrade is therefore not an increase
in vertex order but an increase in the structure of the
self-consistency itself: from a scalar causal function to a
constrained hierarchy of correlated spectral objects, with the
response hierarchy of Eq.~\eqref{eq:resphier} and the seed hierarchy
of Sec.~\ref{sec:num} as its two internal truncation axes.

The fourth remark completes the verification logic: the three gates
of Sec.~\ref{sec:num} test three \emph{different}
properties---fixed-point stability, causal admissibility, and
microscopic correctness---and no one of them implies the others,
each with its own instrument: the closure-stability margin
$\chi_{\mathrm{cl}}\approx0.04$--$0.07$ (Sec.~\ref{sec:num}), the
two-variable Hilbert-transform test at the real boundary
(few-percent residual, on the exact measure and on the closure
kernel alike), and the moment-field/Feshbach residuals.  The
distinction is not academic: in the single-resolvent program the
moment-closed spectrum possesses locally stable but non-causal
roots, and the selection among them is made by analyticity and
spectral positivity---the causality gate---not by the Jacobian
\cite{SRletter}.  The MRSC upgrade carries the same lesson one
resolvent multiplicity higher: a stable fixed point with the wrong
Hilbert dictionary is a parameterization of the correct equations in
the wrong function space.

Within this verification logic the two benchmark laboratories divide
the labor.  The deformed free-fermion family anchors the microscopic
content: the $\lambda=0$ anchor is reproduced, and the basis term
vanishes at the anchor by construction (the sector classes are exact
for the free-fermion benchmark).  The disordered Ising chain
of Sec.~\ref{sec:genbench} tests the generic machinery: the class
expansion Eq.~\eqref{eq:classseed} and its census are generic
(finite and algorithmically constructible for every fixed
$(r,\ell_{\max},d_{\max})$, Proposition~\ref{prop:finite}), and the
four-term closure with the level-1 cross-response kernel of
Sec.~\ref{sec:crossresponse} tracks the exact two-resolvent cumulant
at the sub-percent-to-percent level across the interaction grid with
no sector basis anywhere in the pipeline.

The benchmark also quantifies \emph{where the error of the closure
lives}: Check~B carries this message in the deformation
direction---the microscopic seed content becomes \emph{increasingly}
valuable away from the free anchor, with the $1.9\times$ advantage
at $\lambda=2^{-4}$---which is what makes the free-fermion benchmark
a controlled microscopic anchor rather than a trivial consistency
check, while the higher-cumulant back-reaction
$\varepsilon_{\rm high}$ of Eq.~\eqref{eq:deficitdecomp} is the one
term not separately measured here, retained as unresolved in
Sec.~\ref{sec:conclusion}.

The failed $\omega$-only experiment of Sec.~\ref{sec:num} is, in
retrospect, the most informative single negative result of the
construction.  Its
lesson is not that the causal-ansatz program fails but that
\emph{finite moments alone are insufficient: the ansatz space must
carry the connectivity geometry}.  The physical image behind the two
branches is single: \emph{the moments tell the resolvent where it
must go at infinity; the Feshbach feedback tells it how to get there
through the bulk}---local microscopic constraints fixing the
deep-plane asymptotics, global resolvent feedback supplying the bulk
route.

What is generic in the construction should be stated with the same
precision.  \emph{Generic already:} the Feshbach identity, the
Stieltjes representation, the moment-field identity, the ladder
organization, the DCA functional architecture, the
response-hierarchy organization of the closure deficit, and the
self-consistency principle---none of these assumes any ensemble
structure.  \emph{Not yet generic:} the pair
densities $\rho_m^{(2)}$, the explicit seed coefficients, and the
numerical fixed-point realization---these are the benchmark
realizations of the generic architecture.  Stated as the organizing
principle of the framework: the architecture is generic, the
microscopic basis is not.
What the benchmarks change is the microscopic realization; what
survives is the architecture of information recovery.  The load-bearing
objects---the Feshbach identity, the causal representation, the
moment-field identity, and the self-consistency equation---are defined
before any microscopic basis is chosen, and the basis-dependent residue
is carried entirely by the seeds, which inherit the coupling-order
structure of the Hamiltonian and thereby organize the closure deficit
into a named truncation axis (Sec.~\ref{sec:num}).  The
framework is thereby a closure between the exact hierarchy of
Ref.~\cite{MResR} and the on-shell statistics of
Refs.~\cite{Deutsch1991,Srednicki1994,Dalessio2016,Magan2016}---not
an exact solution of the former, not a fit to the latter.

The open directions mirror the hierarchy's own outlook
\cite{MResR}.  (i)~Closing Eq.~\eqref{eq:MRSCeq} beyond the moment
level at $\lambda>0$ with the deep-plane loop rows, the exact anchor
and sector-resolved response of Ref.~\cite{MResR} as boundary data.
(ii)~The numerical realization of the connectivity basis of
Sec.~\ref{sec:conn}: the response-hierarchy ladder and its
seed-resolution rungs are quantified in Sec.~\ref{sec:num}, and whether
the higher vertices admit a low-complexity continued-fraction
representation remains open, with the sector moment tensors of
Ref.~\cite{MResR} as the natural data.  The eigenstate-free
determination of $G^{(2)}_{mn}$ from the moment fields is delimited in
Sec.~\ref{sec:num} and Appendix~\ref{app:krylov}: it fixes the
coefficients to sub-percent in the asymptotic data, and the in-band
plateau is the rank of the joint weight (the connectivity coordinate)
rather than a missing moment or a violated (Stieltjes) positivity.  (iii)~The non-Gaussian coupling case, where
the pairwise Wick seeds are replaced by higher-order coupling
cumulants: Proposition~\ref{prop:nonGauss} fixes the $r=2$ instance and
Table~\ref{tab:e2nonGauss} verifies it on two test beds, so what
remains is the higher-body hypergraph census and the $r\ge3$ instance
of the same promotion.  (iv)~The unitary
and symplectic symmetry classes, testing which parts of the
parameter self-consistency are ensemble-universal.

On the observable side, three directions follow from
Appendix~\ref{sec:eth}: the boundary values of $\cC^{(r)}$ generate the
multi-time spectral-weight correlations and the nonlinear response
functions $\chi^{(r)}$ of the full-ETH program
\cite{Pappalardi2022,Pappalardi2025}, with the local free cumulants
of Ref.~\cite{VFP25} identified in Appendix~\ref{sec:LRI} as
window-coincident spectral cumulants; the higher matrix-element
correlations carried by $C^{(r)}$ enter out-of-time-order
correlators and non-Gaussian ETH matrix models
\cite{FoiniKurchan2019,Jafferis2023}; and whether the
response-kernel radius $\rho(K_X)$ distinguishes thermalizing from
localized regimes is posed here as a question, not a criterion: the
measured $\rho(K_X)\le0.044$ quantifies the feedback only on the
tested grid, against the background of many-body localization
\cite{NandkishoreHuse2015}.

Finally, the relation between the closure and the diagnostics
assembled by this article.  They meet at the conceptual interface only, as
delimited in Appendix~\ref{sec:LRI}: the fixed-point machinery of the
closure produces the closed multi-resolvent content, while the window
diagnostics consume the measured spectral data of the benchmarks
directly, as a second, independent consumer of the microscopic input;
what the two programs share is the microscopic input, not the closure loop,
so their difference is a test rather than an assumption.  What the
diagnosis establishes is delimited accordingly: a computable
finite-order instrument that measures and distinguishes window
defects---not a demonstration that any specific MRSC vertex or
feedback mechanism causes the observed rotational structure.  The
tracing of defects to channel correlations and microscopic feedback
requires an independent prediction or a response under parameter
variation and is left open here; the drift-kernel identity and the
transferability census themselves are carried out in
Appendix~\ref{sec:LRI}.  The common open layer of the two
programs is the status ladder of Appendix~\ref{sec:LRIstatus}: the within-window
isotropy step (iv) and the thermodynamic-limit control of the
inter-window correlations and errors of steps (v)--(vi) remain open
problems in the closed hierarchy, and no Haar-limit,
thermodynamic-scaling, or universal anisotropy-suppression statement
follows from the present results.

\section{Conclusion}
\label{sec:conclusion}

The question posed in Sec.~\ref{sec:intro} was how the irreducible
multi-resolvent vertices of a generic interacting system are determined from
microscopic data and closed into a finite set of causal parameters.  This
article answers it with a demonstrated route: three construction
layers---microscopic closure, response closure, and causal
self-consistency---assemble the constructive step;
Sec.~\ref{sec:errorprop} turns the construction into verification and
error localization; and the finite-order diagnostics of Appendix~\ref{sec:LRI}
form an adjacent statistical framework, not a fifth co-equal pillar.  In the
constructive step the coupling graph and its walk census of
Sec.~\ref{sec:conn} supply the class amplitudes $\Theta_{\mathfrak c}$ and
the coupling-disorder seed; the Feshbach feedback and cross-response
resummation dress the seeds into the cumulants, and the causal constraints
with the predictive fixed point $\boldsymbol\theta_*=F[\boldsymbol\theta_*;H]$
of Sec.~\ref{sec:predict} close the cycle on Hamiltonian-derived
one-resolvent and coupling input alone.  The seeds are the microscopic residue that no self-consistency
equation can generate; the ladder is the dressing that no seed can do
without.

Verification and error localization are built into the construction:
the closure deficit decomposes into named terms by construction level
(Eq.~\eqref{eq:deficitdecomp}); the isolatable contributions are
measured at the anchor, and the higher-cumulant back-reaction is
retained as an unresolved contribution.  The propagation bound
Eq.~\eqref{eq:errorbound} converts each upgrade into the same reduction of
the fixed-point error, the audited inverse-Jacobian norm
$\|(I-\bar J)^{-1}\|_2\approx1.4$ is weakly amplifying, and the term-by-term
hierarchy locates the dominant remaining deficit in the omitted microscopic
response channel, not the highest-order vertex---the anchor-row residual
falls $1.2\times10^{-3}\to1.2\times10^{-5}$ entirely through the $\kappa_3$
dressing.  On the free-fermion benchmark the Hamiltonian-only moments
reproduce the coincident-sector content of the two-resolvent cumulant at
the few-percent level ($5.4\%,2.8\%$); on the
disordered Ising chain of Sec.~\ref{sec:genbench} the blind protocol---with
every measured higher-resolvent and environment-side two-point quantity
withheld---predicts the exact two-resolvent cumulant to $0.35\%$--$0.97\%$
over the tested probes, and at $\lambda=0$ reduces to the exact identity
$\cC^{(2)}=\kappa_2[X,X]$ of the free-fermion sector algebra \cite{MResR},
reproduced to machine precision.  The response rung carries both: the
level-1 kernel restores the sign the DCA level gets wrong and reduces the
closure deficit by factors $4.1$--$31$ ($\varepsilon_0=4$--$24\%$ to
$\varepsilon_1=0.3$--$5.4\%$).

At the observable level this content is physical, not merely formal: the
two-resolvent sector supplies the spectral-weight correlations that enter
operator-specific ETH functions after projection,
and Appendix~\ref{sec:eth} illustrates the connection for a local operator in
the interacting benchmark, where the two-resolvent sector---evaluated at
the observable level with the exact two-resolvent data---provides a
systematic, order-one, sign-correct correction to the single-resolvent
baseline for the ETH variance of $X_1$---observable corrections absent
from any single-resolvent closure
\cite{Deutsch1991,Srednicki1994,Dalessio2016,Magan2016}; the end-to-end
transport of the closure prediction to that observable is structured in
two bounded steps---the projection transport, bounded by the content
share of Eq.~\eqref{eq:transportbound}, and the kernel transport, an
explicit linear operator whose bound is Eq.~\eqref{eq:bdybound}---with
the remaining error isolated as the model error of the sector image
Eq.~\eqref{eq:bdytransport} and the tail of the seed truncation
(Appendix~\ref{sec:eth}).  The window statistics that this projection
samples are examined separately in the LRI diagnostics of
Appendix~\ref{sec:LRI}, which supply the reference moments and measure
the finite-order defects; the two tests share the microscopic input
but establish different statements (the five-level evidence map of Appendix~\ref{sec:LRI}).

The scope boundary declared in Sec.~\ref{sec:intro} holds throughout.
The construction determines the finite-dimensional causal parameters
$\boldsymbol\theta$, while recovery of the full bulk correlator
additionally requires the mixed and higher-insertion terms of the
functional---a gap the free-fermion anchor quantifies in
Sec.~\ref{sec:num}---and the two tasks are kept distinct rather than
identified.  The skewness promise of Sec.~\ref{sec:intro} is likewise
closed by the $r=3$ sector: the parity-mixing content of $\cC^{(3)}$
is the skewness-generating channel, and the skewness selection rule of
Ref.~\cite{SRletter} follows as its corollary
(Sec.~\ref{sec:num}).

What remains open is structured by the status ladder of
Appendix~\ref{sec:LRIstatus}.  For the closure itself, the $r=1$ corner recovers
the Faddeeva single-resolvent closure of Ref.~\cite{SRletter}, $r=2$ is
fully quantitative, $r=3$ is carried as a demonstration of extensibility
rather than a complete closure (the parity structure and moment dictionary
of Sec.~\ref{sec:num} show the construction continues beyond $r=2$), and
arbitrary $r$ is a constructive formalism,
$\cC^{(r)}=\sum_{\mathfrak c}\Theta_{\mathfrak c}^{(r)}\cB_{\mathfrak c}^{(r)}$
with $\Gamma^{(r)}=\Gamma_{\rm coup}^{(r)}+\Gamma_{\rm res}^{(r)}[\cC]$ fixed
by the seeds of Eq.~\eqref{eq:classseed}; the deferred systematic layer
(class-resolved dressing at higher truncation order, the
connectivity-conditioned polynomial basis) leaves the model-specific seed
layer as the explicitly named input.  For the LRI diagnostics the
finite-order instruments of Appendix~\ref{sec:LRI} establish a working
diagnostic---a second, independent consumer of the microscopic input
rather than a downstream product of the closure---now with an exact
kernel identity (Eq.~\eqref{eq:K1}, machine-checked), a proven
identifiability boundary (Proposition~\ref{prop:nonident}), a
closure-boundary census (Tables~\ref{tab:censusA} and~\ref{tab:censusB}), and a quantitative
transferability boundary (Appendix~\ref{sec:transfer}).  They do not by
themselves establish the emergence of Haar statistics from a general
Hamiltonian.  That step is now split into a proved part and a named
hypothesis: Proposition~\ref{prop:LRIcond} shows that, within the
retained basis, the reference relations follow from the sector-collapse
property (H1) together with its companion weight condition (H1$'$) and
the window realization (H2).  In the free-fermion class (H1) and
(H1$'$) hold exactly; otherwise they are measured, not derived.
Deriving them from $H$ for a general chaotic Hamiltonian remains open,
with the item-by-item status on the ladder of
Appendix~\ref{sec:LRIstatus}.

\begin{acknowledgments}
This work is supported by NSFC Grant No. 12305035.
The numerical verification scripts and the machine-readable output data
underlying the figures, tables, and quoted numbers of this article are
archived at DOI~\href{https://doi.org/10.5281/zenodo.23497355}{10.5281/zenodo.23497355}.
\end{acknowledgments}

\appendix

\section{Cumulants of products}
\label{app:cum}

For centered random variables $\{Y_i\}$ grouped into $r$ families
$I_1,\dots,I_r$, the mixed cumulant of the family products is
\begin{equation}
\kappa_r\Bigl(\prod_{i\in I_1}Y_i,\dots,\prod_{i\in I_r}Y_i\Bigr)
=\sum_{\pi}\prod_{B\in\pi}\kappa_{|B|}\bigl(\{Y_i\}_{i\in B}\bigr),
\label{eq:cumprod}
\end{equation}
the sum running over set partitions $\pi$ of $\cup_jI_j$ into blocks
of size $\ge2$ whose block hypergraph connects the $r$ families
\cite{Lehner2004}.  \emph{Proof sketch:} the moment generating
function of the family products is
$\E\exp[\sum_j\lambda_j\prod_{i\in I_j}Y_i]$; taking the logarithm and
differentiating in all $\lambda_j$ produces exactly the connected
partitions, with single-element blocks excluded because
$\E[Y_i]=0$.  Eq.~\eqref{eq:cumprod} specializes to
Eq.~\eqref{eq:partexp} for $Y=\delta\cG$ and to the vertex-functional
expansion for products of couplings and resolvents.

The centering is a hypothesis, and in the applications it is met by
construction.  Every factor of a product is split into a deterministic
part and a centered fluctuation before the formula is used: the
resolvents through $\cR_a=\bar\cR_a^0+\Delta_a$, with $\Delta_a$ a
series in $\delta\cG_a=\cG_a-\E_H[\cG_a]$
(Eqs.~\eqref{eq:geom}--\eqref{eq:XD}), and the coupling products through
$V^{(\ell)}=\langle V^{(\ell)}\rangle+\delta V^{(\ell)}$.  The
deterministic parts contribute coefficients in front of the cumulants,
so every term of Eq.~\eqref{eq:vertexF} is a deterministic kernel built
from coupling means and $\bar\cR^0$ times a cumulant of centered
fluctuation fields.  The realization-centered difference
$\delta\cR_a=\cR_a-\cR_{\rm tr}$ of Eq.~\eqref{eq:dRdef} is \emph{not}
ensemble-centered---$\E_H[\delta\cR_a]\neq0$, because $\cR_{\rm tr}$
fluctuates with the realization---and Eq.~\eqref{eq:cumprod} is
therefore never applied to it: the resolvent cumulants are first
expanded through $X_a+\Delta_a$, which uses only multilinearity and the
translation invariance of cumulants, and only the resulting $\delta\cG$
monomials enter the product formula.  Where a mean is itself supplied
by a closure, the residual mean is a definitional shift and is carried
by the closure deficit $\delta F$ of Sec.~\ref{sec:errorprop}.

\section{The moment-field identity}
\label{app:moment}

For every channel $a$ and integer $m\ge1$ the spectral identity
$\sum_n p_n^a\lambda_n^m=\bra{\varphi_a}H^m\ket{\varphi_a}$ holds
exactly per realization, with
$p_n^a=|\braket{\varphi_a|\psi_n}|^2$ the participation weights; hence
$\sum_n\delta p_n^a\lambda_n^m=s_m(a)$ with $s_m(a)$ of
Eq.~\eqref{eq:momfield}, and the centered moments of
Eq.~\eqref{eq:moments} follow.  At $r=1$ this gives
Eq.~\eqref{eq:mf1} and the leading self-energy coefficient
$v_a=\sum_{b\neq a}|V_{ab}|^2$ used in Eq.~\eqref{eq:Xdeep}.

\section{Explicit \texorpdfstring{$r=3$}{r=3} ladder}
\label{app:ladder3}

Writing $X_i=X_{a_i}(z_i)$, $\Delta_i=\Delta_{a_i}(z_i)$,
Eq.~\eqref{eq:tower} at $r=3$ reads
\begin{equation}
\begin{aligned}
\cC^{(3)}_{abc}={}&\kappa_3(X_a,X_b,X_c)
+\kappa_3(\Delta_a,X_b,X_c)\\
&+\kappa_3(X_a,\Delta_b,X_c)+\kappa_3(X_a,X_b,\Delta_c)\\
&+\kappa_3(\Delta_a,\Delta_b,X_c)+\kappa_3(\Delta_a,X_b,\Delta_c)\\
&+\kappa_3(X_a,\Delta_b,\Delta_c)
+\kappa_3(\Delta_a,\Delta_b,\Delta_c),
\end{aligned}
\label{eq:ladder3explicit}
\end{equation}
and the vertex sector
\begin{equation}
\begin{aligned}
\kappa_3(\Delta_a,\Delta_b,\Delta_c)
={}&\bigl(\bar{\cR}_a^0\bar{\cR}_b^0\bar{\cR}_c^0\bigr)^2\Gamma^{(3)}_{abc}\\
&+\bigl(\bar{\cR}_a^0\bigr)^3\bigl(\bar{\cR}_b^0\bar{\cR}_c^0\bigr)^2
\bigl[2\Gamma^{(2)}_{ab}\Gamma^{(2)}_{ac}\\
&\qquad\qquad+\Gamma^{(2)}_{aa}\Gamma^{(2)}_{bc}\bigr]
+\text{perms}+\cdots,
\end{aligned}
\label{eq:ladder3vertex}
\end{equation}
the $r=3$ instance of Theorem~\ref{thm:ladder}
(Eqs.~\eqref{eq:ladder3explicit} and \eqref{eq:ladder3vertex}); in the random
free-fermion benchmark the $X$-tower is generically nonzero---the
bare propagator is the harmonic-mean object
$\bar{\cR}_a^0\neq\E_U[\cR_a]=\cR_{\rm tr}$---and carries part of the
deep-plane coefficient $K_{211}=\cM_{211}$ of
Eq.~\eqref{eq:moments} through the moment fields of
Eq.~\eqref{eq:Xdeep}, as verified in Sec.~\ref{sec:num}.

\section{Supplementary LRI calibration details}
\label{app:supplLRI}

This appendix collects the calibration and audit details supporting
the finite-order diagnostics of Appendix~\ref{sec:LRI}: the hand-checked
pairing-Gram reference, the distinct-index closed forms and the
partition-lattice audit, the two-scale fit of the order-two plateau,
the block-level exchangeable null, the free-fermion defect
calibration and drift--diagonal split, and the three-path
certification with order-by-order contents.

\subsection{Hand check of the pairing-Gram reference}
\label{app:D3hand}

For the reduced reference of Eq.~\eqref{eq:wgdistinct}, the
reduction admits an elementary hand check at $D=3$.  The Gram of
Eq.~\eqref{eq:wein} is the $3\times3$ matrix with diagonal entries
$9$ and off-diagonal entries $3$ (two and one connected components of
the union graph), whose inverse carries the weights
$\mathrm{Wg}(\mathrm{id})=2/15$ and $\mathrm{Wg}(t)=-1/30$; each
pairing contributes the product $D\bar q_{|c|}$ over the connected
components $c$ of its index-identification graph, $|c|$ counting the
distinct window indices the component covers, so that the
within-column pairing $(13)(24)$ splits into two one-point components
and is killed by $\bar q_1=0$, while the two surviving pairings
$(14)(23)$ and $(12)(34)$ each form a single two-point component and
contribute $D\bar q_2=3\bar q_2$, so
$M_2^{\rm Wg}=3\bar q_2(2/15-1/30)=3\bar q_2/10$, which at
$\bar q_2=1/2$ [$q=(1,-1/2,-1/2)$] returns $0.15$, the elementary
$O(3)$ moment computed directly from Eq.~\eqref{eq:wein}.

\subsection{Distinct-index closed forms and the partition-lattice audit}
\label{app:T4audit}

The distinct-index protocol of Appendix~\ref{sec:LRIdefect} uses the
falling weights
$\cF_2=1-\sum_n w_n^2$,
$\cF_3=1-3\sum_n w_n^2+2\sum_n w_n^3$,
$\cF_4=1-6\sum_n w_n^2+8\sum_n w_n^3+3\bigl(\sum_n w_n^2\bigr)^2
-6\sum_n w_n^4$, and the sums $\cT_r$ evaluated by set-partition
M\"obius subtraction of the coincident-index terms,
$\cT_2=\operatorname{tr}B^2-\sum_n w_n^2O_{nn}^2$,
$\cT_3=\operatorname{tr}B^3-3\sum_n w_n^2O_{nn}Q_n
+2\sum_n w_n^3O_{nn}^3$ with $Q_n=\sum_l w_lO_{nl}^2$, and
\begin{equation}
\begin{aligned}
\cT_4=\operatorname{tr}B^4
&+8\sum_n w_n^3O_{nn}^2Q_n
+2\sum_{n,l}w_n^2w_l^2O_{nn}O_{ll}O_{nl}^2\\
&+\sum_{n,l}w_n^2w_l^2O_{nl}^4
-4\sum_n w_n^2O_{nn}[OCO]_{nn}\\
&-2\sum_n w_n^2Q_n^2
-6\sum_n w_n^4O_{nn}^4,
\end{aligned}
\label{eq:T4eval}
\end{equation}
where $B_{nl}=w_n^{1/2}w_l^{1/2}O_{nl}$ and $C_{nl}=w_nw_lO_{nl}$ in
Eq.~\eqref{eq:T4eval};
each $\cT_r$ was verified against brute-force enumeration of the
distinct-index sum to machine precision.  The attribution is
exhaustive by construction: every index pattern belongs to exactly
one set-partition class, the all-distinct class carries M\"obius
weight $\mu=1$ and forms the reference numerator $\cT_r$, and each
coincidence class is counted once in $M_r$ and removed from $\cT_r$
with its M\"obius weight $\mu(\pi)$---integer lattice weights, not
probabilities ($\mu\in\{-6,+2,+1,-1\}$ at $r=4$)---so no pattern is
omitted or double-counted, and the M\"obius-inverted sums keep the
same unnormalized window-sum normalization as $M_r$, the division by
$\cF_r$ occurring only at the reference step
$M_r^{\rm micro}=\cT_r/\cF_r$.  The polynomial forms of $\cF_r$ are
exact counting weights by their combinatorial definition, and the
verification against direct distinct-index sums to $\sim10^{-11}$ is
a numerical check of the implementation---the residual is
finite-precision accumulation in the polynomial route, a different
regime from the $10^{-19}$ agreement of the three derivation paths
in Appendix~\ref{sec:LRIdefect}.

The coincidence-sector content the M\"obius
subtraction removes reads
$M_r-\cT_r=(1.38\times10^{-3},1.43\times10^{-4},9.7\times10^{-6})$
on the benchmark---$12.5\%$ of $M_2$, and exceeding the net moment at
$r=3,4$ where the distinct-index class is negative---and its
completeness was audited class by class: at $r=4$ all fifteen set
partitions of the coincidence sector are enumerated and each is
subtracted exactly once, the closed forms agreeing with the
brute-force sums to machine precision.  This is a finite-order,
partition-by-partition verification of the
moment identities of Appendix~\ref{sec:LRIdict}, not a proof of the
general theorem: the structural enumeration, the M\"obius-weighted
closed forms, and the direct index summation provide cross-checks
with partially independent implementations---each with its own loop
and bookkeeping logic---while sharing the underlying matrix $B$ and
the window weights.  The exhaustiveness itself is not a numerical
finding: it follows from the partition-lattice correspondence---each
index pattern realizes exactly one coincidence pattern, hence
exactly one set-partition class---established by definition and
combinatorial argument, the numerics corroborating only the
implementation of that bookkeeping, while the no-omission,
no-double-count statement rests on this partition-lattice
exhaustiveness, a combinatorial argument, not on numerical
agreement.

\subsection{Two-scale fit of the order-two plateau: identifiability and sensitivity}
\label{app:twoscale}

The plateau analysis of Appendix~\ref{sec:LRIdefect} rests on the
following fit and sensitivity details.  The unified model
$\varepsilon_2(N,D_I)=b_N+a_ND_I^{-p}$ with a common exponent fits
the sixteen $(N,\nu)$ entries at $p\approx1.55$ with a residual an order of
magnitude below the global two-parameter fits.  The identifiability of
the two scales is limited, and the plateaus $b_N$ are intercepts of
this specified model rather than directly measured quantities: the
exponent profile is shallow (the residual varies by a factor $6.5$
across $p\in[0.3,1.6]$, with a two-fold residual interval
$[1.0,1.6]$), the inferred $b_N$ rise with the fitted $p$---at $p=0.7$
the $N=10$ plateau falls to $3\times10^{-5}$ and the $N=13$ one turns
slightly negative, while at $p=1.55$ they read $2.9$, $2.6$, $2.3$,
and $1.6$ (in units of $10^{-4}$) at $N=10$--$13$---and the fit is endpoint-driven
(dropping $\nu=6$ moves $p$ to $1.60$, dropping $\nu=48$ to $1.25$,
the plateaus staying positive in both).  Realization-level resampling
enlarges the nominal standard errors ($2$--$3\times10^{-5}$) by a
factor of three to four: the $N=10$--$12$ plateaus stay positive at
the $90\%$ bootstrap level, while the $N=13$ bootstrap interval
$[-0.15,2.5]\times10^{-4}$ includes zero, and the exponent itself is
not resolved at the realization level (bootstrap $90\%$ range
$[0.8,1.6]$).  The system-size dependence trades off against the
fitted exponent---at fixed $p=1.1$ the plateaus are nearly
$N$-independent ($\approx2.0$--$2.2\times10^{-4}$), while the best
fit returns a log-log slope $\approx-2$---and growing the window at
fixed $N$ leaves the residual intact, an observed finite-size
phenomenon of the present data rather than an independent physical
mechanism: the size and the $N$-dependence of the residual term
remain substantially degenerate, and with $b_N$ and $p$ strongly
correlated, model misspecification can absorb a slowly decaying term
into the intercept.  The between-realization standard deviation of
$\delta_2$, $1.1\times10^{-3}\to1.5\times10^{-4}$ across the scan,
carries the estimation error $\xi_r^{\rm samp}$ of
Eq.~\eqref{eq:momentdecomp} together with the
realization-to-realization variation of the profile itself (the
cross term); a single realization resolves the sign of $\delta_2$
in only $58$--$92\%$ of realizations and $62$--$93\%$ of individual
windows, and at $\nu\ge24$ the
between-realization spread exceeds the within-realization
center-to-center spread, the profile variation across disorder draws dominating the
center-to-center variation.

\subsection{Block-level exchangeable null: construction and calibrations}
\label{app:blocknull}

The block-level null of Appendix~\ref{sec:LRIdefect} is implemented as
eight independent blocks of twelve five-window Gaussian draws
recentered per realization, the randomization unit being the five
window values of one realization permuted inside its block, and the
$95\%$ single-pair band $[-0.727,0.371]$ is fixed by $2\times10^{5}$
null draws; the null single-pair distribution has mean $-0.24$, sd
$0.29$, and the observed block-level scatter (observed sd $0.017$
versus null sd $0.008$, $z\approx2.0$) matches the null mean $-0.24$
within its block-level scatter.  The observed per-block counts of
pairs outside the $95\%$ band
are $[2,2,3,0,2,2,1,1]$, and two further calibrations that
match the null's block-level variance and block-level mean to the
observed blocks leave the excess intact ($P=9\times10^{-4}$ and
$8\times10^{-4}$).  The observed distribution is broader-tailed than
the null in both directions: sd $0.373$ versus $0.287$, excess
kurtosis $-0.75$ versus $-0.23$, and a $0.275$ fraction of $|x|>0.5$
versus $0.203$.  Neither alternative null models the intra-block
dependence itself---at twelve realizations per block no faithful
parametric model of it is available---so the excess cannot be
attributed to scale or location structure alone.

\subsection{Free-fermion defect calibration and the drift--diagonal split}
\label{app:ffcalib}

The defect values quoted in Appendix~\ref{sec:LRIdefect} are obtained as
follows: evaluated against the window-independent reference of
Eq.~\eqref{eq:wgdistinct} at the benchmark's parameters ($D=91$,
$D_I\approx23$, band center), the defect reads
$\delta=(-5.2\times10^{-4},+1.1\times10^{-6},+5.6\times10^{-7})$
[$\varepsilon=(5.2\times10^{-4},1.1\times10^{-6},5.6\times10^{-7})$].
In moment units the defect is small in absolute terms; relative to
the reference it stands at $4.9\%$ of $|M_2^{\rm Wg}|$, $4.3\%$ of
$|M_3^{\rm Wg}|$, and $65\%$ of $|M_4^{\rm Wg}|$.  The large
fourth-order ratio is amplified by the small reference moment and
must not be read as a correspondingly large absolute defect or as a
stronger fourth-order LRI violation: the absolute and relative
defects answer different questions---the former compares drift
scales, the latter measures the deviation against its own
target---so the scan quotes absolute moment-unit defects and
compares them within each order across windows and system sizes,
never mixing relative metrics across orders or treating the three
orders as a common scale.
The residual is a \emph{window-content drift}, not an isotropy
violation: the reference is built from the spectrally flat $\bar q_r$
of Eq.~\eqref{eq:qbar}, while the measured moment weights the falling
$r$-content by the window profile.  The defect closes per order as
$\delta_r=\delta_r^{\rm drift}+\delta_r^{\rm diag}$, with
$\delta^{\rm drift}=(-7.6\times10^{-4},-5.2\times10^{-6},+9.3\times10^{-7})$
the difference between the window-resolved and the flat falling
$r$-content---the band-center $r=2$ content falls $7\%$ below its
flat value---and
$\delta^{\rm diag}=(+2.3\times10^{-4},+6.3\times10^{-6},-3.7\times10^{-7})$
the deviation of the realized diagonal content from the Haar-model
content the reference assumes; the decomposition is verified per
order to machine precision.  Neither component is an isotropy
defect: LRI requires only the in-window isotropy part of
$\delta M_r^{\rm sys}$ to vanish, and separating it from the drift
requires a window-resolved conditional reference---a window-indexed
Weingarten contraction built from the window's own $q$-moments---posed
as an open item in Appendix~\ref{sec:LRIstatus}.  The benchmark thereby
fixes the calibration baseline---the defect floor the reference
choice imposes,
\begin{equation}
\begin{aligned}
\varepsilon_r^{\rm floor}\equiv\varepsilon_r\big|_{\rm bench}
&=\frac{|M_r^{\rm micro}-M_r^{\rm Wg}|_{\rm bench}}
{1+|M_r^{\rm Wg}|_{\rm bench}}\\
&=(5.2\times10^{-4},1.1\times10^{-6},5.6\times10^{-7}),
\end{aligned}
\label{eq:deffloor}
\end{equation}
the table's own normalized defect of Eq.~\eqref{eq:LRIdefect}---the
absolute difference $|M_r^{\rm micro}-M_r^{\rm Wg}|$ divided by the
stabilization denominator $1+|M_r^{\rm Wg}|$---evaluated at the
benchmark's parameters ($D=91$, $D_I\approx23$, band center,
recentered weights), so that the floor and the table entries are one
quantity with one normalization and the ratios below are
dimensionless; it is computed from the benchmark's own measured
moments, its own window weights, and the same centered $q$-moments
of Eq.~\eqref{eq:qbar} as the table---the identical pipeline
applied to a known system---a computed calibration, not a fitted
tolerance and not a statistical lower bound on any defect,
comprising both the drift and the diagonal components of the split.
Against this floor the Ising entries of
Table~\ref{tab:isingdefect} are read as the ratios
$|\varepsilon_r|/\varepsilon_r^{\rm floor}$, with the baseline
strictly positive at every order so that each ratio is well
defined: the small-window $r=2$ entries
exceed the floor by up to a factor $2.1$ and the large-window ones
fall below it, at $r=3$ only the smallest-window entries reach the
floor, and at $r=4$ all entries sit well below it---the transfer of
the benchmark floor to the Ising parameters is indicative rather than
exact.

\subsection{Three-path certification and order-by-order contents}
\label{app:threepath}

The calibration is a cross-check of independent
paths rather than a closed algebraic loop: the raw moments $M_r$
(path~1, direct traces), the cumulants $K_r$ (path~2, the
moment-field M\"obius inversion with its own coefficient system), and
the corrections $c_r$ (path~3, the explicit $\sigma_0$-power
polynomials of Eq.~\eqref{eq:coincdict}) are independently
determined: the path-2 inversion coefficients follow from the
M\"obius inversion of the sector moments, and the path-3 coefficient
set follows from a structural enumeration of the merge patterns at
zero $K$- and $M$-input, reproducing the merge combinatorics of the
Weingarten block structure, so the two coefficient systems are fixed
by separate derivations and only their numerical evaluation shares
the realized $K_r$ values; the three paths meet at machine precision,
$M_r=K_r+c_r$ with residuals below $10^{-19}$, which certifies the
implementation self-consistency of the identification, not the
independence of the derivations.  At the benchmark center the
contents read $|K_1|=0.095$, $|K_2|=1.0\times10^{-2}$,
$|K_3|=7.0\times10^{-5}$, $|K_4|=2.0\times10^{-5}$ with
$c_2=5.8\times10^{-4}$, $c_3=1.9\times10^{-4}$,
$c_4=2.9\times10^{-5}$ ($\sigma_0=6.3\times10^{-2}$), and the
relative sensitivities $|d\ln c_r/d\ln K_r|$ are $O(1)$---$2.0$ for
$c_2\leftarrow K_1$, $0.98$ for $c_3\leftarrow K_2$, and $2.01$,
$0.058$ for $c_4\leftarrow K_2,K_3$---so no small-denominator
amplification is present.
The order-by-order contents at the benchmark center read
$c_2/K_2=+0.055$, $c_3/K_3=-2.8$, $c_4/K_4=-1.5$: at order two the
correction and the cumulant share their sign, while at orders three
and four they are opposite in sign and the correction
order-dominates---$K_3$ and $K_4$ are anomalously small, so the
finite-$D$ correction is a small \emph{moment} correction but a large
\emph{cumulant-relative} correction, and the moments reconstruct
exactly even where the cumulant-relative corrections exceed unity.
The same sign structure persists across $D=55$--$105$ ($N=11$--$15$,
$c_2/K_2>0$, $c_3/K_3$ and $c_4/K_4$ negative with
$|c_3/K_3|=1.2$--$2.8$); these values are calibration points, and no
$D$-scaling of the corrections is inferred from them.  The estimation
error is a
distinct object---for a Haar orthogonal window model at the same
dimension $D_I=91$ (observable $q=D_I^{-1}\sum_n O_{nn}^4$) the
single-draw fluctuation rms is $\approx0.33$ of the reference value
(scale $D_I^{-1/2}$) for this observable and normalization, fifteen
times larger than the $\approx2\%$
deterministic finite-$D$ correction (scale $D_I^{-1}$); the
deterministic finite-$D$ correction remains after ensemble
averaging, although its relative magnitude vanishes as $D\to\infty$
in this toy model, and averaging removes only the estimation
fluctuation---the fifteen-fold separation is specific to this
observable, normalization, and ensemble, not a general law.  At
order four both routes of Appendix~\ref{sec:LRIdict} reconstruct the
moment $M_4$ identically, while the cumulant assignments differ by
the route split of Eq.~\eqref{eq:r4split},
$K_4^{\rm NC}-K_4=\Delta_4=1.5\times10^{-5}$ at the benchmark's band
center.  The order-4 reading is therefore a
dictionary-consistency calibration, not an LRI violation: it
quantifies the second-orientation content the merge route omits,
Eq.~\eqref{eq:r4orient}, and it
vanishes identically when $K_2=0$, when $M_2\to0$, or in the
separated limit $\sigma_0\to0$.  The probe reads exactly zero at
$r=2,3$ at the moment level, and its $r=4$ content is fully
accounted for by Eq.~\eqref{eq:r4split}---on the benchmark the probe
is a calibrated instrument, and a residual $\delta_r$ beyond these
known contents and the calibrated drift floor of
Appendix~\ref{sec:LRIdefect} signals systematic non-Haar content.  Three levels of
the order-4 reading should be kept apart: at the \emph{algebraic}
level $\delta_4=0$ is an identity, both routes reconstructing the
same moment; at the \emph{dictionary} level $\delta_4$ equals the
cumulant reassignment \eqref{eq:r4split}, a known, exactly calibrated
bookkeeping content; and only at the third, \emph{real-system} level,
where the sector algebra is approximate and $M_4^{\rm micro}$ is
measured on physical eigenstates, does a residual $\delta_4$ beyond
these contents constitute LRI evidence.  The defect scan below is a
third-level test.

\section{Drift-kernel data and the transferability archive}
\label{app:driftdata}

This appendix collects the machine checks, the pre-registered
protocols, the gate compendium, and the verdict registry behind the
instruments of Appendices~\ref{sec:driftkernel} and \ref{sec:transfer} and
the census of Tables~\ref{tab:censusA} and~\ref{tab:censusB}.

\emph{Kernel identity.}  On the ring TFIM of Sec.~\ref{sec:genbench}
at $N=10$ ($D=1024$), $\nu=24$, band center, the right-hand side of
Eq.~\eqref{eq:K1} equals $2.527306510\times10^{-4}$ and the
moment difference on the left equals $2.527306476\times10^{-4}$
(relative difference $1.3\times10^{-8}$), with
$\sum_{n\neq m}K_{nm}$ at the $10^{-18}$ level (zero mass,
machine precision).  The flat and diagonal references are the closed
single-state forms $M_2^{\rm flat}=\bar q_2/(D-1)$ and
$M_2^{\rm Wg}=D\bar q_2/[(D-1)(D+2)]$, whose difference
$\delta_2^{\rm diag}=2\bar q_2/[(D-1)(D+2)]$ carries no pair
content.

\emph{Identifiability example.}  For $D=4$ let $P$ and $\tilde P$ be
the two fixed-point-free involutions on four sites, acting on the
centered channel by conjugation.  Both realize identical single-body
data: spectrum $\{+1,+1,-1,-1\}$, zero diagonal, $\bar q_2=1$,
$\bar q_3=0$, $\bar q_4=1$, $M_2^{\rm flat}=1/3$,
$M_2^{\rm Wg}=2/9$.  With any window weights satisfying $w_1=w_2$
the drift of Eq.~\eqref{eq:K1} reads $+2.160240837\times10^{-1}$
for $P$ and $-1.080120419\times10^{-1}$ for $\tilde P$---opposite
signs, ratio exactly $-2$.  The same input set therefore
underdetermines the drift; the proposition asserts the
underdetermination for this input set, not for every conceivable
single-state input.

\emph{Transferability protocols.}  All transferability claims are
pre-registered blind tests: the training set (six realizations), the
validation set (two), the test set (twelve fresh realizations, never
touched in training), the bin width ($0.5$, selected on validation),
the seven band positions
$E/\|H\|=0.10,0.30,0.40,0.50,0.60,0.70,0.90$, and every gate were
frozen before the test runs.  The profile object is the
two-dimensional binned mean
$\bar f(|E_n-E_m|,(E_n+E_m)/2)$ of the pair content over the
training realizations, and the prediction is
$\hat\delta_2=\sum_{n\neq m}K_{nm}\bar f(\cdot)$.  Reported numbers:
aggregate signed error $8.9\%$ (factorized baseline $98.5\%$);
per-center aggregate ratios $[0.73,1.27]$ at all seven centers;
sign accuracy $59/84$; per-realization sign patterns
$(7,7,7,7,7,7,7,0,0,7,1,2)$ hits out of seven over the twelve test
realizations; realization-level bootstrap ($B=20000$, resampling
realizations) confidence intervals: sign rate $[45.2\%,92.9\%]$,
five-center shape correlation $[-0.1,1.0]$, signed error
(median) $0.364$; unsigned per-cell error $91.4\%$.  The variance
audit: independent-pair variance model underestimates the
between-realization standard deviation by $5$--$8\times$; the
global-amplitude model by $250$--$500\times$; the diagonal share of
the realized variance is $0.102$.  The coupling scan: profile shape
transfer to $g=0.5$ and $g=2.0$ with five-center correlations
$0.8$, $0.6$ and signed errors $9.9\%$, $19.3\%$ against
recalibrated factorized baselines $100.8\%$, $97.3\%$; the full
protocol replicated at $g=2.0$ passes every gate (sign $78.6\%$,
ratios $0.80$--$1.34$, correlation $0.8$, signed error $9.7\%$);
at $g=0.5$ the mid-band ensemble signal $1.206\times10^{-4}$ falls
below the realization-level scatter ($\sim10^{-3}$) and the
replication fails.  The conditional-LRI scan:
$R_2=M_2^{\rm micro}/M_2^{\rm Wg}=0.926,1.397,2.073$ and
$\eta_2=|R_2-1|=7.4\%,39.7\%,107.3\%$ at $g=0.5,1.0,2.0$.

\emph{Census sources.}  Each row of Tables~\ref{tab:censusA} and~\ref{tab:censusB} cites a
pre-registered protocol with frozen gates; the gate definitions, the
tested values, and the verdicts are collected in
Appendix~\ref{app:gates} below.  The census is the compressed
form, and the distinction between a narrow refutation of the tested
closure and a non-refutation of adjacent claims is stated row by row
in the table.  Two check-design
limitations apply to the moment rounds behind the census: the
Monte-Carlo $\sigma$-scaled engine self-check is ill-posed for
near-degenerate anchors (their MC variance collapses), so the engine
is validated by the second-moment channels ($\le3.2\%$) and the
frozen same-seed replication ($\le0.27\%$); and the $M_4$ median is
heavy-tailed with $20$--$40\%$ sampling drift at $n_{\rm real}=8$,
so $M_4$-based ratios are reported as medians with their
pre-registered band, not as point estimates.

\subsection{Gate compendium and verdict registry}
\label{app:gates}

Each closure route of Tables~\ref{tab:censusA} and~\ref{tab:censusB} was tested against a
frozen gate; the gate definitions, the tested values, and the
outcomes are collected in Table~\ref{tab:gates}.  Every gate was
pre-registered before the corresponding run, and the two check-design
limitations stated above apply to the moment rounds.

\begin{table*}[t]
\caption{Frozen gates and outcomes behind the closure census of
Tables~\ref{tab:censusA}--\ref{tab:censusB} and the moment rounds of
Appendices~\ref{sec:driftkernel} and \ref{sec:transfer}; the lettered
verdicts are defined in the registry below.  Each outcome carries the
decisive number; the full bands and tested values are quoted in the
corresponding subsection.}
\label{tab:gates}
\footnotesize
\begin{ruledtabular}
\begin{tabular}{p{4.0cm}p{6.0cm}p{7.2cm}}
\hline\hline
route & frozen gate & outcome (decisive value)\\
\hline
factorized pair content & drift reproduced at the D2 level (band-center magnitude, displaced-center signs) & fail: $21.8\%$ of the drift with wrong edge signs (free fermion); $6.4$--$33.4\times$ too small mid-band, edge ratios $-16,-25$ (Ising)\\
conditional reference at $D_I$ & window-approximated moment & fail (narrow): $M_2^{\rm cond}$ off by $4\times$, drift by $57\times$ with opposite sign\\
colored block-Haar & reproduction $<1\%$; scaling $<10^{-12}$; $M_2\in[0.9,1.1]$; $s_{\rm nd}<50\%$; trimmed/full $\in[1/3,3]$ & pass at $k\le2$ (S2 = A); fails to extend: $M_4/M_4^{\rm BH}=7890/738/177$\\
CPA cavity dressing & $m_{r8}/\mathrm{exact}\in[0.85,1.15]$ all windows; non-convergence $\le5\%$ & fail: best $0.7513$ ($\lambda=3/32$); non-convergence $9.11\%\to96.55\%$\\
irreducible-ring closure & $\mathrm{med}(\Gamma_1/\Delta\Sigma)\in[0.5,1.5]$; $\ge10\%$ of $\mu_0$ with $\ge2\times$ reduction & fail (escape hatch): $0.9416\to0.3135$; reduction $6.8\%/3.4\%$\\
Route B spectral bound & per-window coverage $100\%$; $\mathrm{med}\le3$; $\rho<1$ & vacuous (declared unclosed): $\rho<1$ on $0.05\%$/$0\%$ of samples\\
diagonal resolvent closure & $\varepsilon_1$ median $<0.5$; $R_{4p}\in[0.85,1.15]$ & fail (structural): $\varepsilon_1=0.981/0.983/0.977$; $R_{4p}=0.0075/0.0026/0.0046$\\
spectrum-level parent matching & rate $\ge0.8$; matched subset $R_{4p}\in[0.85,1.15]$ & fail (rates): $3.0/0.7/0.4\%$ bare; pass on subset: $1.008/0.998/1.003$\\
static single-state variables & isolation $\rho>0.3$, collapse $\ge4/6$, ratio $\in[0.5,2]$ & fail: $\rho=0.148/0.123/0.052$, collapse $0/6$\\
two-state covariance & $M_2\le2$; $s_2\in[0.5,1.5]\wedge\rho_2>0.3$; $|b_c-(b_{M4}-2b_{M2})|\le0.15$ & fail as mechanism: $M_2=2.000$ flat, $\Sigma_{xy}\approx0$\\
single-realization drift (T4) & sign $\ge70\%$; ratios $\in[0.5,2]$ at $\ge5/7$ centers; shape $\rho\ge0.6$; $e_{\rm rel}$ below the D2 baseline & D: $70.2\%$; $7/7$ ($0.73$--$1.27$); shape $0.5$; $8.9\%<98.5\%$; $2/12$ global sign flips\\
realization-level variance (T4b) & $\hat\sigma/\sigma_{\rm emp}\in[0.5,2]$ at $\ge5/7$; $|\hat P-P_{\rm emp}|\le0.25$ at $\ge4/5$ & C: $0/7$ ($5$--$8\times$ low), $0/7$ ($250$--$500\times$ low); diagonal share $0.102$\\
conditional LRI & $R_2\in[0.95,1.05]$; fourth order $\in[0.9,1.1]$ & C: $R_2=0.926/1.397/2.073$; fourth order $7890/738/177$; A, B pass at the aggregate level ($g=2.0$)\\
\hline\hline
\end{tabular}
\end{ruledtabular}
\end{table*}

\emph{Verdict registry.}  The lettered verdicts used in the gate
compendium and the transferability section: the colored block-Haar
audit S2 = A; the transferability test T4 = D (aggregate transferable,
realization-level not); the realization-level variance audit T4b = C;
the conditional-LRI scan = C; the finite-size scan T5 = A; the
cross-operator test T6 = B for $X_1X_2$ / A for $X_4$.  For the
conditional-LRI scan, proposition A (the closed hierarchy expresses
the specified statistic exactly), proposition B (under explicit
conditions the deviation has a computable bound or asymptotic
estimate), and proposition C (under stronger conditions the deviation
vanishes, yielding LRI at that order) are evaluated independently:
A and B hold at the aggregate level, C fails.  The unified-framework
exit condition---the level-3 cross-sample concentration prediction
passes at both tested couplings together with the S2 = A audit---is
satisfied.

\subsection{Size-scan and cross-operator protocols}
\label{app:sizescan}

\emph{Finite-size scan (T5).}  On the ring TFIM at $N=12,14$,
$\lambda=3/32,3/4,2$, $q=0.3$, $\nu=24$, $n_{\rm real}=8$, with the
anchor-pair protocol of Appendix~\ref{sec:driftkernel}, three frozen
hypotheses.  H1 (stability): the $M_4/M_4^{\rm BH}$, $M_{22}^c$, and
$M_2$ ratios lie in $[0.5,2]$ at every cell---all $18/18$ cells pass,
with ratio bands $0.675$--$1.07$ ($N=12$) and $1.01$--$1.63$
($N=14$).  H2 (structure): $M_4/M_4^{\rm BH}\ge100$---$8422/565/119$
($N=12$), $12840/1095/283$ ($N=14$), so the excess grows with $N$
(at $\lambda=3/32$: $7890\to8422\to12840$); and
$M_2\in[0.9,1.1]$---$1.0003$--$1.0016$.  H3 (ordering): the
$\lambda$-ordering of the excess is preserved---$8422>565>119$
($N=12$) and $12840>1095>283$ ($N=14$) at
$\lambda=3/32>3/4>2$.  Verdict: T5 = A.

\emph{Cross-operator test (T6).}  At $\lambda=1$, $N=10$,
$q\in\{0.10,\dots,0.90\}$, $n_{\rm real}=12$, the pair-flip operator
$X_1X_2$ (sites 0--1) and the single-flip operator $X_4$ (site 3).
H4 (formula): the drift-kernel formula of Eq.~\eqref{eq:K1} holds at
machine level for each operator's drift---$1.11\times10^{-16}$
($X_1X_2$), $1.67\times10^{-16}$ ($X_4$).  H5 (sign structure): the
U-shaped sign structure passes for $X_4$; it fails for $X_1X_2$,
whose $q=0.9$ edge is slightly negative, a block-level drift
$\sim10^{-7}$ against the all-$D$ $X_1$ drift
$2.4\times10^{-4}$---the noise floor of the check-design limitations
of Appendix~\ref{app:gates}.  H6 (excess): the fourth-order excess
over the block-Haar reference is $\ge100$ at every $q$ for both
operators---$165$--$344$ ($X_1X_2$), $166$--$349$ ($X_4$).  Verdict:
T6 = B ($X_1X_2$) / A ($X_4$).

\section{Supplementary benchmark verification details}
\label{app:benchver}

\subsection{Free-fermion anchor attribution and deformation data}
\label{app:benchfree}

\emph{Check A (anchor precision) fails, with clean attribution.}  The genuinely
nonlinear kernel moves the $\lambda=0$ anchor by only $0.3$ points
($20.1\%\to19.8\%$), because the ladder is $\bar{\cR}^4$-suppressed:
at the exact amplitudes $\bar{\cR}^4\Gamma_{\mathrm{CC}}^{\rm
lad}(\boldsymbol\theta_{\rm exact})=-4.7\times10^{-5}-2.8\times
10^{-5}\mathrm i$ against $\cC^{(2)}_{\rm exact}=2.7\times10^{-3}
+5.7\times10^{-4}\mathrm i$ at $z_1$ ($\sim2\%$).  The dominant
anchor deficit is the diagonal-fluctuation cross term: the DCA
linear system of Sec.~\ref{sec:predict} gives
$2\kappa_2(X_a,\Delta_b)^{\rm DCA}=-1.9\times10^{-3}-5.3\times
10^{-4}\mathrm i$ while the exact partition
$\cC^{(2)}=\kappa_2[XX]+2\kappa_2[X,\Delta]+\kappa_2[\Delta\Delta]$
of Eq.~\eqref{eq:XD} returns $2\kappa_2[X,\Delta]=-4.5\times
10^{-4}+1.1\times10^{-5}\mathrm i$---nearly fivefold too large in
modulus, with a large imaginary part where the exact value is
real-dominated, and this error carries the bulk of the anchor
deficit through the census projection.  The cross-response kernel of
Sec.~\ref{sec:crossresponse} attributes the deficit term by term.
The level-1 closure
$\Xi^{(1)}=(1-K_X)^{-1}(\Xi^0+T^{(1)})$ of
Eq.~\eqref{eq:XiLevel1} moves the anchor entry to
$2\kappa_2(X_a,\Delta_b)^{(1)}=-2.0\times10^{-4}+1.3\times
10^{-4}\mathrm i$, an error of $2.8\times10^{-4}$ in modulus against
the DCA's $1.5\times10^{-3}$---a $5.5\times$ reduction---and the
anchor-row residual of the kernel equation itself drops from
$1.2\times10^{-3}$ to $1.2\times10^{-5}$, two orders of magnitude,
\emph{entirely} through the $\kappa_3$ dressing $T^{(1)}$: the
per-pair $X$--$X$ covariance and per-row loop weights alone (level
$1$ without $T^{(1)}$) leave the entry at
$-2.0\times10^{-3}-6.0\times10^{-4}\mathrm i$, so the DCA's flat
closure is exactly the level-0 truncation of
Eq.~\eqref{eq:XiLevel1} and its fivefold deficit is the dropped
$\kappa_3(X_a,(\bar{\cR}_b^0)^2,X_c)$ content, not the vertex.  The
residual level-1 error is carried by the neighbor rows, whose own
residuals stay at $2.2$--$2.8\times10^{-4}$ per row: the dropped
distance-2 channels and the $\Delta$-reentry of the $\kappa_3$
tower---the channel-space truncation shared by every closure level.
The kernel content is $\lambda$-insensitive to a few percent across
the grid (it is carried by the diagonal content), so this
attribution holds at every $\lambda$.  The bottleneck at
$\lambda=0$ is therefore the $\kappa_3$-dressed diagonal-fluctuation
cross response, whose leading discrepancy is accounted for by the
level-1 kernel of Eq.~\eqref{eq:XiLevel1}, with the channel
truncation identified as the next upgrade axis---not the
nonlinearity: the cycle-ladder
kernel is implemented exactly and its effect is quantified as
subleading.  \emph{Check B (deformation advantage) passes at the stated $1.9\times$, and is closure-level insensitive.}  The moment branch degrades monotonically,
$0.5\%,8.6\%,16.6\%,29.9\%$, while the D1 fixed point tracks
$19.8\%,15.4\%,13.2\%,15.8\%$---the two branches cross between
$\lambda=2^{-6}$ and $2^{-5}$, and at $\lambda=2^{-4}$ the fixed
point is $1.9\times$ closer to the exact sector amplitudes than the
moment branch ($1.8\times$ already at the D0 level), where its
$\theta_2$ matches the exact value to $1\%$ against the moment
branch's $27\%$: \emph{the Feshbach feedback restores the
interaction-induced content the moments cannot see, without any
measured input}---the
microscopic-to-self-consistent route
$H\to F[\boldsymbol\theta;H]\to\boldsymbol\theta_*$ of
Sec.~\ref{sec:predict}, demonstrated.  The fixed-point amplitudes
are therefore not fitting coefficients: through
Eq.~\eqref{eq:sectoransatz} they are the parameters of the two-fold
Stieltjes kernel itself, and the closure output is a causal spectral
kernel constructed from Hamiltonian data alone---the test
below checks precisely this causal form at the real boundary.  The
near-independence of the
Check-B ratio on the closure level is itself diagnostic: the
$\lambda>0$ improvement is carried by the Hamiltonian-only seed
content (the sector profiles sharpen under the deformation), while
the $\theta$-dressed kernel---exact but $\bar{\cR}^4$-suppressed at
this benchmark scale---acts at the percent level.  The bulk
normalization of
the projected map is $\cC^{(2)'}[\boldsymbol\theta_*]\approx
0.28\,\cC^{(2)}_{\rm exact}$ at the probes for the D1 maps (D0:
$0.16$), and the fraction is $\lambda$-stable ($0.273\to0.281$
across the grid): the omitted mixed
terms and higher insertions carry the
remaining content, and the sector sums' large-cancellation
structure ($\cG2_0\approx-259$, $\cG2_2\approx-2.4$ at $z_1$)
makes the amplitudes the robust output of the fixed point while the
bulk cumulant is the delicate one---the truncation content of the
projected map, quantified rather than assumed.

\subsection{Response-hierarchy rung data on the Ising chain}
\label{app:benchrung}

The response hierarchy of
Eq.~\eqref{eq:resphier} is realized term by term on this laboratory.
With the level-$L$ deficit realized as the normalized closure
deviation at the exact input,
$|C'_L(\boldsymbol\theta_{\rm ex})-C_{\rm ex}|/|C_{\rm ex}|$, the DCA
level sits at $4.0$--$23.6\%$ across the grid
($5.1\%,4.0\%,4.5\%$ at $\lambda=2^{-4}$;
$15.6\%,13.3\%,14.3\%$ at $\lambda=2^{-2}$;
$23.6\%,21.5\%,22.2\%$ at $\lambda=2^{-1}$), the first response rung
removes it by factors of $4.1$--$31$
($\varepsilon_1=0.4\%,0.6\%,0.7\%$ at $\lambda=2^{-4}$;
$0.8\%,2.0\%,2.6\%$ at $\lambda=2^{-2}$;
$4.6\%,4.6\%,5.4\%$ at $\lambda=2^{-1}$), and the $\Delta$-reentry
rung of Eq.~\eqref{eq:Xieq} leaves
$\varepsilon_2\approx\varepsilon_1$ without monotone improvement:
its measured contribution
$2\Xi^{(2)}-2\Xi^{(1)}=-7.8\times10^{-5}+3.0\times10^{-5}\mathrm i$
at $z_1$ is of the same order as the remaining kernel deficit
$2(\Xi_{\rm ex}-\Xi^{(1)})=+8.9\times10^{-5}-1.5\times
10^{-5}\mathrm i$ but of the wrong sign---the second rung is
exhausted on the retained channel system $\{a_0,f_1,\dots,f_N\}$.
The distance-2 diagnosis is decisive: promoting the representative
double-flip channel $f_1\oplus f_2$ to a full row of the response
system moves the kernel by
$2\Xi^{(2d)}-2\Xi^{(1)}=-3.2\times10^{-7}-9.0\times10^{-8}\mathrm
i$, and the full $N(N-1)/2=36$-channel double-flip class by
$2\Xi^{(2f)}-2\Xi^{(1)}=-1.2\times10^{-5}-3.3\times10^{-6}\mathrm
i$---$13\%$ of the remaining kernel deficit and of the wrong
direction---so the star truncation $\{a_0,f_1,\dots,f_N\}$ is
already adequate for the two-resolvent cumulant at this coupling:
the first response rung identifies the dominant missing microscopic
information, and the remaining error is associated primarily with
the second-order class seeds within the retained channel basis,
rather than with the omitted response channels or with an
uncontrolled increase in vertex order.  The same data verify the resummation itself:
the response-kernel spectral radius
$\rho(K_X)=0.007,0.014,0.026$ at
$\lambda=2^{-4},2^{-3},2^{-2}$ (zero at $\lambda=0$; $0.012,0.024,0.044$ with the full distance-2 class) lies far below
one at every probe, so the Neumann series of
Eq.~\eqref{eq:XiResum} converges---the Ising-side companion of the
closure-stability margin $\chi_{\rm cl}\approx0.04$--$0.07$ measured
on the free-fermion map.

\subsection{Hilbert-form boundary test}
\label{app:benchhilbert}

\emph{Causal-form test at the real boundary.}  The dispersion
relation Eq.~\eqref{eq:disp} makes the closure output a testable
object: is the fixed-point kernel
$\cC^{(2)'}(z_1,z_2)=\sum_m\theta_*^m\cG_2^m(z_1,z_2)$ the boundary
value of a two-fold Stieltjes kernel, or merely a resolvent
polynomial?  For real $S^{(2)}$ the two-variable Hilbert dictionary
of Eq.~\eqref{eq:disp} composes to the involutions
$H_1H_2[\mathrm{Re}\,\cC^{(2)}]=-\mathrm{Re}\,\cC^{(2)}$ and
$H_1H_2[\mathrm{Im}\,\cC^{(2)}]=-\mathrm{Im}\,\cC^{(2)}$ at fixed
$\eta$, tested here at $\eta=1$ on the real grid $[-64,64]$ (step
$0.25$ on $|E|<8$, step $1$ outside; residuals on the central
square $|E_1|,|E_2|<6$).  Three objects are tested at three
levels.  (i) The exact channel-resolved kernel evaluated
pointwise---$8281$ point masses whose total mass vanishes: residual
$(5.2,12.0)$---a \emph{resolution floor}, not a causal violation:
the discrete transform carries a ${\sim}1\%$ quadrature error per
point mass, amplified by the total-variation-to-cancellation ratio
${\sim}10^3$ of the point-mass sum.  (ii) The same exact measure
smoothed onto $40\times40$ boxes: $(0.03,0.03)$---the numerical
machinery verified on the exact object.  (iii) The fixed-point
kernel in its centered sector form, $\cG_2^m-N_m\bar g(z_1)\bar
g(z_2)$ with $\bar g$ the trace-mean resolvent (the exact measure is
centered, $\sum_n\delta p_n=0$; the solver's mass row, held to
$7.5\times10^{-6}$ relative, is subtracted analytically):
$(0.035,0.031)$ at every $\lambda$, with the three sector kernels
passing individually at $2$--$4\%$.  The closure output therefore
\emph{passes the causal-form test at the few-percent level}: the
object assembled by the map Eq.~\eqref{eq:predictmap} is a
legitimate two-fold Stieltjes boundary kernel---constructively so,
since it is a sector combination of the Stieltjes transforms
$\cG_2^m$---built from
Hamiltonian data alone, and no boundary data entered its
construction; the Hilbert test is then a numerical consistency check
of that form, while the content gap
between (iii) and (ii) is the truncation deficit quantified above:
the reconstructed spectral covariance
$S_{\mathrm{pred}}=\sum_m\theta_*^m\rho_m^{(2)}$ reaches overlap
$0.97$ with deviation $0.25$ against the exact binned measure, and
the diagonal boundary comparison of the closure kernel against the
exact channel cumulant reaches deviation $0.89$ (D1) on $[-8,8]$,
consistent with the $0.28$ bulk normalization quoted above.

\subsection{Itemized checks of the free-fermion anchor}
\label{app:anchorchecks}

\emph{Deep-plane seeds.}  Table~\ref{tab:dp} shows the approach of
$t^2\Gamma^{(2)}(\mathrm it,\mathrm it)\to2e_{ab}/\mathrm i^2=-2e_{ab}$
and $t^4\cC^{(2)}(\mathrm it,\mathrm it)\to K_{11}=2\ell-2k^2/N$:
the corrections are $O(1/t)$ with coefficients of many-body
bandwidth scale, and by $t=64$ the seeds are approached to a few
percent ($-45.4$ vs $-48$; $3.31$ vs $3.43$).  The three-point seed
$t^3\Gamma^{(3)}\to 8e_{abc}/\mathrm i^3=8\mathrm i e_{abc}$ converges
more slowly ($143.7\mathrm i$ vs $192\mathrm i$ at $t=64$ for the
coincident triple), consistent with the $O(t^{-1})$ corrections
carrying the third moments of the many-body bandwidth.

\begin{table}[tbp]
\caption{Deep-plane approach to the microscopic seeds at
$(N,k)=(14,2)$, $n_{\mathrm{real}}=4000$.  Targets: $t^2\Gamma^{(2)}
\to-2e_{ab}$, $t^4\cC^{(2)}\to K_{11}=2\ell-2k^2/N$,
$t^3\Gamma^{(3)}\to8\mathrm i\,e_{abc}$.}
\label{tab:dp}
\footnotesize
\begin{ruledtabular}
\begin{tabular}{c@{\quad}cccc}
\hline\hline
rep & $t$ & $t^2\Gamma^{(2)}$ & $t^4\cC^{(2)}$ & $\mathrm{Im}\,t^3\Gamma^{(3)}$\\
\hline
$\ell{=}2$ ($-48$; $3.43$) & $16$ & $-29.7+0.19\mathrm i$ & $2.09$ & $52.9$\\
 & $32$ & $-41.0+0.14\mathrm i$ & $2.98$ & $113.7$\\
 & $64$ & $-45.4+0.08\mathrm i$ & $3.31$ & $143.7$\\
$\ell{=}1$ ($-24$; $1.43$) & $16$ & $-14.4-0.17\mathrm i$ & $0.90$ & $35.8$\\
 & $32$ & $-20.3-0.15\mathrm i$ & $1.26$ & $57.7$\\
 & $64$ & $-22.6-0.09\mathrm i$ & $1.40$ & $68.0$\\
$\ell{=}0$ ($-8$; $-0.57$) & $16$ & $-4.08-0.47\mathrm i$ & $-0.37$ & $-1.46$\\
 & $32$ & $-6.43-0.33\mathrm i$ & $-0.51$ & $-1.18$\\
 & $64$ & $-7.41-0.18\mathrm i$ & $-0.57$ & $-0.75$\\
\hline\hline
\end{tabular}
\end{ruledtabular}
\end{table}

\emph{Causality and sum rules.}  Hermiticity
$\cC^{(2)}(z_1^*,z_2^*)=\cC^{(2)}(z_1,z_2)^*$ holds exactly
(difference at machine precision).  The one-variable dispersion relation in its
smoothed form, $\cC^{(2)}(E+2\mathrm i\varepsilon,z_2)
=\int\rho_\varepsilon(E')/(E+\mathrm i\varepsilon-E')\dd E'$ with
$\rho_\varepsilon(E')=-(1/\pi)\,\mathrm{Im}_1\cC^{(2)}(E'+\mathrm
i\varepsilon,z_2)$, is verified at $E=-0.5,0.4$
($\varepsilon=0.05$) to $2.6\times10^{-5}$--$9.4\times10^{-5}$ against
values $\sim2\times10^{-3}$, the residual being grid and quadrature
limited.  The diagonal mass sum rule gives
$\sum_{a,n}\E[(\delta p_n^a)^2]=3.562$ against the exact Selberg
value $q_2-1=3.55$ ($0.3\%$).

\emph{Moment-field dictionary.}  The microscopic moment
$\cM_{11}=\kappa_2[s_1(a),s_1(b)]$ measured directly from the
channel Hamiltonian reproduces the closed form $K_{11}=2\ell-2k^2/N$
to $0.4\%$ ($3.443$ vs $3.429$ at $\ell=2$); the next moments
$\cM_{22}=\kappa_2[s_2(a),s_2(b)]$ and
$\cM_{31}=\kappa_2[s_3(a),s_1(b)]$ take the values
$(64.5,17.2,-6.8)$ and $(242.2,103.3,-39.5)$ for $\ell=2,1,0$, the
microscopic inputs of the closure.

\emph{Explicit $r=2$ functional.}  The block decomposition of
Eq.~\eqref{eq:r2functional} is verified term by term at every probe
point.  The dressing block equals its closed form
$\sum_{e,f}\cC^{(2)}_{c(e),d(f)}$ to machine precision
($0.264579773410957$ vs $0.264579773410958$ at $z_1$), the seed block
equals $2\sum_{e\in E_a\cap E_b}\hat{\cR}_{c(e)}\hat{\cR}_{d(e)}$ to
the sampling accuracy of the chi-square averages ($\sim2\%$), and the
mixed blocks take their predicted $O(z^{-4})$ values.  The
mean-field-level residual---the cross-correlated-channel deficit of
the functional---decays monotonically with the deep-plane scale,
$\cE_{\rm MF}(z)=|\Gamma^{(2)}-\Gamma^{(2)}_{\rm MF}|/|\Gamma^{(2)}|
=6.2,5.0,1.67,0.67,0.22,0.06,0.02$ at $z_1,z_2,\mathrm it$
($t=4,8,16,32,64$) for the coincident pair: the functional is exact
in the deep plane, and the $O(1)$ bulk deficit is precisely the
cross-correlated channel of Sec.~\ref{sec:vertex}, quantified here as
a function of the broadening.

\subsection{Krylov and $r=3$ extensibility: technical details}
\label{app:krylov}
\emph{The eigenstate-free Krylov basis.}  The first step of the
eigenstate-free route is exercised: the two-resolvent Krylov basis
$\psi_m(z)=D^{-1}\sum_n P_m(\lambda_n)/(z-\lambda_n)$, with $P_m$
the orthonormal polynomials of the mean spectral density built from
the trace moments alone, and the $K\times K$ coefficients
$G^{(2)}_{mn}$ determined by the diagonal moment dictionary through
the asymptotic matching
$\cM_{pq}=\sum_{mn}G_{mn}c_{mp}c_{nq}$ ($c_{mp}$ the
$z^{-p-1}$ coefficient of $\psi_m$)---no rank labels, no
eigenstate overlaps anywhere, in the spirit of the Krylov-basis
program \cite{Nandy2025}.  Two features of this determination decide
how far the route can be pushed, and both are worth recording.
First, the moment rows $p,q\ge1$ alone do not determine $G$: the
identity $\sum_n\delta p_n=0$ forbids any weight at the leading
$z^{-1}$ order, so every row with $p=0$ or $q=0$ vanishes identically,
and the square system on $p,q\ge1$ left after dropping them is so
ill-conditioned ($\kappa=7\times10^{5}$ at $K=3$ and
$1\times10^{7}$ at $K=5$) that its solution is numerically
meaningless in-band, with reconstruction errors of order $10^{4}$.
Restoring the exactly vanishing rows turns the determination into an
overdetermined linear problem, solved here in row-weighted least
squares.  Second, with that completion the determination is stable,
and the reconstruction error at the two bulk probes is $16\%/20\%$ at
$K=2$, $23\%/21\%$ at $K=3$, $12.5\%/12.4\%$ at $K=4$ and
$9.5\%/7.3\%$ at $K=5$, while on the deep plane ($z=4i,8i,16i$) it
reaches $(0.3\%,0.4\%,0.1\%)$ at $K=5$.  The in-band error therefore
\emph{plateaus} near $10\%$ instead of decreasing with $K$, against
the sector basis's few-percent (which uses the rank labels): the
spectral-polynomial basis captures the $z$-separable content the rank
measurement of Sec.~\ref{sec:num} promised, and the residual plateau
is precisely the connectivity content the sector labels carry---the
quantitative trade between the eigenstate-free basis and the
connectivity coordinate.  What a determination at in-band probes lacks
is not more moments, whose asymptotic content is already reproduced to
sub-percent.  The natural remaining candidate is the causal (Stieltjes)
positivity of Sec.~\ref{sec:causal}, probed at the real boundary by
Sec.~\ref{sec:hilbert}, and it is verified here to be
\emph{inactive}.  The joint spectral weight of the reconstruction is
the covariance $\mathbb E[\delta p_n\delta p_{n'}]$ of the spectral
weights---positive semidefinite by Cauchy--Schwarz, with signed
off-diagonals---so the causal condition on the coefficients reads
$\mathrm{sym}(G)\succeq0$; the row-weighted determination already lies
on that cone at every $K$ (smallest eigenvalue of
$\mathrm{sym}(G)$ equal to $2\times10^{-2}$, $3\times10^{-8}$,
$6\times10^{-5}$ and $2\times10^{-10}$ for $K=2,3,4,5$, against a
largest eigenvalue $\simeq1.2\times10^{-1}$), and enforcing the
constraint by a positive-semidefinite parametrization $G=LL^{\top}$
changes the in-band reconstruction by less than $0.02\%$.  What the
pure energy coordinate truncates is instead the \emph{rank} of that
covariance: the exact joint weight has participation ratio $82$ of the
$91$ modes and its five largest directions carry only $43\%$ of the
Frobenius norm, while the reconstruction's weight is rank-$K$ by
construction and differs from the exact covariance by $93\%$ in
Frobenius norm at every $K$ from $2$ to $5$.  The in-band plateau is
therefore a rank deficiency of the energy coordinate---the
connectivity content---and not a positivity violation, and what the
synthesis below requires is the connectivity-conditioned polynomial
basis itself, not a restored moment or a repaired causal form.  The synthesis is the
product basis
$\Phi_{\mathfrak c,\alpha\beta}(z_1,z_2)
=\sum_{n,n'}P_\alpha(\lambda_n)P_\beta(\lambda_{n'})\,
\rho_{\mathfrak c}(n,n';z_1,z_2)$, with $\rho_{\mathfrak c}$ the
class kernel Eq.~\eqref{eq:Bclass} evaluated on the eigenpair
$(n,n')$ and $P_\alpha$ the Krylov polynomial of degree $\alpha$:
the connectivity class supplies the microscopic coordinate and the
Krylov polynomial the energy coordinate of the \emph{same} basis.
The Krylov route is therefore not an alternative basis but the
spectral-coordinate tool of the class basis of
Sec.~\ref{sec:conn}---the form in which
Eq.~\eqref{eq:genericansatz} is realized numerically---and the
in-band deficits above are interpreted as a pure-energy-coordinate
truncation of that same closure, diagnosing missing connectivity
information rather than defining a competing closure.

\emph{$r=3$ sector closure and the skewness sector.}  The
eight-class ansatz Eq.~\eqref{eq:sectoransatz3} is exercised with
the even-degree moment set of Sec.~\ref{sec:sectorclosure3} on the
coincident channel triple.  The Hamiltonian-only moment dictionary
is verified at $r=3$: the exact Weingarten sector values of
Ref.~\cite{MResR} satisfy the moment equations to $0.3\%$
($\cM_{211}$), $1.6\%$ ($\cM_{411}$), $2.0\%$ ($\cM_{321}$),
$4.8\%$ ($\cM_{332}$), $5.0\%$ ($\cM_{521}$)---$12$--$16\%$ for
the noisiest high-degree entries---while every odd-degree moment
vanishes on both sides: $\kappa_3[s_p,s_q,s_r]=0$ for $p+q+r$ odd
in the reflection-symmetric Gaussian class, the moment form of the
selection rule.  The eight-class expansion itself is verified
exactly: an independent classification of every ordered level triple
by its mask pattern reproduces the eight class counts
$(91,6552,24024,2184,360360,72072,18018,270270)$, summing to $D^3$,
and the class sums reproduce the brute-force sum over all
$7.5\times10^{5}$ ordered triples to $10^{-13}$---the representation
is coverage-complete.  Reconstructing $\cC^{(3)}$ from the exact
sector table is then limited by the comparison rather than by the
representation: the third power of the resolvent fluctuation is
heavy-tailed (per-realization coefficient of variation $34$ and $24$
at the two probes), so the directly accumulated cumulant carries a
standard error of $54\%$ and $38\%$ at $4\times10^{3}$ realizations,
and the exact-$\theta$ reconstruction agrees with it to $0.8$ paired
standard errors at both probes---the $12$--$15\%$ agreement is inside
that floor and does not by itself bound the closure.  The closure's
own error is deterministic and separate: at the same sample size the
eight-class inversion reproduces $\cC^{(3)}$ to $36$--$38\%$, the
residual being carried by the three small shared-mode sectors, whose
closure values exceed the exact ones by factors up to $87$, while the
four resolved sectors are recovered at $0.8$--$5.8\%$.  The purpose of
this exercise is extensibility, not closure quality: it demonstrates
that the formal ladder, the vertex-functional construction and the
moment dictionary of Secs.~\ref{sec:ladder}--\ref{sec:system}
can be instantiated at $r=3$, while the quantitative $r=3$ closure
level is bounded by the sector resolution reported below.  The
$8\times8$
Hamiltonian-only inversion has condition number $9.4\times10^3$:
the large-amplitude sectors are resolved ($0.8\%$, $3\%$, $5.8\%$,
$20\%$ for $[0000],[2000],[1000],[2111]$), while the small
shared-mode and chain sectors ($\theta\sim10^{-7}$) lie below the
moment resolution at this size---the SVD spectrum resolves four to
five of the eight directions.  The parity content is preserved by
the closure: the odd-moment predictions remain at the noise level.
The skewness-generating content of the hierarchy is thereby
identified: it is the odd-degree moment sector of $\cC^{(3)}$,
microscopically the $\kappa_3[s_p,s_q,s_r]$ with $p+q+r$ odd, which
the selection rule sets to zero for reflection-symmetric kernels
(the free-fermion benchmark, verified here) and which carries the
$\mathrm{Im}(\cF\cF)_{\rm odd}$ interference of
Ref.~\cite{SRletter} in parity-mixing systems.

\section{Proof of the multi-resolvent ladder theorem}
\label{app:proofladder}

Proof of Theorem~\ref{thm:ladder}.  Equation \eqref{eq:tower} is multilinearity of cumulants;
Eq.~\eqref{eq:vertexTower} is Eq.~\eqref{eq:XD} term by term;
Eq.~\eqref{eq:partexp} is the product formula
Eq.~\eqref{eq:cumprod} of
Appendix~\ref{app:cum} applied to the centered fields $\delta\cG_{a_i}$
(the $s_i=1$ term is the vertex itself).  The deep-plane bookkeeping
splits accordingly.  (i) For the insertion series: in the joint scaling
$z_i=t\zeta_i$ each additional factor $\bar{\cR}_a^0\delta\cG_a$ in
the product $\prod_i(\bar{\cR}_{a_i}^0)^{s_i+1}$ costs $t^{-2}$ (from
$\delta\cG_a(z)=\delta v_a/z+O(z^{-2})=O(z^{-1})$, the self-energy
fluctuation of Eq.~\eqref{eq:Gdef}, together with
$\bar{\cR}_a^0=O(z^{-1})$), so every insertion beyond the leading one
raises the total degree in $t$ by at least two.  (ii) For the $X$-tower: the
individual $X$-fields start at $O(z^{-2})$ through
Eq.~\eqref{eq:Xdeep}, while their connected $r$-point cumulants
acquire the stronger $3r-2$ degree bound under the connectivity
assumptions of Sec.~\ref{sec:causal}---$X_a\sim z^{-2}$ but
$\cC^{(r)}_{\rm conn}\sim z^{-(3r-2)}$---and in
the Gaussian class the leading single-site tower terms vanish,
$\kappa_r(X_a,\dots,X_a)=0$ for $r\ge3$, raising the effective onset one
degree further.  These two statements control the two classes (i) and
(ii) of the theorem independently.

\section{Observable-level test: the ETH variance}
\label{sec:eth}

The loop that runs from the Hamiltonian to the two-resolvent
cumulant is closed in Secs.~\ref{sec:genbench} and \ref{sec:num};
this section carries its output to a measured observable and asks
what the higher-resolvent hierarchy buys there.  Two statistical
questions are in play, and this section addresses the first of them.
The ETH hypothesis concerns the energy-resolved statistics of the
eigenbasis matrix elements themselves, and local rotational
invariance (LRI) is one concrete statistical description of those
data (Appendix~\ref{sec:LRI}).  The test carried out here is at the
level of a single observable projection: whether the retained
multi-resolvent content reproduces the energy-resolved off-diagonal
variance of a local operator---one projection of the underlying
statistics, not the statistics themselves.  Whether the window
moments themselves satisfy the LRI reference relations is a distinct
test, addressed in Appendix~\ref{sec:LRI}.  On the Ising chain
of Eq.~\eqref{eq:ising} we take the transverse field of site one,
$O=X_1$---a local term of $H$ itself---and split its energy-resolved
off-diagonal variance
\begin{multline}
V_O(\bar E,\omega)
=\E_H\Bigl[\,|O_{nm}|^2\,\Big|\,\tfrac{\lambda_n+\lambda_m}{2}
\in\bar E\pm\tfrac{\Delta E}{2},\\
|\lambda_n-\lambda_m|\in\omega\pm\tfrac{\Delta\omega}{2}\Bigr],
\label{eq:Vdef}
\end{multline}
binned with $\Delta E=\Delta\omega=0.5$ over the eigenpairs $n\ne m$
of each realization.  The observable couples exactly the $2^{N-1}$
channel pairs connected by the flip move of site one,
$(a,a\oplus e_1)$, so its matrix elements are
\begin{equation}
O_{nm}=\sum_{\mu=1}^{2^{N-1}}
\bigl[u_{n,2\mu}^*u_{m,2\mu+1}+u_{n,2\mu+1}^*u_{m,2\mu}\bigr],
\label{eq:Onm}
\end{equation}
with $u_{n,a}=\braket{\varphi_a}{n}$ the $Z$-basis eigenvector
amplitudes---the channel expansion of the observable projection
anticipated in Sec.~\ref{sec:genbench}, executed for a concrete
operator.  The observable level is where the two-resolvent content
must be visible: the single-resolvent environment fixes at most the
factorized part of $\E|O_{nm}|^2$, and everything beyond it is
correlation.  All profiles and tables of this section are estimated
from $n_{\mathrm{real}}=2\times10^{4}$ disorder realizations; the
resolution of the binned comparisons is set by the $3.7\%$ bin-crossing
floor of the $\lambda=0$ row of Table~\ref{tab:eth} (see its caption),
which exceeds the estimation scatter of the ensemble averages.

At the pair level $|O_{nm}|^2$ splits exactly,
\begin{equation}
|O_{nm}|^2=V^{(1\mathrm R)}_{nm}+\Delta^{(2)}_{nm}+R_{nm},
\label{eq:splitpair}
\end{equation}
into three terms.  The single-resolvent baseline
\begin{equation}
V^{(1\mathrm R)}_{nm}=\sum_{\mu}
\bigl[q^+_{2\mu}(\lambda_n)\,q^+_{2\mu+1}(\lambda_m)
+q^+_{2\mu+1}(\lambda_n)\,q^+_{2\mu}(\lambda_m)\bigr],
\label{eq:V1R}
\end{equation}
with $q^+_a(E)=\E_H[p_n^a\,|\,\lambda_n=E]$ the mean channel
participation profile, is the maximal factorized content available
from exact one-resolvent data: it factorizes the two channel
weights at the two energies.  Built from exact one-resolvent
profiles, it provides the strongest single-resolvent baseline
within this decomposition: any $r=1$ closure can at most
reproduce this contribution.  The remaining shortfall therefore
cannot be removed by improving the $r=1$ description alone.  The class-organized two-resolvent term
\begin{equation}
\Delta^{(2)}_{nm}=\sum_{\mu}
\Bigl[\Cov\bigl(p_n^{2\mu},p_m^{2\mu+1}\bigr)
+\Cov\bigl(p_n^{2\mu+1},p_m^{2\mu}\bigr)\Bigr],
\label{eq:Delta2}
\end{equation}
is the connected channel-pair covariance---the observable projection
$P_O[\cS^{(2)}]$ of the spectral cumulant
Eq.~\eqref{eq:Sdef} onto the family of channel pairs coupled by
$O$, the projection of exactly the two-resolvent sector that
Secs.~\ref{sec:conn}--\ref{sec:closure} construct.  The remainder
$R_{nm}$ collects the off-diagonal channel pairings: the
$\mu\ne\mu'$ contractions of Eq.~\eqref{eq:Onm} and the
cross term, the same two-resolvent sector's remaining channel
content.  The split is an identity pair by pair and survives the
binning, so
\begin{equation}
V_O=V^{(1\mathrm R)}+\Delta^{(2)}+R,
\label{eq:VOsplit}
\end{equation}
and the two-resolvent sector supplies the entire correction
$V_O-V^{(1\mathrm R)}=\Delta^{(2)}+R$ by construction; the split
quantifies how much of it the class-organized projection carries.

The remainder is not an unspecified leftover: it has an exact
microscopic expression in the channel basis of Eq.~\eqref{eq:Onm}.
With $a_\mu=u_{n,2\mu}u_{m,2\mu+1}$ and
$b_\mu=u_{n,2\mu+1}u_{m,2\mu}$, so that
$O_{nm}=\sum_\mu(a_\mu+b_\mu)$, squaring separates the coincident
channel pairs from the inter-pair terms,
\begin{equation}
\begin{aligned}
|O_{nm}|^2={}&\sum_\mu\bigl(p_n^{2\mu}p_m^{2\mu+1}
+p_n^{2\mu+1}p_m^{2\mu}\bigr)
+2\sum_\mu c_n^{\mu}c_m^{\mu}\\
&+\sum_{\mu\ne\mu'}(a_\mu+b_\mu)(a_{\mu'}+b_{\mu'}),
\end{aligned}
\label{eq:Rmicro}
\end{equation}
with $c_n^\mu\equiv u_{n,2\mu}u_{n,2\mu+1}$ the pair-coherence
amplitude of eigenstate $n$ on channel pair $\mu$.  The first group
is the coincident product content whose ensemble average is the
baseline plus the projected two-resolvent term,
$V^{(1\mathrm R)}+\Delta^{(2)}$, by the definitions
\eqref{eq:V1R}--\eqref{eq:Delta2}; the remainder therefore splits
into two named objects,
\begin{equation}
R=R^{\rm coh}+R^{(4)},\qquad
R^{\rm coh}_{nm}=2\sum_\mu c_n^{\mu}c_m^{\mu},
\label{eq:Rsplit}
\end{equation}
where $R^{(4)}$ is the last group of Eq.~\eqref{eq:Rmicro}.  Neither
object is reachable from the channel-diagonal weight $p_n^a$ alone:
$R^{\rm coh}$ is the two-point function of the pair-coherence
field---an off-diagonal-in-channel, two-energy object---and $R^{(4)}$
is four-channel inter-pair content.  This is the structural reason
the class-organized projection carries $\Delta^{(2)}$ and not $R$:
every kernel of Secs.~\ref{sec:conn}--\ref{sec:closure} is diagonal
in the channel index, so it fixes the weights and never the
coherences.  At $\lambda=0$ the eigenbasis is the channel basis
itself, every $c_n^\mu$ vanishes identically, and with it the entire
remainder---the exact product-state limit the $\lambda=0$ row of
Table~\ref{tab:eth} realizes.  In the interacting benchmark the split
is measured rather than modeled (Table~\ref{tab:Rsplit}, same
grids and $n_{\mathrm{real}}=1200$, with the two weakest couplings
added so that the approach to $\lambda=0$ is resolved): neither part
is coupling-selected.  For the single-flip operator the four-channel
term grows from $|R^{(4)}|/V_O=0.18$ at $\lambda=2^{-6}$ to
$0.53$ at $2^{-3}$ and falls again at strong coupling ($0.30$ at
$2^{-1}$), while the coherence correlator is non-monotone---it carries
$49\%$ of $|R|$ at $\lambda=2^{-6}$, comparable to the four-channel
part there, drops to $10$--$17\%$ for $2^{-5}\le\lambda\le2^{-2}$, and
changes sign between $\lambda=2^{-3}$ and $2^{-2}$.  For the
distance-two operator the two pieces stay comparable at every tested
coupling ($|R^{\rm coh}|/V_O=0.03$--$0.06$ against
$|R^{(4)}|/V_O=0.004$--$0.06$).  The weak-coupling behaviour is thus
not a limit in which one kernel dominates: a prediction of $R$ at
small $\lambda$ needs \emph{both} the channel-coherent and the
four-channel kernels, which is why the remainder is quoted here as
a measured quantity rather than as a closed object.  The identity
\eqref{eq:Rmicro} closes to $10^{-17}$ in every bin, and both terms
vanish exactly at $\lambda=0$.  This fixes what the next closure
order would have to supply, in the order of the measured weights: a
channel-coherent two-resolvent kernel for $R^{\rm coh}$, and $r=4$
channel content for $R^{(4)}$.  For a random-phase matrix-element
structure the four-channel contractions are removed by the
Weingarten algebra \cite{MResR}, so a nonzero $R^{(4)}$ is a
coherence-phase diagnostic rather than a noise term; the free-fermion
content of Eq.~\eqref{eq:Rsplit} is fixed by the single-particle
overlaps; the same pair content is the object of the drift-kernel
diagnostics of Appendix~\ref{sec:driftkernel}.

\begin{table}[t]
\begin{ruledtabular}
\begin{tabular}{lccc}
$\lambda$ & $R^{\rm coh}/V_O$ & $R^{(4)}/V_O$ & $R/V_O$\\
\hline
\multicolumn{4}{l}{$O=X_1$}\\
$2^{-6}$ & $-0.176$ & $-0.184$ & $-0.360$\\
$2^{-5}$ & $-0.140$ & $-0.309$ & $-0.449$\\
$2^{-4}$ & $-0.094$ & $-0.467$ & $-0.561$\\
$2^{-3}$ & $-0.031$ & $-0.532$ & $-0.563$\\
$2^{-2}$ & $+0.021$ & $-0.488$ & $-0.467$\\
$2^{-1}$ & $+0.042$ & $-0.298$ & $-0.256$\\
\multicolumn{4}{l}{$O=X_1X_2$}\\
$2^{-6}$ & $-0.034$ & $-0.004$ & $-0.038$\\
$2^{-5}$ & $-0.047$ & $-0.010$ & $-0.057$\\
$2^{-4}$ & $-0.056$ & $-0.023$ & $-0.079$\\
$2^{-3}$ & $-0.061$ & $-0.042$ & $-0.103$\\
$2^{-2}$ & $-0.050$ & $-0.058$ & $-0.108$\\
$2^{-1}$ & $-0.033$ & $-0.044$ & $-0.077$\\
\end{tabular}
\end{ruledtabular}
\caption{Measured splitting Eq.~\eqref{eq:Rsplit} of the interference
remainder into its channel-coherence and four-channel parts at
$\lambda=2^{-6},\ldots,2^{-1}$, band
center $\bar E\in[-0.5,0.5]$, $N=9$, $n_{\rm real}=1200$;
$V_O$ is the exact off-diagonal variance of the same row.  Both
columns vanish identically at $\lambda=0$ and the pair identity
\eqref{eq:Rmicro} closes to $10^{-17}$ at every entry.}
\label{tab:Rsplit}
\end{table}

\begin{figure}[t]
\begin{tikzpicture}
\begin{groupplot}[
 group style={group size=1 by 2, vertical sep=0.35cm,
   x descriptions at=edge bottom},
 width=\columnwidth, height=4.4cm,
 xlabel={$\omega$},
 xmin=-11.5, xmax=11.5,
 tick label style={font=\scriptsize},
 label style={font=\small},
]
\nextgroupplot[
 ylabel={variance ($10^{-3}$)}, ymin=0, ymax=0.018,
 mark size=1.2pt,
 legend style={at={(0.02,0.98)},anchor=north west,font=\scriptsize,
   legend cell align=left},
]
\addplot[only marks, mark=*, color=black]
 coordinates {(-10.75,0.00700)(-9.75,0.00463)(-8.75,0.00391)(-7.75,0.00320)
 (-6.75,0.00258)(-5.75,0.00215)(-4.75,0.00182)(-3.75,0.00159)
 (-2.75,0.00142)(-1.75,0.00131)(-0.75,0.00125)(0.25,0.00123)
 (1.25,0.00127)(2.25,0.00135)(3.25,0.00149)(4.25,0.00169)
 (5.25,0.00197)(6.25,0.00234)(7.25,0.00281)(8.25,0.00354)
 (9.25,0.00436)(10.25,0.00543)};
\addlegendentry{$V_O$}
\addplot[only marks, mark=square*, color=blue]
 coordinates {(-10.75,0.01703)(-9.75,0.01037)(-8.75,0.00841)(-7.75,0.00692)
 (-6.75,0.00566)(-5.75,0.00461)(-4.75,0.00384)(-3.75,0.00329)
 (-2.75,0.00290)(-1.75,0.00266)(-0.75,0.00253)(0.25,0.00250)
 (1.25,0.00258)(2.25,0.00277)(3.25,0.00308)(4.25,0.00354)
 (5.25,0.00421)(6.25,0.00512)(7.25,0.00632)(8.25,0.00769)
 (9.25,0.00939)(10.25,0.01253)};
\addlegendentry{$V^{(1\mathrm R)}$}
\addplot[only marks, mark=triangle*, color=red]
 coordinates {(-10.75,0.00933)(-9.75,0.00657)(-8.75,0.00536)(-7.75,0.00443)
 (-6.75,0.00368)(-5.75,0.00303)(-4.75,0.00259)(-3.75,0.00226)
 (-2.75,0.00203)(-1.75,0.00188)(-0.75,0.00180)(0.25,0.00177)
 (1.25,0.00183)(2.25,0.00194)(3.25,0.00215)(4.25,0.00241)
 (5.25,0.00284)(6.25,0.00337)(7.25,0.00404)(8.25,0.00492)
 (9.25,0.00598)(10.25,0.00753)};
\addlegendentry{$V^{(1\mathrm R)}+\Delta^{(2)}$}
\nextgroupplot[
 ylabel={correction ($10^{-3}$)}, ymin=-0.011, ymax=0,
 mark size=1.2pt,
 legend style={at={(0.02,0.02)},anchor=south west,font=\scriptsize,
   legend cell align=left},
]
\addplot[only marks, mark=*, color=black]
 coordinates {(-10.75,-0.01003)(-9.75,-0.00574)(-8.75,-0.00450)(-7.75,-0.00372)
 (-6.75,-0.00308)(-5.75,-0.00246)(-4.75,-0.00202)(-3.75,-0.00170)
 (-2.75,-0.00148)(-1.75,-0.00135)(-0.75,-0.00128)(0.25,-0.00127)
 (1.25,-0.00131)(2.25,-0.00142)(3.25,-0.00159)(4.25,-0.00185)
 (5.25,-0.00224)(6.25,-0.00278)(7.25,-0.00351)(8.25,-0.00415)
 (9.25,-0.00503)(10.25,-0.00710)};
\addlegendentry{$\Delta^{\mathrm{ex}}=V_O-V^{(1\mathrm R)}$}
\addplot[only marks, mark=triangle*, color=red]
 coordinates {(-10.75,-0.00770)(-9.75,-0.00380)(-8.75,-0.00305)(-7.75,-0.00249)
 (-6.75,-0.00198)(-5.75,-0.00158)(-4.75,-0.00125)(-3.75,-0.00103)
 (-2.75,-0.00087)(-1.75,-0.00078)(-0.75,-0.00073)(0.25,-0.00073)
 (1.25,-0.00075)(2.25,-0.00083)(3.25,-0.00093)(4.25,-0.00113)
 (5.25,-0.00137)(6.25,-0.00175)(7.25,-0.00228)(8.25,-0.00277)
 (9.25,-0.00341)(10.25,-0.00500)};
\addlegendentry{$\Delta^{(2)}=P_O[\cS^{(2)}]$}
\end{groupplot}
\end{tikzpicture}
\caption{Energy-resolved off-diagonal variance of $O=X_1$ at
$\lambda=2^{-2}$---the intermediate coupling at which the deformation
advantage of Sec.~\ref{sec:response} is fully developed---band center
$\bar E\in[-0.5,0.5]$ ($2\times10^{4}$
realizations, $\Delta\omega=0.5$ bins).  Top: exact
$V_O$, single-resolvent baseline $V^{(1\mathrm R)}$ of
Eq.~\eqref{eq:V1R}, and the two-resolvent reconstruction
$V^{(1\mathrm R)}+\Delta^{(2)}$ of Eq.~\eqref{eq:Delta2}; the
baseline misses both the magnitude and the $\omega$ shape, the
reconstruction restores the shape.  Bottom: the exact missing
correction $\Delta^{\mathrm{ex}}=V_O-V^{(1\mathrm R)}$ and the
class-organized two-resolvent projection
$\Delta^{(2)}=P_O[\cS^{(2)}]$ evaluated with the exact $\cS^{(2)}$,
which
carries $57$--$77\%$ of it at every resolved $\omega$
($57$--$68\%$ in the interior, exceeding $70\%$ only in the two
far-edge bins) with the
correct sign; the gap is the interference remainder $R$.}
\label{fig:eth}
\end{figure}
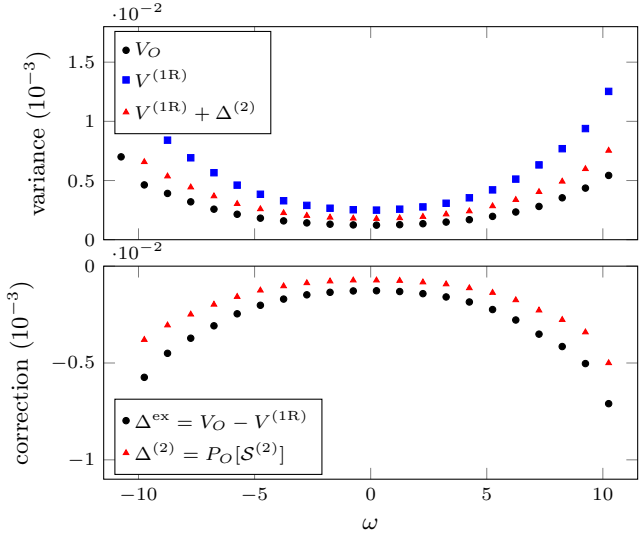

Displayed in Eq.~\eqref{eq:Delta2} in its ensemble form, the
two-resolvent term is evaluated pair by pair as the profile-subtracted
channel product---each channel weight with its mean participation
profile removed---and then averaged over realizations: both operations
are exact, and the pair-level identity Eq.~\eqref{eq:splitpair} closes
before any averaging.
The decomposition is exact algebra: all three terms entering
Table~\ref{tab:eth} and Fig.~\ref{fig:eth} are computed from the
exact eigenstates of each realization, nothing is fitted, and the
pair-level identity Eq.~\eqref{eq:splitpair} closes to machine
precision at every grid point.  The MRSC content enters in three
steps.  First, the closure is blind: the parameters of
Secs.~\ref{sec:conn}--\ref{sec:closure} are determined from
Hamiltonian-only input---moments, seeds, and connectivity
data---and the blind protocol of Sec.~\ref{sec:num} verifies the
resulting cumulant $\cC^{(2)}$ to
$0.35\%$--$0.97\%$ on the tested probes; no observable-level
datum enters the closure anywhere.  Second, the projection is
exact: the map $P_O$ in Eq.~\eqref{eq:Delta2} is an exact linear
identity, applied identically to any candidate cumulant, so no
accuracy is introduced or lost in passing to the observable.
Third, the observable output is fixed directly:
Table~\ref{tab:eth} and Fig.~\ref{fig:eth} determine the
observable content the two-resolvent sector must supply,
evaluated with the exact $\cS^{(2)}$.  The probe-level fidelity is
not a relative bound at the observable level, and the two differ by a
content ratio that can be stated exactly.  For any linear projection a
cumulant perturbation is projected as
$\delta(P_O\cC)=P_O\delta\cC$, so
\begin{equation}
\varepsilon_O\equiv\frac{|\delta(P_O\cC)|}{|P_O\cC|}
\le\frac{\varepsilon_C}{\rho_O},
\qquad
\rho_O\equiv\frac{|P_O\cC|}{|\cC|},
\label{eq:transportbound}
\end{equation}
with $\varepsilon_C=|\delta\cC|/|\cC|$ the relative closure error on
the cumulant: the exact projection preserves the numerator but
renormalizes the denominator, so a content share $\rho_O\ll1$
amplifies relative errors by $1/\rho_O$.  Both extremes occur in the
construction.  If the closure error is proportional to the cumulant,
$\delta\cC\propto\cC$, then $\varepsilon_O=\varepsilon_C$ and no
fidelity is lost in the projection; if instead the error lies where
the projection cancels, the full factor $1/\rho_O$ appears.  On the
benchmark the share is measured rather than assumed: the diagonal
(trivial) content carries $94\%$ of the cumulant amplitude at the
tested point (Sec.~\ref{sec:num}), so $\rho_O\simeq0.06$ and the
amplification is at most $\sim17$.  Transporting the blind closure
bound $0.35\%$--$0.97\%$ through Eq.~\eqref{eq:transportbound} places
the end-to-end observable accuracy between the coherent value
($0.4\%$--$1.0\%$) and the incoherent bound ($6\%$--$16\%$): the
projection adds no error of its own, but it reweights relative ones.
Which limit applies is decided by the boundary transport of the
kernel---the closure is validated at complex probes while the
observable requires the on-shell energy-resolved content.  On the
retained branch that transport is a finite-dimensional operator
rather than an interpolation.  By the sector ansatz
Eq.~\eqref{eq:sectoransatz} the closed cumulant is the linear image of
the amplitude vector, so the probe data and the observable content are
linear functionals of the same amplitudes: the probes fix
$\boldsymbol\theta$ through the matrix $M_p$ of sector sums $\cG2_m$
evaluated at the blind-test points---a Hamiltonian-only object---and
the binned content is the image
\begin{equation}
\Delta^{(2)}_{\rm pred}(E,\omega)=\sum_m\theta_m\,w_m(E,\omega),
\qquad w_m=P_O[\rho_m](E,\omega),
\label{eq:bdytransport}
\end{equation}
of the same vector, with $w_m$ built from the level sums and the
profile-subtracted sector densities.  Two consequences follow.
First, the transport bound is explicit,
\begin{equation}
\begin{aligned}
\bigl|\delta\Delta^{(2)}(E,\omega)\bigr|
&\le\|w(E,\omega)\|\,\|\delta\boldsymbol\theta\|,\\
\|\delta\boldsymbol\theta\|
&\le\|M_p^{+}\|\,\bigl|\delta\cC\bigr|_{\rm probe},
\end{aligned}
\label{eq:bdybound}
\end{equation}
both factors Hamiltonian-only and computable before any eigenstate is
touched: no interpolation between probe and on-shell content is
involved on the retained branch, because the same amplitude vector
carries both.  Second, the residual error is isolated rather than
absorbed.  It is the model error of the ansatz---the part of the exact
binned content outside $\operatorname{span}\{w_m\}$---whose
probe-level size the benchmark has already measured (the level-one
kernel captures $92\%$ of the exact kernel, with a total deviation of
$0.23\%$ against the $94\%$ diagonal carry of
Sec.~\ref{sec:num}), and whose binned-level evaluation on the
flip-pair channel is left open here.  The seed truncation is bounded on the same footing
rather than merely measured: the deep-plane content of
Table~\ref{tab:dp} falls by the consecutive ratio
$\varkappa\simeq0.65$ between adjacent orders ($5.4\%$ and $3.5\%$
deficits at the two measured orders), so the omitted tail is at most
$\varkappa/(1-\varkappa)\simeq1.9$ times the last retained order,
placing $\delta F_{\rm seed}$ within a factor $\lesssim2$ of its
measured $0.42\%$--$0.64\%$ of the kernel.  With this the two items
named above are in final form: the projection transport is bounded by
Eq.~\eqref{eq:transportbound}, the kernel transport is a computable
linear operator with its model error isolated by
Eqs.~\eqref{eq:bdytransport}--\eqref{eq:bdybound}, and the seed
truncation is bounded a posteriori by the measured consecutive-order
ratio.  The exact-projection comparison of Table~\ref{tab:eth} fixes
every ingredient on the near side of that chain.

Thus the
observable-level comparison provides an out-of-sample
characterization of the retained two-resolvent sector, with
closure-level errors isolated by construction; what a fully
end-to-end closure test at the observable level still requires is the
evaluation of the model error of Eq.~\eqref{eq:bdytransport} on the
flip-pair channel itself, which fixes the last unspecified factor of
the chain; on the benchmark the binned comparison is resolution-limited
by the bin-crossing floor of Table~\ref{tab:eth}, and the sub-floor
attribution remains open here.

Table~\ref{tab:eth} gives the band-center values
$\bar E\in[-0.5,0.5]$.  The single-resolvent baseline overshoots the
exact variance by a factor $1.7$--$2.3$ across
$\lambda=2^{-4},2^{-3},2^{-2},2^{-1}$: the correction missing from
the single-resolvent description is not a refinement but the
dominant content.  The class-organized projection $\Delta^{(2)}$
supplies $56$--$68\%$ of that correction across the tested
coupling range, with the correct (negative) sign throughout, and
the interference term the remainder; the
deterministic limit collapses both to the exact product-state
result---at $\lambda=0$ the eigenstates are the channel states
themselves, $V^{(1\mathrm R)}_{nm}=|O_{nm}|^2$ pair by pair with
exact profiles, $\Delta^{(2)}=R=0$ exactly, and the binned estimator
reproduces the vanishing correction only to its bin-crossing floor: the
$+0.12$ entry of the $\lambda=0$ row is $3.6\%$ of $V_O$ there, equal
to the narrow-window floor $3.7\%$ ($1.9\%$ in the wider window), and
$R=0$ in that row is exact rather than binned---the profile-subtracted
pair product and the exact single-resolvent content coincide
identically, so the entire residual is the binning of the profiles and
not a violation of the pair-level identity.

\begin{table}[t]
\begin{ruledtabular}
\begin{tabular}{ccccccc}
$\lambda$ & $V_O$ & $V^{(1\mathrm R)}$ & $\Delta^{(2)}$ & $R$
& $V^{(1\mathrm R)}/V_O$ & $\Delta^{(2)}/(V_O-V^{(1\mathrm R)})$\\
\hline
$0$     & $3.29$ & $3.17$ & $+0.12$ & $0$      & $0.96$ & $1.00$\\
$2^{-4}$ & $1.43$ & $3.17$ & $-0.97$ & $-0.76$ & $2.21$ & $0.56$\\
$2^{-3}$ & $1.39$ & $3.18$ & $-1.02$ & $-0.76$ & $2.28$ & $0.57$\\
$2^{-2}$ & $1.52$ & $3.17$ & $-0.98$ & $-0.66$ & $2.08$ & $0.60$\\
$2^{-1}$ & $1.79$ & $3.09$ & $-0.89$ & $-0.41$ & $1.73$ & $0.68$
\end{tabular}
\end{ruledtabular}
\caption{Band-center ($\bar E\in[-0.5,0.5]$) off-diagonal ETH
variance of $O=X_1$ on the Ising benchmark of
Sec.~\ref{sec:genbench}, split by Eq.~\eqref{eq:VOsplit}; all
entries in units of $10^{-3}$ per qualifying pair, averaged over
$n_{\mathrm{real}}=2\times10^{4}$ disorder realizations and over eigenpairs.  The last column is the fraction of the
missing correction $V_O-V^{(1\mathrm R)}$ carried by the
class-organized two-resolvent projection.  At $\lambda=0$ the exact
correction vanishes; the $3.7\%$ residual is the bin-crossing floor
of the binned profiles and is carried identically by the projection
($R=0$ exactly).}
\label{tab:eth}
\end{table}

Figure~\ref{fig:eth} resolves the $\omega$ dependence at
$\lambda=2^{-2}$ along the band center.  The exact variance is
suppressed at small $\omega$---$1.2\times10^{-3}$ at $\omega=0$
against $1.7\times10^{-3}$ at $|\omega|\approx4$ and
$5.4\times10^{-3}$ at the band edge---while the single-resolvent
baseline sits a factor $\sim2$ above it at every resolved
$\omega$.  The lower panel compares the exact missing correction
$\Delta^{\mathrm{ex}}=V_O-V^{(1\mathrm R)}$ with the projected
$\Delta^{(2)}$, evaluated with the exact $\cS^{(2)}$: the latter
carries $57$--$77\%$ of the former at every resolved $\omega$
($57$--$68\%$ in the interior, exceeding $70\%$ only in the two
far-edge bins) with the correct sign, reproducing the
correction's full energy-resolved shape rather than a global
normalization, and the interference remainder $R$ completes the
identification.  The single-resolvent baseline fails not only in
magnitude but in the entire energy-resolved shape of the observable
correction.

The same decomposition extends beyond the tested window and beyond the
single-flip operator (Table~\ref{tab:ethstrong}).  Three features
survive and one changes.  First, the overshoot of the
single-resolvent baseline is not a weak-coupling artifact: it persists
to $\lambda=1$ ($V^{(1\mathrm R)}/V_O=1.33$) and disappears only at
$\lambda=2$ ($1.04$), where the flip deformation becomes comparable to
the disorder scale $h,J\sim1$.  Second, the class-organized projection
keeps carrying the correction with the correct sign up to
$\lambda=1$, where it supplies $102\%$ of $V_O-V^{(1\mathrm R)}$ with
the interference remainder at $|R|/V_O=0.6\%$.  Third, at $\lambda=2$
the total correction itself becomes small and splits between two
nearly cancelling contributions: $\Delta^{(2)}$ overshoots it by a
factor $5.5$ while the interference term, now $19\%$ of $V_O$ and
\emph{positive}, cancels most of it---the identity
$\Delta^{(2)}+R=V_O-V^{(1\mathrm R)}$ holds at every entry, so the
two-resolvent content is still order one while no longer being
identifiable term by term in this regime.  The operator dependence
probes the same structure: for the two-flip operator $O=X_1X_2$ the
single-resolvent baseline is far better ($V^{(1\mathrm
R)}/V_O=1.00$--$1.20$ over the whole coupling range), the projection
supplies $40$--$64\%$ of the much smaller correction, and
$|R|/V_O\le13\%$---distance-two channel pairs factorize more strongly
than single-flip pairs, as expected, since their participation
products pair states differing by two site flips.  Translation
invariance of the ensemble provides the control: $X_1$ and $X_2$
agree within $3$--$7\%$ in every column, the residual difference being
the open boundary of the chain.  The $\omega$ resolution is intact at
strong coupling: at $\lambda=1$ the projection is negative at every
resolved $\omega$ and deepens from $-0.5\times10^{-3}$ at
$\omega\simeq0$ to $-3.1$ and $-3.8\times10^{-3}$ near the two band
edges, the shape measured at weak coupling in Fig.~\ref{fig:eth}.

\begin{table}[t]
\begin{ruledtabular}
\begin{tabular}{ccccccc}
$\lambda$, $O$ & $V_O$ & $V^{(1\mathrm R)}$ & $\Delta^{(2)}$ & $R$
& $V^{(1\mathrm R)}/V_O$
& $\Delta^{(2)}/\Delta^{\mathrm{ex}}$\\
\hline
$1$, $X_1$    & $2.23$ & $2.95$ & $-0.74$ & $+0.01$ & $1.33$ & $1.02$\\
$2$, $X_1$    & $2.64$ & $2.75$ & $-0.62$ & $+0.51$ & $1.04$ & $5.50$\\
$2^{-3}$, $X_2$ & $1.39$ & $3.17$ & $-1.04$ & $-0.75$ & $2.29$ & $0.58$\\
$2^{-1}$, $X_2$ & $1.70$ & $3.09$ & $-0.90$ & $-0.49$ & $1.82$ & $0.65$\\
$2^{-3}$, $X_1X_2$ & $1.19$ & $1.39$ & $-0.08$ & $-0.12$ & $1.16$ & $0.40$\\
$2^{-1}$, $X_1X_2$ & $1.10$ & $1.32$ & $-0.14$ & $-0.08$ & $1.20$ & $0.64$\\
$2$, $X_1X_2$ & $1.22$ & $1.22$ & $-0.15$ & $+0.15$ & $1.00$ & ---
\end{tabular}
\end{ruledtabular}
\caption{Strong-coupling and multi-operator extension of
Table~\ref{tab:eth}, same decomposition Eq.~\eqref{eq:VOsplit}, same
binning ($\Delta E=\Delta\omega=0.5$, band center
$\bar E\in[-0.5,0.5]$, $\omega$ summed over the full grid), units of
$10^{-3}$ per qualifying pair, estimated from
$n_{\mathrm{real}}=1.5\times10^{3}$ realizations.  Here
$\Delta^{\mathrm{ex}}\equiv V_O-V^{(1\mathrm R)}$.  The
$\lambda=2^{-3}$
and $2^{-1}$ rows of $X_2$ reproduce the corresponding $X_1$ rows of
Table~\ref{tab:eth} within $2\%$ and serve as a cross-check of the two
runs.  The last column is left open where $V_O-V^{(1\mathrm R)}$ is
comparable to the bin-crossing floor and the ratio is not resolved.}
\label{tab:ethstrong}
\end{table}

The projection is organized by the class machinery of
Sec.~\ref{sec:conn}: the pairs $(a,a\oplus e_1)$ are the one-flip
family generated by the $g_1X_1$ vertex of $H$ itself---the
observable's own coupling class, whose anchor-level member is the
one-shared-move class of the census of Sec.~\ref{sec:genbench} at
amplitude $\Theta=2\lambda^2$.  The chain
\begin{multline}
\Theta_{\mathfrak c}\;\longrightarrow\;\cC^{(2)}
\;\longrightarrow\;\cS^{(2)}\\
\longrightarrow\;P_O[\cS^{(2)}]=\Delta^{(2)}
\;\longrightarrow\;V_O=V^{(1\mathrm R)}+\Delta^{(2)}+R
\label{eq:chainpair}
\end{multline}
is therefore completed with its last three steps exact identities
verified here, and its first step the content of the blind protocol
of Sec.~\ref{sec:num}, which predicts the class-resolved cumulant
at $0.35$--$0.97\%$ over the tested probes: the observable payload
of that sector content is quantified below.  The
interference term $R$ is by definition the residual beyond the
retained two-resolvent projection.  It has no random-phase
counterpart---for
a fully delocalized random basis every four-channel contraction
with four distinct channel indices vanishes by the orthogonal
Weingarten pairings \cite{Weingarten,CollinsSniady} for
$n\ne m$---which suggests that part of the residual is associated
with eigenbasis coherence not captured by the retained
two-resolvent projection: the measured $R$, comparable to
$\Delta^{(2)}$ rather than a small residue, records the
structured phases surviving
the disorder average.  This picture is consistent with the
observation that physical matrix elements retain correlations beyond
the simplest random-matrix description even where standard
thermalization indicators already hold \cite{Richter2020}.

Two boundaries are respected.  The two-resolvent output is not
itself an operator-specific ETH function: it supplies the
spectral-weight correlations that enter such a function after the
projection $P_O$ onto a chosen observable, exactly as delimited in
Sec.~\ref{sec:genbench}, and the $r=1$ sector supplies only the
disconnected baseline $V^{(1\mathrm R)}$.  What the hierarchy buys
at the observable level is concrete and order-one: dressed
multi-resolvent correlations determine the observable ETH
corrections---the sign, magnitude, and energy-resolved shape of the
correction missing from any single-resolvent baseline for the ETH
variance, organized by the coupling census and dressed by the
self-consistent loop.

It remains to state precisely what this section does and does not
establish.  It establishes an observable-level fact: the retained
two-resolvent sector supplies the dominant part of the correction
that the single-resolvent baseline misses, with the correct sign and
energy-resolved shape, and the projection that carries the sector
content into the observable is exact.  It does not establish that the
underlying energy-window statistics obey local rotational invariance.
A single observable samples one projection $P_O$ of a distribution,
so a distribution can pass this test while violating the window
moment relations---and conversely, agreement with an LRI reference
would not by itself fix the value of this correction at a given
coupling.  The pair-level split of Eq.~\eqref{eq:splitpair} is the
instrument that Appendix~\ref{sec:LRI} reuses as its microscopic
diagnostic---the split is an exact identity and carries no model
input---but the question posed there is the different one: whether
the microscopic window moments satisfy the Weingarten reference
relations, and by how much they fail to.

\section{Microscopic diagnostics of local rotational invariance}
\label{sec:LRI}

The preceding section measured one projection of the correlation
content; the question posed here concerns the statistics it projects
from.  The ETH-variance test of Appendix~\ref{sec:eth} establishes an
observable-level fact about one operator---correlated channel
contributions dominate the correction to the single-resolvent
baseline---and says nothing by itself about the window moment
relations behind it.  LRI supplies the statistical reference for
exactly those relations, and the microscopic closure supplies an
independent route to the same moments, so that the comparison is a
test rather than a consistency check of a shared assumption.

The question whether the energy-local random structure of full ETH
can be \emph{derived} rather than postulated is posed for the exact
hierarchy in the companion paper \cite{HOETH}; the derivation
requires six steps: a statistical structure for the coupling
fluctuations, an effective distribution for the eigenstate rotation,
energy-window localization, approximate isotropy within each window,
sufficiently weak correlations between windows, and error control in
the dense-spectrum limit.  The exact hierarchy of \cite{HOETH} is
constructed per Hamiltonian and carries no statistical content, so it
supplies the window bookkeeping but cannot complete the derivation.
The closure of the present work supplies the missing statistical
layer, and this section assembles the derivation program around it,
with its status step by step.  The closure and the diagnostics are
linked at the conceptual interface only: the fixed-point machinery of the closure produces
the closed multi-resolvent content, while the window diagnostics
below consume the measured spectral data of the benchmarks
directly---the defect scan of Appendix~\ref{sec:LRIdefect} is a second,
independent consumer of the microscopic input rather than a
downstream product of the closure, and the propagation of the
parameter-space bound $\varepsilon_{\rm cl}$ to the moment level is
bounded explicitly in Eqs.~\eqref{eq:momtrans}--\eqref{eq:momprop}
(Appendix~\ref{sec:LRIstatus}).  The deeper structural
development of the pair channel---the dressed pair-statistics
mechanism behind the drift kernel, the realization-level
transferability mechanics, and the higher-order spectral-weight
structure---is not attempted here; the instruments are used only as
diagnostics, and the census of Appendix~\ref{sec:census} records the
verdicts they produce.
The five levels of the evidence map are then the following, and no
level implies the one above or below it: the same microscopic input
feeds both tests, and each level carries its own boundary.
\begin{itemize}
  \item \emph{Microscopic identity.}  The pair split of the fixed
    spectral cumulant, Eq.~\eqref{eq:splitpair}, is exact per
    realization and closes before any averaging; it implies no ensemble
    closure and no accuracy statement for a projected observable.
  \item \emph{MRSC closure.}  The closed cumulant $\cC^{(2)}$ at the
    fixed point is a blind prediction at $0.35$--$0.97\%$ on the tested
    probes (Sec.~\ref{sec:num}), and its projection and boundary
    transport are bounded by
    Eqs.~\eqref{eq:transportbound}--\eqref{eq:bdybound}; no
    Hamiltonian-only bound on the ansatz model error---the binned
    content outside the sector image, Eq.~\eqref{eq:bdytransport}---is
    implied.
  \item \emph{ETH observable.}  The two-resolvent sector supplies an
    order-one, sign-correct, energy-resolved correction to the
    off-diagonal variance of Eq.~\eqref{eq:Vdef} (Appendix~\ref{sec:eth});
    the model error of Eq.~\eqref{eq:bdytransport} and the seed tail
    beyond the measured consecutive-order ratio remain open.
  \item \emph{LRI reference.}  The window dictionary against the
    Weingarten reference $M_r^{\rm Wg}$ is exact, with finite-order
    defects quantified (Appendix~\ref{sec:LRIdefect}); matching at
    $r=2$ is not complete LRI, and the thermodynamic control of the
    residuals is open.
  \item \emph{LRI microscopic origin.}  The same multi-resolvent
    statistics yield the exact kernel identity, the identifiability
    boundary, and the closure-boundary census
    (Table~\ref{tab:LRIstatus}); a derivation of Haar window statistics
    from a general Hamiltonian remains open.
\end{itemize}

\subsection{Microscopic-to-LRI moment dictionary}
\label{sec:LRIdict}

The target is the \emph{local rotational invariance} (LRI) of
Ref.~\cite{VFP25}, which makes the statistical content of full ETH
precise: the energy-resolved matrix data of an observable $A$ is
modeled by the invariance $A\sim U_{\rm loc}^{\dagger}AU_{\rm loc}$
with $U_{\rm loc}=\bigoplus_kU^{(I_k)}$ an independent Haar rotation
within each energy window $I_k$ carrying $D_k\simeq e^{S(E_k)}$
states.  The consequences derived in Ref.~\cite{VFP25} are the
factorization of window-averaged matrix-element moments into local
free cumulants $K_n(E_1,\dots,E_n)$ with a non-crossing (cactus)
structure, subleading corrections generated by merging two blocks of
the non-crossing partition at coinciding energy arguments, and the
identification of the ensemble average with the energy-window
smoothing of a single realization.

The dictionary below is accordingly a \emph{conditional prediction},
not an emergence theorem: it states what the window moments must be
\emph{given} the LRI structure, while the microscopic closure of
Secs.~\ref{sec:closure}--\ref{sec:predict} supplies the independent
route through which the same moments are computed from $H$, so that
their difference is a test rather than an assumption.

The difference itself is carried by an exact object, organized in
three layers with no jumps between them.  \emph{Layer 1 (exact
representation):} the channel data entering the order-two defect is
the two-state joint intensity
$S_X^{(2)}(E,E')=\sum_{n\neq m}|X_{nm}|^2\,\delta(E-E_n)\delta(E'-E_m)$,
which the double-resolvent trace of Eq.~\eqref{eq:doubleres}
represents exactly,
\begin{equation}
\operatorname{Tr}\bigl[X R(z) X R(z')\bigr]
=\iint \frac{S_X^{(2)}(E,E')\,dE\,dE'}{(z-E)(z'-E')},
\label{eq:doubleres}
\end{equation}
and the splitpair identity of Appendix~\ref{sec:eth} writes each pair
content $|X_{nm}|^2=V^{(1R)}_{nm}+\Delta^{(2)}_{nm}+R_{nm}$ as the
one-resolvent projection, the projected order-2 spectral cumulant, and
the off-diagonal pairing remainder.  \emph{Layer 2 (statistical
closure):} whether the pair content is decided by lower-order
(hierarchy-level) objects is the closure question---the exact
representation states nothing about it.  \emph{Layer 3 (computable
estimation):} closed formulas or controlled approximations with named
error sources.  Exactness at Layer~1 does not imply closure at
Layer~2, and closure does not imply computability at Layer~3; the
merit of the double-resolvent identity is that it converts the LRI
window conditions into a bookkeeping problem whose failure modes are
themselves testable, as Appendix~\ref{sec:LRIdefect} shows.  Three
tiers of hypothesis are used, and stated explicitly.
(i)~\emph{Random structure}: the rotation class is the orthogonal
Haar group within each window, carrying the orthogonal Weingarten
weights of Sec.~\ref{sec:conn}; this tier fixes the
finite-dimensional combinatorial form of the reference moments.
(ii)~\emph{Window structure}: the window rotations $U^{(I_k)}$ are
independent across distinct windows, and the coincidence limit is the
$\delta_\Delta(E_i-E_j)/D_i$ contraction of
Eq.~\eqref{eq:coincdict}, one merge per $\sigma$-power; this tier
decides which moment terms merge and which keep distinct window
labels.  The merge rule is exact within the model class---the
$\sigma_0$-powers of Eq.~\eqref{eq:coincdict} are its $r=2,3,4$
instances, with combinatorial weights fixed by the Weingarten block
structure---and outside the model class the merge rule becomes a computable,
order-by-order testable window condition exercised in
Appendix~\ref{sec:LRIdefect}: the bookkeeping observable is the residual route split
$K_4^{\rm NC}-K_4-X_4$ read at partial coincidence
$(E_A,E_A,E_B,E_B)$, and the pass criterion is that it reduces to the
smoothness-mixing content rather than an unaccounted
orientation content (Prop.~\ref{prop:r4sep} below).  The order-2
instance of that content is the closed-form route-difference
identity Eq.~\eqref{eq:r2mix}, which holds by definition between the
two merge routes and therefore carries no test content itself; the
physical test is the defect $\delta_r$ of
Appendix~\ref{sec:LRIdefect}, the residual between the measured moments
and the conditional predictions.  (iii)~\emph{Finite-separation
approximation}: the separated projection \eqref{eq:projection} is
exact in the separated limit---where ``separation'' denotes three
distinct regimes: beyond the window scale ($|E_i-E_j|\gg w$, where the
kernel $\sigma_{ij}$ decays), outside the populated band (where the
window has no support), and the decoupling regime $\sigma_{ij}\to0$ in
which the two energy arguments become independent; only the third is
used in the limit statements---while at finite separation the merged
reading holds to smoothness-mixing accuracy
only, as quantified below; the chain of objects is explicit---the
exact object is the separated projection, the approximate object is
the coincidence dictionary read at finite separation, and the error
source is the $\sigma$-weighted smoothness mixing of the window
kernels, exhibited in closed form at Eq.~\eqref{eq:r2mix} below.
No statement of this section asserts
that a general microscopic Hamiltonian produces Haar window
statistics; the assertions are exact inside the model class and
elsewhere become the computable window conditions that the defect
scan of Appendix~\ref{sec:LRIdefect} tests.

For a channel-diagonal observable
$A=\sum_aA_a\ket{\varphi_a}\bra{\varphi_a}$ the matrix-element moments
factorize through the participation weights $p_n^a$ alone, so the
local free cumulants are observable projections of the spectral
cumulants \eqref{eq:Sdef} taken in their window-coincidence
structure.  At order two the identification is already explicit:
\begin{equation}
\begin{gathered}
K_1(E_i)=\sum_aA_a\,\langle p^a\rangle_{E_i}^{\rm w},\\[2pt]
K_2(E_i,E_j)=\sum_{a,b}A_aA_b\,S^{\rm w}_{ab}(E_i,E_j),
\end{gathered}
\label{eq:LRIbridge}
\end{equation}
where $\langle\,\cdot\,\rangle^{\rm w}$ denotes the window projection
of the on-shell boundary value---the $\eta$-coarse-grained reading of
\eqref{eq:stieltjes}---and $S^{\rm w}_{ab}$ is the corresponding
window-projected two-point spectral cumulant of channels $a$ and $b$.
The diagonal element of \eqref{eq:LRIbridge} at coinciding energies is
the diagonal ETH variance: by \eqref{eq:diagmass} the coincidence
limit of $S^{(2)}$ is $(q_2-1)/D$, and the LRI statement that the
diagonal variance equals the off-diagonal zero-frequency value
\cite{VFP25} is the merge-two-blocks rule at order two, realized per
channel by \eqref{eq:diagmass} and per observable by the two-resolvent
projection $\Delta^{(2)}$ of Appendix~\ref{sec:eth}, which supplies
$56$--$68\%$ of the observed deviation with the correct sign.  At
general $n$ the LRI moment--cumulant relation
$\Tr(\hat\sigma_{E_1}A\cdots\hat\sigma_{E_n}A)
=\sum_{\mu\in\mathrm{NC}(n)}K_\mu\,\sigma_{\mu^*}$, with
$\hat\sigma_E$ the microcanonical window projector, \cite{VFP25}
mirrors the hierarchical structure of the spectral cumulants: each
$K_\mu$ is the observable projection of the window-coincident
$S^{(m)}$, and the subleading corrections obtained by merging two
blocks at equal energies are the coincidence limits of the
multi-energy arguments---the same limits that organize the diagonal
masses \eqref{eq:diagmass} at every order.

The partition-level content of the bridge admits the following
projection form, which upgrades the analogy with
Ref.~\cite{VFP25} from a correspondence to a dictionary.
\begin{proposition}[Conditional window-projection dictionary]
\label{prop:windict}
Under the orthogonal-Haar window model and the stated
coincidence/merger rules, for a channel-diagonal observable
$A=\sum_aA_a\ket{\varphi_a}\bra{\varphi_a}$ let
\begin{multline}
\mathcal P_A^{(r)}\left[S^{(r),\rm w}\right](E_1,\dots,E_r)\\
=\sum_{a_1,\dots,a_r}A_{a_1}\cdots A_{a_r}
\,S^{(r),\rm w}_{a_1\cdots a_r}(E_1,\dots,E_r)
\label{eq:channelproj}
\end{multline}
be the channel-weight projection of the window-projected $r$-point
spectral cumulant.  For energy arguments localized in distinct
windows the local free cumulants of Ref.~\cite{VFP25} factorize into
the block projections of Eq.~\eqref{eq:channelproj},
\begin{equation}
K_\pi(E_1,\dots,E_n)
=\prod_{B\in\pi}
\mathcal P_A^{(|B|)}\left[S^{(|B|),\rm w}\right]
\bigl(\{E_i\}_{i\in B}\bigr),
\label{eq:projection}
\end{equation}
and every window coincidence contributes the merger corrections of
the inversion pair of Ref.~\cite{VFP25}: each power of
$\sigma_{ij}=\delta_\Delta(E_i-E_j)/D_i$ contracts one coincident
pair of arguments, merging the two blocks of $\pi$ that carry them.
The merged content is Eq.~\eqref{eq:coincdict} below; the separated
projection \eqref{eq:projection} is exact in the separated limit,
and at finite separation the two routes differ by the
$\sigma$-weighted smoothness mixing of the route-difference identity
Eq.~\eqref{eq:r2mix}, whose
benchmark magnitude at order two is
$|K_2^{\rm NC}-K_2|/|M_2|=1.0\times10^{-3}$ ($\Delta=2\bar d$) and
$2.8\times10^{-3}$ ($\Delta=4\bar d$): the mixing carries the
coincidence link, which decays with separation, times the mean shift
$M_1(E)-M_1(F)$, which on the benchmark grows across the pair, so the
relative mixing is not monotone in $\Delta$.
\end{proposition}

The bridge is anchored at the algebraic level by the sector algebra
of Ref.~\cite{MResmerge}, which computes the same spectral cumulants
$S^{(r)}$ exactly for the orthogonal free-fermion class in the
sector basis
$S^{(r)}=\sum_s N_s\,\mathrm{Cov}_r(s)\,\rho^{(r)}_s$, where $N_s$
is the class degeneracy, $\mathrm{Cov}_r(s)$ the exact Weingarten
content of class $s$, and the fully coincident class the
single-channel moment projection.  Since the closure computes the
same $S^{(r)}$ from $H$ through Eq.~\eqref{eq:Sdef}, the two routes
meet at the coincidence limits, where the inversion pair of
Ref.~\cite{VFP25} takes a closed form.
\begin{proposition}[Coincidence/merge dictionary]
\label{prop:coincdict}
Setting all energy arguments to $E$ and writing
$\sigma_0=\delta_\Delta(0)/D_E$, the fully coincident window moments
close on the local free cumulants as
\begin{equation}
\begin{gathered}
M_2=K_2+K_1^2\sigma_0,\\[2pt]
M_3=K_3+3K_1K_2\sigma_0+K_1^3\sigma_0^2,\\[2pt]
M_4=K_4+4K_1K_3\sigma_0+4K_2^2\sigma_0\\[2pt]
\qquad{}+8K_1^2K_2\sigma_0^2+K_1^4\sigma_0^3,
\end{gathered}
\label{eq:coincdict}
\end{equation}
with $M_n(E)=\Tr(\hat\sigma_EA)^n$ the fully coincident window
moment and all $K$'s evaluated at $(E,\dots,E)$; each power of
$\sigma_0$ contracts one coincident pair---one merge of two blocks of
the non-crossing partition, with the combinatorial weight of its
block structure.
\end{proposition}
The dictionary is
exact within the model class of Appendix~\ref{sec:LRIdict}; on the
free-fermion benchmark of Sec.~\ref{sec:bench} it is verified to
machine precision at all twelve tested window centers,
reproducing the multi-resolvent Weingarten results of
Ref.~\cite{MResmerge} at the moment level without importing their
sector tables; the $r=3$ line is the coincidence content of the
eight-class sector closure \eqref{eq:sectorsolve3}, whose basis is
the sixteen-class $(m_{12},m_{13},m_{23},t)$ census of the replica
hierarchy.  The merge-two-blocks regrouping of Ref.~\cite{VFP25}---
the $(1-\sigma_0)$-weighted recombination that generates the
subleading corrections---agrees with the dictionary through $n=3$
identically, while at $n=4$ the two routes differ by
$2\sigma_0M_2(M_2-M_1^2\sigma_0)$.  The tripartite
correspondence---Weingarten contraction merging of the sector
algebra, energy-window coincidence of the spectral cumulants, and
LRI block merging---is therefore quantitative: it is exact at
$r=2,3$, and at $r=4$ the sector algebra of
Ref.~\cite{MResmerge} remains the exact anchor against which the LRI
subleading structure is calibrated.

At $n=4$ the split has a partition-level reading that fixes its
sign, magnitude, and limits.  At full coincidence the merge route of
Ref.~\cite{VFP25} counts each of the two non-crossing double
pairings $\{12|34\}$ and $\{14|23\}$ once---one $K_2^2\sigma_0$
position and its $K_1^2K_2\sigma_0^2$ decorations, six positions in
all---whereas the exact dictionary \eqref{eq:coincdict} counts each
pairing twice, once per orientation of its coincidence link, giving
four $K_2^2\sigma_0$ and eight $K_1^2K_2\sigma_0^2$ positions.  The
four missing positions are precisely the second orientations,
\begin{equation}
\begin{aligned}
X_4&\equiv K_2(12)K_2(34)\sigma_{13}
+K_2(23)K_2(41)\sigma_{24}\\
&\quad{}+K_1(1)K_1(3)K_2(24)\sigma_{12}\sigma_{34}\\
&\quad{}+K_1(2)K_1(4)K_2(31)\sigma_{23}\sigma_{41},
\end{aligned}
\label{eq:r4orient}
\end{equation}
degenerating at full coincidence to
$X_4=2K_2^2\sigma_0+2K_1^2K_2\sigma_0^2$.  The candidate attribution
of the excess to the crossing pairing $\{13|24\}$ is excluded on two
grounds: its content carries no coincidence link---it is the single
pairing with no $\sigma_0$ factor, so it survives the separated
limit $\sigma_0\to0$ in which the split vanishes---and evaluated
numerically it returns a value of the split's own magnitude but not
equal to it, while $X_4$ reproduces $\Delta_4$ to machine precision.
The two route contents differ by
\begin{equation}
\begin{aligned}
\Delta_4^{\rm pair}
\equiv M_4^{\rm NC}-M_4
&=2\sigma_0K_2(K_2+K_1^2\sigma_0)\\
&=2\sigma_0K_2M_2
=2\sigma_0M_2(M_2-M_1^2\sigma_0),
\end{aligned}
\label{eq:r4split}
\end{equation}
and the same $\Delta_4$ separates the cumulant assignments,
$K_4^{\rm NC}-K_4=\Delta_4$, $K_4^{\rm NC}$ denoting the cumulant of
the merge route and $K_4$ that of the inversion route; the two routes
therefore reconstruct
the \emph{same} moment $M_4$, the split being a redistribution
between the order-4 cumulant and the pairing content, not a
moment-level disagreement.  Equivalently, merging the coincidence
pairings before the M\"obius inversion (the route of
Ref.~\cite{VFP25}) and inverting before merging (the dictionary
route) do not commute at order four: $K_4^{\rm NC}$ is the cumulant
of the merge route, $K_4$ that of the moment-field inversion
\eqref{eq:coincdict}, and the two are distinct order-4 cumulants of
the same moment related by the split.  Four limits calibrate it: (i)
$K_2=0$ ($M_2=M_1^2\sigma_0$) gives $\Delta_4=0$; (ii) $M_1=0$ gives
$\Delta_4=2\sigma_0M_2^2$, so the split is mean-independent and
survives centering; (iii) $\sigma_0\to0$ gives $\Delta_4\to0$, the
separated routes agreeing exactly; (iv) $M_2\to0$ gives
$\Delta_4\to0$.  On the benchmark the identity
$K_4^{\rm NC}-K_4=X_4=2\sigma_0M_2(M_2-M_1^2\sigma_0)$ holds to
machine precision at every tested center; away from coincidence an
additional smoothness-mixing correction of the
$\sigma_{ij}K_1(i)[M_2(j,k)-M_2(i,k)]$ class---the same mixing that
distinguishes the $r=2$ dictionaries \eqref{eq:coincdict} and
Ref.~\cite{VFP25} at partial coincidence---enters at relative order
a few percent at the tested near-coincident points, and vanishes at
full coincidence and full separation.  A minimal partial-coincidence
example quantifies the approach to these limits on the $N=14$, $k=2$
benchmark ($w=4\bar d$): pinning one coincident pair at
$(E,E,E+2w,E+4w)$ lowers the split from $1.46\times10^{-5}$ at full
coincidence (the band-center value $\Delta_4=1.5\times10^{-5}$ of
Proposition~\ref{prop:r4coinc}) to $1.0\times10^{-6}$, $X_4$ tracking
it within a mixing
correction of $1.4\%$ of the split; recentering the channel weights,
which suppresses $K_1$, moves that fraction to $6.7\%$ at the pinned
pair and lowers the absolute mixing by an order of magnitude at the
larger separations, consistent with the mixing carrying one factor of
$K_1$.  At four separated anchors the split falls to $1\times10^{-8}$
---the routes re-converge at general position---while $X_4$ retains a
$2.4\times10^{-7}$ value that the mixing cancels, confirming that the
orientation-doubling content is a coincidence-limit object whose
general-domain residue is the smoothness mixing itself.  A minimal
analytic example makes the two domains explicit in closed form.
On a small band of levels with channel weights $a_n=\pm1$ and
diagonal $A$, the window weights
$w_{E,n}=G_w(E-\lambda_n)/D_E$---$G_w$ the Gaussian window kernel of
width $w$ and $D_E$ the density of states at $E$---close the moments
elementarily---$M_1(E)=\sum_n a_nw_{E,n}$,
$M_2(E,F)=\sum_n w_{E,n}w_{F,n}$, and
$M_4(E_A,E_A,E_B,E_B)=\sum_n w_{A,n}^2w_{B,n}^2$---and the order-2
route difference of the two definitions
$K_2^{\rm NC}\equiv M_2(E,F)-\sigma(E,F)M_1(E)M_1(F)$ and
$K_2\equiv M_2(E,F)-\sigma(E,F)M_1(E)^2$, which share the same
moment, is exactly
\begin{equation}
K_2^{\rm NC}-K_2
=\sigma(E,F)\,M_1(E)\bigl[M_1(E)-M_1(F)\bigr],
\label{eq:r2mix}
\end{equation}
with $\sigma(E,F)=\delta_\Delta(E-F)/D_E$: the smoothness mixing is
the coincidence link times the mean shift across the pair, vanishing
at coincidence ($F=E$) and at separation ($\sigma\to0$), and on the
benchmark Eq.~\eqref{eq:r2mix} reads
$|K_2^{\rm NC}-K_2|/|M_2|\approx10^{-3}$ at the tested separations.
At order four the partially coincident configuration
$(E_A,E_A,E_B,E_B)$ separates the two contents cleanly: at
$E_B=E_A$ the route split $K_4^{\rm NC}-K_4$ equals $\Delta_4$ of
Eq.~\eqref{eq:r4split} and $X_4$ to machine precision with vanishing
mixing, while across the populated band ($E_B=E_A+w$, $2w$) the
split decays ($4.3\times10^{-2}\to1.2\times10^{-2}$) and $X_4$
retains a finite residue approaching
$K_1(E_A)K_1(E_B)K_2(E_A,E_B)\sigma_0^{A}\sigma_0^{B}$ as the cross
link $\sigma_{AB}$ vanishes, the mixing difference
$K_4^{\rm NC}-K_4-X_4$ closing the gap.  The orientation-doubling
reading of the split is therefore operative on the coincidence limit
alone, order by order; no all-order generalization is asserted.
A reconstruction built from
$K_4$ with the merge pairing form misses $M_4$ by exactly
$-\Delta_4$: the same bookkeeping fact stated twice.  The
distinct-index protocol of Appendix~\ref{sec:LRIdefect} uses a
window-independent reference in which no $\sigma_0$-pairing enters,
so the route split does not propagate into the defect scan.
\begin{proposition}[Route split at full coincidence]
\label{prop:r4coinc}
For the diagonal observable of the six-level example, at full
coincidence $(E,E,E,E)$ the route split equals the
orientation-doubling content,
$K_4^{\rm NC}-K_4=X_4=\Delta_4$ with $X_4$ of
Eq.~\eqref{eq:r4orient} and $\Delta_4$ of Eq.~\eqref{eq:r4split}.
The identity is exact algebra, verified pointwise to machine
precision---the six-level value $0.0434$ with the mixing
residual $K_4^{\rm NC}-K_4-X_4$ below $5\times10^{-17}$ and
$X_4-\Delta_4=0$; the magnitude is example-specific, and the
band-center value of the $N{=}14$ benchmark,
$\Delta_4=1.5\times10^{-5}$, belongs to a different model, weight
profile, and normalization and is not to be compared across
examples---and the coincidence reduction
$X_4\to2K_2^2\sigma_0+2K_1^2K_2\sigma_0^2$ is verified to fifteen
digits at every tested center.
\end{proposition}
\begin{proposition}[Finite-separation bias and its status]
\label{prop:r4sep}
At partial coincidence $(E_A,E_A,E_B,E_B)$ with $E_A\neq E_B$ the
split is no longer the orientation content:
$K_4^{\rm NC}-K_4-X_4\neq0$.  The band sweep above quantifies the
deviation---split $4.3\times10^{-2}\to1.2\times10^{-2}$ across
$E_B=E_A\to E_A+2w$, $X_4$ residue approaching
$K_1(E_A)K_1(E_B)K_2(E_A,E_B)\,\allowbreak\sigma_0^{A}\sigma_0^{B}$,
and the
mixing crossing zero between $w$ and $1.5w$ beyond
$E_A$ ($-1.4\times10^{-3}$ to $+6.4\times10^{-3}$); on the benchmark pinned
pair the mixing is $1.4\%$ of the split ($6.7\%$ for the recentered
weights).  The reading of the missing positions as \emph{second
orientations} is therefore a coincidence-limit characterization---at
general energy configurations the dictionary's four pair positions
$M_2(ab)\sigma(ac)M_2(ad)$, one per cyclic anchor, and the merge
route's two partition-shaped positions
$K_2(ab)K_2(cd)\sigma(ac)$ are positionally distinct and degenerate
onto common values only at full coincidence---and the
finite-separation statements are numerical observations at the tested
points; a stronger claim, uniqueness of the missing content over the
entire parameter domain, is not asserted.
\end{proposition}
\begin{proposition}[Order-2 route-difference identity]
\label{prop:r2mix}
Equation~\eqref{eq:r2mix} holds for any operator $A$ and any window
kernel: it is the difference of the two route definitions introduced
above, which share the same
moment---pure algebra, with no diagonal assumption (the diagonal
model closes $M_1,M_2$ elementarily, but the identity does not
require it).  Its coincidence limit vanishes when $M_1$ is
continuous---the mean shift is linear in $F-E$ with slope
$\sigma M_1M_1'$---and its separation limit vanishes when the
kernel decays, $\sigma\to0$, with the relevant moments bounded.
Both conditions can fail: a kernel
with a nonzero tail keeps the separation limit finite at the tail
value times the mean shift, and a discontinuous (box) window breaks
the coincidence limit at window edges where $M_1$ jumps.  The
mixing vanishes identically whenever $M_1(E)=M_1(F)$.  Three
distinct vanishing conditions are therefore in play and must not be
conflated under a single centering label: window-resolved centering
($M_1(E)=0$ throughout the tested window domain, for which
$M_1(E)=0$ at one endpoint already suffices), window-mean equality
($M_1(E)=M_1(F)$ without centering), and the global mean subtraction
the benchmark's recentered weights realize, which guarantees
neither.  On the benchmark the global subtraction suppresses $M_1$
from $0.095$ to $0.018$ at the band center and cuts the mixing
fivefold ($2.1\times10^{-4}$ and $5.5\times10^{-4}$ of $|M_2|$ at
$\Delta=2\bar d$ and $4\bar d$)---a benchmark-specific numerical
observation rather than a general statement about recentering---and
$M_1$ stays at the $10^{-2}$ level across the tested pair.  On the benchmark the magnitude is the
$10^{-3}$-of-$|M_2|$ level quoted in
Prop.~\ref{prop:windict}---the $\sigma$-weighted route-difference
content of the tier-(ii) merge rule of Appendix~\ref{sec:LRIdict}, a
definitional identity; the physical test is the defect of
Appendix~\ref{sec:LRIdefect}.
\end{proposition}

The dictionary so far states what follows \emph{if} the window model
holds.  The microscopic question is the converse: under what property
of the retained amplitudes does the whole chain---sector algebra,
window projection, coincidence merge---close?  Two hypotheses answer
it, and their separation is the logical content of the statement: one
gives the \emph{factorization} of the class sum over overlap sectors,
the other identifies the collapse ratios with the \emph{weights} of
the distinct-index reference.  Both are exactly the structures the
sector algebra of Ref.~\cite{MResmerge} realizes in the free-fermion
class; neither is assumed elsewhere.

\begin{proposition}[Conditional form of the LRI dictionary]
\label{prop:LRIcond}
Let the retained basis carry the class amplitudes
$\cC^{(r)}=\sum_{\mathfrak c}\Theta_{\mathfrak c}^{(r)}
\mathfrak B_{\mathfrak c}^{(r)}$ of Eq.~\eqref{eq:classseed}, and
suppose that
\begin{itemize}
\item[(H1)] \emph{sector collapse.}  For every retained class
$\mathfrak c$ and every augmentation $\mathfrak c\cup\{c\}$ the
amplitude ratio
$\Theta^{(r+1)}_{\mathfrak c\cup\{c\}}/\Theta^{(r)}_{\mathfrak c}$
depends only on the overlap sector $\ell(c)$ of the augmentation and
on its degree, not on the remaining leaf structure of
$\mathfrak c$;
\item[(H1$'$)] \emph{reference weights.}  The common ratio of (H1) is
the per-index weight of the distinct-index reference,
$\Theta^{(r+1)}_{\mathfrak c\cup\{c\}}/\Theta^{(r)}_{\mathfrak c}
=q_{\ell(c)}D$ for every augmentation of overlap $\ell(c)$, with
$q_\ell$ the centered $q$-moments of Eq.~\eqref{eq:qbar};
\item[(H2)] \emph{window realization.}  The retained channel content
projects onto the window structure through
Eq.~\eqref{eq:channelproj}, with the merge rule of
Prop.~\ref{prop:windict} valid inside each window up to the
smoothness mixing of Eq.~\eqref{eq:r2mix}.
\end{itemize}
Then for every $r\le r_{\max}$ and every window tuple the window
moments of the retained basis satisfy
\begin{equation}
\bigl|M_r^{\rm cond}-M_r^{\rm Wg}\bigr|
\le\varepsilon_{\rm iso}^{(r)}
+L_r\Lambda_r\,\varepsilon_{\rm cl}
+\bigl|{\rm mix}\bigr|_r,
\label{eq:LRIcondbound}
\end{equation}
with $\varepsilon_{\rm iso}^{(r)}$ the within-window anisotropy
defect of Appendix~\ref{sec:LRIdefect}---measured from the pair-content
identity alone, without reference to $M_r^{\rm Wg}$---$\varepsilon_{\rm
cl}$ the closure residual of the retained parameter vector in the norm
of Eq.~\eqref{eq:momprop} (a model-deficit quantity, not the solver
residual $r_0$ of Sec.~\ref{sec:errorprop}), and the last term the route
mixing of Eq.~\eqref{eq:r2mix}.  All three terms on the right-hand side
of Eq.~\eqref{eq:LRIcondbound} are window-moment units: the first and
the third are measured in that space directly, and the second is the
parameter-space residual of Eq.~\eqref{eq:momprop} transmitted by
$L_r\Lambda_r$; in particular the reference
relations of the window model---and with them the LRI moment--cumulant
dictionary---hold as \emph{identities on the retained basis} whenever
the three terms vanish.
\end{proposition}

\begin{proof}[Proof]
Under (H1) the class sum of a retained cumulant factors over overlap
sectors: every augmentation of a class by a leaf of overlap $\ell$
multiplies the parent amplitude by the same ratio, so the summed
content of a class depends on its leaves only through their overlap
populations.  Factorization alone does not fix the coefficient with
which a given cycle structure enters---the reference's Weingarten
weights, its Gram inverse, and its normalization all sit in that
coefficient---and this is precisely what (H1$'$) supplies: each
augmentation carries the reference weight $q_\ell D$, so a class of
cycle structure $\tau$ contributes
$\mathrm{Wg}(D,\sigma_*\tau^{-1})\prod_{c\in\tau}q_{|c|}D$, and the
class sum of any retained cumulant reproduces the term-by-term content
of Eq.~\eqref{eq:wgdistinct}.  The residual of the retained block is
the anisotropy content: it is measured from the pair-content identity
Eq.~\eqref{eq:splitpair} alone, without reference to
$M_r^{\rm Wg}$, so the bound compares two independently obtained
quantities rather than restating the defect.  Under (H2) each sector
term is projected onto the window indices by
Eq.~\eqref{eq:channelproj}, and the coincidence limits of the
projected moments close on Eq.~\eqref{eq:coincdict} up to the route
mixing of Eq.~\eqref{eq:r2mix}.  The amplitude vector itself is fixed
by the closure only to within its own converged residual, and that
parameter-space error reaches the moments through the transmission
bound of Eq.~\eqref{eq:momprop}, giving the second term of
Eq.~\eqref{eq:LRIcondbound}.
\end{proof}

Three statements of different status are separated here, and the
separation is the point.  Hypothesis (H1) and its companion
(H1$'$) are \emph{exact} in the
free-fermion class, where the class amplitudes reduce to the
$\chi^2$ cumulants $\Theta^{(2)}=2$, $\Theta^{(3)}=8$
of Eq.~\eqref{eq:G2seed} and carry no leaf dependence, and the
dictionary was verified there to machine precision at all twelve
tested window centres.  In a generic-interaction system (H1) and
(H1$'$) are \emph{measured, not derived}: the defect magnitudes
$0.02$--$0.35$ ($r=3$) and $0.03$--$0.70$ ($r=4$) of the spread in
Appendix~\ref{sec:LRIdefect} are exactly the failure of the collapse and
of the reference weights, so the LRI defect becomes a coordinate on
how far a given Hamiltonian
sits from the sector-collapsed structure---a falsifiable statement in
place of an external statistical assumption.  What remains open, and
is the microscopic step of Appendix~\ref{sec:LRIstatus}, is a derivation
of (H1) and (H1$'$) from $H$ for a general chaotic Hamiltonian: the
free-fermion
class obtains it from the Weingarten algebra, and no analogous
mechanism is known outside that class.  With
Eq.~\eqref{eq:LRIcondbound} the logical role of the LRI reference in
this article is therefore fixed: it is not assumed as a property of
eigenstate statistics at large, but derived from the collapse
property of the retained amplitudes, with the deviation from
collapse carried by the three named terms.

\emph{What (H1) does and does not require.}  The collapse is not a
consequence of Gaussian statistics of the couplings alone, and this
can be verified exactly.  Take the Gaussian coupling class with leaves
of support up to three sites, the amplitude of a leaf being the
$\chi^2$-type object $\prod_{i\in S}g_i^2$; the class amplitude is the
joint cumulant of the leaf monomials and is given in closed form by
Wick's theorem.  At $r=2$ the amplitudes are $2$, $8$, and $26$ for
overlaps one, two, and three---the first of these is exactly the
free-fermion amplitude $\Theta^{(2)}=2$---so the $r=2$ sector already
reproduces the class content of the benchmark.  Augmenting a class by
one further leaf, however, does not produce a ratio fixed by the
overlap sector: at fixed $\ell(c)=1$ the ratio of the augmented to the
parent amplitude takes the values $2$ and $4$ according to whether
the augmentation falls on a single-site or on a multi-site mother; at
$\ell(c)=2$ it takes the values $8$, $10$, $14.5$, $16$, $24.3$, and
$25$; at $\ell(c)=3$ the values $52$, $58$, $79$, and $127$.  The
collapse is therefore a property of the \emph{leaf structure} of the
retained basis, not of the coupling statistics: it is exact in the
free-fermion class because there every retained leaf is a
single-entry object, so the mother carries no leaf structure for the
ratio to depend on, and it must be measured---and can fail---as soon
as the retained basis contains multi-site leaves with connected
overlaps.  In the two benchmarks of this article the retained leaves
are of the trivial type, single-entry in the free-fermion class and
factorized in the census of the ring TFIM, which is consistent with
the collapse being observed there and with its failure appearing only
as the small window-level defect of Appendix~\ref{sec:LRIdefect} rather
than as a breakdown of the dictionary.

\subsection{Finite-order LRI defect and residuals}
\label{sec:LRIdefect}

The dictionary of Appendix~\ref{sec:LRIdict} converts the LRI target
from a measure-level Haar statement into a finite set of window
conditions on the spectral cumulants.  Their failure is carried by
the order-$r$ \emph{LRI defect}
\begin{equation}
\delta_r(I)
=M_r^{\rm micro}(I)-M_r^{\rm Wg}(I),
\label{eq:LRIdefect}
\end{equation}
the difference between the window moment measured on the
Hamiltonian's eigenstates and the value the sector algebra
prescribes.  The measured moment is an estimator, and it decomposes as
\begin{equation}
\widehat M_r^{\rm micro}
=M_r^{\rm Wg}+\delta M_r^{\rm sys}+\xi_r^{\rm samp},
\label{eq:momentdecomp}
\end{equation}
The decomposition is an identity by definition: let
$M_r^{\rm target}(I)$ be the exact model-level moment the estimator
targets---the microcanonical ensemble value at window $I$, with the
window specified by its Gaussian kernel $G_w$ of width $w=4\bar d$
($\bar d$ the mean level spacing),
$w_{E_I,n}=G_w(E_I-\lambda_n)/\sum_mG_w(E_I-\lambda_m)$,
normalized to
$\sum_n w_{E_I,n}=1$, evaluated at the tested window centers, and with
the channel weights $a_n$ entering through the participation data
$X$ of Sec.~\ref{sec:bench}.  Then set
$\xi_r^{\rm samp}\equiv\widehat M_r^{\rm micro}-M_r^{\rm target}$,
$\delta M_r^{\rm sys}\equiv M_r^{\rm target}-M_r^{\rm Wg}$.
The identity carries no error content: the smallness of
$\delta M_r^{\rm sys}$ is a separate claim, established here only by
independent microscopic computation (the benchmark calibration below)
or by statistical control---never by the decomposition itself.
The three objects then have the following precise contents.
$M_r^{\rm Wg}$ is the distinct-index Weingarten
reference of Eq.~\eqref{eq:wgdistinct}---a window-independent scalar
obtained by contracting the centered $q$-moments of
Eq.~\eqref{eq:qbar} with the orthogonal Weingarten weights---in
which no $\sigma_0$-pairing enters.  The contraction is performed in
the full Hilbert dimension $D$: the window matrix is a submatrix of
the $D$-dimensional orthogonal matrix, not an independent Haar
rotation of the window population $D_I$, so the reference carries no
$D_I$ argument.  (The \emph{conditional} reference of
Appendix~\ref{sec:LRIdict}, deliberately evaluated at $D_I$, is a
different object: it is the window-approximated moment used to test
whether the defect could be absorbed by a window-dimension
convention, and the census records that it cannot.)  What the window does carry is the measured moment
$M_r^{\rm micro}(I)$, and with it the defect $\delta_r(I)$; the
numerical size of the difference between the two conventions is
quantified in Appendix~\ref{sec:LRIdefect}.  The $\sigma_0$-powers of
Eq.~\eqref{eq:coincdict} live in the \emph{coincidence sector} of
the dictionary of Appendix~\ref{sec:LRIdict}, where they are
deterministic finite-$D_I$ reference content at window coincidence,
with combinatorial weights fixed by the merger coefficients; the
distinct-index protocol removes that sector from both measured and
reference moments through the falling weights $\cF_r$, so the route
split of Appendix~\ref{sec:LRIdict} does not propagate into the defect
scan.  Both objects are reference content and do not vanish as more
samples are taken.  $\delta M_r^{\rm sys}$ is the
non-Haar systematic defect, the difference between the target moment
and the reference; it need not vanish with the sample size.
$\xi_r^{\rm samp}$ is the estimation error, and only its statistical
properties are fixed by form: it has ensemble mean zero under the
specified randomization scheme, and its scale is set by the
\emph{effective} number of independent samples and their dependence
structure---correlated windows, shared realizations, or uneven
weighting raise it above the nominal per-draw value---the sampling
error is controlled by the variance of the specified estimator and by
its effective independent sample size: a $D_I^{-1/2}$ scale is a
benchmark-specific expectation under the stated variance assumptions,
not a universal consequence of the moment decomposition---it holds
per independent draw ($D_I^{-1/2}/\sqrt{n_{\rm eff}}$ after
$n_{\rm eff}$ effective draws) only when the estimator variance
scales as $D_I^{-1}$, and it is defined only once the
randomization scheme and sampling unit are specified.  The two roles do not commute: the
reference content---the distinct-index contraction and, in the
coincidence sector, the $\sigma_0$-powers---is deterministic,
unaffected by the amount of sampling, while $\xi_r^{\rm samp}$ is
sampling noise, removed by averaging but carrying no LRI information.  On the
free-fermion benchmark the sums are exact, so
$\xi_r^{\rm samp}=0$ identically and the route identity
Eq.~\eqref{eq:ffcalibration} holds to machine precision, while the
defect of Eq.~\eqref{eq:LRIdefect} there is the drift-calibrated
quantity of the calibration below; on the Ising scan
the twelve-realization mean is the estimator, and its
between-realization spread carries $\xi_r^{\rm samp}$ together with
the disorder variation of $\delta M_r^{\rm sys}$ (the cross term).
The defect of Eq.~\eqref{eq:LRIdefect} is correspondingly
$\delta_r=\delta M_r^{\rm sys}+\xi_r^{\rm samp}$, and LRI requires
only the systematic part to vanish (Eq.~\eqref{eq:LRItarget}),
\begin{equation}
\delta M_r^{\rm sys}\to 0,
\label{eq:LRItarget}
\end{equation}
in the dense-window regime
$\Delta_{\rm level}\ll\eta\ll E_{\rm macro}$.  Without the
separation \eqref{eq:momentdecomp} the deterministic finite-$D$
corrections of the reference, or the sampling fluctuation of a
finite-window estimator, would each be misread as LRI violation.

The free-fermion laboratory of Sec.~\ref{sec:bench} calibrates the
instrument at two levels.  At the route level the sector algebra is
exact and the reconstruction closes---with $M_r$ the measured input
and $K_r,c_r$ reconstructed from it by the two routes below,
\begin{equation}
M_r=K_r+c_r
\label{eq:ffcalibration}
\end{equation}
to machine precision at all twelve tested window centers: the
machine-precision statement is the route identity of
Appendix~\ref{sec:LRIdict}, not the defect of
Eq.~\eqref{eq:LRIdefect}.  At the defect level the benchmark is not trivial: against the window-independent reference the defect reads $\delta=(-5.2\times10^{-4},+1.1\times10^{-6},+5.6\times10^{-7})$, closing per order as a window-content drift plus a diagonal component; the benchmark defect then defines the floor (Eq.~\eqref{eq:deffloor}) $\varepsilon_r^{\rm floor}=(5.2\times10^{-4},1.1\times10^{-6},5.6\times10^{-7})$, against which the Ising entries of Table~\ref{tab:isingdefect} are read as ratios---the small-window $r=2$ entries exceed it by up to $2.1\times$, at $r=3$ only the smallest windows reach it, and at $r=4$ all entries sit below it.  The calibration values and the drift--diagonal split are given in Appendix~\ref{app:ffcalib}.  The two roles in Eq.~\eqref{eq:momentdecomp} are
quantitatively disjoint there.  The calibration is a cross-check of three independent paths---raw moments, moment-field cumulants, and the explicit $\sigma_0$-power corrections---meeting at machine precision (residuals below $10^{-19}$) with no small-denominator amplification, an order-by-order sign structure that persists across $D=55$--$105$, and a deterministic finite-$D$ correction well separated from the estimation fluctuation at this dimension; the order-4 route split $\Delta_4=1.5\times10^{-5}$ is fully accounted for as dictionary bookkeeping.  The path contents and sensitivities are collected in Appendix~\ref{app:threepath}.

On the interacting chain the probe is applied to the fluctuating
sector of the channel, with the ETH mean content removed first.  The
diagonal profile $\braket{n|X_1|n}$ carries the $S^{(1)}$ mean: its
spectral rms is $0.05$--$0.13$ in the tests below, absorbing
$7$--$48\%$ of $\operatorname{Tr}X_1^2$ at $N=10$--$13$, so the raw
window moment is dominated by the $\sigma_0$-content that
Eq.~\eqref{eq:coincdict} already attributes to $K_1$.  The test
therefore acts on the centered channel
$X=O-\operatorname{diag}(O_{nn})$ in the eigenbasis, keeps all
window indices distinct, and builds the reference from the centered
$q$-moments
\begin{equation}
\begin{aligned}
\bar q_1&=0,\qquad
\bar q_2=1-\langle O_{nn}^2\rangle,\\
\bar q_3&=-\frac{3\operatorname{tr}(X^2 d)-2\sum_n O_{nn}^3}{D},\\
\bar q_4&=1-\frac{4\operatorname{tr}(X^3 d)
-6\sum_n O_{nn}^2(X^2)_{nn}+3\sum_n O_{nn}^4}{D},
\end{aligned}
\label{eq:qbar}
\end{equation}
with $d=\operatorname{diag}(O_{nn})$ and $\langle\cdot\rangle$ the
spectral mean $D^{-1}\operatorname{tr}$.  Under the distinct-index
constraint only the cyclic shift $\sigma_*(i)=i-1\pmod r$ survives
the Weingarten contraction, so the reference collapses to a
window-independent scalar,
\begin{equation}
M_r^{\rm Wg}
=\sum_{\tau\in S_r}\mathrm{Wg}\bigl(D,\sigma_*\tau^{-1}\bigr)
\prod_{c\in\tau}\bar q_{|c|}\,D,
\label{eq:wgdistinct}
\end{equation}
with $\mathrm{Wg}$ the orthogonal Weingarten weights of
Sec.~\ref{sec:conn}, evaluated in the pairing-Gram convention of
Eq.~\eqref{eq:wein}: $\mathrm{Wg}(D,\sigma_*\tau^{-1})$ is the
reduced orthogonal weight of the surviving pairing class, and at
$r=2$ the reduction is elementary---of the three pairings the
within-column one, whose blocks pair each index with itself, is
killed by $\bar q_1=0$, and the surviving class carries the net
weight $1/[(D-1)(D+2)]$, so that
$M_2^{\rm Wg}=D\bar q_2/[(D-1)(D+2)]$; the reduction admits an elementary hand check at $D=3$ returning the elementary $O(3)$ moment, worked through in Appendix~\ref{app:D3hand}.  The measured moment
$M_r^{\rm micro}=\cT_r/\cF_r$ is the distinct-index cyclic sum of
the window, with the explicit closed forms of the falling weights
$\cF_r$ and of the M\"obius-subtracted sums $\cT_r$ collected in
Appendix~\ref{app:T4audit}.  The falling weights $\cF_r$ are counting weights, not
projection operators: they equal the probability that $r$
independent draws from the window profile $w_n$ land on distinct
indices, and the protocol \emph{divides} by them, excluding no index
pair a priori.  The coincidence-sector content removed by the M\"obius subtraction, $M_r-\cT_r=(1.38\times10^{-3},1.43\times10^{-4},9.7\times10^{-6})$ ($12.5\%$ of $M_2$, exceeding the net moment at $r=3,4$), is audited class by class in Appendix~\ref{app:T4audit}, where the closed forms agree with the brute-force sums to machine precision and the exhaustiveness rests on the partition-lattice correspondence.

\begin{table*}[t]
\caption{Finite-order LRI defect on the ring TFIM of
Sec.~\ref{sec:genbench} at full coupling, $O=X_1$, extended to
$N=10$--$13$ ($D=1024$--$8192$):
$\varepsilon_r=|M_r^{\rm micro}-M_r^{\rm Wg}|/(1+|M_r^{\rm Wg}|)$
from Eq.~\eqref{eq:LRIdefect}, with $D_I=\rho(E)\eta$ the window
dimension, twelve disorder realizations per entry, five centers
$E/\|H\|=0.3,0.4,\ldots,0.7$ averaged.
$\varepsilon_{\rm iso}$ is the signed residual fraction
$R/\langle|O_{nm}|^2\rangle$ of Eq.~\eqref{eq:splitpair} over the
distinct-index window pairs.}
\label{tab:isingdefect}
\footnotesize
\begin{ruledtabular}
\begin{tabular}{c@{\quad}c@{\quad}c@{\quad}c@{\quad}c@{\quad}c@{\quad}c@{\quad}c}
\hline\hline
$N$ & $D$ & $\nu$ & $D_I$ & $\varepsilon_2$ & $\varepsilon_3$ & $\varepsilon_4$ & $\varepsilon_{\rm iso}$\\
\hline
10 & 1024 & 6 & 31.1 & $1.09\times10^{-3}$ & $1.84\times10^{-6}$ & $2.52\times10^{-8}$ & $1.86$\\
 &  & 12 & 61.4 & $5.45\times10^{-4}$ & $6.56\times10^{-7}$ & $4.11\times10^{-9}$ & $0.36$\\
 &  & 24 & 122.1 & $3.62\times10^{-4}$ & $3.98\times10^{-7}$ & $1.79\times10^{-9}$ & $0.28$\\
 &  & 48 & 239.4 & $3.61\times10^{-4}$ & $2.18\times10^{-7}$ & $1.02\times10^{-9}$ & $0.19$\\
11 & 2048 & 6 & 33.0 & $6.70\times10^{-4}$ & $1.14\times10^{-6}$ & $7.16\times10^{-9}$ & $2.82$\\
 &  & 12 & 65.3 & $4.16\times10^{-4}$ & $3.68\times10^{-7}$ & $2.22\times10^{-9}$ & $0.98$\\
 &  & 24 & 129.1 & $2.67\times10^{-4}$ & $1.60\times10^{-7}$ & $7.27\times10^{-10}$ & $0.47$\\
 &  & 48 & 255.6 & $3.10\times10^{-4}$ & $1.27\times10^{-7}$ & $5.87\times10^{-10}$ & $0.04$\\
12 & 4096 & 6 & 34.5 & $3.78\times10^{-4}$ & $6.39\times10^{-7}$ & $1.63\times10^{-9}$ & $1.52$\\
 &  & 12 & 68.0 & $3.00\times10^{-4}$ & $2.12\times10^{-7}$ & $5.37\times10^{-10}$ & ---\\
 &  & 24 & 135.6 & $2.69\times10^{-4}$ & $1.05\times10^{-7}$ & $4.43\times10^{-10}$ & ---\\
 &  & 48 & 270.8 & $2.04\times10^{-4}$ & $1.21\times10^{-7}$ & $3.21\times10^{-10}$ & ---\\
13 & 8192 & 6 & 36.2 & $6.76\times10^{-4}$ & $9.84\times10^{-7}$ & $9.83\times10^{-9}$ & ---\\
 &  & 12 & 71.5 & $3.93\times10^{-4}$ & $2.54\times10^{-7}$ & $1.38\times10^{-9}$ & ---\\
 &  & 24 & 141.8 & $2.28\times10^{-4}$ & $1.03\times10^{-7}$ & $6.65\times10^{-10}$ & ---\\
 &  & 48 & 281.5 & $1.48\times10^{-4}$ & $8.15\times10^{-8}$ & $2.64\times10^{-10}$ & ---\\
\hline\hline
\end{tabular}
\end{ruledtabular}
\end{table*}

At the table's dimensions ($D\ge1024$) the pairing-Gram convention
does not alter the reported entries: the reduced orthogonal weights
of Eq.~\eqref{eq:wein} differ from the unreduced (cycle-counting)
weights by less than $0.1\%$ of the reference moment, below the
two-significant-digit rounding unit of the quoted $\varepsilon_r$,
so the entries stand as printed under either convention---a property
of the present dimensions and display precision, not a
convention-independence of the framework, and the theoretical
reference remains the reduced weights of Eq.~\eqref{eq:wein}
throughout.

Two structures emerge.  (i)~The $r=2$ defect dominates and
saturates in $D_I$: $\varepsilon_2$, the normalized residual
$|M_2^{\rm micro}-M_2^{\rm Wg}|/(1+|M_2^{\rm Wg}|)$ of
Eq.~\eqref{eq:LRIdefect}, falls from $\sim10^{-3}$ at
$N=10$, $\nu=6$ to $1.5\times10^{-4}$ at $N=13$, $\nu=48$, as a
finite-$N$ plateau rather than a $D_I^{-\alpha}$ decay (per-$N$ fits
$\alpha_2\approx0.55$, $0.40$, $0.29$, $0.75$ at $N=10$--$13$, noisy
at the smallest windows).  A unified two-scale model
$\varepsilon_2(N,D_I)=b_N+a_ND_I^{-p}$ with a common exponent fits
all sixteen entries at $p\approx1.55$ with a residual an order of
magnitude below the global two-parameter fits, but the two scales
are only weakly identified: the inferred plateaus depend on the
fitted exponent, the fit is endpoint-driven, and realization-level
resampling does not resolve the exponent; the identifiability,
bootstrap, and system-size sensitivity analyses are collected in
Appendix~\ref{app:twoscale}.  Over the accessible system sizes the
window dependence of $\varepsilon_2$ is therefore consistent with a
two-scale description containing a positive size-dependent residual
term, whose thermodynamic scaling cannot be resolved for $N\le13$.
What the data do establish is that this residual is not a sampling
artifact: the between-realization scatter of $\delta_2$
($1.1\times10^{-3}\to1.5\times10^{-4}$ across the scan, carried by
$\xi_r^{\rm samp}$ together with the profile variation) bounds the
statistical error of each entry above by
$\sigma_\delta/\sqrt{12}$, i.e.\ by $3.2\times10^{-4}$ at the smallest
windows and $4.3\times10^{-5}$ at the largest, and every reported
$\varepsilon_2$ exceeds that bound by a factor $2$--$8$.  The
plateau is thus systematic rather than statistical, and the surviving
open question is its size dependence---not the estimation noise.  An
ensemble-level bootstrap makes both statements quantitative.  At
$N=10$ and $11$, where twenty-four realizations are affordable,
resampling the realizations four thousand times gives
$\varepsilon_2$ confidence intervals whose half-widths are
$20$--$30\%$ of the means ($N=11$: $6.9\times10^{-4}$ with $95\%$
interval $[5.0,9.0]\times10^{-4}$ at $\nu=6$, falling to
$2.8\times10^{-4}$ with $[1.9,3.8]\times10^{-4}$ at $\nu=48$); a free
power law in $D_I$ over the four window sizes gives
$\alpha_2=0.44$ at $N=11$ with a $95\%$ interval $[0.33,0.56]$, which
straddles the counting exponent $1/2$ and therefore cannot separate
the two laws at these sizes, while a fit constrained to that exponent
returns a plateau that changes sign between the two sizes
($-2.8\times10^{-5}$ at $N=10$, $+1.8\times10^{-5}$ at $N=11$), so the
constrained extrapolation is not consistent across $N$.  The variance
decomposition locates the residual: the between-realization scatter
$\sigma_b$ of the per-realization mean exceeds the within-realization
center-to-center scatter $\sigma_w$ by a factor growing from $1.4$ at
the widest windows to $5.3$ at the narrowest, and the mean residual at
the narrowest window is of order $\sigma_b$ itself
($\varepsilon_2\simeq1.1$--$1.4\,\sigma_b$).  The deficit is therefore
a block-to-block scale rather than an average over centers: it is
reduced by sampling more realizations, not by combining more windows
within a realization, and its $N$ dependence remains the open
question.  The
D$_I$-scaling of the statistical part is in any case steeper than the
naive $D_I^{-1/2}$ of independent window drawing, since the $D_I$
windows of one realization are correlated through the same profile;
the measured per-$N$ exponents ($\alpha_2\approx0.55$--$1.9$,
endpoint-driven) reflect that correlation and cannot be used to
subtract a statistical floor.  The benchmark calibration of
Appendix~\ref{sec:LRIdefect} locates this range against the
reference-induced floor $\varepsilon_2^{\rm floor}=5.2\times10^{-4}$:
the small-window entries exceed the floor by up to a factor $2.1$ and
the large-window entries fall below it, so the excess at small
$(N,\nu)$ is the part that survives the reference artifact---the
floor being a benchmark calibration whose transfer to the Ising
parameters is indicative rather than exact.  With
$\delta_2>0$ at every tested window---the band-center enhancement of
the variance profile over its spectral mean---the bookkeeping
$\delta_2=\delta_2^{\rm profile}+\delta_2^{\rm fluct}$ separates the
two: the positive mean, common to all $80$ ($N,\nu$,center) entries,
is the profile content, while the between-realization standard
deviation carries the estimation error $\xi_r^{\rm samp}$ of
Eq.~\eqref{eq:momentdecomp} together with the
realization-to-realization variation of the profile itself.  Aggregated first within each realization over its
five windows, the ensemble mean of $\delta_2$ is positive in all
sixteen $(N,\nu)$ entries with $t=2.0$--$5.0$ (twelve realizations
per entry), while a single realization resolves the sign only
marginally---robust at the ensemble level, marginal at the
realization level (the sign-resolution fractions and the spread
crossover are in Appendix~\ref{app:twoscale}).  The
sixteen $(N,\nu)$ entries share one system family and one estimation
pipeline, and at fixed $N$ the four window entries share the same
twelve disorder realizations---the pairing structure of
Table~\ref{tab:isingdefect}---so they are not sixteen independent
replications; the reported $t$-values are per-entry effect sizes at
the realization level rather than an aggregate significance claim.  The profile
content matches the isotropy diagnostic in order and sign.
The signed residual fraction of Eq.~\eqref{eq:splitpair} measures the
anisotropy as a fraction of the local off-diagonal intensity
$\langle|O_{nm}|^2\rangle$, not of the Weingarten reference, so its
natural scale is order one rather than the $\varepsilon_r$ scale:
$\varepsilon_{\rm iso}=1.5$--$2.8$ at $\nu=6$, falling to
$0.36$--$0.98$ at $\nu=12$, $0.28$--$0.47$ at $\nu=24$, and
$0.04$--$0.19$ at $\nu=48$ (the $N=12$--$13$ entries at
$\nu\ge12$ are not resolved above the residual scatter and are left
blank in Table~\ref{tab:isingdefect}); the attribution is consistent
at the level
the sample supports, while a residual below the between-realization
spread is not excluded.  (ii)~The higher-order defects are consistent
with zero at the current resolution: the means of $\delta_3$ and
$\delta_4$ over realizations and centers sit at $0.02$--$0.35$ and
$0.03$--$0.70$ of their between-realization standard deviations
across all sixteen $(N,\nu)$ entries, so no significant systematic
non-Haar content is detected at $r=3,4$ for $N\le13$ at these
statistics; the standard
deviations themselves, $1.8\times10^{-7}$--$3.5\times10^{-6}$ at
order three and $8.9\times10^{-10}$--$6.4\times10^{-8}$ at order
four, set the present resolution; the benchmark drift floor,
$\varepsilon_3^{\rm floor}=1.1\times10^{-6}$ and
$\varepsilon_4^{\rm floor}=5.6\times10^{-7}$, brackets the table's
$r=3,4$ range from above---only the smallest-window $r=3$ entries
reach the floor---so no $r=3,4$ non-Haar content can be resolved
beyond the reference artifact at the current construction.

\emph{Class attribution of the higher-order defect.}  The
centered-channel protocol fixes which classes can carry a defect at
all.  With $\bar q_1=0$, every permutation in the reference sum of
Eq.~\eqref{eq:wgdistinct} that has a fixed point contributes nothing,
so the surviving classes are the fixed-point-free ones: at $r=2$ the
single pairing, at $r=3$ the two $3$-cycles, at $r=4$ the six
$4$-cycles together with the three double transpositions.  A window
defect can therefore enter through exactly two structural
classes---the \emph{drift} class, obtained by replacing the global
$q$-moments in the reference by their window-weighted values, and the
\emph{non-Haar} class, the remainder of the measured moment beyond
that reconstruction.  Both are evaluated on the same windows used for
the table (the window-weighted moments average only the diagonal
profile, so the drift class costs no additional eigenvector
contraction): at $r=2$ the drift class accounts for only $2$--$7\%$ of
$\delta_2$, the bulk being the off-diagonal profile content of the
band-center enhancement; at $r=3$ it is a positive systematic of
$5\times10^{-8}$ at the narrowest windows, falling to
$1.5\times10^{-8}$ at $\nu=48$ ($N=9$, two to three standard errors
per entry) and below $10^{-8}$ at $N=10$, an order of magnitude below
the table's resolution; and at $r=4$ it is at most $2\%$ of
$\delta_4$.  The attribution therefore rules out the drift origin of
the higher-order defect.  It also sharpens statement (ii): with
thirty-six realizations instead of twelve, the $r=4$ defect is no
longer consistent with zero.  It is negative---$-2.2\times10^{-8}$ to
$-5.2\times10^{-9}$ at $N=9$ and $-1.7\times10^{-9}$ at
$N=10,\nu=48$---with a significance growing from two standard errors
at the narrowest window to seven at $\nu=48$, and, since the drift
class is two orders of magnitude smaller, it localizes in the
non-Haar class at the unnormalized $10^{-9}$ level.  That level is
itself two orders of magnitude below the benchmark floor
$\varepsilon_4^{\rm floor}=5.6\times10^{-7}$, so it lies outside the
range in which an LRI violation could be claimed.  The $r=3$ entries
remain unresolved at the same statistics (means within two standard
errors of zero at $N=10$), so the higher-order defect is resolved
only at $r=4$, only in the negative direction, and only below the
reference floor.

\emph{Operator genericity.}  The attribution above is made for the
single-site operator $O=X_1$.  Repeating the same scan with the
two-site operators $O=X_1X_2$ and $O=X_1X_3$ changes the diagonal
profile, the $q$-moments, and the reference, while leaving the class
census untouched; it gives the same picture.  At $N=10$, $\nu=48$,
thirty-six realizations, $\delta_2$ stays positive and of the same
order ($+4.0$, $+2.9$, $+2.2$ in units of $10^{-4}$ for $X_1$,
$X_1X_2$, $X_1X_3$), with the drift class at $1$--$2\%$ of it;
$\delta_3$ stays at the resolution floor ($+1.4$, $-9.2$, and
$-0.7$ in units of $10^{-8}$, against standard errors of $3$--$4$);
and $\delta_4$ stays negative for all three operators,
$-1.7\times10^{-9}\pm0.4$, $-1.1\times10^{-9}\pm0.3$, and
$-1.0\times10^{-9}\pm0.2$, while the drift class remains at most
$2\%$ of it.  The same ordering holds at $N=9$ ($\delta_4$ negative
at four to seven standard errors for all three operators).  The
higher-order defect is therefore not an artifact of the single-site
operator: it belongs to the non-Haar class of every operator tested,
with the drift class negligible throughout.

\emph{Reference dimension.}  The reference of
Eq.~\eqref{eq:wgdistinct} is contracted in the full dimension $D$.
Choosing instead the window population $D_I$ would move it by
$M_r^{\rm Wg}(D)-M_r^{\rm Wg}(D_I)$: at $N=10$ this shift is
$-4.2\times10^{-3}$, $-1.8\times10^{-3}$, $-6.4\times10^{-4}$, and
$-1.3\times10^{-4}$ at $\nu=6,12,24,48$ for $r=2$, and
$+5.9\times10^{-9}$ for $r=4$ at $\nu\ge12$.  The measured defects
are $+4.0\times10^{-4}$ ($r=2$) and $-1.7\times10^{-9}$ ($r=4$) at
$\nu=48$: the reference-dimension shift is of the same order as the
defects---which is why the convention has to be stated---but for
$r=4$ it has the opposite sign and a larger magnitude, so it neither
produces nor cancels the negative systematic.  The full-$D$
contraction is also the correct one: the window matrix is a
submatrix of the $D$-dimensional orthogonal matrix, not an
independent Haar rotation of dimension $D_I$.  The $r=4$ residual is
therefore not a reference-convention artifact; it survives the three
bookkeeping explanations tested here---drift, operator choice, and
reference dimension.

The statement (ii) is channel-specific and must not be read as the
absence of higher-order structure elsewhere: it concerns the
\emph{distinct-index cyclic} moments $M_r^{\rm micro}$ of
Eq.~\eqref{eq:wgdistinct}, whose loop contractions average over
pairings.  The \emph{pair-content channel}---the fourth moment of the
off-diagonal intensity itself, $\sum_{n\neq m}w_nw_m|X_{nm}|^4$
against its block-Haar reference---carries an order-$10^3$ excess
($\sim$$7890\times$ at $\lambda=3/32$ down to $\sim$$177\times$ at
$\lambda=2$ on the ring TFIM of Sec.~\ref{sec:genbench}), and its
quartic fine structure is exactly a quartic functional of the
single-state spectral-weight profiles
(Appendices~\ref{sec:driftkernel} and \ref{sec:transfer}).  The cyclic
defects being small while the pair-content fourth moment is not is
not a contradiction: the loop contraction of $M_4^{\rm micro}$ mixes
the structured pair channel with the flat content it is normalized
against, while $M_4^{\rm pair}$ reads the channel alone.  The
higher-order development of the pair channel and of the spectral-weight
structure is left open here.  (iii)~The inter-window diagnostic
\eqref{eq:windecorr} returns raw correlations
$\rho_{\rm win}^{(2)}\approx0.7$--$0.98$ at all tested centers and
sizes, but this coherence is dominated by the realization
mean: recentering each realization's defect over its five windows
reduces the correlations to scatter over $[-0.85,0.70]$ with mean
$-0.22$---the shift explains the mean, while the residual scatter
keeps significant non-exchangeable content.
Against an exchangeable-noise null---$m=5$ centered window
values whose pair correlations then carry the known
$-1/(m-1)=-0.25$ centering bias---the observed mean matches the null
mean within its block-level scatter.  The outside-count excess
survives the block-level calibration of Appendix~\ref{app:blocknull},
which preserves the observed centers, windows, and weights and shares
realizations within blocks: the observed per-block outside counts
total $13$ against a null total of $3.9\pm2.2$ over eight blocks
($95\%$ band $[0,9]$) give a one-sided $P=0.0011$ with Monte Carlo
standard error $7\times10^{-5}$, with no single block dominating,
and two further calibrations that match the null's block-level
variance and block-level mean to the observed blocks leave the
excess intact ($P=9\times10^{-4}$ and $8\times10^{-4}$).  The centered residuals therefore carry, under the
constructed block-level exchangeable null, significant
non-exchangeable content beyond the finite-sample null---residual
finite-$N$ window structure survives recentering---while the shared
disorder shift fully explains the mean; the thermodynamic control of
this excess is the remaining open $N\to\infty$ task.

The remaining question is how the closure error transmits to
$\delta_r$.  The propagation is monitored through four complementary
diagnostics---the microscopic anisotropy $\varepsilon_{\rm iso}$, the
inter-window dependence $\rho_{\rm win}^{(r)}$ of
Eq.~\eqref{eq:windecorr}, the closure residual $\varepsilon_{\rm cl}$,
and the finite-size floor $\varepsilon_{\rm finite}$---defined next.
These diagnostics are not assumed to form an additive decomposition
of the defect: no relation of the form
$\varepsilon_{\rm LRI}^{(r)}=\varepsilon_{\rm iso}+\varepsilon_{\rm win}
+\varepsilon_{\rm cl}+\varepsilon_{\rm finite}$ (or a $\le$ version of
it) is asserted, and the four terms are not objects of one type.
They are defined as follows.  The microscopic anisotropy
$\varepsilon_{\rm iso}$ is the signed residual ratio
carried by the residual $R$ of the decomposition
\eqref{eq:splitpair} of Appendix~\ref{sec:eth}---a diagnostic, with
values up to $2.8$ in Table~\ref{tab:isingdefect}, not a norm; the
inter-window correlation
\begin{equation}
\rho_{\rm win}^{(r)}(I,J)
=\frac{\left|\operatorname{Cov}\left[\mathcal M_r(I),\mathcal M_r(J)\right]\right|}
{\sqrt{\operatorname{Var}\mathcal M_r(I)\,
\operatorname{Var}\mathcal M_r(J)}},
\label{eq:windecorr}
\end{equation}
a dedicated LRI diagnostic, distinct from the response-kernel radius
$\rho(K_X)$ and the closure margin $\chi_{\rm cl}$ of
Secs.~\ref{sec:num} and~\ref{sec:closure}; under the
independent-window statistical hypothesis adopted here the residual
correlation between distinct energy windows must be controlled, and
its thermodynamic fate is an open problem---LRI constrains the
within-window statistics, while the decorrelation of distinct
windows is an additional statistical assumption rather than a
consequence of the invariance; the closure
truncation $\varepsilon_{\rm cl}$, fed by the deficit decomposition
\eqref{eq:deficitdecomp} and transmitted by the contraction bound
\eqref{eq:errorbound} as
$\varepsilon_{\rm cl}\le\|\delta F\|/(1-\chi_{\rm op})$---a
parameter-space contraction bound on the fixed point; its
transmission to the moment level is bounded explicitly in
Eqs.~\eqref{eq:momtrans}--\eqref{eq:momprop} below---so that
the microscopic closure supplies an explicit parameter-space bound on
the fixed-point error rather than a bare numerical statement, and the
propagation to the LRI moment constraints is bounded rather than
merely asserted; and the finite-size floor
$\varepsilon_{\rm finite}$ of the discrete spectrum, which carries
the estimation content $\xi_r^{\rm samp}$ of
Eq.~\eqref{eq:momentdecomp}---a $D_I^{-1/2}$ scale under the stated
variance assumptions, not a universal consequence of the
decomposition.  The four diagnostics are
heterogeneous---$\varepsilon_{\rm iso}$ is a signed ratio,
$\rho_{\rm win}^{(r)}$ a correlation coefficient rather than a
defect norm, $\varepsilon_{\rm cl}$ a parameter-space contraction
bound, and $\varepsilon_{\rm finite}$ a scale---and they share cross
terms, so the three
physical classes of defect content are defined through
their observables rather than as additive components of
$\delta_r$, and they organize the diagnostics without being asserted
to exhaustively or uniquely decompose the measured defect: the
finite-size floor $\Delta_{\rm finite}$ (the
estimation content $\xi_r^{\rm samp}$ of Eq.~\eqref{eq:momentdecomp},
diagnosed by $\varepsilon_{\rm finite}$ and scaling as
$D_I^{-1/2}$
under the stated variance assumptions), the residual coherent structure
$\Delta_{\rm LRI}$ (window correlations surviving recentering, probed
by $\rho_{\rm win}^{(r)}$ and diagnosed by the block-level null test),
and the closure-transmitted profile $\Delta_{\rm closure}$ (the
$\varepsilon_{\rm cl}$ content of Eq.~\eqref{eq:errorbound}).  In
particular, the recentered inter-window correlation is a probe of
$\Delta_{\rm LRI}$ rather than $\Delta_{\rm LRI}$ itself: its excess
over the block-level exchangeable null is consistent with a nonzero
$\Delta_{\rm LRI}$, but no equality between the diagnostic and the
defect is asserted.  Promoting the diagnostics to a quantitative
error identity or bound would
require absolute or norm definitions of all four terms and a
probability level for the finite-size term; the Lipschitz constant
propagating the parameter-space contraction to the moment level is
fixed by Eqs.~\eqref{eq:momtrans}--\eqref{eq:momprop}, so the
remaining open items are the norm definitions and the probability
level.  The scaling ansatz
$\varepsilon_r\sim D_I^{-\alpha_r}$ separates the two: the
intrinsic estimation fluctuation $\xi_r^{\rm samp}$ sets an
$\alpha_r=\tfrac12$ floor under the stated variance assumptions of
Appendix~\ref{sec:LRIdefect}, and a
systematic defect appears as deficient convergence above it---the
$\delta M_r^{\rm sys}$ of Eq.~\eqref{eq:momentdecomp}---so the
$D_I$ scan is a sharp probe of non-Haar content.

The transmission of the closure truncation to the moment level is
explicit because the LRI probe is centered: the test acts on
$X=O-\operatorname{diag}(O_{nn})$
(Appendix~\ref{sec:LRIdefect}), so $K_1=0$ by construction and the
coincidence dictionary \eqref{eq:coincdict} collapses to
\begin{equation}
M_2=K_2,\qquad M_3=K_3,\qquad M_4=K_4+4\sigma_0K_2^2 .
\label{eq:momtrans}
\end{equation}
The moment map $G_r$ is therefore the identity at orders two and
three, with a single quadratic term at order four, so~\eqref{eq:momtrans}
supplies its Lipschitz constant on the audited ball,
\begin{equation}
L_2=L_3=1,\qquad L_4=1+8\sigma_0|K_2| ,
\label{eq:Lipschitz}
\end{equation}
with $|K_2|$ bounded by the fixed-point content.  The parameter step
is linear as well: the sector dictionary predicts the moments as
$M_{pq}=\sum_m\theta_mL_{pq}(m)$ with the level sums of
Eq.~\eqref{eq:sectorsolve}, hence
$|\delta K_r|\le\Lambda_r|\delta\boldsymbol\theta|$, with
$\Lambda_r$ the corresponding coefficient norm fixed by those sums.
Combining the two steps with the contraction bound
\eqref{eq:errorbound} gives
\begin{equation}
|\delta M_r|\le L_r\Lambda_r\,\varepsilon_{\rm cl},
\qquad
\varepsilon_{\rm cl}\le\frac{\|\delta F\|}{1-\chi_{\rm op}} .
\label{eq:momprop}
\end{equation}
The relative version does not follow: an exact linear map need not
preserve relative errors---the same caveat as at the observable level
in Appendix~\ref{sec:eth}---so \eqref{eq:momprop} bounds absolute moment
errors at fixed scale, and the defect ratio $\delta_r/M_r^{\rm Wg}$
inherits it only up to the scale ratio it carries.  The norms are
compatible by construction: $\|\delta F\|$ is the Euclidean norm on the
retained amplitude vector of Sec.~\ref{sec:errorprop}, $\Lambda_r$ is
the operator norm of the sector dictionary induced by that norm, and
$|\delta M_r|$ is the absolute value at each fixed window tuple; the
same conventions apply to all three terms of
Eq.~\eqref{eq:LRIcondbound}.

\subsection{The drift kernel and the exact order-two decomposition}
\label{sec:driftkernel}

The order-two defect admits an exact split whose two terms carry
exactly named contents.  For the centered channel $X$ of
Eq.~\eqref{eq:qbar} and the Gaussian window kernel of
Eq.~\eqref{eq:momentdecomp}, the measured moment and its flat
reference differ by
\begin{multline}
\delta_2=M_2^{\rm micro}-M_2^{\rm flat}
=\sum_{n\neq m} K_{nm}\,|X_{nm}|^2,\\
K_{nm}\equiv\frac{w_nw_m}{1-\sum_l w_l^2}-\frac{1}{D(D-1)},
\label{eq:K1}
\end{multline}
with $\delta_2^{\rm diag}\equiv M_2^{\rm flat}-M_2^{\rm Wg}$ the
closed diagonal correction of Eq.~\eqref{eq:wgdistinct}.  The
identity is bookkeeping---no statistical hypothesis enters---and the
kernel has zero mass, $\sum_{n\neq m}K_{nm}=0$, so a constant pair
content produces no defect.  The identity is machine-checked: on the
ring TFIM of Sec.~\ref{sec:genbench} at $N=10$, $\nu=24$, band
center, the kernel sum and the moment difference agree to
$2.527306510\times10^{-4}$ versus $2.527306476\times10^{-4}$
(relative difference $1.3\times10^{-8}$).  The flat piece of
Eq.~\eqref{eq:wgdistinct} is closed in the single-state marginals;
the drift piece is a genuine two-state functional of the pair content.
The sign structure of the drift is a quantitative law of the pair
channel, not a generic property: at the tested couplings the drift is
positive at every one of the eighty tested windows; across the band it
follows a U-shape---at $\lambda=1$ the mean $\delta_2$ reads
$1.030\times10^{-3}$ at pair-flip strength $q=0.10$,
$2.917\times10^{-4}$ at $q=0.50$,
and $9.385\times10^{-4}$ at $q=0.90$ ($q$ the deformation strength of
Appendix~\ref{app:sizescan})---and against the coupling it is
U-shaped as well with a minimum near $g\approx0.8$ and a value
$\varepsilon_2(g{=}0)=8.13\times10^{-4}$ at the free-fermion
limit.  The joint profile $|X_{nm}|^2$ itself decays with the level
separation $|E_n-E_m|$ (a factor $\sim\!2.6$ from the band center to
the edges at the tested dimensions) and carries the anisotropic
content measured by $\varepsilon_{\rm iso}$.

The drift of Eq.~\eqref{eq:K1} is \emph{not} decided by any
single-state observable, as the following proven statement records.
\begin{proposition}[Single-state non-identifiability of the drift]
\label{prop:nonident}
On $D=4$ with two fixed-point-free
involutions $P$ and $\tilde P$ the single-body data coincide
identically---spectrum $\{+1,+1,-1,-1\}$, zero diagonal,
$\bar q_2=1$, $\bar q_3=0$, $\bar q_4=1$, $M_2^{\rm flat}=1/3$,
$M_2^{\rm Wg}=2/9$---while the drift of
Eq.~\eqref{eq:K1} takes the values
\begin{multline}
\delta_2^{\rm drift}=+2.160240837\times10^{-1}\quad\text{versus}\\
-1.080120419\times10^{-1},
\label{eq:nonident}
\end{multline}
their ratio being exactly $-2$ whenever $w_1=w_2$.  The statement is
restricted to this input set---spectrum, diagonal, window weights,
and the centered $q$-moments---and does not assert that no richer
single-state input can ever decide the drift; it asserts the negative
content of the present construction: the two-state information in
$\delta_2^{\rm drift}$ is not reconstructible from the single-state
layer of the hierarchy at the order tested.
\end{proposition}  The boundary is realized
in the benchmark data as well: at the $N=10$, $\nu=24$, band-center
window the conditional reference evaluated at the window dimension
$D_I$ (the window-approximated moment of
Appendix~\ref{app:gates}) reads $M_2^{\rm cond}=4.06\times10^{-2}$
against $M_2^{\rm micro}=1.01\times10^{-2}$, off by $4\times$, and
its drift $\delta^{\rm ref}=+3.0\times10^{-2}$ overshoots the
measured defect by $57\times$ with the opposite sign.  The same boundary
reappears as a structural failure of the diagonal resolvent closure
(Appendix~\ref{sec:census}): the self-consistency equation for the
spectral weights admits only single-row localized fixed points, so
the profile's loop-level tail---the part of the pair content that
carries the fourth-order structure---lies outside its solution space.

\subsection{Transferability and the realization-level boundary}
\label{sec:transfer}

The drift-kernel content can be predicted across disorder
realizations and couplings, but only at a specific level, and the
boundary of that level is quantitative.  A blind protocol fixes the
profile law $\bar f(|E_n-E_m|,(E_n+E_m)/2)$---the
realization-averaged pair content, binned in two dimensions, learned
on six training realizations and validated on two---and predicts the
drift of twelve fresh realizations at seven band positions through
$\hat\delta_2=\sum_{n\neq m}K_{nm}\bar f(\cdot)$.  The aggregate
signed error is $8.9\%$ against the factorized baseline's $98.5\%$,
the per-center aggregate ratios lie in $[0.73,1.27]$ at all seven
centers, and the sign accuracy is $59/84=70.2\%$.  The
realization-level audit then fixes exactly what this does and does
not buy.  The sign accuracy is a per-realization phenomenon: eight of
the twelve realizations are predicted with all seven signs correct,
two are predicted with a global sign flip, and two are mixed; the
bootstrap confidence interval of the aggregate rate over
realization-level resampling is $[45.2\%,92.9\%]$, which does not
separate the measured rate from chance.  The absolute (unsigned)
per-cell error is $91.4\%$---the aggregate $8.9\%$ is the signed
cancellation of order-one per-cell errors around a correctly
transferred ensemble mean---and the five-center shape correlation
($\rho=0.5$) has a realization-level interval $[-0.1,1.0]$.  The
variance of the drift across realizations has no
profile-moment structure either: the independent-pair model
underestimates the between-realization standard deviation by $5$--$8$
times and the global-amplitude model by $250$--$500$ times, and the
diagonal share of the realized variance is $0.102$---the
realization-level fluctuations are $90\%$ correlated and not carried
by any tested moment functional of the profile.  The transferability
law therefore has a sharp boundary: \emph{the ensemble-mean profile
transfers across realizations and couplings, and the realization
level does not.}

Across couplings the same boundary persists.  The profile learned at
$g=1$ transfers in shape to $g=0.5$ and $g=2.0$ (five-center
Spearman correlations $0.8$ and $0.6$, signed errors $9.9\%$ and
$19.3\%$ against the recalibrated factorized baselines $100.8\%$ and
$97.3\%$), and the full blind protocol replicated at $g=2.0$ passes
every gate (sign $78.6\%$, per-center ratios $0.80$--$1.34$,
shape correlation $0.8$, signed error $9.7\%$), while at $g=0.5$ the
mid-band ensemble signal ($1.206\times10^{-4}$) collapses below the
realization-level scatter ($\sim\!10^{-3}$), a signal-to-fluctuation
ratio of $\approx\!0.1$ versus $\approx\!0.5$--$0.6$ at $g=1,2$, and
the replication fails.  The coupling range of the law is set by the
same realization-level fluctuation object, now dominating the signal
as the coupling decreases.

The same boundary closes the conditional-LRI question.  Testing
whether the Weingarten reference
$M_2^{\rm Wg}$ of Eq.~\eqref{eq:wgdistinct} reproduces the measured
moment at couplings $g=0.5,1,2$ gives the ratio
$R_2=M_2^{\rm micro}/M_2^{\rm Wg}=0.926,\;1.397,\;2.073$---the
reference is correct within $5\%$ at none of the tested couplings,
and the normalized deviations $\eta_2=|R_2-1|$ read $7.4\%$,
$39.7\%$, $107.3\%$.  The reference is a baseline, not a conditional
prediction: the drift of Eq.~\eqref{eq:K1} is the leading systematic
content the reference misses, and its aggregate-level
transferability is what remains of the LRI program at the order-two
level.  The finite-size scan ($N=12,14$) and the cross-operator test
($X_1X_2$ versus $X_4$) behind these statements are collected in
Appendix~\ref{app:sizescan}.

\subsection{The closure-boundary census}
\label{sec:census}

Each route below was tested against pre-registered gates on the same
ring-TFIM family; the table records for every route the narrow claim
its failure refutes and the claim it leaves untouched.  The rows fall
into four groups, each anchored in the main text.  The two
pair-content routes (rows 1--2) test what the order-two defect of
Appendix~\ref{sec:driftkernel} can be reduced to: a factorized
single-state form, or a reference evaluated at the window dimension
$D_I$ rather than the global $D$.  The reference route (row~3) tests
the colored block-Haar reference of Appendix~\ref{sec:LRIdict} beyond
second order.  The five dressing and closure routes (rows 4--8) test
the loop-level organization of Sec.~\ref{sec:crossresponse}: cavity
dressing, the irreducible-ring closure, the Route-B spectral-radius
bound, the diagonal resolvent closure, and spectrum-level parent
matching.  The five mechanism routes (rows 9--13) test candidate
mechanisms for the fourth-moment concentration and for the drift
transfer of Appendices~\ref{sec:driftkernel} and \ref{sec:transfer}.
Where a route is not developed further in the main text, the row
states the tested claim and the verdict.  The gate definitions, the
tested values, and the verdicts behind each row are collected in
Appendix~\ref{app:gates}; the table is the compressed form
(Tables~\ref{tab:censusA} and~\ref{tab:censusB}).

\begin{table*}[t]
\caption{Closure-boundary census, part A: the reference and closure
routes (rows 1--8 of the grouping in the text).  Each entry is a
closure route tested against pre-registered gates; ``refutes'' states
the narrow claim the measured failure excludes, ``leaves open'' the
claim the failure does not touch.  All rows on the ring TFIM of
Sec.~\ref{sec:genbench}; $\lambda$ is the transverse-field strength.
Part B collects the mechanism and transfer routes.}
\label{tab:censusA}
\footnotesize
\begin{ruledtabular}
\begin{tabular}{p{4.5cm}p{4.5cm}p{4.15cm}p{4.15cm}}
\hline\hline
route & measured failure & refutes (narrowly) & leaves open\\
\hline
factorized pair content $|X_{nm}|^2\approx f_nf_m$ &
anisotropy residual $6$--$430\times$ the gate &
pair content factorizes into single-state factors &
any genuine two-state description\\
conditional reference at the window dimension $D_I$ &
$M_2^{\rm cond}$ off by $4\times$, drift by $57\times$, opposite sign &
the window-local reference dimension &
global-$D$ references\\
colored block-Haar reference &
$M_4/M_4^{\rm BH}=7890/738/177$, $M_{22}^c/M_{22}^{c,\rm BH}=0.648/0.867/0.936$ ($\lambda=3/32,3/4,2$) &
reference correctness beyond second order at the tested block class &
the informative $k\ge3$ sectors\\
CPA cavity dressing of the pair content &
non-convergence $9.1\%\to96.5\%$; $V_4$ gate $0.75\to0.13$--$0.22$ &
cavity-level closure of the dressing &
loop-level dressing\\
irreducible-ring closure probe &
mechanism ratio $0.9416\to0.3135$; error reduction $6.8\%/3.4\%<10\%$ &
the tested ring closure at production $\lambda$ &
weaker ring-level statements\\
Route B spectral-radius bound &
$\rho<1$ on $0.05\%$/$0\%$ of samples; truncation error $O(1)$ relative at every $N$ &
uniform spectral-radius control at production $\lambda$ &
the bound at stronger disorder\\
diagonal resolvent closure of the spectral weights &
$\varepsilon_1=0.981/0.983/0.977$; $R_{4p}=0.0075/0.0026/0.0046$ &
(structural) profiles with a loop-level tail lie outside the closure's fixed-point space &
non-diagonal (loop) self-consistency\\
spectrum-level parent matching of the localization basis &
match rates $3.0\%/0.7\%/0.4\%$ (bare) and $15.2\%/1.5\%/0.3\%$ (second order) &
the localization basis is spectrum-identifiable &
eigenvector-level identification\\
\hline\hline
\end{tabular}
\end{ruledtabular}
\end{table*}

\begin{table*}[t]
\caption{Closure-boundary census, part B: the mechanism and transfer
routes (rows 9--13).  Conventions as in part A.}
\label{tab:censusB}
\footnotesize
\begin{ruledtabular}
\begin{tabular}{p{4.5cm}p{4.5cm}p{4.15cm}p{4.15cm}}
\hline\hline
route & measured failure & refutes (narrowly) & leaves open\\
\hline
static single-state variables for the fourth-moment concentration &
correlations $\rho\le0.15$; $F$-collapse $0/6$ bins &
any tested static single-state variable decides the concentration &
path-level objects\\
two-state covariance as the concentration mechanism &
$M_2=2.000$ flat (Cauchy--Schwarz bound), $\Sigma_{xy}\approx0$ &
the covariance-channel mechanism &
the compound law (concentration $\approx(\mathrm{IPR}_x\,\mathrm{IPR}_y)/2$)\\
single-realization drift prediction &
shape $\rho=0.5$ with interval $[-0.1,1.0]$; $4/12$ global sign flips; per-cell scatter $\sim3\times$ &
realization-level transfer of the profile law &
the ensemble-mean transfer (signed error $8.9\%$)\\
realization-level variance closure &
both models fail every center; diagonal share $0.102$ &
the profile-moment structure of the realization-level variance &
the correlation structure of the fluctuations\\
conditional LRI prediction &
$R_2=0.926/1.397/2.073$ at $g=0.5/1/2$ &
the Weingarten reference as a conditional prediction within $5\%$ &
the deviation theory (aggregate level)\\
\hline\hline
\end{tabular}
\end{ruledtabular}
\end{table*}

The census reads as one consistent boundary.  The routes that fail do
so at the same point where the main text locates the failure: the
factorized and window-local references miss the two-state content
(rows 1--2), the block-Haar reference stops being informative beyond
second order (row~3), the dressing and closure routes fail at
production $\lambda$ or at the fixed-point space (rows 4--8), the
static and covariance mechanisms miss the concentration law
(rows 9--10), and the realization-level transfer and variance closure
fail exactly where the transferability boundary of
Appendix~\ref{sec:transfer} sits (rows 11--13).  Each row's ``leaves
open'' column records what the failure does not touch, which is what
keeps the ensemble-mean transfer of Appendix~\ref{sec:transfer} and the
aggregate-level conditional statements of Appendix~\ref{sec:LRIdefect}
intact.

\subsection{Status of the derivation}
\label{sec:LRIstatus}

The six steps of the derivation program acquire concrete status
within the closure; Table~\ref{tab:LRIstatus} summarizes them in the
evidence vocabulary of Sec.~\ref{sec:levels}.  Read against
Prop.~\ref{prop:LRIcond}, steps (i)--(ii) are hypotheses on the
retained amplitudes---the sector-collapse property and the window
realization---while steps (iv)--(vi) are the measured deviations from
them, so the table states where the derivation stands rather than
what eigenstate statistics are assumed.

\begin{table*}[t]
\begin{ruledtabular}
\begin{tabular}{p{3.5cm}p{3.1cm}p{8.4cm}}
\hline\hline
Step & Status & Evidence and open item\\
\hline
(i) coupling statistics & established &
connectivity census and coupling disorder fix the class amplitudes and
$\Gamma_{\rm coup}^{(r)}$ (Secs.~\ref{sec:conn}, \ref{sec:bench});
non-Gaussian promotion verified at $r=2$ (Proposition~\ref{prop:nonGauss});
higher-body census open\\
(ii) effective distribution & moment level &
$S^{(r)}$ and the moment-field identity fix the retained moments from
$H$; no measure-level statement is made\\
(iii) window localization & native &
the $\eta$-coarse-graining of \eqref{eq:stieltjes} \emph{is} the window
projection (off-shell/on-shell distinction of Sec.~\ref{sec:disc})\\
(iv) within-window isotropy & benchmark-exact; generic open &
free fermions exact sector by sector; Ising defects $0.02$--$0.35$
($r=3$) and $0.03$--$0.70$ ($r=4$) of their spreads
(Table~\ref{tab:isingdefect}); general isotropy theorem open\\
(v) inter-window correlations & quantified, not proven &
$\rho(K_X)\le0.044$, $\chi_{\rm cl}\approx0.04$--$0.07$;
exchangeable-noise mean with non-exchangeable scatter after recentering
($P=0.0011$, Appendix~\ref{sec:LRIdefect}); thermodynamic control of
the realization-shared shift open\\
(vi) dense-spectrum error control & established at fixed $N$ &
deficit decomposition \eqref{eq:deficitdecomp}, propagation bound
\eqref{eq:errorbound}, and moment transmission \eqref{eq:momprop}
locate every truncation error at fixed $N$; $N\to\infty$ open\\
\hline\hline
\end{tabular}
\end{ruledtabular}
\caption{Status of the six derivation steps of Appendix~\ref{sec:LRIstatus} in the evidence vocabulary of Sec.~\ref{sec:levels}.  No entry claims a generic emergence theorem; Proposition~\ref{prop:LRIcond} states the conditional form in which steps (i)--(ii) imply the reference relations, and the open items of steps (iv)--(vi) are conditions on the closed hierarchy, so progress in the closure directly advances the derivation.}
\label{tab:LRIstatus}
\end{table*}

The instruments carry explicit evidence labels throughout:
Eq.~\eqref{eq:r2mix} and the route split
$K_4^{\rm NC}-K_4=X_4=\Delta_4$ are exact identities verified
algebraically; the window-projection dictionary of
Appendix~\ref{sec:LRIdict} is a conditional proposition under the stated
Haar model and coincidence rules; the closure error bound of
Eq.~\eqref{eq:errorbound} is a parameter-space propagation bound; and
the $\rho_{\rm win}$ excess of Appendix~\ref{sec:LRIdefect} is a
numerical observation under the constructed block-level null.

What is established, therefore, is not a proof that arbitrary
many-body spectra produce exactly Haar window statistics, but a
constructive reduction of that statement: the closure computes from
$H$ every object the LRI expansion requires, so LRI is converted from
a postulate into a finite set of computable window conditions on the
spectral cumulants---realized exactly at the route level on the
Weingarten benchmark, calibrated at the defect level by the drift
floor of Appendix~\ref{sec:LRIdefect}, and quantified on the Ising
architecture by the defect scan of Table~\ref{tab:isingdefect}.  The two
steps that would complete the derivation---the generic within-window
isotropy theorem of step (iv) and the thermodynamic-limit control of
inter-window correlations and errors of steps (v)--(vi)---are posed
here as open problems, now with precise formulations in the language
of the closed hierarchy.  \emph{A roadmap, not a bare limitation:}
each of the open steps (iv)--(vi) is formulated as a condition on the
closed hierarchy, so progress in the closure directly advances the
derivation: higher $r$ extends the finite-order dictionary of
Appendix~\ref{sec:LRIdict} and the resolution of step (iv), larger $N$
and the window-resolved conditional reference of
Appendix~\ref{sec:LRIdefect} sharpen the quantitative control of steps
(v)--(vi), and the Lipschitz constant of the moment map, fixed by
Eq.~\eqref{eq:Lipschitz}, carries the closure error of
Eq.~\eqref{eq:errorbound} to the moment level through
Eq.~\eqref{eq:momprop}.  In particular, if the connectivity-class amplitudes of
a general chaotic Hamiltonian collapse onto the overlap sectors in
the manner of Theorem~\ref{thm:ffred}, and the microscopic share of
the three-source decomposition of Eq.~\eqref{eq:momentdecomp} then
reproduces the Haar-sector moments, LRI follows as a fixed-point
property of the closure map $F$ of Eq.~\eqref{eq:fixedpoint} rather
than as an external assumption.  The established content is thereby fixed:
the finite-order dictionary between microscopic spectral statistics
and window moments of Appendix~\ref{sec:LRIdict}, the fourth-order
orientation-doubling distinction between the merge and inversion
routes of Eq.~\eqref{eq:r4orient} together with the moment-level
equivalence those routes share, the quantified defect
decomposition of Appendix~\ref{sec:LRIdefect}, the exact drift-kernel
identity of Eq.~\eqref{eq:K1} with its machine-checked kernel, the
identifiability boundary of Proposition~\ref{prop:nonident}, and the
transferability census of Appendix~\ref{sec:transfer}---the ensemble
mean transfers, the realization level does not; the emergent
claims---isotropy for arbitrary microscopic bases and the
thermodynamic control of the residuals---remain the open layer.  The
division of labor with
Ref.~\cite{MResmerge} is correspondingly precise: the sector algebra
supplies what the window statistics must be when the local
rotational structure holds exactly, and the present closure computes
the defect $\delta_r=M_r^{\rm micro}-M_r^{\rm Wg}$, the difference
between the measured window moment and the value the sector algebra
prescribes, at the moment level of Eq.~\eqref{eq:coincdict}, with
the three-source separation \eqref{eq:momentdecomp} and the
diagnostic framework of Appendix~\ref{sec:LRIdefect}.

\end{document}